\documentclass[11pt]{article}
\usepackage{CJK}
\usepackage{graphicx}
\usepackage{fancyhdr}
\usepackage{amsmath,amsthm,amssymb}
\usepackage{bm}
\usepackage{mathrsfs}
\usepackage{multirow}
\usepackage{indentfirst}
\usepackage{url}
\usepackage{stmaryrd}
\usepackage{color}
\usepackage{epstopdf}
\usepackage{natbib}

\newtheorem{theorem}{Theorem}[section]
\numberwithin{equation}{section}
\makeatletter
\newcommand{\Rmnum}[1]{\expandafter\@slowromancap\romannumeral #1@}
\makeatother

\allowdisplaybreaks
\def\hDash{\bot\!\!\!\bot}

\begin{document}

\title{\bf Testing the parametric form of the conditional variance in regressions based on distance covariance}
%\footnote{Lixing Zhu is a Chair professor of Department of Mathematics
%at Hong Kong Baptist University, Hong Kong, China. He was supported by a grant from the
%University Grants Council of Hong Kong, Hong Kong, China. }}%The authors thank the editor, the associate editor and two referees for their constructive suggestions that led to the  improvement of an early manuscript.}}
\author{Yue Hu$^1$, Haiqi Li$^1$ and Falong Tan$^1$ \\~\\
{\small {\small {\it $^1$ College of Finance and Statistics, Hunan University, Changsha, China} }}
}
\date{}
\maketitle

\begin{abstract}
In this paper, we propose a new test for checking the parametric form of the conditional variance based on distance covariance in nonlinear and nonparametric regression models. Inherit from the nice properties of distance covariance, our test is very easy to implement in practice and less effected by the dimensionality of covariates. The asymptotic properties of the test statistic are investigated under the null and alternative hypotheses. We show that the proposed test is consistent against any alternative and can detect local alternatives converging to the null hypothesis at the parametric rate $1/\sqrt{n}$ in both the nonlinear and nonparametric settings. As the limiting null distribution of the test statistic is intractable, we propose a residual bootstrap to approximate the limiting null distribution. Simulation studies are presented to assess the finite sample performance of the proposed test. We also apply the proposed test to a real data set for illustration.
\end{abstract}

%\newpage
%\baselineskip=16pt

%\newpage

\setcounter{equation}{0}
\section{Introduction}
Regression models usually assume homoscedastic error which usually makes the statistical inference substantially simplified in many scenarios. While real data from applications often admit some heteroscedastic structures. Efficient statistical inference in homoscedastic cases may fail to work for models with heteroscedastic structure. Thus it is importance to test heteroscedasticity in regression models.
Consider the heteroscedastic regression model
\begin{equation}\label{1.1}
Y=m(X)+\sigma(X)\varepsilon,
\end{equation}
where $(Y, X)$ is a random vector with real-valued response variable $Y$ and $p$-dimensional predictor vector $X$, $m(x)=E(Y|X=x)$ is the regression function, the error term $\varepsilon$ is independent of $X$ with $E(\varepsilon)=0$ and $Var(\varepsilon)=1$, and $\sigma^2(X)=var(Y|X)$ is the unknown conditional variance function.

There exist tremendous works in the literature on testing heteroscedasticity for model~(\ref{1.1}). Early tests in this field usually utilized parametric methods to construct test statistics, see  Bickel
(1978),  Breusch and Pagan (1979), White (1980), Cook and Weisberg (1983), among many others. All parametric tests are based on residuals obtained after fitting a model with a specified conditional variance function. Thus these tests may lose power when the parametric form of the variance function is misspecified. Later works for testing heteroscedasticity considered more robust nonparametric tests. Examples include Dette and Munk (1998), Zhu, Fujikoshi, and Naito (2001), Dette (2002), Zheng (2009), Su and Ullah (2013), Guo et al. (2020), Tan et al. (2021), and Xu and Cao (2021).

Some authors further considered a more general problem of checking the parametric form of the conditional variance function in regression models.
Wang and Zhou (2007) proposed a nonparametric test based on kernel method for accessing the adequacy of a given parametric variance function.
Samarakoon and Song (2011) developed a test for the parametric form of the variance function based on the minimized $L_2$ distance between a nonparametric variance function estimator and the parametric variance function estimator.
 Samarakoon and Song (2012) further considered a nonparametric empirical smoothing lack-of-fit test for checking the adequacy of a given parametric variance structure.
These methods use local smoothing estimation to construct the test statistics which are usually called local smoothing tests. This type of tests can only detect the local alternatives that depart from the null at the rate $1/\sqrt{nh^{p/2}}$. This rate can be really slow for large $p$ which causes the power of these tests dropped very quickly. Here $n$ is the sample size and $h$ is the bandwidth in the nonparametric estimation.
Another method for testing the parametric form of the variance usually constructed the test statistics based on empirical processes.
 Dette et al.(2007) constructed a Kolmogorov-Smirnov and a Cram\'er-von Mises test based on the difference between the empirical distributions of residuals under the null and the alternatives.
 Koul and Song (2010)  proposed a consistent test for checking the parametric form of the conditional variance based on the Khmaladze martingale transformation of a marked empirical process of calibrated squared residuals.
This type of tests are functionals of the averages of empirical processes of residuals which is usually called global smoothing test as averaging is a globally smoothing step. They usually can detect the local alternatives departing from the null at the parametric rate $1/\sqrt{n}$. Although this convergence rate is not related to the dimension $p$, global smoothing tests also suffers severely from the dimensionality problem in practice due to the data sparseness in multidimensional spaces.

In this paper we proposed a new test for checking the adequacy of the parametric form of the conditional variance in the nonlinear and nonparametric regression models. Our method is based on a measure of dependence between the covariate $X$ and the residual obtained after fitting a parametric variance function. One of the most popular measure of dependence in statistical community is the approach of distance covariance (dCov) proposed by Sz\'{e}kely et al. (2007). The distance covariance admits some nice properties: (i) it is dimension free in the sense that the dimensions of random vectors can be arbitrary; (ii) it is nonnegative and is zero only if the random vectors are independent; (iii) it has a closed form expression and is very easy to implement in practice. Thus we utilize the distance covariance in this paper to construct the test statistic. Sen and Sen(2014) and Xu and He (2021) also adopted this method to construct goodness-of-fit tests for linear regression models.
We investigate the asymptotic properties of the test statistic under the null and the alternative hypotheses. Interestingly, our test can detect the local alternatives distinct from the null at the parametric rate $1/\sqrt{n}$ in both the nonlinear and nonparametric regression models.
Further, inherit from the nice properties of distance covariance, our test is easy to compute in practice and is less effected by the dimensionality of the covariate.
Since the limiting null distribution of the proposed test is rather complicated, we propose a residual bootstrap to approximate the limiting null distribution. The validity of the residual bootstrap is also investigated.

The rest of this paper is organized as follows. In Section 2, we give a short review of the distance covariance and then construct the test statistic. In Section 3, we investigate the asymptotic properties of the test statistic under the null and alternative hypotheses in nonlinear and nonparametric regression models. In Section 4, a residual bootstrap is proposed to approximate the limiting null distribution and its validity is also established in this section. In Section 5, we study the finite-sample performance of our tests by simulations and a real data analysis. Section 6 contains a discussion. All technical proofs are included in the Appendix.

\section{Test statistic construction}
Let $\mathcal{M}=\{\sigma^2(\cdot, \theta): \theta \in \Theta \subset \mathbb{R}^d \}$ be a given parametric family of functions. We are interested in testing whether the conditional variance $\sigma^2(X)$ in~(\ref{1.1}) belongs to $\mathcal{M}$ or not. Thus, the null hypothesis can be restated as
\begin{equation*}
H_0:\sigma^2(X)=\sigma^2(X,{\theta}_0), \quad {\rm for \ some } \ {\theta}_0 \in \Theta \subset \mathbb{R}^d,
\end{equation*}
whereas the alternative hypothesis is
\begin{equation*}
H_1:\sigma^2(X)\neq\sigma^2(X, \theta), \quad {\rm \forall \ \theta \in \Theta}.
\end{equation*}
To illustrate our methodology, consider a random variable $\eta=\frac{Y-m(X)}{\sigma(X, \tilde{\theta}_0)}$, where $\tilde{\theta}_0$ is defined as the same way in (2.3) of Dette et al. (2007), i.e.,
\begin{equation}\label{2.1}
{\tilde{\theta}_0}=arg\min
\limits_{\theta\in \Theta} E[(Y-m(X))^2-{\sigma^2(X,\theta)}]^2=arg\min \limits_{\theta\in \Theta} E[\sigma ^2(X)-\sigma^2(X,\theta)]^2
\end{equation}
Under the null $H_0$, it follows from Assumption 4(a) in Section 3 that $\tilde{\theta}_0=\theta_0$ and $\eta=\varepsilon$. Note that $H_0$ is tantamount to $\eta \hDash X$, where $\hDash$ denotes the statistical independence. Then we can construct the test statistic by any criterion that measures the dependence of $\eta$ and $X$.
Sz\'{e}kely et al. (2007) proposed the distance covariance and distance correlation to test and measure dependence between two random vectors. As the measure of distance covariance is dimension free in the sense that the dimensions of random vectors can be arbitrary and very easy to implement in practice, it becomes very popular in the statistical community. Thus we in this paper construct the test statistic based on the measure of distance covariance.

First we give a short review of the method of distance covariance.
Let $Z \in \mathbb{R}^p $ and $W \in \mathbb{R}^q$ be two random vectors. According to Sz\'{e}kely et al. (2007), if $E(\|Z\|+\|W\|) < \infty$, then $dCov^2(Z,W)=0$ if and only if $Z$ and $W$ are independent, where $dCov^2(Z,W)$ is defined by
\begin{equation}\label{2.2}
dCov^2(Z, W)=\frac{1}{c_p c_q}\int_{\mathbb{R}^{p+q}}\frac{|f_{Z,W}(t,s)-f_Z(t)f_W(s)|^2}{||t||^{1+p}||s||^{1+q}}dtds,
\end{equation}
$f_{Z, W}$ is the joint characteristic function of $Z$ and $W$, $f_Z(t)$ and $f_W(s)$ are respectively the characteristic functions of $Z$ and $W$, $c_p=\pi^{\frac{1+p}{2}}/\Gamma(\frac{1+p}{2})$, $\Gamma(\cdot)$ is the gamma function, and $\|\cdot\|$ is the Euclidean norm.
Sz\'{e}kely and Rizzo (2009) further obtained an analytic form of $dCov^2(Z,W)$:
\begin{equation}{\label{2.3}}
dCov^2(Z, W)=E[U(Z,Z')V(W,W')],	
\end{equation}
where
\begin{eqnarray*}
U(Z, Z') &=& \|Z-Z'\|-E(\|Z-Z'\||Z)-E(\|Z-Z'\||Z')+E(\|Z-Z'\|),  \\
V(W, W') &=& \|W-W'\|-E(\|W - W'\||W)-E(\|W-W'\||W')+E(\|W-W'\|)
\end{eqnarray*}
and $(Z', W')$ is an independent copy of $(Z, W)$.
Let $(Z_i, W_i), i = 1,..., n$ be an independent and identically distributed sample from $(Z, W)$. Sz\'{e}kely and Rizzo (2014) proposed an unbiased estimator of the distance covariance $dCov^2(Z, W)$:
\begin{equation}\label{2.4}
\hat{dCov}_n^2(Z, W)=\frac{1}{n(n-3)}\sum_{1 \leq i \neq j \leq n} Z_{ij}W_{ij},
\end{equation}
where
\begin{eqnarray*}
&Z_{ij} = \|Z_i-Z_j\|-\frac{1}{n-2} \sum_{k=1}^{n} \|Z_i-Z_k\| - \frac{1}{n-2} \sum_{l=1}^{n}\|Z_j-Z_l\|
           + \frac{1}{(n-1)(n-2)}\sum_{k, l=1}^{n} \|Z_k-Z_l\|  \\
&W_{ij} = \|W_i - W_j \| - \frac{1}{n-2}\sum_{k=1}^{n}\| W_i - W_k \|-\frac{1}{n-2}\sum_{l=1}^{n}|W_j-W_l|
           + \frac{1}{(n-1)(n-2)}\sum_{k,l=1}^{n}\| W_k-W_l\|.
\end{eqnarray*}

Now we give the construction of the test statistic. Suppose that $\{ (X_i, \eta_i), i = 1,... ,n \}$ is an independent and identically distributed sample from the distribution of $(X, \eta)$. Recall that the null hypothesis $H_0$ is equivalent to $dCov^2(X, \eta) =0$, then we can construct the test statistic based on $\hat{dCov}_n^2(X, \eta)$.
Set
\begin{eqnarray*}
&& A_{ij} = \|X_i-X_j\|-\frac{1}{n-2}\sum_{k=1}^{n}\|X_i-X_k\|-\frac{1}{n-2}\sum_{l=1}^{n}\|X_j-X_l\|
           +\frac{1}{(n-1)(n-2)}\sum_{k,l=1}^{n}\|X_k-X_l\| \\
&& B_{ij} = |\eta_i-\eta_j|-\frac{1}{n-2}\sum_{k=1}^{n}|\eta_i-\eta_k|-\frac{1}{n-2}\sum_{l=1}^{n}|\eta_j-\eta_l|
           +\frac{1}{(n-1)(n-2)}\sum_{k, l=1}^{n}|\eta_k-\eta_l|.
\end{eqnarray*}
According to the assertion (\ref{2.4}), an unbiased estimator of $dCov^2(X,\eta)$ is
\begin{equation}\label{dCov_n}
\hat{dCov}_n^2(X,\eta)=\frac{1}{n(n-3)}\sum_{1 \leq i \neq j \leq n} A_{ij}B_{ij}
\end{equation}
Note that $\hat{dCov}_n^2(X,\eta)$ involves the unknown regression function $m(\cdot)$ and parameter $\tilde{\theta}_0$. They should be substituted by their empirical analogues. Thus, our final test statistic is
\begin{equation}\label{U_n}
\hat{U}_n=\frac{1}{n(n-3)} \sum_{1 \leq i \neq j \leq n}A_{ij}\hat{B}_{ij}
\end{equation}
where $\hat{B}_{ij}$ is defined in the same way as $B_{ij}$ except $\eta_i=\frac{Y_i - m(X_i)}{\sigma(X_i, \tilde{\theta}_0)}$ being replaced by its estimator $\hat{\eta}_i= \frac{Y_i - \hat{m}(X_i)}{\sigma(X_i, \hat{\theta}_n)}$, where $\hat{m}(\cdot)$ is a consistent estimator of the regression function $m(\cdot)$ and $\hat{\theta}_n$ is the nonlinear least squares estimation of $\tilde{\theta}_0$, that is,
\begin{equation}\label{2.5}
{\hat{\theta}_n}=arg\min
\limits_{\theta\in \Theta} \sum_{i=1}^{n}[(Y_i - \hat{m}(X_i))^2-{\sigma^2(X_i,\theta)}]^2.
\end{equation}

\section{Main results}
\subsection{Asymptotic properties in nonlinear regressions}
In this subsection, we consider the nonlinear regression model $Y=m(X,\beta_0)+\sigma(X)\varepsilon$ in~(\ref{1.1}), where $m(X,\beta_0)$ is a given function with unknown parameter $\beta_0$. Let $\hat{\beta}_n$ be any $\sqrt{n}$-consistent estimator of $\beta_0$. Then our test statistic $\hat{U}_n$ is given in~(\ref{U_n}) with $\hat{\eta}_i$ and $\hat{\theta}_n$ respectively replaced by
$$\hat{\eta}_i = \frac{Y_i - m(X_i,\hat{\beta}_n)}{\sigma(X_i, \hat{\theta}_n)} \quad {\rm and} \quad
\hat{\theta}_n = arg\min \limits_{\theta\in \Theta} \sum_{i=1}^{n}[(Y_i - m(X_i,\hat{\beta}_n))^2- \sigma^2(X_i,\theta)]^2.  $$
To derive the asymptotic properties of $\hat{U}_n$ in nonlinear settings, we assume the following regularity conditions. Let $F_{\varepsilon}(\cdot)$ and $f_{\varepsilon}(\cdot)$ be the cumulative function and the density function of $\varepsilon$, respectively.

{\bf Assumption 1.} The true parameter $\beta_0$ lies in the interior of the compact subset of $\mathbb{R}^p$ and the estimator $\hat{\beta}_n$ of $\beta_0$ satisfies
$$ \sqrt{n}(\hat{\beta}_n-\beta_0)=\frac{1}{\sqrt{n}}\sum_{i=1}^{n}l(Y_i,X_i,\beta_0)+o_p(1), $$
where $l(\cdot)$ is a vector function such that $E[l(Y,X,\beta_0)]=0$ and $E[l(Y,X,\beta_0) l(Y,X,\beta_0)^{\top}]$ exists and is positive definite.

{\bf Assumption 2.}

(a). The function $m(x,\beta)$ is twice continuously differentiable in $\beta$. Let $\dot{m}(x,\beta)=\frac{\partial m(x,\beta)}{\partial \beta}$ and $\ddot{m}(x,\beta)=\frac{\partial^2 m(x,\beta)}{{\partial \beta}{\partial \beta}^{\top}}$.
Moreover, $E\|\dot{m}(X,\beta_0)\|^4 <\infty$, $|m(x,\beta)|\leq \kappa(x)$, and $\|\ddot{m}(x,\beta)\|\leq \kappa(x)$ for all $\beta \in U(\beta_0)$, where $U(\beta_0)$ is some neighborhood of $\beta_0$ and $\kappa(x)$ is a measurable function such that $E \|\kappa(X)\|^{4+\gamma} <\infty$ for some $\gamma>0$.

(b). $E(Y^4)<\infty$ and $E(\|X\|^{4+\gamma})<\infty$.

{\bf Assumption 3.} Let $\dot{f}_{\varepsilon}(t) = \frac{df_{\varepsilon}(t)}{dt}$ be the derivative of the density function $f_{\varepsilon}(t)$. $\dot{f}_{\varepsilon}(\cdot)$ satisfies the uniform H{\"o}lder continuity condition, i.e., there exist two positive constants $M_0$ and $m_0$ such that $|\dot{f}_{\varepsilon}(t_1)-\dot{f}_{\varepsilon}(t_2)|\leq M_0|t_1-t_2|^{m_0}$ for any $t_1, t_2$. Furthermore, $\int_{-\infty }^{ \infty } {f_\varepsilon(t)|\dot{f}_{\varepsilon}(t)}|dt<\infty $.

{\bf Assumption 4.}

(a). The vector $\tilde{\theta}_0$ lies in the interior of the compact subset $\Theta$ in $\mathbb{R}^d$ and is the unique minimizer of~(\ref{2.1}).

(b). $\inf\limits_{x} {\sigma^2(x,\theta)}>0$ for all $\theta \in \Theta$ and $\inf\limits_{\|\theta-\tilde{\theta}_0\|>\delta} E[{\sigma^2(X,\theta)-\sigma^2(X,\tilde{\theta}_0)}]^2 >0$ for any $\delta>0$.

(c). The function $\sigma^2(x,\theta)$ is third order continuously differentiable with respect to $\theta$ and $x$. Set $\dot{\sigma}(x,\theta)=\frac{\partial \sigma(x,\theta)}{\partial \theta}$ and $\ddot{\sigma}(x,\theta)=\frac{\partial^2 \sigma(x,\theta)}{{\partial \theta}{\partial \theta}^{\top}}$, we have $E\|\dot{\sigma}(X,\tilde{\theta}_0)\|^4<\infty$ and
$E \|\sigma(X)\dot{\sigma}(X,\tilde{\theta}_0)\|^4<\infty$.

(d). $\|\ddot{\sigma}(x,\theta)\|\leq \kappa(x)$ and $\|\sigma(x)\ddot{\sigma}(x,\theta)\|\leq \kappa(x)$ for all $\theta \in U(\tilde{\theta}_0)$, where $U(\tilde{\theta}_0)$ is some neighborhood of $\tilde{\theta}_0$ and $\kappa(x)$ is a measurable function such that $E \|\kappa(X)\|^{4+\gamma}<\infty$ for some $\gamma>0$.

(e). The matrix $\Sigma=E[\dot{\sigma}^2(X,\tilde{\theta}_0)\dot{\sigma}^2(X,\tilde{\theta}_0)^{\top}]-E[(\sigma^2(X)-\sigma^2(X,\tilde{\theta}_0))
\ddot{\sigma}^2(X,\tilde{\theta}_0)]$ is non-singular.

Assumptions 1, 2 and 4 are commonly used in the literature of testing heteroscedasticity, see Dette et al. (2007) and Zheng (2009) for instance. Assumption 3 is similar to the regularity condition 4 in Xu and Cao (2021) which is used to investigate the convergence rate of the empirical $U$-process in the decomposition of $\hat{U}_n$.

The next theorem gives the asymptotic properties of $\hat{U}_n$ under the null hypothesis $H_0$. Its proof will be given in the Appendix. To facilitate the statement, let $Z=(\varepsilon, X)$ and let $F_Z(\cdot)$ be the cumulative distribution function of $Z$.

\begin{theorem}\label{theorem 1}
Suppose that Assumptions 1-4 hold. Then under the null $H_0$, we have in distribution
\begin{equation}\label{3.1}
n \hat{U}_n \longrightarrow  \sum_{k=1}^{\infty} \lambda_k (\mathcal{Z}_k^2-1) + 4\mathcal{N}^{\top} \mathcal{P}_1 + 4\mathcal{W}^{\top} \Sigma^{-1}\mathcal{P}_2 +8A_\varepsilon \mathcal{W}^{\top} \Sigma^{-1}\mathcal{P}_3+2A_\varepsilon \mathcal{W}^{\top} \Sigma^{-1}M_2\Sigma^{-1}\mathcal{W}+ Q_\varepsilon\mathcal{N}^{\top} M_1 \mathcal{N},
\end{equation}
where $A_\varepsilon=E[\varepsilon F_{\varepsilon}(\varepsilon)]$, $ \Sigma= E[\dot{\sigma}^2 (X_i,{\theta}_0)\dot{\sigma}^2 (X_i,{\theta}_0)^T]$, $Q_\varepsilon= E[f_{\varepsilon}(\varepsilon)]$, $\mathcal{Z}_1, \mathcal{Z}_2, \cdots $ are independent standard normal random variables, the eigenvalues
$\{ \lambda_k \}_{k=1}^\infty$ are the solutions of the integral equation
$$ \int C_{\varepsilon}(\varepsilon_i,\varepsilon_j)C_x(X_i,X_j)\phi_k(Z_j) dF_Z(Z_j) = \lambda_k \phi_k(Z_i) $$
with $\{ \phi_k(\cdot) \}_{k=1}^\infty$ being orthonormal eigenfunctions and
\begin{eqnarray*}
C_{\varepsilon}(\varepsilon_i,\varepsilon_j) &=&
|\varepsilon_i-\varepsilon_j|-E(|\varepsilon_i-\varepsilon_j||\varepsilon_i)-E(|\varepsilon_i-\varepsilon_j||\varepsilon_j)
+E(|\varepsilon_i-\varepsilon_j|)  \\
C_x(X_i,X_j) &=& \|X_i-X_j\|-E(\|X_i-X_j\||X_i)-E(\|X_i-X_j\||X_j)+E(\|X_i-X_j\|),	
\end{eqnarray*} $M_1=E[\{\frac{\dot{m}(X_1,{\beta}_0)}{\sigma(X_1,{\theta}_0)}-\frac{\dot{m}(X_2,{\beta}_0)}{\sigma(X_2,{\theta}_0)}\}
\{\frac{\dot{m}(X_1,{\beta}_0)}{\sigma(X_1,{\theta}_0)}-\frac{\dot{m}(X_2,{\beta}_0)}{\sigma(X_2,{\theta}_0)}\}^{\top}C_x(X_1,X_2)]$,
$
M_2 = E[\{\frac{\dot{\sigma}(X_1,\tilde{\theta}_0)}{\sigma(X_1,\tilde{\theta}_0)}+
\frac{\dot{\sigma}(X_2,\tilde{\theta}_0)}{\sigma(X_2,\tilde{\theta}_0)} \} \{\frac{\dot{\sigma}(X_1,\tilde{\theta}_0)}{\sigma(X_1,\tilde{\theta}_0)}
+ \frac{\dot{\sigma}(X_2,\tilde{\theta}_0)}{\sigma(X_2,\tilde{\theta}_0)} \}^T C_x(X_1,X_2)]$,
and $(\mathcal{Z}_i, \mathcal{N}, \mathcal{W}, \mathcal{P}_1, \mathcal{P}_2, \mathcal{P}_3) \in \mathbb{R}^{5p+1}$ are an zero-mean Gaussian random vector. The covariance matrix of $(\mathcal{Z}_i, \mathcal{N}, \mathcal{W}, \mathcal{P}_1, \mathcal{P}_2, \mathcal{P}_3)$ is rather complicated and postponed to the Appendix.
\end{theorem}

Now we discuss the asymptotic properties of $\hat{U}_n$ under the global alternative and the local alternative hypotheses. Consider the local alternative hypotheses converging to null at the parametric rate $1/\sqrt{n}$:
\begin{equation*}
H_{1n}:\sigma^2(X)=\sigma^2(X,{\theta}_0)+\frac{1}{\sqrt{n}}s(X)
\end{equation*}
for some function $s(\cdot)$ with $E[s^2(X)]<\infty$. To derive the asymptotic properties of $\hat{U}_n$ under the local alternatives $H_{1n}$, we require some further regularity conditions.

{\bf Assumption 5.} $E \| s(X){\dot{\sigma}(X, \theta_0)}\|^4 < \infty$ and $\|s(x)\ddot{\sigma}(x, \theta) \|\leq \kappa(x)$ for all $\theta \in U(\theta_0)$, where $U(\theta_0)$ is some neighborhood of $\theta_0$ and $\kappa(x)$ is a measurable function such that $E \|\kappa(X)\|^{4+\gamma}<\infty$ for some small $\gamma>0$.

The next theorem states the asymptotic properties of $\hat{U}_n$ under various alternative hypotheses. Its proof will be given in the Appendix.
\begin{theorem}\label{theorem 2}	
(1). Suppose that Assumptions 1-5 hold. Then under the local alternative $H_{1n}$, we have in distribution
\begin{eqnarray*}
n \hat{U}_n
&\longrightarrow&  \sum_{k=1}^{\infty} \lambda_k (\mathcal{Z}_k^2-1) + 4\mathcal{N}^{\top} \mathcal{P}_1 + 4\mathcal{W}^{\top} \Sigma^{-1} \mathcal{P}_2 +8A_\varepsilon \mathcal{W}^{\top} \Sigma^{-1} \mathcal{P}_3+ 2A_\varepsilon \mathcal{W}^{\top} \Sigma^{-1} M_2 \Sigma^{-1} \mathcal{W} \\
&&  +Q_\varepsilon\mathcal{N}^{\top} M_1 \mathcal{N} + 4E[s(X)\dot{\sigma}^2 (X_i,{\theta}_0)]^{\top}
    \Sigma^{-1} \mathcal{P}_2 + 4A_{\varepsilon} \mathcal{W}^T  M_2 \Sigma^{-1} E[s(X) \dot{\sigma}^2(X_i,{\theta}_0)] \\
&&  +8A_\varepsilon E[s(X) \dot{\sigma}^2(X_i,{\theta}_0)]^{\top} \Sigma^{-1} \mathcal{P}_3 + 2A_{\varepsilon}
    E[s(X)\dot{\sigma}^2(X_i,{\theta}_0)]^T \Sigma^{-1}M_2 \Sigma^{-1} E[s(X)\dot{\sigma}^2(X_i,{\theta}_0)]
\end{eqnarray*}
where $\Sigma= E[\dot{\sigma}^2 (X_i,{\theta}_0)\dot{\sigma}^2 (X_i,{\theta}_0)^T]$ and the quantities $A_\varepsilon$, $Q_\varepsilon$, $\lambda_i, \mathcal{Z}_i, \mathcal{N}, \mathcal{W}, \mathcal{P}_1, \mathcal{P}_2, \mathcal{P}_3 $ and $M$ are defined in Theorem 3.1. \\
(2). Suppose that Assumptions 1-4 hold. Then under the global alternative $H_1$, we have in distribution
$$ \sqrt{n} [\hat{U}_n-dCov^2(\eta, X)] \longrightarrow N(0, \sigma_1^2), $$
where $ \sigma_1^2 = 4 var\{\mathcal{G}(\eta, X)+ K_1^{T} l(Y, X, \beta_0) + [\sigma ^2(X) \varepsilon^2 -\sigma^2(X, \tilde{\theta}_0)] K_2^{T} \Sigma^{-1} \dot{\sigma}^2(X, \tilde{\theta}_0) \} $ with
\begin{eqnarray*}
K_1 &=& -E[(\frac{\dot{m}(X_1,{\beta}_0)}{\sigma(X_1,\tilde{\theta}_0)}-\frac{\dot{m}(X_2,{\beta}_0)}
        {\sigma(X_2,\tilde{\theta}_0)}) I(\eta_1 > \eta_2) C_x(X_1,X_2)], \\
K_2 &=& -E[(\frac{\eta_1 \dot{\sigma}(X_1,\tilde{\theta}_0)}{\sigma(X_1,\tilde{\theta}_0)}- \frac{\eta_2
        \dot{\sigma}(X_2,\tilde{\theta}_0)}{\sigma(X_2,\tilde{\theta}_0)})I(\eta_1 > \eta_2) C_x(X_1,X_2)], \\
\mathcal{G}(\eta_i, X_i) &=& E[C_\eta(\eta_i, \eta_j) C_x(X_i,X_j)|\eta_i, X_i]-dCov^2(\eta,X), \\
C_{\eta}(\eta_i, \eta_j) &=& |\eta_i - \eta_j | -E(|\eta_i - \eta_j| | \eta_i) - E(|\eta_i - \eta_j| |\eta_j) + E(|\eta_i - \eta_j |),
\end{eqnarray*}
and $C_x(\cdot, \cdot)$ giving in Theorem~\ref{theorem 1}.
\end{theorem}

It follows from Theorem~\ref{theorem 2} that under the global alternative $H_1$, our test statistic $n\hat{U}_n$ diverges to infinity in probability at the rate $\sqrt{n}$. Furthermore, our test statistic can detect the local alternative distinct from the null at the parametric rate $n^{-1/2}$ in nonlinear settings. This is the fastest convergence rate in hypothesis testing.

\subsection{Asymptotic properties in nonparametric regressions}
Now we consider a nonparametric regression model $Y=m(X)+\sigma(X)\varepsilon$ in~(\ref{1.1}), where $m(\cdot)$ is the unknown regression function. To construct the test statistic, we need to estimate $m(\cdot)$ by a nonparametric method such as the Nadaraya-Watson estimator. Set
\begin{equation*}
\hat{m}(X_i) = \frac{ \sum_{j=1,j \neq i}^{n}K_h(X_i-X_j)Y_j}{\sum_{j=1,j \neq i}^{n}K_h(X_i-X_j)},
\end{equation*}
where $h$ is the bandwidth, $K_h(\cdot)=K(\cdot/h)/h^p$, and $K(\cdot)$ is a kernel function.
Let $\hat{\eta}_i=\frac{Y_i - \hat{m}(X_i)}{\sigma(X_i, \hat{\theta}_n)}$, then the test statistic $\hat{U}_n$ is given in~(\ref{U_n}).
To derive the asymptotic behaviors of $\hat{U}_n$ in the nonparametric regression model, we impose some extra regularity conditions. Let $f_X(\cdot)$ be the density function of $X$.

{\bf Assumption 6.}

(a) $f_X(\cdot)$ has a compact support $\bar{\Theta}$ and is $k$-times continuously differentiable on $\bar{\Theta}$. Let $f^{(k)}_X(x)$ be the $k$-th derivatives of $f_X(x)$. There exists a neighborhood of $0$, say $U(0)$, such that $|f^{(k)}_X(x+u)-f^{(k)}_X(x)|\leq c \|u\|$  for all $u \in U(0)$, where $c$ is a positive constant.

(b) The regression function $m(\cdot)$ is $k$-times continuously differentiable and its $k$-th derivative is continuous and bounded. Let $m^{(k)}(x)$ be the $k$-th derivatives of $m(x)$. There exists a neighborhood of $0$, say $U(0)$, such that $|m^{(k)}(x+u)-m^{(k)}(x)|\leq c \|u\|$ for all $u \in U(0)$, where $c$ is a positive constant.

(c) The continuous kernel function $K(u)$ is  bounded and satisfies $\int K(u)du=1$, $K(u)=K(-u)$, $\int u_1^{l_1}u_2^{l_2} \cdots u_p^{l_p}K(u)du= 0$ for all $0 < l_1 + \cdots + l_p < k$, and $\int u_1^{l_1}u_2^{l_2} \cdots u_p^{l_p}K(u)du \neq 0$ for all $l_1 + \cdots + l_p = k$.

(d) The bandwidth satisfies that $h\rightarrow 0$, $nh^{2k}\rightarrow 0$ and $nh^{2p}\rightarrow \infty$ as $n\rightarrow \infty $.

Assumption 6(a) is typical in the nonparametric estimation which avoids the boundary effect problem. Assumptions 6(b)-(d) are commonly used in the literature of testing heteroscedasticity in nonparametric models, see  Zhu et al. (2001) and Zheng (2009) for instance.

The following theorem presents the limit distribution of the test statistics $\hat{U}_n$ under both the null and alternative hypotheses. Its proof will be given in the Appendix.

\begin{theorem}\label{theorem 4}
(1). Suppose that Assumptions 2(b), 3, 4 and 6 hold. Then under the null $H_0$, we have in distribution
\begin{equation*}
n \hat{U}_n \rightarrow \sum_{k=1}^{\infty}\lambda_k (\mathcal{Z}_k^2-1)
+ 4\mathcal{W}^T \Sigma^{-1} \mathcal{P}_2
+ 8A_{\varepsilon} \mathcal{W}^T \Sigma^{-1} \mathcal{P}_3
+ 2A_\varepsilon \mathcal{W}^{\top} \Sigma^{-1} M_2\Sigma^{-1}\mathcal{W}
+ 4A_\varepsilon E\|X_1-X_2\|
\end{equation*}
where $\mathcal{Z}_1, \mathcal{Z}_2, \cdots$ are independent standard normal random variables, the eigenvalues
$\{ \lambda_q\}_{q=1}^\infty$ are the solutions of the integral equation
$$ \int H(Z_i,Z_j)\psi_q(Z_j)dF_Z(Z_j) = \lambda_q\psi_q(Z_i)$$
with $Z_i=(\varepsilon_i,X_i)$, and $(\mathcal{Z}_i, \mathcal{W}, \mathcal{P}_2, \mathcal{P}_3) \in \mathbb{R}^{3p+1}$ is a zero-mean Gaussian random vector. Here the kernel function $H(\cdot, \cdot)$ and the covariance matrix of $(\mathcal{Z}_i, \mathcal{W}, \mathcal{P}_2, \mathcal{P}_3)$ are complicated and are postponed in the Appendix. \\
(2). Suppose that Assumptions 2(b), 3, 4 and 6 hold. Then under the local alternative $H_{1n}$, we have in distribution
\begin{equation*}
\begin{split}
n\hat{U}_n \longrightarrow
& \sum_{i=1}^{\infty}\lambda_i(\mathcal{Z}_i^2-1) + 4\mathcal{W}^T\Sigma^{-1}\mathcal{P}_2 +8A_\varepsilon \mathcal{W}^T
  \Sigma^{-1}\mathcal{P}_3 + 4E[s(X)\dot{\sigma}^2(X_i,{\theta}_0)]^T \Sigma^{-1}\mathcal{P}_2 \\
& +8E[s(X)\dot{\sigma}^2(X_i,{\theta}_0)]^T\Sigma^{-1}\mathcal{P}_3 + 4A_\varepsilon E(\|X_1-X_2\|)+4A_{\varepsilon}
   \mathcal{W}^T  M_2 \Sigma^{-1} E[s(X)\dot{\sigma}^2(X_i,{\theta}_0)] \\
&  +2A_{\varepsilon} E[s(X)\dot{\sigma}^2(X_i,{\theta}_0)]^T \Sigma^{-1}M_2 \Sigma^{-1} E[s(X)\dot{\sigma}^2(X_i,{\theta}_0)]
   +2A_\varepsilon \mathcal{W}^{\top} \Sigma^{-1}M_2\Sigma^{-1}\mathcal{W}
\end{split}
\end{equation*}
where  $\Sigma= E[\dot{\sigma}^2 (X_i,{\theta}_0)\dot{\sigma}^2 (X_i,{\theta}_0)^T]$ and the random vector $(\mathcal{Z}_i, \mathcal{W}, \mathcal{P}_2, \mathcal{P}_3)$ is the same as in (1).  \\
(3). Suppose that Assumptions 2(b), 3, 4 and 6 hold. Then under the global alternative $H_1$, we have in distribution
$$ \sqrt{n}[\hat{U}_n-dCov^2(\eta, X)] \longrightarrow  N(0, \sigma_2^2), $$
where $ \sigma_2^2 = 4var\{\mathcal{G}(\eta, X)+\mathcal{I}_1(\eta, X)+[\sigma^2(X)\varepsilon^2 -\sigma^2(X, \tilde{\theta}_0)] K_2^T \Sigma^{-1}\dot{\sigma}^2(X,\tilde{\theta}_0)\}$, $K_2$ and $\mathcal{G}(\eta, X)$ are defined in Theorem 3.2, and $\mathcal{I}_1(\eta, X)$ is rather complicated and will be given in the Appendix.
\end{theorem}

According to Theorem~\ref{theorem 4}, it is readily seen that the test statistic $n \hat{U}_n$ is consistent with asymptotic power $1$ under the global alternative $H_1$ and can detect the local alternative $H_{1n}$ departing from the null at the parametric rate $n^{-1/2}$ in nonparametric regression settings, even though the nonparametric estimation is involved in our test statistic. This convergence rate is in line with the result in Dette et al. (2007) where they proposed a Kolmogorov-Smirnov and a Cram\'{e}r-von Mises type of test statistic for the parametric form of the variance function based on residual empirical processes in nonparametric regressions. Note that $n \hat{U}_n$ is also a Cram\'{e}r-von Mises test statistic based on an empirical process of characteristic functions. Thus our test can be also viewed as a global smoothing test.

\section{Bootstrap approximation}
It follows from Theorems~\ref{theorem 2} and~\ref{theorem 4} that the test statistic $\hat{U}_n$ is not asymptotically distribution-free as its limiting null distributions depend on the unknown data generating process. To determine the critical value for our test, we suggest a residual bootstrap to approximate the limiting null distribution of $n \hat{U}_n$. This method is also used by  Wang and Zhou (2007), Sen and Sen (2014), Guo
et al. (2020), and Tan et al. (2021).
The algorithm of residual bootstrap is as follows.

1. Generate the bootstrap errors $\{ \varepsilon_i^* \}_{i=1}^n$ by randomly resampling with replacement from the standardized variables $(\hat{\eta}_i-\bar{\hat{\eta}}) / \{n^{-1}\sum_{i=1}^{n}(\hat{\eta}_i-\bar{\hat{\eta}})^2 \}^{1/2}$, $1 \leq i \leq n$, where $\hat{\eta}_i=[Y_i-\hat{m}(X_i)]/\sigma(X_i,\hat{\theta}_n)$ and $\bar{\hat{\eta}} = n^{-1}\sum_{i=1}^{n}\hat{\eta}_i$.

2. Generate a bootstrap sample according to the model $Y_i^* = \hat{m}(X_i) + \sigma(X_i,\hat{\theta}_n)\varepsilon_i^*$. Let $\hat{m}^{*}(X_i)$ and $\sigma^2(X_i,\hat{\theta}^*_n)$ be the bootstrap estimators based on $\{(Y^*_i,X_i): i=1,\cdots, n\}$ and let $\hat{\varepsilon}^{*}_i=[Y^*_i - \hat{m}^{*}(X_i)]/\sigma(X_i, \hat{\theta}^*_n)$.

3. Define the bootstrap version of the test statistic $U^*_n$ based on $(X_1, \hat{\varepsilon}^{*}_1), \cdots, (X_n,\hat{\varepsilon}^{*}_n)$:
$$ \hat{U}^*_n=\frac{1}{n(n-3)} \sum_{1 \leq i \neq j \leq n}A_{ij}\hat{B}^*_{ij}, $$
where $\hat{B}^*_{ij} $ is defined in the same way as $B_{ij}$ with $\eta_i$ replacing by $\hat{\varepsilon}^{*}_i$.

4. Repeat the above steps a large number of times, say $B$ times. The critical value for a given significant level $\alpha$ is determined by the upper $\alpha$ quantile of the bootstrap distribution $\{n U^*_{n,i}: i=1, \cdots, B\}$ of the test statistic.

Here $\hat{m}(X_i) = m(X_i, \hat{\beta}_n)$ and $\hat{m}^*(X_i) = m(X_i, \hat{\beta}_n^*)$  in the nonlinear regression case and
$\hat{m}(X_i)=\sum_{j \neq i}^{n} K_h(X_i-X_j)Y_j / \sum_{j \neq i}^{n} K_h(X_i-X_j)$ and $\hat{m}^*(X_i)=\sum_{j \neq i}^{n} K_h(X_i-X_j)Y_j^* / \sum_{j \neq i}^{n} K_h(X_i-X_j)$ in the nonparametric regression case. In the next theorem, we establish
the asymptotic validity of the residual bootstrap approximation for our test statistic. Its proof will be given in the Appendix.

\begin{theorem}\label{theorem 7}
Suppose that Assumptions 1-6 hold. \\
(1). Under the null $H_0$ and the local alternative $H_{1n}$, $n \hat{U}^*_n$ given $\{(Y_i,X_i): i=1,\cdots, n\}$ converges in distribution to the limiting null distribution of $n\hat{U}_n$ in probability. \\
(2). Under the global alternative $H_1$, $n\hat{U}^*_n$ given $\{(Y_i,X_i): i=1,\cdots, n\}$ converges in distribution to the limiting null distribution of $n\hat{U}_n$ in probability, except for $Z=(\varepsilon, X)$ replacing by $Z=(\eta, X)$.	
\end{theorem}

\section{Numerical studies}
\subsection{Simulations}
In this subsection, we conduct detailed simulation studies to investigate the finite sample performance of the proposed test statistics $\hat{U}_n$. We also make a comparison between our test, the Dette et al. (2007) test $T_n^{CvM}$, and
the Wang and Zhou (2007) test $T_n^{WZ}$ in both the nonlinear and nonparametric regression settings. The Cram\'er-von Mises (CvM) test statistic of Dette et al. (2007) is defined as
$$ T_n^{CvM} = n \int [ \hat{F}_{\hat{\varepsilon}}(y)-\hat{F}_{\hat{\eta}}(y)]^2d\hat{F}_{\hat{\varepsilon}}(y), $$
where $\hat{F}_{\hat{\varepsilon}}(y)=\frac{1}{n}\sum_{i=1}^{n}I(\hat{\varepsilon}_i \leq y)$ with $\hat{\varepsilon}_i = [Y_i - \hat{m}(X_i)]/\hat{\sigma}(X_i)$, $\hat{F}_{\hat{\eta}}(y)=\frac{1}{n}\sum_{i=1}^{n}I(\hat{\eta}_i \leq y)$ with $\hat{\eta}_{i} =[Y_i - \hat{m}(X_i)]/ \sigma(X_i, \hat{\theta}_n)$, and $\hat{\sigma}(\cdot)$ is a nonparametric estimator of $\sigma(\cdot)$. The critical value of $T_n^{CvM} $ is determined by the smooth residual bootstrap as suggested by Dette et al. (2007). The test statistic of Wang and Zhou (2007) is
$$ T_n^{WZ}=\frac{1}{n(n-1)h^p} \sum_{1 \leq i\neq j \leq n} K(\frac{X_i-X_j}{h}) \{[Y_i -\hat{m}(X_i)]^2-\sigma^2(X_i,\hat{\theta}_n)\} \{[Y_j- \hat{m}(X_j)]^2-\sigma^2(X_j,\hat{\theta}_n)\}, $$
where $K(\cdot)$ is the kernel functions and $h$ is the bandwidth. For the kernel function and the bandwidth selection in the nonparametric estimations of Dette et al. (2007) and Wang and Zhou (2007), one can refer their papers for details.

In the following simulation examples, $a=0$ and $a\neq 0$ correspond to the null hypothesis and the alternative hypotheses respectively. The significant level is set to be $0.05$. The simulation results are based on the average of $1000$ replications and the bootstrap approximation of $B=500$ samples. The sample sizes are $50$ and $100$ in each model. The dimension $p$ of covariates is set to be $2, 4, 8$ to see how the performance of these tests is affected by the dimensionality.

For our test in nonlinear regression models, we use the least square method to estimate the unknown parameter $\beta_0$ in the regression function. In the nonparametric regression case, we adopt the standard normal density function as the kernel function. To assess the effect of bandwidth $h$ for our test $\hat{U}_n$ in nonparametric cases, we conduct a simple simulation for a large bunch of bandwidths. This strategy was also adopted by Zhu, Fujikoshi, and Naito (2001), Wang and Zhou (2007), Khmaladze and Koul (2009), among many others. Let $h=c n^{-1/(p+4)}$ where $c$ varies from $0.6$ to $1.4$.
Figures 1-2 give the empirical sizes and powers of our test in nonparametric settings for model $H_{21}: Y = \beta_0^{T}X + |1+ (\theta_0^{T}X)^2 + a \sin(\theta_0^{T}X )|^{1/2} \varepsilon$. Here $\beta_0 = \theta_0 = (1,1,...1)^T/\sqrt{p}$ and $X \sim N(0, I_p)$ independent of the standard normal error term $\varepsilon$. Simulation results for other models in the following are very similar. Thus we omit them for brevity. From these figures, we can see that when the bandwidth is relatively small, our test $\hat{U}_n$ is conservative and have very low empirical powers. While if the bandwidth is too large, our test can not control the significant level either. Thus, we adopt $h=1.2n^{-1/(p+4)}$ in nonparametric settings in the following simulation examples.
$$ \rm{Figures \ 1-2 \ about \ here} $$

{\bf Example 1}. The data are generated from the following regression models:
\begin{eqnarray*}
H_{11}: Y &=& \beta_0^{T}X + (1+ |\theta_0^{T}X+ a \exp(\theta_0^{T}X )|  )^{1/2} \varepsilon \\
H_{12}: Y &=& \beta_0^{T}X + (1+ |\theta_1^{T}X|+ a (\theta_1^{T}X) )^{1/2} \varepsilon,
\end{eqnarray*}		
where $X\sim N(0, I_p)$ independent of the standard normal error term $\varepsilon$, $\beta_0=(1,1,\cdots, 1)^T/\sqrt{p}$, $\theta_0=(1,1,\cdots, 1)^T/\sqrt{p}$, and $\theta_1=(\underbrace{1,\cdots,1}_{p/2},0, \cdots ,0)^T/\sqrt{p/2}$.

The simulation results are reported in Tables $1-4$. We can see that all these tests can maintain the significance level very well in both the nonlinear and nonparametric regression models when the dimension $p=2, 4$. While the empirical sizes of these tests become unstable for large dimension ($p=8$). For the empirical power, it is easy to see that our test works better than the other two competitors in most cases. However, all these tests have bad power performance when the dimension becomes large, especially in nonparametric settings. This may be due to the curse of dimensionality resulting from the nonparametric estimate of the regression function.
$$ \rm{Tables \ 1-4 \ about \ here} $$

In the next simulation, we consider regression models with a slightly more complicated variance functions under the null hypotheses.

{\bf Example 2}. The data are generated from the following regression models:
\begin{eqnarray*}	
H_{21}: Y &=& \beta_0^{T}X + |1+ (\theta_0^{T}X)^2 + a \sin(\theta_0^{T}X )|^{1/2} \varepsilon, \\
H_{22}: Y &=& \beta_0^{T}X + (1+ |\sin(\theta_0^{T}X)|+a \exp(\theta_0^{T}X ) )  \varepsilon,
\end{eqnarray*}
where the predictor vector $X$, the error term $\varepsilon$ and the parameters $\beta_0$ and $\theta_0$ are the same as in Example 1.

The simulation results are presented in Tables 5-8. We can observe that for model $H_{21}$, all the tests can control the empirical size very well in both the nonlinear and nonparametric cases. The empirical powers of our test for model $H_{21}$ grow very quickly in most cases. While $T_n^{CvM}$ has the worst power performance in this model. For model $H_{22}$, all the tests can maintain the significant level very well for $p=2$. When $p$ is relatively large, the empirical sizes of $T_n^{CvM}$ and $T_n^{WZ}$ are slightly far away from the significant level. For the empirical power in model $H_{22}$, $T_n^{CvM}$ beats the other two competitors and the Wang and Zhou (2007) test $T_n^{WZ}$ performs the worst, especially in nonparametric cases.
$$ \rm{Tables \ 5-8 \ about \ here} $$

\subsection{A real data example}
In this subsection we apply the proposed test to the Esterase count data set that is obtained from Example 2.2 in Carroll and Ruppert (1988). This data set contains $108$ observations obtained from a calibration experiment for measuring the concentration of an enzyme esterase with the radioimmunoassay (RIA) counts as the response variable $Y$ and the concentration of esterase as the predictor variable $X$.
%\cite{carroll2017transfor}
Carroll and Ruppert (1988) first suggested to fit this data set by a linear regression model $Y=\beta_0 + \beta_1 X +\varepsilon$. To see whether there exist heteroscedasticity in this model, we give a scatter plot of $Y-\hat{\beta}_0 - \hat{\beta}_1 X$ against the fitted values $\hat{Y}$ in Figures 3(a), where $\hat{Y}=\hat{\beta}_0 + \hat{\beta}_1 X$. This plot clearly indicates the existence of heteroscedasticity.
Wang and Zhou (2007) further analysed the Esterase count data set and suggested to use a heteroscedasticity regression model for data fitting:
\begin{equation}\label{5.1}
Y_i=\beta_0+\beta_1X_i+\sigma(\beta_0+\beta_1X_i)^{\theta}\varepsilon_i, \quad i=1, \cdots, 108.
\end{equation}
When applying our tests to model~(\ref{5.1}), we find that the $p$-value is about $0.968$. This indicates that the model~(\ref{5.1}) may be plausible to fit the Esterase count data set. To further visualize this fit, Figure 3(b) gives the scatter plot of $\hat{\eta}$ against the predictor variable $X$ where $\hat{\eta}=\frac{Y-\hat{\beta}_0-\hat{\beta}_1 X}{\hat{\sigma}(\hat{\beta}_0 + \hat{\beta}_1 X)^{\hat{\theta}}}$. This plot also shows that there exists no trend between the residual $\hat{\eta}$ and the predictor $X$. Thus the heteroscedasticity model~(\ref{5.1}) is reasonable for fitting the Esterase count data.
$$ \rm{Figure \ 3 \ about \ here} $$

\section{Discussion}
In this paper we proposed a new test based on distance covariance to check the adequacy of the parametric form of the variance function in nonlinear and nonparametric regression models. Inherit from the nice properties of distance covariance, the proposed test is very easy to implement in practice and is less sensitive to the dimension of covariates. The asymptotic properties of the test statistic is investigated under the null and the alternative hypotheses. It has been shown that the proposed test is consistent against any alternative and can detect the local alternatives distinct from the null at a parametric rate $1/\sqrt{n}$, even in nonparametric regression settings.
As the proposed test is not asymptotically distribution-free, we suggested a residual bootstrap to approximate the limiting null distribution of the test statistic. Simulation results show that our test can control the nominal level very well and has good power performance.

\begin{figure}\label{Figure 1}
\centering
\includegraphics[width=12cm,height=8cm]{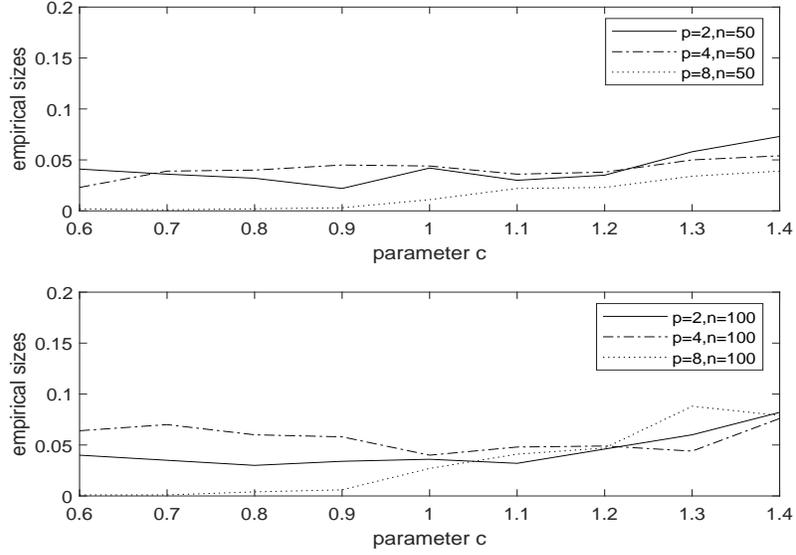}
\caption{The empirical sizes of $\hat{U}_n$ in nonparametric cases against different bandwidths for model $H_{21}$.}
\end{figure}

\begin{figure}
\centering
\includegraphics[width=12cm,height=8cm]{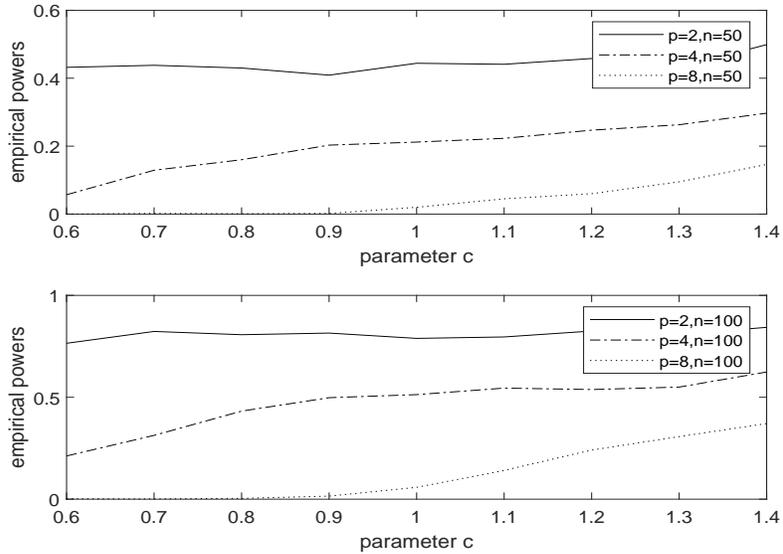}
\caption{The empirical powers of $\hat{U}_n$ in nonparametric cases against different bandwidths when $a=2$ for model $H_{21}$.}\label{Figure 3}
\end{figure}

\clearpage
\begin{table}[ht!]\caption{Empirical sizes and powers of $\hat{U}_n$, $T_n^{CvM}$, and $T_n^{WZ}$ for $H_{11}$ in nonlinear cases
in Example 1. }
\centering
{\small\scriptsize\hspace{12.5cm}
\renewcommand{\arraystretch}{1}\tabcolsep 0.25cm
\begin{tabular}{cccccccccccc}
\hline
&\multicolumn{1}{c}{a} &\multicolumn{2}{c}{$\hat{U}_n$}  &\multicolumn{2}{c}{$T_n^{CvM}$} &\multicolumn{2}{c}{$T_n^{WZ}$}  \\
&&\multicolumn{1}{c}{n=50}&\multicolumn{1}{c}{n=100} &\multicolumn{1}{c}{n=50}&\multicolumn{1}{c}{n=100} &\multicolumn{1}{c}{n=50}&\multicolumn{1}{c}{n=100}\\
\hline
$ p=2$  &0.0       &0.037 &0.040 &0.041 &0.051 &0.049 &0.050\\
        &0.5       &0.102 &0.202 &0.095 &0.109 &0.112 &0.216\\
        &1.0       &0.210 &0.466 &0.109 &0.164 &0.222 &0.449\\
        &1.5       &0.322 &0.629 &0.119 &0.219 &0.290 &0.625\\
        &2.0       &0.379 &0.727 &0.168 &0.239 &0.374 &0.698\\
        &2.5       &0.452 &0.798 &0.170 &0.255 &0.378 &0.738\\	
\hline
$ p=4$  &0.0    &0.054 &0.044  &0.058 &0.061   &0.058 &0.052 \\
		&0.5    &0.091 &0.139  &0.087 &0.120   &0.093 &0.135\\
		&1.0    &0.154 &0.314  &0.113 &0.161   &0.115 &0.227\\
		&1.5    &0.215 &0.464  &0.120 &0.192   &0.155 &0.347 \\
		&2.0    &0.255 &0.556  &0.124 &0.191   &0.197 &0.359 \\
		&2.5    &0.285 &0.652  &0.168 &0.205   &0.205 &0.398  \\
\hline
$ p=8$  &0.0    &0.046 &0.055  &0.087 &0.095  &0.051 &0.043 \\
		&0.5    &0.111 &0.126  &0.117 &0.136  &0.073 &0.103 \\
		&1.0    &0.115 &0.241  &0.133 &0.196  &0.076 &0.093 \\
		&1.5    &0.155 &0.346  &0.145 &0.206  &0.089 &0.120\\
		&2.0    &0.186 &0.402  &0.147 &0.184  &0.082 &0.141\\
		&2.5    &0.218 &0.481  &0.134 &0.180  &0.103 &0.160\\
\hline
\end{tabular}}
\end{table}

\begin{table}[ht!]\caption{Empirical sizes and powers of $\hat{U}_n$, $T_n^{CvM}$, and $T_n^{WZ}$ for $H_{11}$ in nonparametric cases in Example 1. }
\centering{
\small\scriptsize\hspace{12.5cm}
\renewcommand{\arraystretch}{1}\tabcolsep 0.25cm
\begin{tabular}{cccccccccccc}
\hline
&\multicolumn{1}{c}{a} &\multicolumn{2}{c}{$\hat{U}_n$}&\multicolumn{2}{c}{$T_n^{CvM}$} &\multicolumn{2}{c}{$T_n^{WZ}$}  \\
&&\multicolumn{1}{c}{n=50}&\multicolumn{1}{c}{n=100} &\multicolumn{1}{c}{n=50}&\multicolumn{1}{c}{n=100}
&\multicolumn{1}{c}{n=50}&\multicolumn{1}{c}{n=100}\\
\hline
$ p=2$      &0.0       &0.036 &0.053 &0.039 &0.048 &0.032 &0.044\\
			&0.5       &0.085 &0.185 &0.097 &0.119 &0.079 &0.111\\
			&1.0       &0.159 &0.435 &0.149 &0.182 &0.119 &0.285\\
			&1.5       &0.268 &0.636 &0.183 &0.260 &0.208 &0.378\\
			&2.0       &0.327 &0.755 &0.193 &0.318 &0.234 &0.481\\
			&2.5       &0.415 &0.843 &0.225 &0.378 &0.287 &0.569\\
\hline
$ p=4$      &0.0    &0.041 &0.052 &0.048 &0.045 &0.033 &0.053\\
			&0.5    &0.044 &0.093 &0.088 &0.114 &0.053 &0.067\\
			&1.0    &0.081 &0.174 &0.134 &0.163 &0.061 &0.111\\
			&1.5    &0.084 &0.291 &0.152 &0.222 &0.089 &0.157\\
			&2.0    &0.127 &0.375 &0.165 &0.258 &0.099 &0.192\\
			&2.5    &0.153 &0.470 &0.153 &0.266 &0.123 &0.227\\
\hline
$ p=8$      &0.0    &0.032 &0.066 &0.021 &0.052  &0.026 &0.043\\
			&0.5    &0.039 &0.092 &0.051 &0.087  &0.031 &0.039\\
			&1.0    &0.055 &0.149 &0.066 &0.122  &0.033 &0.055\\
			&1.5    &0.060 &0.185 &0.067 &0.158  &0.038 &0.068\\
			&2.0    &0.063 &0.257 &0.084 &0.190  &0.047 &0.066\\
			&2.5    &0.080 &0.256 &0.081 &0.195  &0.045 &0.080\\
			\hline
\end{tabular}}
\end{table}

\begin{table}[ht!]\caption{Empirical sizes and powers of $\hat{U}_n$, $T_n^{CvM}$, and $T_n^{WZ}$ for $H_{12}$ in nonlinear cases
		in Example 1.}
	\centering{
		\small\scriptsize\hspace{12.5cm}
		\renewcommand{\arraystretch}{1}\tabcolsep 0.25cm
		\begin{tabular}{cccccccccccc}
			\hline
			&\multicolumn{1}{c}{a} &\multicolumn{2}{c}{$\hat{U}_n$}  &\multicolumn{2}{c}{$T_n^{CvM}$} &\multicolumn{2}{c}{$T_n^{WZ}$}  \\
			&&\multicolumn{1}{c}{n=50}&\multicolumn{1}{c}{n=100} &\multicolumn{1}{c}{n=50}&\multicolumn{1}{c}{n=100} &\multicolumn{1}{c}{n=50}&\multicolumn{1}{c}{n=100}\\
			\hline
			$ p=2$      &0.0       &0.049 &0.048 &0.049 &0.043 &0.029 &0.049\\
			&0.2       &0.050 &0.047 &0.049 &0.059 &0.042 &0.069    \\
			&0.4       &0.062 &0.091 &0.052 &0.058 &0.045 &0.083 \\
			&0.4       &0.079 &0.162 &0.048 &0.080 &0.093 &0.128  \\
			&0.6       &0.135 &0.287 &0.086 &0.088 &0.111 &0.248 \\
			&1.0       &0.219 &0.474 &0.075 &0.143 &0.153 &0.383   \\	
			\hline
			$ p=4$      &0.0      &0.043 &0.045 &0.052 &0.051 &0.047 &0.045    \\
			&0.2    &0.047 &0.052 &0.067 &0.067 &0.059 &0.050    \\
			&0.4    &0.058 &0.071 &0.055 &0.084 &0.058 &0.069    \\
			&0.6    &0.089 &0.133 &0.081 &0.081 &0.057 &0.115     \\
			&0.8    &0.102 &0.242 &0.103 &0.114 &0.085 &0.152     \\
			&1.0    &0.185 &0.396 &0.120 &0.145 &0.113 &0.203      \\
			\hline
			$ p=8$      &0.0    &0.029 &0.050 &0.083 &0.075 &0.055 &0.057\\
			&0.2    &0.052 &0.061 &0.100 &0.099 &0.045 &0.053 \\
			&0.4    &0.073 &0.076 &0.086 &0.096 &0.060 &0.065  \\
			&0.6    &0.074 &0.120 &0.114 &0.135 &0.077 &0.072  \\
			&0.8    &0.102 &0.176 &0.117 &0.176 &0.095 &0.083 \\
			&1.0    &0.140 &0.320 &0.151 &0.211 &0.087 &0.090 \\
			\hline
	\end{tabular}}
\end{table}

\begin{table}[ht!]\caption{Empirical sizes and powers of $\hat{U}_n$, $T_n^{CvM}$, and $T_n^{WZ}$ for $H_{12}$ in nonparametric cases in Example 1.}
	\centering
	{\small\scriptsize\hspace{12.5cm}
		\renewcommand{\arraystretch}{1}\tabcolsep 0.25cm
		\begin{tabular}{cccccccccccc}
			\hline
			&\multicolumn{1}{c}{a} &\multicolumn{2}{c}{$\hat{U}_n$}  &\multicolumn{2}{c}{$T_n^{CvM}$} &\multicolumn{2}{c}{$T_n^{WZ}$}  \\
			&&\multicolumn{1}{c}{n=50}&\multicolumn{1}{c}{n=100} &\multicolumn{1}{c}{n=50}&\multicolumn{1}{c}{n=100}
			&\multicolumn{1}{c}{n=50}&\multicolumn{1}{c}{n=100}\\
			\hline
			$ p=2$      &0.0       &0.053 &0.072 &0.047 &0.051 &0.044 &0.042\\
			&0.2       &0.070 &0.098 &0.051 &0.057 &0.040 &0.066 \\
			&0.4       &0.074 &0.108 &0.053 &0.079 &0.046 &0.079   \\
			&0.6       &0.101 &0.161 &0.067 &0.097 &0.062 &0.134\\
			&0.8       &0.129 &0.230 &0.082 &0.140 &0.093 &0.217\\
			&1.0       &0.163 &0.350 &0.086 &0.196 &0.112 &0.329\\
			\hline
			$ p=4$      &0.0    &0.049 &0.063 &0.046 &0.050 &0.040 &0.048\\
			&0.2    &0.053 &0.067 &0.058 &0.072 &0.042 &0.039\\
			&0.4    &0.053 &0.075 &0.046 &0.070 &0.045 &0.057\\
			&0.6    &0.070 &0.106 &0.058 &0.098 &0.046 &0.075\\
			&0.8    &0.103 &0.128 &0.077 &0.125 &0.062 &0.093\\
			&1.0    &0.105 &0.200 &0.126 &0.161 &0.062 &0.134\\
			\hline
			$p=8$      &0.0    &0.047 &0.077 &0.015 &0.048 &0.018 &0.039 \\
			&0.2    &0.044 &0.059 &0.018 &0.033 &0.023 &0.029\\
			&0.4    &0.038 &0.080 &0.013 &0.039 &0.034 &0.035\\
			&0.6    &0.047 &0.075 &0.025 &0.051 &0.025 &0.038\\
			&0.8    &0.043 &0.072 &0.033 &0.078 &0.034 &0.040\\
			&1.0    &0.054 &0.085 &0.045 &0.116 &0.033 &0.061\\
			\hline
		\end{tabular}
	}
\end{table}

\begin{table}[ht!]\caption{Empirical sizes and powers of $\hat{U}_n$, $T_n^{CvM}$, and $T_n^{WZ}$ for $H_{21}$ in nonlinear cases in Example 2. }
\centering
{\small\scriptsize\hspace{12.5cm}
\renewcommand{\arraystretch}{1}\tabcolsep 0.25cm
\begin{tabular}{cccccccccccc}
\hline
&\multicolumn{1}{c}{a} &\multicolumn{2}{c}{$\hat{U}_n$}  &\multicolumn{2}{c}{$T_n^{CvM}$} &\multicolumn{2}{c}{$T_n^{ZH}$}  \\
&&\multicolumn{1}{c}{n=50}&\multicolumn{1}{c}{n=100} &\multicolumn{1}{c}{n=50}&\multicolumn{1}{c}{n=100}
&\multicolumn{1}{c}{n=50}&\multicolumn{1}{c}{n=100}\\
\hline
$ p=2$      &0.0     &0.052 &0.047 &0.042 &0.053  &0.037 &0.051\\
			&0.5     &0.053 &0.090 &0.053 &0.047  &0.064 &0.067  \\
			&1.0     &0.079 &0.163 &0.065 &0.065  &0.087 &0.164 \\
			&1.5     &0.176 &0.409 &0.066 &0.084  &0.164 &0.339\\
			&2.0     &0.422 &0.777 &0.098 &0.132  &0.289 &0.617\\
			&2.5     &0.666 &0.960 &0.140 &0.205  &0.400 &0.749  \\ 	
\hline
$ p=4$      &0.0  &0.045 &0.040 &0.065 &0.048  &0.048 &0.043 \\
			&0.5  &0.052 &0.065 &0.049 &0.047  &0.060 &0.053\\
			&1.0  &0.061 &0.124 &0.057 &0.073  &0.074 &0.083\\
			&1.5  &0.129 &0.320 &0.083 &0.086  &0.114 &0.158 \\
			&2.0  &0.301 &0.622 &0.101 &0.109  &0.130 &0.269 \\
			&2.5  &0.480 &0.903 &0.115 &0.119  &0.190 &0.369  \\
\hline
$ p=8$      &0.0  &0.050 &0.048 &0.077 &0.065  &0.057 &0.044 \\
			&0.5  &0.061 &0.058 &0.076 &0.078  &0.074 &0.050 \\
			&1.0  &0.077 &0.084 &0.097 &0.107  &0.068 &0.065 \\
			&1.5  &0.115 &0.201 &0.123 &0.137  &0.096 &0.103\\
			&2.0  &0.191 &0.434 &0.145 &0.146  &0.091 &0.127\\
			&2.5  &0.386 &0.729 &0.162 &0.157  &0.137 &0.167\\
\hline
\end{tabular}}
\end{table}

\begin{table}[ht!]\caption{Empirical sizes and powers of $\hat{U}_n$, $T_n^{CvM}$, and $T_n^{WZ}$ for $H_{21}$ in nonparametric cases in Example 2. }
\centering
{\small\scriptsize\hspace{12.5cm}
\renewcommand{\arraystretch}{1}\tabcolsep 0.25cm
\begin{tabular}{cccccccccccc}
\hline
&\multicolumn{1}{c}{a} &\multicolumn{2}{c}{$\hat{U}_n$}  &\multicolumn{2}{c}{$T_n^{CvM}$} &\multicolumn{2}{c}{$T_n^{ZH}$}  \\
&&\multicolumn{1}{c}{n=50}&\multicolumn{1}{c}{n=100} &\multicolumn{1}{c}{n=50}&\multicolumn{1}{c}{n=100}
&\multicolumn{1}{c}{n=50}&\multicolumn{1}{c}{n=100}\\
\hline
$ p=2$      &0.0     &0.032 &0.048 &0.043 &0.056 &0.035 &0.050\\
			&0.5     &0.046 &0.079 &0.054 &0.063 &0.046 &0.076  \\
			&1.0     &0.103 &0.194 &0.046 &0.077 &0.065 &0.180 \\
			&1.5     &0.209 &0.453 &0.079 &0.132 &0.117 &0.413\\
			&2.0     &0.440 &0.805 &0.171 &0.320 &0.222 &0.652\\
			&2.5     &0.711 &0.971 &0.302 &0.632 &0.312 &0.789\\
\hline
$ p=4$      &0.0  &0.034 &0.042 &0.042 &0.044  &0.037 &0.045 \\
			&0.5  &0.042 &0.053 &0.031 &0.041  &0.040 &0.055\\
			&1.0  &0.060 &0.104 &0.051 &0.057 &0.043 &0.083\\
			&1.5  &0.109 &0.243 &0.075 &0.101 &0.056 &0.124 \\
			&2.0  &0.252 &0.524 &0.115 &0.167 &0.100 &0.231 \\
			&2.5  &0.443 &0.808 &0.199 &0.310 &0.132 &0.326  \\
\hline
$ p=8$      &0.0  &0.023 &0.059 &0.025 &0.023 &0.029 &0.038\\
			&0.5  &0.041 &0.065 &0.028 &0.033 &0.034 &0.025\\
			&1.0  &0.046 &0.082 &0.035 &0.037 &0.029 &0.031\\
			&1.5  &0.049 &0.129 &0.038 &0.087 &0.043 &0.051\\
			&2.0  &0.073 &0.226 &0.088 &0.110 &0.051 &0.077\\
			&2.5  &0.129 &0.420 &0.151 &0.218 &0.072 &0.085\\
\hline
\end{tabular}}
\end{table}

\begin{table}[ht!]\caption{Empirical sizes and powers of $\hat{U}_n$, $T_n^{CvM}$, and $T_n^{WZ}$ for $H_{22}$ in nonlinear cases in Example 2. }
\centering
{\small\scriptsize\hspace{12.5cm}
\renewcommand{\arraystretch}{1}\tabcolsep 0.25cm
\begin{tabular}{cccccccccccc}
\hline
&\multicolumn{1}{c}{a} &\multicolumn{2}{c}{$\hat{U}_n$}  &\multicolumn{2}{c}{$T_n^{CvM}$} &\multicolumn{2}{c}{$T_n^{ZH}$}  \\
&&\multicolumn{1}{c}{n=50}&\multicolumn{1}{c}{n=100} &\multicolumn{1}{c}{n=50}&\multicolumn{1}{c}{n=100}
&\multicolumn{1}{c}{n=50}&\multicolumn{1}{c}{n=100}\\
\hline
$ p=2$      &0.0       &0.047 &0.044 &0.047 &0.046 &0.056 &0.053\\
			&0.5       &0.319 &0.632 &0.778 &0.970 &0.334 &0.733\\
			&1.0       &0.601 &0.901 &0.993 &1.000 &0.610 &0.843\\
			&1.5       &0.754 &0.988 &0.999 &1.000 &0.692 &0.839\\
			&2.0       &0.860 &0.993 &1.000 &1.000 &0.679 &0.810\\
			&2.5       &0.890 &0.998 &1.000 &1.000 &0.685 &0.799\\
\hline
$p=4$       &0.0    &0.062 &0.043 &0.114 &0.165 &0.094 &0.133\\
			&0.5    &0.237 &0.471 &0.669 &0.950 &0.066 &0.316\\
			&1.0    &0.426 &0.789 &0.986 &1.000 &0.232 &0.572\\
			&1.5    &0.586 &0.928 &1.000 &1.000 &0.315 &0.607 \\
			&2.0    &0.676 &0.974 &1.000 &1.000 &0.342 &0.587 \\
			&2.5    &0.728 &0.981 &1.000 &1.000 &0.348 &0.549 \\
\hline
$p=8$       &0.0    &0.059 &0.043 &0.104 &0.166 &0.077 &0.112\\
			&0.5    &0.108 &0.278 &0.541 &0.940 &0.019 &0.081\\
			&1.0    &0.222 &0.596 &0.972 &1.000 &0.051 &0.223\\
			&1.5    &0.287 &0.732 &0.999 &1.000 &0.114 &0.344\\
			&2.0    &0.337 &0.835 &1.000 &1.000 &0.171 &0.388\\
			&2.5    &0.401 &0.888 &1.000 &1.000 &0.207 &0.406\\
\hline
\end{tabular}}
\end{table}

\begin{table}[ht!]\caption{Empirical sizes and powers of $\hat{U}_n$, $T_n^{CvM}$, and $T_n^{WZ}$ for $H_{22}$ in nonparametric cases in Example 2. }
\centering
{\small\scriptsize\hspace{12.5cm}
\renewcommand{\arraystretch}{1}\tabcolsep 0.25cm
\begin{tabular}{cccccccccccc}
\hline
&\multicolumn{1}{c}{a} &\multicolumn{2}{c}{$\hat{U}_n$}  &\multicolumn{2}{c}{$T_n^{CvM}$} &\multicolumn{2}{c}{$T_n^{ZH}$}  \\
&&\multicolumn{1}{c}{n=50}&\multicolumn{1}{c}{n=100} &\multicolumn{1}{c}{n=50}&\multicolumn{1}{c}{n=100}
&\multicolumn{1}{c}{n=50}&\multicolumn{1}{c}{n=100}\\
\hline
$p=2$       &0.0     &0.044 &0.055 &0.047 &0.065 &0.040 &0.045\\
			&0.5     &0.169 &0.405 &0.551 &0.927 &0.227 &0.601\\
			&1.0     &0.362 &0.751 &0.964 &1.000 &0.500 &0.762\\
			&1.5     &0.527 &0.905 &0.998 &1.000 &0.623 &0.797\\
			&2.0     &0.632 &0.958 &1.000 &1.000 &0.676 &0.800\\
			&2.5     &0.698 &0.979 &1.000 &1.000 &0.666 &0.790\\	
\hline
$ p=4$      &0.0    &0.052 &0.056 &0.084 &0.097 &0.070 &0.098\\
			&0.5    &0.085 &0.217 &0.312 &0.815 &0.016 &0.170\\
			&1.0    &0.194 &0.465 &0.893 &1.000 &0.089 &0.425\\
			&1.5    &0.257 &0.685 &0.997 &1.000 &0.176 &0.518 \\
			&2.0    &0.317 &0.775 &1.000 &1.000 &0.261 &0.565 \\
			&2.5    &0.404 &0.832 &1.000 &1.000 &0.336 &0.623 \\
\hline
$ p=8$      &0.0    &0.020 &0.038 &0.032 &0.073 &0.053 &0.093\\
			&0.5    &0.053 &0.114 &0.023 &0.202 &0.004 &0.009\\
			&1.0    &0.062 &0.196 &0.401 &0.970 &0.009 &0.047\\
			&1.5    &0.077 &0.273 &0.827 &0.998 &0.029 &0.095\\
			&2.0    &0.077 &0.355 &0.967 &1.000 &0.062 &0.177\\
			&2.5    &0.125 &0.384 &0.996 &1.000 &0.121 &0.259\\
\hline
\end{tabular}}
\end{table}

\begin{figure}
\centering
\includegraphics[width=15cm,height=6cm]{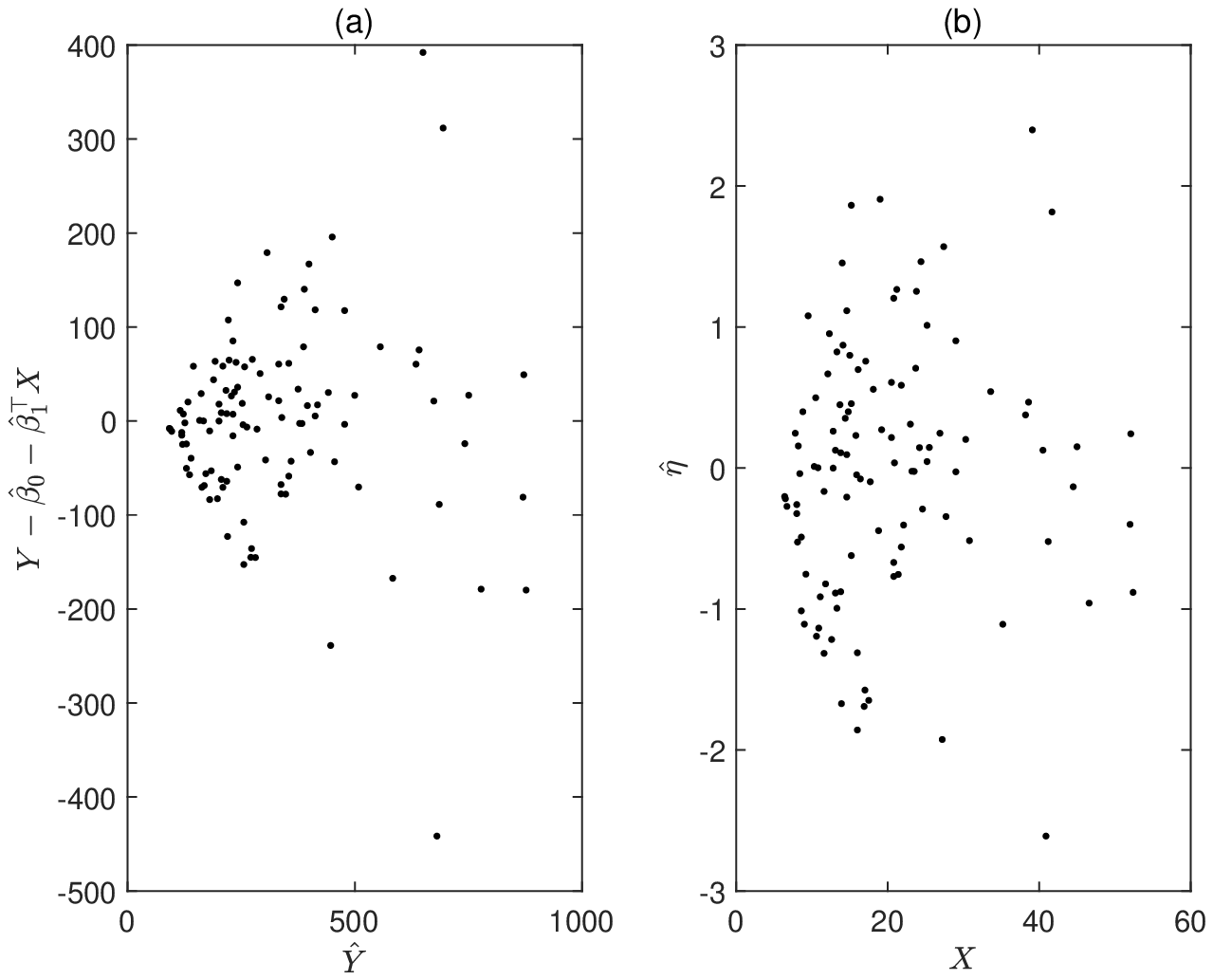}
\caption{(a) is the scatter plot of $Y- \hat{\beta}_0 - \hat{\beta}_1 X$ against the fitted values $\hat{Y}$ and (b) is the scatter plot of $\hat{\eta}$ against the predictor variable $X$ for the Esterase count data set.}\label{Figure 5}
\end{figure}

\section*{Acknowledgement}
Falong Tan was supported by the National Natural Science Foundation of China (12071119) and Haiqi Li was support of the National Natural Science Foundation of China (72171076) and Hunan Provincial Natural Science Foundation of China for Distinguished Young Scholars (2021JJ10026).

\clearpage
%\bibliographystyle{Chicago}
%\bibliography{mybibfile}

\section*{References}
[1] Bickel, P. J. (1978). Using residuals robustly i: Tests for heteroscedasticity, nonlinearity. The Annals of Statistics 6, 266–291.

[2] Breusch, T. S. and A. R. Pagan (1979). A simple test for heteroscedasticity and random coefficient variation. Econometrica 46, 1287–1294.

[3] Carroll, R. J. and D. Ruppert (1988). Transformation and Weighting in Regression. New York: Wiley.

[4] Cook, R. D. and S. Weisberg (1983). Diagnostics for heteroscedasticity in regression. Biometrika 70, 1–10.

[5] Dette, H. (2002). A consistent test for heteroscedasticity in nonparametric regression based on the kernel method. Journal of Statistical Planning and Inference 103(1-2), 311–329.

[6] Dette, H. and A. Munk (1998). Testing heteroscedasticity in nonparametric regression. Journal of the Royal Statistical Society: Series B (Statistical Methodology) 60(4), 693–708.

[7] Dette, H., N. Neumeyer, and I. V. Keilegom (2007). A new test for the parametric form of the variance function in non-parametric regression. Journal of the Royal Statistical Society: Series B (Statistical Methodology) 69(5), 903–917.

[8] Guo, X., X. Jiang, S. Zhang, and L. Zhu (2020). Pairwise distance-based heteroscedasticity test for regressions. Science China Mathematics 63(12), 2553–2572.

[9] Knight, K. (1998). Limiting distributions for l1 regression estimators under general conditions. Annals of Statistics 26(2), 755–770.

[10] Koul, H. L. and W. Song (2010). Conditional variance model checking. Journal of Statistical Planning and Inference 140(4), 1056–1072.

[11] Pakes, A. and D. Pollard (1989). Simulation and the asymptotics of optimization estimators. Econometrica 57, 1027–1057.

[12] Samarakoon, N. and W. Song (2011). Minimum distance conditional variance function checking in heteroscedastic regression models. Journal of multivariate analysis 102(3), 579–600.

[13] Samarakoon, N. and W. Song (2012). Empirical smoothing lack-of-fit tests for variance function. Journal of Statistical Planning and Inference 142(5), 1128–1140.

[14] Sen, A. and B. Sen (2014). Testing independence and goodness-of-fit in linear models. Biometrika 101(4), 927–942.

[15] Serfling, R. J. (2009). Approximation theorems of mathematical statistics. John Wiley $\And$ Sons, New York.

[16] Sherman, R. P. (1994). Maximal inequalities for degenerate u-processes with applications to optimization estimators. The Annals of Statistics 22, 439–459.

[17] Su, L. and A. Ullah (2013). A nonparametric goodness-of-fit-based test for conditional heteroskedasticity. Econometric Theory 29(1), 187–212.

[18] Sz\'{e}kely , G. J. and M. L. Rizzo (2009). Brownian distance covariance. The Annals of Applied Statistics 3(4), 1236–1265.

[19] Sz\'{e}kely , G. J. and M. L. Rizzo (2014). Partial distance correlation with methods for dissimilarities. The Annals of Statistics 42(6), 2382–2412.

[20]. Sz\'{e}kely , G. J., M. L. Rizzo, and N. K. Bakirov (2007). Measuring and testing dependence by correlation of distances. The Annals of Statistics 35(6), 2769–2794.

[21] Tan, F., X. Guo, and L. Zhu (2022). Weighted residual empirical processes, martingale transformations and model checking for regressions. arXiv preprint arXiv:2201.12537.

[22] Tan, F., X. Jiang, X. Guo, and L. Zhu (2021). Testing heteroscedasticity for regression models based on projections. Statistica Sinica 31, 625–646.

[23] Wang, L. and X.-H. Zhou (2007). Assessing the adequacy of variance function in heteroscedastic regression models. Biometrics 63(4), 1218–1225.

[24] White, H. (1980). A heteroskedasticity-consistent covariance matrix estimator and a direct test for heteroskedasticity. Econometrica 48, 817–838.

[25] Xu, K. and M. Cao (2021). Distance-covariance-based tests for heteroscedasticity in nonlinear regressions. Science China Mathematics 64(10), 2327–2356.

[26] Xu, K. and D. He (2021). Omnibus model checks of linear assumptions through distance covariance. Statistica Sinica 31, 1055–1079.

[27] Yao, S., X. Zhang, and X. Shao (2018). Testing mutual independence in high dimension via distance covariance. Journal of the Royal Statistical Society: Series B (Statistical Methodology) 80(3), 455–480.

[28] Zheng, X. (2009). Testing heteroscedasticity in nonlinear and nonparametric regressions. Canadian Journal of Statistics 37(2), 282–300.

[29]  Zhu, L., Y. Fujikoshi, and K. Naito (2001). Heteroscedasticity checks for regression models. Science in China Series A: Mathematics 44(10), 1236–1252.

\section*{Appendix}
The Appendix contains the proofs of the theoretical results stated in the main context.

{\bf Proof of Theorem 3.1.} Recall that $\eta_i=\frac{Y_i-m(X_i, \beta_0)}{\sigma(X_i, \tilde{\theta}_0)}$ and $\hat{\eta}_i = \frac{Y_i - m(X_i,\hat{\beta}_n)}{\sigma(X_i, \hat{\theta}_n)} $ in nonlinear cases.
It follows that
\begin{equation}\label{6.1}
\hat{\eta}_i-{\eta}_i = -\frac{m(X_i,\hat{\beta}_n)-m(X_i,{\beta}_0)}{\sigma(X_i,\tilde{\theta}_0)}-
\frac{\varepsilon_i(\sigma(X_i,\hat{\theta}_n)-\sigma(X_i,\tilde{\theta}_0))}{\sigma(X_i,\tilde{\theta}_0)}+R_i,
\end{equation}
where
\begin{equation*}
\begin{split}
R_i
=& \frac{\varepsilon_i[\sigma(X_i,\tilde{\theta}_0)-\sigma(X_i,\hat{\theta}_n)]^2}{\sigma(X_i,\tilde{\theta}_0)
   \sigma(X_i,\hat{\theta}_n)}+\frac{m(X_i,{\beta}_0)-m(X_i,\hat{\beta}_n)}{\sigma^2(X_i,\tilde{\theta}_0)}
   [\sigma(X_i,\tilde{\theta}_0)-\sigma(X_i,\hat{\theta}_n)] \\
&  +\frac{m(X_i,{\beta}_0)-m(X_i,\hat{\beta}_n)}{\sigma^2(X_i,\tilde{\theta}_0)}\frac{[\sigma(X_i,\tilde{\theta}_0)
   -\sigma(X_i,\hat{\theta}_n)]^2}{\sigma(X_i,\hat{\theta}_n)}.
\end{split}
\end{equation*}
According to the identity in the proof of Theorem 1 in Knight (1998), if $x \neq 0$, then we have
\begin{equation}\label{6.2}
|x-y|-|x|=-y\{\mathbb{I}(x> 0)-\mathbb{I}(x< 0)\}+2\int_{0}^{y}\{\mathbb{I}(x\leq s)-\mathbb{I}(x \leq 0)\}ds.
\end{equation}
Here $\mathbb{I}(A)$ is the indicator function of the set $A$. Note that $\eta_i = \varepsilon_i$ under the null $H_0$. It follows from (\ref{6.1}) and (\ref{6.2}) that
\begin{equation}\label{6.3}
\begin{split}
&  | \hat{\eta}_i-\hat{\eta}_j|  \\
&= |\varepsilon_i-\varepsilon_j|-\{ \frac{m(X_i,\hat{\beta}_n)-m(X_i,{\beta}_0)}{\sigma(X_i,\tilde{\theta}_0)}-
   \frac{m(X_j,\hat{\beta}_n)-m(X_j,{\beta}_0)}{\sigma(X_j,\tilde{\theta}_0)}+\frac{\varepsilon_i[{\sigma(X_i,\hat{\theta}_n)}
   -{\sigma(X_i,\tilde{\theta}_0)}]}{\sigma(X_i,\tilde{\theta}_0)}\\
&  -\frac{\varepsilon_j({\sigma(X_j,\hat{\theta}_n)}-{\sigma(X_j, \tilde{\theta}_0)})}{\sigma(X_j, \tilde{\theta}_0)}+(R_i-R_j)\}
   [\mathbb{I}(\varepsilon_i>\varepsilon_j)-\mathbb{I}(\varepsilon_i<\varepsilon_j)] \\
&  +2\int_{0}^{\frac{m(X_i,\hat{\beta}_n)-m(X_i,{\beta}_0)}{\sigma(X_i,\tilde{\theta}_0)}-\frac{m(X_i,\hat{\beta}_n)
   -m(X_j,{\beta}_0)}{\sigma(X_j,\tilde{\theta}_0)}+\frac{\varepsilon_i[{\sigma(X_i,\hat{\theta}_n)}-{\sigma(X_i,\tilde{\theta}_0)}]}
   {\sigma(X_i,\tilde{\theta}_0)}-\frac{\varepsilon_j[{\sigma(X_j,\hat{\theta}_n)}-{\sigma(X_j,\tilde{\theta}_0)}]}
   {\sigma(X_j,\tilde{\theta}_0)}+R_i-R_j}\\
&  [\mathbb{I}(\varepsilon_s-\varepsilon_t \leq  z)-\mathbb{I}(\varepsilon_s\leq \varepsilon_t)]dz.
\end{split}
\end{equation}

Recall that $\hat{U}_n=\frac{1}{n(n-3)} \sum_{1 \leq i \neq j \leq n}A_{ij}\hat{B}_{ij}$ with
\begin{eqnarray*}
&& A_{ij} =     \|X_i-X_j\|-\frac{1}{n-2}\sum_{k=1}^{n}\|X_i-X_k\|-\frac{1}{n-2}\sum_{l=1}^{n}\|X_j-X_l\|
                +\frac{1}{(n-1)(n-2)}\sum_{k,l=1}^{n}\|X_k-X_l\| \\
&& \hat{B}_{ij} = |\hat{\eta}_i-\hat{\eta}_j|-\frac{1}{n-2}\sum_{k=1}^{n}|\hat{\eta}_i-\hat{\eta}_k|-\frac{1}{n-2}\sum_{l=1}^{n}
                |\hat{\eta}_j-\hat{\eta}_l| +\frac{1}{(n-1)(n-2)}\sum_{k, l=1}^{n}|\hat{\eta}_k-\hat{\eta}_l|.
\end{eqnarray*}
By Lemma 1 of Yao et al. (2018), we can rewrite $\hat{U}_n$ as
$$ \hat{U}_n=\frac{1}{C_n^4} \sum\limits_{i< j< k< l}\tilde{h}_0(\hat{Z}_i,\hat{Z}_j,\hat{Z}_k,\hat{Z}_l), $$
where
\begin{equation}\label{6.4}
\tilde{h}_0 (\hat{Z}_i,\hat{Z}_j,\hat{Z}_k,\hat{Z}_l)
= \frac{1}{6} \sum_{s< t,u< v}^{(i,j,k,l)} |\hat{\eta}_{st}| (\|X_{st}\| + \|X_{uv}\|)
 -\frac{1}{12} \sum_{(s,t,u)}^{(i,j,k,l)} |\hat{\eta}_{st}| \| X_{su} \|,
\end{equation}
$\hat{Z}_i=(\hat{\eta}_i, X_i)$, $X_{st}=X_s-X_t$, and $\hat{\eta}_{st}=\hat{\eta}_s-\hat{\eta}_t$. Here the summation in (6.4) is over all permutations of the 4-tuples of indices $(i,j,k,l)$. Combining (6.3) and (6.4), we have	
\begin{eqnarray}\label{6.5}
	\hat{U}_n  \nonumber
	&=&   \frac{1}{C_n^4}\sum\limits_{i< j< k< l} \left( \frac{1}{6} \sum_{s< t,u< v}^{(i,j,k,l)} |\varepsilon_{st}|
	(\|X_{st}\|+\|X_{uv}\|) -\frac{1}{12} \sum_{(s,t,u)}^{(i,j,k,l)}|\varepsilon_{st}|\|X_{su}\| \right) \\  \nonumber
	&&    + \frac{1}{C_n^4}\sum\limits_{i< j< k< l} \left( \frac{1}{6} \sum_{s< t,u< v}^{(i,j,k,l)}\delta_{1st}
	(\|X_{st}\|+\|X_{uv}\|) - \frac{1}{12} \sum_{(s,t,u)}^{(i,j,k,l)}\delta_{1st}\|X_{su}\| \right) \\  \nonumber
	&&    + \frac{1}{C_n^4}\sum\limits_{i< j< k< l} \left( \frac{1}{6} \sum_{s< t,u< v}^{(i,j,k,l)}\delta_{2st}
	(\|X_{st}\|+\|X_{uv}\|) - \frac{1}{12} \sum_{(s,t,u)}^{(i,j,k,l)}\delta_{2st}\|X_{su}\| \right) \\
    &=:& \hat{U}_{n0}+\hat{U}_{n1}+\hat{U}_{n2},
\end{eqnarray}
where
\begin{eqnarray*}
\delta_{1st}
&=& -[\frac{\varepsilon_s({\sigma(X_s,\hat{\theta}_n)}-{\sigma(X_s,\tilde{\theta}_0)})}{\sigma(X_s,\tilde{\theta}_0)}
    -\frac{\varepsilon_t({\sigma(X_t,\hat{\theta}_n)}-{\sigma(X_t,\tilde{\theta}_0)})}{\sigma(X_t,\tilde{\theta}_0)}
    +\frac{m(X_s,\hat{\beta}_n)-m(X_s,{\beta}_0)}{\sigma(X_s,\tilde{\theta}_0)} \\
&&  -\frac{m(X_t,\hat{\beta}_n)-m(X_t,{\beta}_0)}{\sigma(X_t,\tilde{\theta}_0)}+(R_s-R_t)]\{\mathbb{I}(\varepsilon_s>\varepsilon_t)
    -\mathbb{I}(\varepsilon_s<\varepsilon_t)\}  \\
\delta_{2st}
&=&  2\int_{0}^{\frac{m(X_s,\hat{\beta}_n)-m(X_s,{\beta}_0)}{\sigma(X_s,\tilde{\theta}_0)}
	 -\frac{m(X_t,\hat{\beta}_n)-m(X_t,{\beta}_0)}{\sigma(X_t,\tilde{\theta}_0)}
	 +\frac{\varepsilon_s{\sigma(X_s,\hat{\theta}_n)}-{\sigma(X_s,\tilde{\theta}_0)})}
    {\sigma(X_s,\tilde{\theta}_0)}-\frac{\varepsilon_t({\sigma(X_t,\hat{\theta}_n)}-{\sigma(X_t,\tilde{\theta}_0)})}
    {\sigma(X_t,\tilde{\theta}_0)}+R_s-R_t} \\
&&   \{\mathbb{I}(\varepsilon_s-\varepsilon_t\leq z)-\mathbb{I}(\varepsilon_s\leq \varepsilon_t)\}dz.
\end{eqnarray*}

First we deal with $\hat{U}_{n1}$.  By Taylor's expansion, we have
\begin{eqnarray*}
 &&  \frac{m(X_s,\hat{\beta}_n)-m(X_s,{\beta}_0)}{\sigma(X_s,\tilde{\theta}_0)}-\frac{m(X_t,\hat{\beta}_n)
    -m(X_t,{\beta}_0)}{\sigma(X_t,\tilde{\theta}_0)}\\
&=& \frac{(\hat{\beta}_n-\beta_0)^T \dot{m}(X_s,{\beta}_0)}{\sigma(X_s,\tilde{\theta}_0)}
    -\frac{(\hat{\beta}_n-\beta_0)^T
    \dot{m}(X_t,{\beta}_0)}{\sigma(X_t,\tilde{\theta}_0)} \\
&&  +2^{-1}(\hat{\beta}_n-\beta_0)^T[\frac{\ddot{m}(X_s,(\hat{\beta}_n-\beta_0)\zeta + \beta_0)}{\sigma(X_s,\tilde{\theta}_0)}
    -\frac{\ddot{m}(X_t,(\hat{\beta}_n-\beta_0)\zeta+\beta_0)}{\sigma(X_t,\tilde{\theta}_0)}] (\hat{\beta}_n-\beta_0),  \\
\end{eqnarray*}
and
\begin{eqnarray*}
&&  \frac{\sigma(X_s,\hat{\theta}_n)-\sigma(X_s,\tilde{\theta}_0)}{\sigma(X_s,\tilde{\theta}_0)}
    -\frac{\sigma(X_t,\hat{\theta}_n)
    -\sigma(X_t,\tilde{\theta}_0)}{\sigma(X_t,\tilde{\theta}_0)}\\
&=& \frac{(\hat{\theta}_n-\tilde{\theta}_0)^T\dot{\sigma}(X_s,\tilde{\theta}_0)}{\sigma(X_s,\tilde{\theta}_0)}
    -\frac{(\hat{\theta}_n-\tilde{\theta}_0)^T \dot{\sigma}(X_t,\tilde{\theta}_0)}{\sigma(X_t,\tilde{\theta}_0)} \\
&& +2^{-1}(\hat{\theta}_n-\tilde{\theta}_0)^T[\frac{\Ddot{\sigma}
	(X_s,(\hat{\theta}_n-\tilde{\theta}_0)\zeta+\tilde{\theta}_0)}
    {\sigma(X_s,\tilde{\theta}_0)}-\frac{\ddot{\sigma}(X_t,(\hat{\theta}_n-\tilde{\theta}_0)\zeta+\tilde{\theta}_0)}
    {\sigma(X_t,\tilde{\theta}_0)}](\hat{\theta}_n-\tilde{\theta}_0),
\end{eqnarray*}
where $\zeta\in (0,1)$. Then we have
\begin{eqnarray}\label{6.6}
\nonumber \hat{U}_{n1}  \nonumber
&=& (\hat{\beta}_n-\beta_0)^T \frac{1}{C_n^4} \sum\limits_{i< j< k< l} h_{11}(Z_i,Z_j,Z_k,Z_l) \\   \nonumber
&&  + 2^{-1}(\hat{\beta}_n-\beta_0)^T  \frac{1}{C_n^4}\sum\limits_{i< j< k< l}h_{12}(Z_i,Z_j,Z_k,Z_l) (\hat{\beta}_n-\beta_0) \\ \nonumber
&&  + 2^{-1}(\hat{\beta}_n-\beta_0)^T \frac{1}{C_n^4}\sum\limits_{i< j< k< l}h_{13}(Z_i,Z_j,Z_k,Z_l) (\hat{\beta}_n-\beta_0)\\   \nonumber
&&  + (\hat{\theta}_n-\tilde{\theta}_0)^T \frac{1}{C_n^4}\sum\limits_{i< j< k< l}h_{14}(Z_i,Z_j,Z_k,Z_l) \\    \nonumber
&&  + 2^{-1}(\hat{\theta}_n-\tilde{\theta}_0)^T \frac{1}{C_n^4}\sum\limits_{i< j< k< l}h_{15}(Z_i,Z_j,Z_k,Z_l)
    (\hat{\theta}_n-\tilde{\theta}_0) \\  \nonumber
&&  + 2^{-1}(\hat{\theta}_n-\tilde{\theta}_0)^T \frac{1}{C_n^4}\sum\limits_{i< j< k< l}h_{16}(Z_i,Z_j,Z_k,Z_l)
    (\hat{\theta}_n-\tilde{\theta}_0) \\    \nonumber
&&  + \frac{1}{C_n^4}\sum\limits_{i< j< k< l}h_{17}(Z_i,Z_j,Z_k,Z_l) \\
&=:&I_{11} + I_{12} + I_{13}+ I_{14}+ I_{15}+ I_{16} + I_{17},
\end{eqnarray}
where $Z_i = (\varepsilon_i, X_i)$,
\begin{eqnarray*}
&&   h_{1m}(Z_i,Z_j,Z_k,Z_l) \\
&=&  -6^{-1}\sum_{s< t,u< v}^{(i,j,k,l)}\delta_{1mst}\{\mathbb{I}(\varepsilon_s>\varepsilon_t)
     -\mathbb{I}(\varepsilon_s<\varepsilon_t)\}(\|X_{st}\|+\|X_{uv}\|)\\
&&   +12^{-1}\sum_{(s,t,u)}^{(i,j,k,l)}\delta_{1mst}\{\mathbb{I}(\varepsilon_s>\varepsilon_t)
     -\mathbb{I}(\varepsilon_s<\varepsilon_t)\}\|X_{su}\|,  \quad {\rm for} \ m=1, \cdots, 7  \\
\end{eqnarray*}
and
\begin{eqnarray*}
\delta_{11st} &=& \frac{\dot{m}(X_s,{\beta}_0)}{\sigma(X_s,\tilde{\theta}_0)}-\frac{\dot{m}(X_t,{\beta}_0)}
                  {\sigma(X_t,\tilde{\theta}_0)} \\
\delta_{12st} &=& \frac{\ddot{m}(X_s,(\hat{\beta}_n-\beta_0)\zeta+\beta_0)}{\sigma(X_s,\tilde{\theta}_0)}
                  -\frac{\ddot{m}(X_s,\beta_0)}{\sigma(X_s,\tilde{\beta}_0)}
                  -[\frac{\ddot{m}(X_t,(\hat{\beta}_n-\beta_0)\zeta+\beta_0)}{\sigma(X_t,\tilde{\theta}_0)}
                  -\frac{\ddot{m}(X_t,\beta_0)}{\sigma(X_t,\tilde{\theta}_0)} ]\\
\delta_{13st} &=&
             \frac{\ddot{m}(X_s,\beta_0)}{\sigma(X_s,\tilde{\theta}_0)}
             -\frac{\ddot{m}(X_t,\beta_0)}{\sigma(X_t,\tilde{\theta}_0)} \\
\delta_{14st} &=& \frac{\varepsilon_s
             \dot{\sigma}(X_s,\tilde{\theta}_0)}{\sigma(X_s,\tilde{\theta}_0)}
              -\frac{\varepsilon_t\dot{\sigma}(X_t,\tilde{\theta}_0)}{\sigma(X_t,\tilde{\theta}_0)} \\
\delta_{15st} &=& \frac{\varepsilon_s \ddot{\sigma}
	          (X_s,(\hat{\theta}_n-\tilde{\theta}_0)\zeta+\theta_0)}{\sigma(X_s,\tilde{\theta}_0)}
              -\frac{\varepsilon_s\ddot{\sigma}(X_s,\tilde{\theta}_0)}{\sigma(X_s,\tilde{\theta}_0)} -[\frac{\varepsilon_t\ddot{\sigma}(X_t,(\hat{\theta}_n-\tilde{\theta}_0)\zeta+\tilde{\theta}_0)} {\sigma(X_t,\tilde{\theta}_0)}
              -\frac{\varepsilon_t \ddot{\sigma}(X_t,\tilde{\theta}_0)}{\sigma(X_t,\tilde{\theta}_0)}] \\
\delta_{16st} &=& \frac{\varepsilon_s \ddot{\sigma}(X_s,\tilde{\theta}_0)}{\sigma(X_s,\tilde{\theta}_0)}-\frac{\varepsilon_t\ddot{\sigma}
                  (X_t,\tilde{\theta}_0)}{\sigma(X_t,\tilde{\theta}_0)} \\
\delta_{17st} &=& R_s-R_t.
\end{eqnarray*}

For the term $I_{12}$, recall that
$$ I_{12} = 2^{-1}(\hat{\beta}_n-\beta_0)^T  \frac{1}{C_n^4}\sum\limits_{i< j< k< l}h_{12}(Z_i,Z_j,Z_k,Z_l) (\hat{\beta}_n-\beta_0). $$
Next we will show that $\frac{1}{C_n^4}\sum\limits_{i< j< k< l}h_{12}(Z_i,Z_j,Z_k,Z_l) =o_p(1).$  For this, set $\mathcal{B} =\{\beta: \sqrt{n}\|\beta - \beta_0\| \leq C\} $, $U_{12}(\beta) = \frac{1}{C_n^4}\sum\limits_{i< j< k< l}h_{12}(\beta, Z_i,Z_j,Z_k,Z_l) $, and
\begin{eqnarray*}
h_{12}(\beta, Z_i,Z_j,Z_k,Z_l)
&=&  -6^{-1}\sum_{s< t,u<v}^{(i,j,k,l)} [\frac{\ddot{m}(X_s,\beta) - \ddot{m}(X_s,\beta_0)}
    {\sigma(X_s, \tilde{\theta}_0)}-\frac{\ddot{m}(X_t,\beta)-\ddot{m}(X_t,\beta_0)}{\sigma(X_t,\tilde{\theta}_0)}]  \\
&&  \times [\mathbb{I}(\varepsilon_s>\varepsilon_t) -\mathbb{I}(\varepsilon_s<\varepsilon_t)] (\|X_{st}\|+\|X_{uv}\|)\\
&&  + 12^{-1}\sum_{(s,t,u)}^{(i,j,k,l)} [\frac{\ddot{m}(X_s,\beta) - \ddot{m}(X_s,\beta_0)}
    {\sigma(X_s, \tilde{\theta}_0)}-\frac{\ddot{m}(X_t,\beta)-\ddot{m}(X_t,\beta_0)}{\sigma(X_t,\tilde{\theta}_0)}] \\
&&  \times [\mathbb{I}(\varepsilon_s>\varepsilon_t) - \mathbb{I}(\varepsilon_s<\varepsilon_t)] \| X_{su}\| .
\end{eqnarray*}
It is easy to see that
$$ \frac{1}{C_n^4}\sum\limits_{i< j< k< l}h_{12}(Z_i,Z_j,Z_k,Z_l) = U_{12}((\hat{\beta}_n-\beta_0)\zeta+\beta_0).  $$
To show that $\frac{1}{C_n^4}\sum\limits_{i< j< k< l}h_{12}(Z_i,Z_j,Z_k,Z_l) =o_p(1) $, since $\sqrt{n}(\hat{\beta}_n-\beta_0)=O_p(1)$, it remains to show that $U_{12}(\beta) = o_p(1) $ uniformly in $\beta \in \mathcal{B}$. By Assumption 2, we have
$$ \| \frac{\ddot{m}(X_s,\beta) - \ddot{m}(X_s,\beta_0)}{\sigma(X_s, \tilde{\theta}_0)} -\frac{\ddot{m}(X_t,\beta)
-\ddot{m}(X_t,\beta_0)}{\sigma(X_t,\tilde{\theta}_0)} \| \leq C[\kappa_1(X_s) + \kappa_1(X_t)], $$
whence
\begin{eqnarray*}
&& \| h_{12}(\beta, Z_i,Z_j,Z_k,Z_l)\|
\leq C(\kappa_1(X_i) +\kappa_1(X_j) + \kappa_1(X_k) + \kappa_1(X_l))(\|X_i\|+\|X_j\|+\|X_k\|+\|X_l\|). \\
&& E[\|h_{12}(\beta, Z_i,Z_j,Z_k,Z_l)\|^2 ]
\leq CE \{ [\kappa^2_1(X) + \kappa^2_1(X)] E \|X\|^2 \}< \infty
\end{eqnarray*}
Note that $ \{U_{12}(\beta):\beta \in \mathcal{B} \}$ is a non-degenerate $U$-process of order 4.
Together with lemma 2.13 of Pakes and Pollard (1989), we obtain that $\{U_{12}(\beta):\beta \in \mathcal{B} \}$ is Euclidean for a squared-integrable envelope. Then it follows from the main corollary and corollary 4 of Sherman (1994) that $U_{12}( \beta )=o_p(1)$ uniformly in $\beta \in \mathcal{B}$. By Slutsky's theorem, we obtain that
$$nI_{12}= \frac{n}{2} (\hat{\beta}_n-\beta_0)^T  \frac{1}{C_n^4}\sum\limits_{i< j< k< l}h_{12}(Z_i,Z_j,Z_k,Z_l) (\hat{\beta}_n-\beta_0)=o_p(1).$$

For the term $I_{13}$, it is readily seen that $\frac{1}{C_n^4}\sum\limits_{i< j< k< l}h_{13}(Z_i,Z_j,Z_k,Z_l)$ is non-degenerate $U$-statistic of order $4$. By the independence between $\varepsilon$ and $X$, we have
$$E\{h_{13}(Z_i,Z_j,Z_k,Z_l)\}=E[(\frac{\ddot{m}(X_1,\beta_0)}{\sigma(X_1,\tilde{\theta}_0)}-\frac{\ddot{m}(X_2,\beta_0)}
{\sigma(X_2,\tilde{\theta}_0)})(\mathbb{I}(\varepsilon_1>\varepsilon_2)-\mathbb{I}(\varepsilon_1<\varepsilon_2))C_x(X_1,X_2)]=0,$$
where $C_x(X_1,X_2) = \|X_1-X_2\|-E(\|X_1-X_2\||X_1)-E(\|X_1-X_2\||X_2)+E(\|X_1-X_2\|)$.
By the law of large numbers for U-statistics, we obtain that
$$ \frac{1}{C_n^4}\sum\limits_{i< j< k< l}h_{13}(Z_i,Z_j,Z_k,Z_l) = o_p(1). $$
Combining this with $\sqrt{n} (\hat{\beta}_n-\beta_0)=O_p(1)$, we have
$$ nI_{13}=n(\hat{\beta}_n-\beta_0)^T \frac{1}{C_n^4}\sum\limits_{i< j< k< l}h_{13}(Z_i,Z_j,Z_k,Z_l) (\hat{\beta}_n-\beta_0)=o_p(1). $$
Similarly to the arguments for $I_{12}$ and $I_{13}$, we can show that
$$ nI_{15} = o_p(1), \quad nI_{16} = o_p(1), \quad {\rm and } \quad nI_{17} = o_p(1). $$
Consequently, we obtain that
\begin{eqnarray}\label{6.7}
  n \hat{U}_{n1}
= n (\hat{\beta}_n-\beta_0)^T \frac{1}{C_n^4} \sum\limits_{i< j< k< l} h_{11}(Z_i,Z_j,Z_k,Z_l) +
   n (\hat{\theta}_n-\tilde{\theta}_0)^T \frac{1}{C_n^4} \sum\limits_{i< j< k< l} h_{14}(Z_i,Z_j,Z_k,Z_l) + o_p(1).
\end{eqnarray}

Next we deal with the term $\hat{U}_{n2}$. Decompose it as
\begin{eqnarray*}
\hat{U}_{n2}
&=&  \frac{1}{C_n^4}\sum\limits_{i< j< k< l}h_{21}(Z_i,Z_j,Z_k,Z_l)
     +\frac{1}{C_n^4}\sum\limits_{i< j< k< l}h_{22}(Z_i,Z_j,Z_k,Z_l) \\
&=:& \hat{U}_{n21} + \hat{U}_{n22},
\end{eqnarray*}
where
\begin{eqnarray*}
h_{21}(Z_i,Z_j,Z_k,Z_l)&=& \frac{1}{6} \sum_{s< t,u< v}^{(i,j,k,l)}E(\delta_{2st}|X_s,X_t) (\|X_{st}\|+\|X_{uv}\|)
                           -\frac{1}{12}\sum_{(s,t,u)}^{(i,j,k,l)}E(\delta_{2st}|X_s,X_t) \|X_{su}\|,  \\
h_{22}(Z_i,Z_j,Z_k,Z_l)&=& \frac{1}{6} \sum_{s< t,u< v}^{(i,j,k,l)}[\delta_{2st}-E(\delta_{2st}|X_s,X_t)](\|X_{st}\|+\|X_{uv}\|)
                           -\frac{1}{12} \sum_{(s,t,u)}^{(i,j,k,l)}[\delta_{2st}-E(\delta_{2st}|X_s,X_t)]\|X_{su}\|.
\end{eqnarray*}

For the term $\hat{U}_{n21}$, similar to the arguments in Theorem 1 of Xu and Cao (2021), we can show that uniformly over $ 1 \leq s, t \leq n$,
\begin{eqnarray*}
&&   E(\delta_{2st}|X_s,X_t) \\
&=&  2E\{ \int_{0}^{\frac{m(X_s,\hat{\beta}_n)-m(X_s,{\beta}_0)}{\sigma(x_s,\tilde{\theta}_0)}-\frac{m(X_t,\hat{\beta}_n)
     -m(X_t,{\beta}_0)}{\sigma(X_t,\tilde{\theta}_0)}+\frac{\varepsilon_s({\sigma(X_s,\hat{\theta}_n)}
     -{\sigma(X_s,\tilde{\theta}_0)})}{\sigma(X_s,\tilde{\theta}_0)}-\frac{\varepsilon_t({\sigma(X_t,\hat{\theta}_n)}
     -{\sigma(X_t,\tilde{\theta}_0)})}{\sigma(X_t,\tilde{\theta}_0)}+R_s-R_t}\\
&&   [\mathbb{I}(\varepsilon_s-\varepsilon_t\leq z)-\mathbb{I}(\varepsilon_s\leq \varepsilon_t)]dz |X_s,X_t\}\\	
&=&  2 E[ \int_{0}^{\frac{m(X_s,\hat{\beta}_n)-m(X_s,{\beta}_0)}{\sigma(X_s,\tilde{\theta}_0)}-\frac{m(X_t,\hat{\beta}_n)
	 -m(X_t,{\beta}_0)}{\sigma(X_t,\tilde{\theta}_0)}} \mathbb{I}(\varepsilon_s - \varepsilon_t \leq  z)
     -\mathbb{I}(\varepsilon_s-\varepsilon_t\leq 0) dz |X_s,X_t]  \\
&&   + 2E[\int_{\frac{m(X_s,\hat{\beta}_n)-m(X_s,{\beta}_0)}{\sigma(X_s,\tilde{\theta}_0)}-\frac{m(X_t,\hat{\beta}_n)
	 -m(X_t,{\beta}_0)}{\sigma(X_t,\tilde{\theta}_0)}}^{\frac{m(X_s,\hat{\beta}_n)-m(X_s,{\beta}_0)}
	 {\sigma(X_s,\tilde{\theta}_0)}-\frac{m(X_t,\hat{\beta}_n)-m(X_t,{\beta}_0)}{\sigma(X_t,\tilde{\theta}_0)}
	 +\frac{\varepsilon_s({\sigma(X_s,\hat{\theta}_n)}-{\sigma(X_s,\tilde{\theta}_0)})}{\sigma(X_s,\tilde{\theta}_0)}
	 -\frac{\varepsilon_t({\sigma(X_t,\hat{\theta}_n)}-{\sigma(X_t,\tilde{\theta}_0)})}{\sigma(X_t,\tilde{\theta}_0)}}\\
&&   [\mathbb{I}(\varepsilon_s-\varepsilon_t \leq  z)-\mathbb{I}(\varepsilon_s-\varepsilon_t\leq 0)]dz|X_s,X_t] \\	
%&=&  2 E[ \int_{0}^{\frac{m(X_s,\hat{\beta}_n)-m(X_s,{\beta}_0)}{\sigma(X_s,\tilde{\theta}_0)}-\frac{m(X_t,\hat{\beta}_n)-m(X_t,{\beta}_0)}{\sigma(X_t,\tilde{\theta}_0)}+R_s-R_t} \mathbb{I}(\varepsilon_s - \varepsilon_t \leq  z) -\mathbb{I}(\varepsilon_s-\varepsilon_t\leq 0) dz |X_s,X_t]  \\
&&   + 2E[\int_{\frac{m(X_s,\hat{\beta}_n)-m(X_s,{\beta}_0)}{\sigma(X_s,\tilde{\theta}_0)}-\frac{m(X_t,\hat{\beta}_n)
     - m(X_t,{\beta}_0)}{\sigma(X_t,\tilde{\theta}_0)}+\frac{\varepsilon_s({\sigma(X_s,\hat{\theta}_n)}-
     {\sigma(X_s,\tilde{\theta}_0)})}{\sigma(X_s,\tilde{\theta}_0)} -\frac{\varepsilon_t({\sigma(X_t,\hat{\theta}_n)}
     -{\sigma(X_t,\tilde{\theta}_0)})}{\sigma(X_t,\tilde{\theta}_0)}}^{\frac{m(X_s,\hat{\beta}_n)-m(X_s,{\beta}_0)}
	 {\sigma(X_s,\tilde{\theta}_0)}-\frac{m(X_t,\hat{\beta}_n)-m(X_t,{\beta}_0)}{\sigma(X_t,\tilde{\theta}_0)}
	 +\frac{\varepsilon_s({\sigma(X_s,\hat{\theta}_n)}-{\sigma(X_s,\tilde{\theta}_0)})}{\sigma(X_s,\tilde{\theta}_0)}
	 -\frac{\varepsilon_t({\sigma(X_t,\hat{\theta}_n)}-{\sigma(X_t,\tilde{\theta}_0)})}{\sigma(X_t,\tilde{\theta}_0)}+R_s-R_t}\\
&&   [\mathbb{I}(\varepsilon_s-\varepsilon_t \leq  z)-\mathbb{I}(\varepsilon_s-\varepsilon_t\leq 0)]dz|X_s,X_t] \\	
&=&  Q_{\varepsilon} \left( [\frac{\dot{m}(X_s,{\beta}_0)}{\sigma(X_s,\tilde{\theta}_0)}-\frac{\dot{m}(X_t,{\beta}_0)}
     {\sigma(X_t,\tilde{\theta}_0)}]^T (\hat{\beta}_n-\beta_0) \right)^2
     + 2A_{\varepsilon}(\hat{\theta}_n-\tilde{\theta}_0)^T [\frac{\dot{\sigma}(X_s,\tilde{\theta}_0)}
     {\sigma(X_s,\tilde{\theta}_0)} +\frac{\dot{\sigma}(X_t,\tilde{\theta}_0)}{\sigma(X_t,\tilde{\theta}_0)}] \\
&&   + 2A_{\varepsilon} \left([\frac{\dot{\sigma}(X_s,\tilde{\theta}_0)}
{\sigma(X_s,\tilde{\theta}_0)}+\frac{\dot{\sigma}(X_t,\tilde{\theta}_0)}{\sigma(X_t,\tilde{\theta}_0)}]^T
(\hat{\theta}_n-\tilde{\theta}_0) \right) ^2 +o_p(\frac{1}{n}),
\end{eqnarray*}
where $Q_{\varepsilon} = E[f_{\varepsilon}(\varepsilon)]$ and $A_{\varepsilon}=E[\varepsilon F_{\varepsilon}(\varepsilon)]$.
Consequently, we obtain that
\begin{eqnarray*}
\hat{U}_{n21}
&=&   Q_{\varepsilon}(\hat{\beta}_n-\beta_0)^T \frac{1}{C_n^4}\sum\limits_{i< j< k< l}h_{211}(Z_i,Z_j,Z_k,Z_l)(\hat{\beta}_n-\beta_0) \\
&&    +2A_{\varepsilon} (\hat{\theta}_n-\tilde{\theta}_0)^T \frac{1}{C_n^4}\sum\limits_{i< j< k< l}h_{212}(Z_i,Z_j,Z_k,Z_l)  \\
&&    +2A_{\varepsilon}(\hat{\theta}_n-\tilde{\theta}_0)^T \frac{1}{C_n^4}\sum\limits_{i< j< k< l}h_{213}(Z_i,Z_j,Z_k,Z_l)(\hat{\theta}_n-\tilde{\theta}_0)  + o_p(\frac{1}{n}), \\
&=:&  Q_{\varepsilon} \hat{U}_{n211} + 2A_{\varepsilon} \hat{U}_{n212} + 2A_{\varepsilon} \hat{U}_{n213}  +o_p(\frac{1}{n}),
\end{eqnarray*}
where
\begin{eqnarray*}
&&  h_{211}(Z_i,Z_j,Z_k,Z_l) \\
&=& 6^{-1}\sum_{s< t,u< v}^{(i,j,k,l)}\{\frac{\dot{m}(X_s,{\beta}_0)}{\sigma(X_s,\tilde{\theta}_0)}
    -\frac{\dot{m}(X_t,{\beta}_0)}{\sigma(X_t,\tilde{\theta}_0)} \} \{\frac{\dot{m}(X_s,{\beta}_0)}{\sigma(X_s,\tilde{\theta}_0)}
    -\frac{\dot{m}(X_t,{\beta}_0)}{\sigma(X_t,\tilde{\theta}_0)}\}^T (\|X_{st}\|+\|X_{uv}\|)   \\
&&  -12^{-1}\sum_{(s,t,u)}^{(i,j,k,l)}\{\frac{\dot{m}(X_s,{\beta}_0)}{\sigma(X_s,\tilde{\theta}_0)}-\frac{\dot{m}(X_t,{\beta}_0)}
    {\sigma(X_t,\tilde{\theta}_0)}\} \{\frac{\dot{m}(X_s,{\beta}_0)}{\sigma(X_s,\tilde{\theta}_0)}-\frac{\dot{m}(X_t,{\beta}_0)}
    {\sigma(X_t,\tilde{\theta}_0)}\}^T \|X_{su}\|, \\
&&  h_{212}(Z_i,Z_j,Z_k,Z_l) \\
&=& 6^{-1}\sum_{s< t,u<v}^{(i,j,k,l)}\{\frac{\dot{\sigma}(X_s,\tilde{\theta}_0)}{\sigma(X_s,\tilde{\theta}_0)}
    + \frac{\dot{\sigma}(X_t,\tilde{\theta}_0)}{\sigma(X_t,\tilde{\theta}_0)} \}(\|X_{st}\|+\|X_{uv}\|)   \\
&&  - 12^{-1}\sum_{(s,t,u)}^{(i,j,k,l)}\{\frac{\dot{\sigma}(X_s,\tilde{\theta}_0)}{\sigma(X_s,\tilde{\theta}_0)}
    + \frac{\dot{\sigma}(X_t,\tilde{\theta}_0)}{\sigma(X_t,\tilde{\theta}_0)} \} \|X_{su}\|, \\
&&  h_{213}(Z_i,Z_j,Z_k,Z_l) \\
&=& 6^{-1}\sum_{s< t,u<v}^{(i,j,k,l)}\{\frac{\dot{\sigma}(X_s,\tilde{\theta}_0)}{\sigma(X_s,\tilde{\theta}_0)}
+ \frac{\dot{\sigma}(X_t,\tilde{\theta}_0)}{\sigma(X_t,\tilde{\theta}_0)} \}\{\frac{\dot{\sigma}(X_s,\tilde{\theta}_0)}{\sigma(X_s,\tilde{\theta}_0)}
+ \frac{\dot{\sigma}(X_t,\tilde{\theta}_0)}{\sigma(X_t,\tilde{\theta}_0)} \}^T(\|X_{st}\|+\|X_{uv}\|)   \\
&&  - 12^{-1}\sum_{(s,t,u)}^{(i,j,k,l)}\{\frac{\dot{\sigma}(X_s,\tilde{\theta}_0)}{\sigma(X_s,\tilde{\theta}_0)}
+ \frac{\dot{\sigma}(X_t,\tilde{\theta}_0)}{\sigma(X_t,\tilde{\theta}_0)} \}\{\frac{\dot{\sigma}(X_s,\tilde{\theta}_0)}{\sigma(X_s,\tilde{\theta}_0)}
+ \frac{\dot{\sigma}(X_t,\tilde{\theta}_0)}{\sigma(X_t,\tilde{\theta}_0)} \}^T \|X_{su}\|,
\end{eqnarray*}

For the term $\hat{U}_{n211}$, it is easy to see that $\frac{1}{C_n^4}\sum\limits_{i< j< k< l}h_{211}(Z_i,Z_j,Z_k,Z_l)$ is a non-degenerate $U$-statistic. By the law of large numbers, we have
\begin{eqnarray*}
\frac{1}{C_n^4}\sum\limits_{i< j< k< l}h_{211}(Z_i,Z_j,Z_k,Z_l) \longrightarrow M_1, \quad {\rm in \ probability,}
\end{eqnarray*}
where
\begin{equation*}
M_1 = E[\{\frac{\dot{m}(X_1,{\beta}_0)}{\sigma(X_1,\tilde{\theta}_0)}-\frac{\dot{m}(X_2,{\beta}_0)}
     {\sigma(X_2,\tilde{\theta}_0)}\} \{\frac{\dot{m}(X_1,{\beta}_0)}{\sigma(X_1,\tilde{\theta}_0)}-\frac{\dot{m}(X_2,{\beta}_0)}
     {\sigma(X_2,\tilde{\theta}_0)}\}^T C_x(X_1,X_2)],
\end{equation*}
where $C_x(X_1,X_2) = \|X_1-X_2\|-E(\|X_1-X_2\||X_1)-E(\|X_1-X_2\||X_2)+E(\|X_1-X_2\|)$.
Consequently, we obtain that
\begin{equation}\label{6.8}
n \hat{U}_{n211} = \sqrt{n}(\hat{\beta}_n-\beta_0)^T M_1 \sqrt{n}(\hat{\beta}_n-\beta_0)+o_p(1).
\end{equation}

For the term $\hat{U}_{n212}$, by the standard theory of $U$-statistics (see Subsection 5.3.1 in  Serfling
(2009),%Serfling 1984,
for instance), we have
\begin{eqnarray}\label{6.9}
  \nonumber n\hat{U}_{n212}   \nonumber
&=& \sqrt{n}(\hat{\theta}_n-\tilde{\theta}_0) \sqrt{n}\frac{1}{C_n^4}\sum\limits_{i< j< k< l} h_{212}(Z_i,Z_j,Z_k,Z_l)  \\  \nonumber
&=& \sqrt{n}(\hat{\theta}_n-\tilde{\theta}_0) \{ \frac{4}{\sqrt{n}}\sum_{i=1}^{n}E[h_{212}(Z_i,Z_j,Z_k,Z_l)|Z_i]+o_p(1) \} \\
&=& \sqrt{n}(\hat{\theta}_n-\tilde{\theta}_0) \frac{1}{\sqrt{n}}\sum_{i=1}^{n}4E[\{\frac{\dot{\sigma}(X_i,{\beta}_0)}
    {\sigma(X_i,\tilde{\theta}_0)} +\frac{\dot{\sigma}(X,{\beta}_0)}{\sigma(X,\tilde{\theta}_0)}\}C_x(X_i,X)|X_i]+o_p(1).
\end{eqnarray}

For the term $\hat{U}_{n213}$, similar to the arguments for $\hat{U}_{n211}$, we have
\begin{eqnarray*}
\frac{1}{C_n^4}\sum\limits_{i< j< k< l}h_{213}(Z_i,Z_j,Z_k,Z_l) \longrightarrow M_2, \quad {\rm in \ probability,}
\end{eqnarray*}
where
\begin{equation*}
M_2 = E[\{\frac{\dot{\sigma}(X_1,\tilde{\theta}_0)}{\sigma(X_1,\tilde{\theta}_0)}+
      \frac{\dot{\sigma}(X_2,\tilde{\theta}_0)}{\sigma(X_2,\tilde{\theta}_0)} \} \{\frac{\dot{\sigma}(X_1,\tilde{\theta}_0)}{\sigma(X_1,\tilde{\theta}_0)}
      + \frac{\dot{\sigma}(X_2,\tilde{\theta}_0)}{\sigma(X_2,\tilde{\theta}_0)} \}^T C_x(X_1,X_2)].
\end{equation*}
Consequently,
\begin{equation}\label{6.10}
n \hat{U}_{n213} =\sqrt{n}(\hat{\theta}_n-\tilde{\theta}_0)^T M_2 \sqrt{n}(\hat{\theta}_n-\tilde{\theta}_0)+o_p(1).
\end{equation}
Thus we obtain that
\begin{eqnarray*}
\begin{split}
n\hat{U}_{n21}
&=  Q_{\varepsilon} \sqrt{n}(\hat{\beta}_n-\beta_0)^T M_1 \sqrt{n}(\hat{\beta}_n-\beta_0) + 2A_{\varepsilon} \sqrt{n}
    (\hat{\theta}_n-\tilde{\theta}_0)^T  M_2 \sqrt{n}(\hat{\theta}_n-\tilde{\theta}_0)   \\
&   + 2A_{\varepsilon} \sqrt{n}(\hat{\theta}_n-\tilde{\theta}_0) \frac{1}{\sqrt{n}}\sum_{i=1}^{n}
    4E[\{\frac{\dot{\sigma}(X_i,{\beta}_0)} {\sigma(X_i,\tilde{\theta}_0)} +\frac{\dot{\sigma}(X,{\beta}_0)}{\sigma(X,\tilde{\theta}_0)}\}C_x(X_i,X)|X_i]+o_p(1).
\end{split}
\end{eqnarray*}

For the term $\hat{U}_{n22}$, following the same line as Theorem 1 inXu and Cao (2021), we can show that
$n \hat{U}_{n22} = o_p(\frac{1}{n}) $.
Altogether we obtain that
\begin{eqnarray}\label{6.11}
\begin{split}
n\hat{U}_n
&= n \hat{U}_{n0} + n (\hat{\beta}_n-\beta_0)^T \frac{1}{C_n^4} \sum\limits_{i< j< k< l} h_{11}(Z_i,Z_j,Z_k,Z_l) \\
&  + n (\hat{\theta}_n-\theta_0)^T \frac{1}{C_n^4} \sum\limits_{i< j< k< l} h_{14}(Z_i,Z_j,Z_k,Z_l)  \\
&  + 2A_{\varepsilon} \sqrt{n}(\hat{\theta}_n-\tilde{\theta}_0) \frac{1}{\sqrt{n}}\sum_{i=1}^{n} 4
   E[h_{212}(Z_i,Z_j,Z_k,Z_l)|Z_i] \\
&  + Q_{\varepsilon} \sqrt{n}(\hat{\beta}_n-\beta_0)^T M_1 \sqrt{n}(\hat{\beta}_n-\beta_0)
    + 2 A_{\varepsilon} \sqrt{n}(\hat{\theta}_n-\tilde{\theta}_0)^T M_2 \sqrt{n}(\hat{\theta}_n-\tilde{\theta}_0) +o_p(1),
\end{split}
\end{eqnarray}
where $E[h_{212}(Z_i,Z_j,Z_k,Z_l)|Z_i] = E\{[\frac{\dot{\sigma}(X_i,{\beta}_0)} {\sigma(X_i,\tilde{\theta}_0)} +\frac{\dot{\sigma}(X,{\beta}_0)}{\sigma(X,\tilde{\theta}_0)}]C_x(X_i,X)|X_i \}$.
For the term $\hat{U}_{n0}$, recall that
\begin{eqnarray*}
\hat{U}_{n0}
&=&  \frac{1}{C_n^4}\sum\limits_{i< j< k< l} \left( \frac{1}{6} \sum_{s< t,u< v}^{(i,j,k,l)} |\varepsilon_{st}|
     (\|X_{st}\|+\|X_{uv}\|) -\frac{1}{12} \sum_{(s,t,u)}^{(i,j,k,l)}|\varepsilon_{st}|\|X_{su}\| \right) \\
&=:& \frac{1}{C_n^4}\sum\limits_{i< j< k< l} h_0(Z_i,Z_j,Z_k,Z_l).
\end{eqnarray*}
It is easy to verify that $U_0$ is degenerate. By Hoeffding decomposition in the technical appendix of Yao et al. (2018), we can show that
$$ E[h_0(Z_i,Z_j,Z_k,Z_l)|Z_i,Z_j]=\frac{1}{6} C_{\varepsilon}(\varepsilon_i,\varepsilon_j)C_x(X_i,X_j),
$$
where
\begin{eqnarray*}
C_{\varepsilon}(\varepsilon_i,\varepsilon_j)&=& |\varepsilon_i-\varepsilon_j|-E(|\varepsilon_i-\varepsilon_j\|\varepsilon_i)-E(|\varepsilon_i-\varepsilon_j\|\varepsilon_j)
+E(|\varepsilon_i-\varepsilon_j|) \\
C_x(X_i,X_j) &=& \|X_i-X_j\|-E(\|X_i-X_j\| |X_i)-E(\|X_i-X_j\| |X_j)+E(\|X_i-X_j\|).
\end{eqnarray*}
It follows from the arguments in Section 5.3 of  Serfling (2009) that
\begin{eqnarray*}
n \hat{U}_{n0}
&=& \frac{6}{n-1}\sum_{i=1}^{n}\sum_{j\neq i}^{n}E\{h_0(Z_i,Z_j,Z_k,Z_l)|Z_i,Z_j \}+o_p(1) \\
&=& \frac{1}{n-1}\sum_{i=1}^{n}\sum_{j\neq i}^{n}C_{\varepsilon}(\varepsilon_i,\varepsilon_j)C_x(X_i,X_j) + o_p(1).
\end{eqnarray*}
For the second and third terms in (\ref{6.11}), set %(\ref{6.7}), set
\begin{eqnarray*}
\hat{U}_{n11} &=& \frac{1}{C_n^4} \sum\limits_{i< j< k< l} h_{11}(Z_i,Z_j,Z_k,Z_l)  \\
\hat{U}_{n14} &=& \frac{1}{C_n^4} \sum\limits_{i< j< k< l} h_{14}(Z_i,Z_j,Z_k,Z_l).
\end{eqnarray*}
By some elementary calculations, we can show that $\hat{U}_{n11}$ and $\hat{U}_{n14} $ are non-degenerate and
\begin{eqnarray*}
E[h_{11}(Z_1,Z_2,Z_3,Z_4)|Z_1]
&=&  -(F_{\varepsilon}(\varepsilon_1)-1/2)E[\{\frac{\dot{m}(X_1,{\beta}_0)}{\sigma(X_1,\tilde{\theta}_0)}
     -\frac{\dot{m}(X_2,{\beta}_0)}{\sigma(X_2,\tilde{\theta}_0)}\}C_x(X_1,X_2)|X_1] \\
E[h_{14}(Z_1,Z_2,Z_3,Z_4)|Z_1 ]
&=&  -\frac{1}{2}E\{[\frac{\varepsilon_1\dot{\sigma}(x_1,\tilde{\theta}_0)}{\sigma(X_1,\tilde{\theta}_0)}
     -\frac{\varepsilon_2\dot{\sigma}(X_2,\tilde{\theta}_0)}{\sigma(X_2,\tilde{\theta}_0)}][\mathbb{I}(\varepsilon_1>\varepsilon_2)
     -\mathbb{I}(\varepsilon_1<\varepsilon_2)] C_x(X_1,X_2)|Z_1 \},
\end{eqnarray*}
where $Z_i = (X_i, \varepsilon_i)$.
By the standard theory of $U$-statistics (see Section 5.3 in Serfling (2009), for instance), we have
\begin{eqnarray*}
\sqrt{n}U_{n11}
&=&  \frac{4}{\sqrt{n}}\sum_{i=1}^{n}E\{h_{11}(Z_i,Z_j,Z_k,Z_l)|Z_i\}+o_p(1)\\
&=&  \frac{1}{\sqrt{n}}\sum_{i=1}^{n}4[1/2-F_{\varepsilon}(\varepsilon_i)]E[\{\frac{\dot{m}(X_i,{\beta}_0)}
     {\sigma(X_i,\tilde{\theta}_0)} -\frac{\dot{m}(X,{\beta}_0)}{\sigma(X,\tilde{\theta}_0)}\}C_x(X_i,X)|X_i]+o_p(1) \\
\sqrt{n}U_{14}
&=&  \frac{4}{\sqrt{n}}\sum_{i=1}^{n}E\{h_{14}(Z_i,Z_j,Z_k,Z_l)|Z_i\}+o_p(1)\\
&=&  -\frac{1}{\sqrt{n}}\sum_{i=1}^{n}2E[\{\frac{\varepsilon_i\dot{\sigma}(X_i,\tilde{\theta}_0)}{\sigma(X_i,\tilde{\theta}_0)}
     -\frac{\varepsilon\dot{\sigma}(X,\tilde{\theta}_0)}{\sigma(X,\tilde{\theta}_0)}\}\{\mathbb{I}(\varepsilon_i>\varepsilon)
     -\mathbb{I}(\varepsilon_i<\varepsilon)\}C_x(X_i,X)|Z_i] + o_p(1).
\end{eqnarray*}
To obtain the limiting distribution of $n\hat{U}_n$, it remains to derive the asymptotic expansion of $\hat{\beta}_n - \beta_0$ and $\hat{\theta}_n - \tilde{\theta}_0$.  By Assumption 1 and Proposition 3 of Tan et al. (2022), we have
\begin{eqnarray*}
\sqrt{n}(\hat{\beta}_n-\beta_0)
&=&  \frac{1}{\sqrt{n}}\sum_{i=1}^{n}l(Y_i,X_i,\beta_0)+o_p(1) \\
\sqrt{n} (\hat{\theta}_n-\tilde{\theta}_0)
&=&  \frac{1}{\sqrt{n}} \sum_{i=1}^{n} [\sigma^2(X_i)\varepsilon_i^2 - \sigma^2(X_i, \tilde{\theta}_0)] \Sigma^{-1} \dot{\sigma}^2(X_i, \tilde{\theta}_0) + o_p(1).
\end{eqnarray*}
Altogether we obtain that
\begin{eqnarray*}
&&   n\hat{U}_n \\
&=&  \frac{6}{n-1}\sum_{i=1}^{n}\sum_{j\neq i}^{n}E\{h_0(Z_i,Z_j,Z_k,Z_l)|Z_i,Z_j\}   \\
&&   +4 \frac{1}{\sqrt{n}}\sum_{i=1}^{n}l(Y_i,X_i,\beta_0)^T  \frac{1}{\sqrt{n}}\sum_{i=1}^{n}
     E[h_{11}(Z_i,Z_j,Z_k,Z_l)|Z_i]  \\
&&   +4 \frac{1}{\sqrt{n}}\sum_{i=1}^{n}[\sigma^2(X_i,\tilde{\theta}_0)(\varepsilon_i^2-1)]\dot{\sigma}^2
     (X_i,\tilde{\theta}_0) ^T \Sigma^{-1} \frac{1}{\sqrt{n}}\sum_{i=1}^{n} E[h_{14}(Z_i,Z_j,Z_k,Z_l)|Z_i] \\
&&   +8 A_{\varepsilon} \frac{1}{\sqrt{n}}\sum_{i=1}^{n}[\sigma^2(X_i,\tilde{\theta}_0)(\varepsilon_i^2-1)]\dot{\sigma}^2
     (X_i,\tilde{\theta}_0)^T \Sigma^{-1} E[h_{212}(Z_i,Z_j,Z_k,Z_l)|Z_i] \\
&&   +2A_{\varepsilon} \frac{1}{\sqrt{n}}\sum_{i=1}^{n}[\sigma^2(X_i,\tilde{\theta}_0)(\varepsilon_i^2-1)]\dot{\sigma}^2
     (X_i,\tilde{\theta}_0)^T\Sigma^{-1} M_2 \Sigma^{-1} \frac{1}{\sqrt{n}}\sum_{i=1}^{n}[\sigma^2(X_i,\tilde{\theta}_0)
     (\varepsilon_i^2-1)]\dot{\sigma}^2(X_i,\tilde{\theta}_0) \\
&&   + Q_{\varepsilon} \frac{1}{\sqrt{n}}\sum_{i=1}^{n}l(Y_i,X_i,\beta_0)^T M_1
     \frac{1}{\sqrt{n}}\sum_{i=1}^{n}l(Y_i,X_i,\beta_0) + o_p(1).
\end{eqnarray*}
Since $E[h_0(Z_i,Z_j,Z_k,Z_l)]^2\leq CE\|X\|^2E(\varepsilon^2)<\infty$, it follows that
\begin{equation*}
n \hat{U}_n \longrightarrow  \sum_{k=1}^{\infty} \lambda_k (\mathcal{Z}_k^2-1) + 4\mathcal{N}^{\top} \mathcal{P}_1 + 4\mathcal{W}^{\top} \Sigma^{-1}\mathcal{P}_2 +8A_\varepsilon \mathcal{W}^{\top} \Sigma^{-1}\mathcal{P}_3+ 2A_\varepsilon \mathcal{W}^{\top} \Sigma^{-1}M_2\Sigma^{-1}\mathcal{W}+Q_\varepsilon\mathcal{N}^{\top} M_1 \mathcal{N},
\end{equation*}
where $\mathcal{Z}_1, \mathcal{Z}_2, \cdots$ are independent standard normal random variables, the eigenvalues $\{ \lambda_q \}_{q=1}^\infty$ are the solutions of the integral equation
$$ \int C_{\varepsilon}(\varepsilon_i,\varepsilon_j)C_x(X_i,X_j)\phi_q(Z_j)dF(Z_j)= \lambda_q\phi_q(Z_i), $$
$\{ \phi_i(\cdot) \}_{i=1}^\infty$ are orthonormal eigenfunctions and $F(\cdot)$ is the cumulative
distribution function of  of $Z$,
and $(\mathcal{Z}_i, \mathcal{N}, \mathcal{W}, \mathcal{P}_1, \mathcal{P}_2, \mathcal{P}_3) \in \mathbb{R}^{5p+1}$ are jointly Gaussian random variables with zero-mean and the covariance matrix satisfying
\begin{eqnarray*}
var(Z_i) &=& 1, \ var(\mathcal{N})   = var(l(Y_i,X_i,\beta_0))\\
var(\mathcal{P}_1) &=& var(E\{h_{11}(Z_i,Z_j,Z_k,Z_l)|Z_i\})\\
var(\mathcal{P}_2) &=& var(E\{h_{14}(Z_i,Z_j,Z_k,Z_l)|Z_i\})\\
var(\mathcal{P}_3) &=& var(E\{h_{212}(Z_i,Z_j,Z_k,Z_l)|Z_i\})\\
var(\mathcal{W})   &=& var([\sigma^2(X_i,{\theta}_0)(\varepsilon_i^2-1)]\dot{\sigma}^2(X_i,{\theta}_0))\\
cov(Z_i,\mathcal{P}_1) &=& cov(\phi_i(Z_i), E\{h_{11}(Z_i,Z_j,Z_k,Z_l)|Z_i\})\\
cov(Z_i,\mathcal{P}_2) &=& cov(\phi_i(Z_i), E\{h_{14}(Z_i,Z_j,Z_k,Z_l)|Z_i\})\\
cov(Z_i,\mathcal{P}_3) &=& cov(\phi_i(Z_i), E\{h_{212}(Z_i,Z_j,Z_k,Z_l)|Z_i\})\\
cov(Z_i,\mathcal{N})   &=& cov(\phi_i(Z_i),l(Y_i,X_i,\beta_0))\\
cov(Z_i,\mathcal{W})   &=& cov(\phi_i(Z_i),[\sigma^2(X_i,{\theta}_0)(\varepsilon_i^2-1)]\dot{\sigma}^2(X_i,{\theta}_0))\\
cov(\mathcal{P}_1,\mathcal{N}) &=& cov(l(Y_i,X_i,\beta_0),E\{h_{11}(Z_i,Z_j,Z_k,Z_l)|Z_i\}\\
cov(\mathcal{P}_2,\mathcal{N}) &=& cov(l(Y_i,X_i,\beta_0),E\{h_{14}(Z_i,Z_j,Z_k,Z_l)|Z_i\}\\
cov(\mathcal{P}_3,\mathcal{N}) &=& cov(l(Y_i,X_i,\beta_0),E\{h_{212}(Z_i,Z_j,Z_k,Z_l)|Z_i\}\\
cov(\mathcal{W},\mathcal{N})   &=& cov([\sigma^2(X_i,{\theta}_0)(\varepsilon_i^2-1)]\dot{\sigma}^2(X_i,{\theta}_0),
                                   l(Y_i,X_i,\beta_0))\\
cov(\mathcal{P}_1,\mathcal{P}_2) &=& cov(E\{h_{11}(Z_i,Z_j,Z_k,Z_l)|Z_i,E\{h_{14}(Z_i,Z_j,Z_k,Z_l)|Z_i\})\\
cov(\mathcal{P}_1,\mathcal{P}_3) &=& cov(E\{h_{11}(Z_i,Z_j,Z_k,Z_l)|Z_i,E\{h_{212}(Z_i,Z_j,Z_k,Z_l)|Z_i\})\\
cov(\mathcal{P}_3,\mathcal{P}_2) &=& cov(E\{h_{212}(Z_i,Z_j,Z_k,Z_l)|Z_i\},E\{h_{14}(Z_i,Z_j,Z_k,Z_l)|Z_i\})\\
cov(\mathcal{W},\mathcal{P}_1)   &=& cov([\sigma^2(X_i,{\theta}_0)(\varepsilon_i^2-1)]\dot{\sigma}^2(X_i,{\theta}_0),
                                     E\{h_{11}(Z_i,Z_j,Z_k,Z_l)|Z_i\}) \\
cov(\mathcal{W},\mathcal{P}_2)   &=& cov([\sigma^2(X_i,{\theta}_0)(\varepsilon_i^2-1)]\dot{\sigma}^2(X_i,{\theta}_0),
                                     E\{h_{14}(Z_i,Z_j,Z_k,Z_l)|Z_i\}) \\	
cov(\mathcal{W},\mathcal{P}_3)   &=& cov([\sigma^2(X_i,{\theta}_0)(\varepsilon_i^2-1)]\dot{\sigma}^2(X_i,{\theta}_0),
                                     E\{h_{212}(Z_i,Z_j,Z_k,Z_l)|Z_i\}).
\end{eqnarray*}
Hence we complete the proof of Theorem 3.1.   \hfill$\Box$

{\bf Proof of Theorem 3.2.} (1) Under the local alternatives $H_{1n}$, recall that
$ \sigma^2(X)=\sigma^2(X,{\theta}_0)+\frac{1}{\sqrt{n}}s(X)$ and
%$\eta_i=\frac{Y_i-m(X_i, \beta_0)}{\sigma(X_i, \tilde{\theta}_0)}$ and
$\hat{\eta}_i = \frac{Y_i - m(X_i,\hat{\beta}_n)}{\sigma(X_i, \hat{\theta}_n)} $ in nonlinear cases.
It follows from (\ref{6.2}) in the proof of Theorem 3.1 that
\begin{eqnarray}\label{6.12}
&&  | \hat{\eta}_i-\hat{\eta}_j| \nonumber \\
&=& |\varepsilon_i-\varepsilon_j| \nonumber \\
&&  -\{P_{ij}+ \frac{\varepsilon_i s(X_i) [\sigma(X_i,\hat{\theta}_n)- \sigma(X_i, \theta_0)]}{2 \sqrt{n}
    \sigma^3(X_i, \theta_0)}-\frac{\varepsilon_j s(X_j)[ \sigma(X_j,\hat{\theta}_n) - \sigma(X_j, \theta_0)]}
    {2\sqrt{n}\sigma^3(X_j, \theta_0)}\} [\mathbb{I}(\varepsilon_i>\varepsilon_j)-\mathbb{I}(\varepsilon_i <\varepsilon_j)] \nonumber \\
&&  +2\int_{0}^{P_{ij} + \frac{\varepsilon_i s(X_i)[{\sigma(X_i, \hat{\theta}_n)}-{\sigma(X_i, \theta_0)}]}
    {2\sqrt{n}\sigma^3(X_i, \theta_0)}-\frac{\varepsilon_j s(X_j)[{\sigma(X_j,\hat{\theta}_n)}-{\sigma(X_j, \theta_0)}]}
    {2\sqrt{n}\sigma^3(X_j, \theta_0)}} \{\mathbb{I}(\varepsilon_i-\varepsilon_j \leq z)-\mathbb{I}(\varepsilon_i \leq \varepsilon_j)\}dz,
\end{eqnarray}
where
\begin{eqnarray*}
P_{ij}
&=& \frac{m(X_i,\hat{\beta}_n)-m(X_i,{\beta}_0)}{\sigma(X_i,\theta_0)}-\frac{m(X_j,\hat{\beta}_n)-m(X_j,{\beta}_0)}
    {\sigma(X_j, \theta_0)}+\frac{\varepsilon_i({\sigma(X_i,\hat{\theta}_n)}-{\sigma(X_i, \theta_0)})}
    {\sigma(X_i, \theta_0)}  \\
&&  -\frac{\varepsilon_j({\sigma(X_j, \hat{\theta}_n)}-{\sigma(X_j, \theta_0)})}{\sigma(X_j, \theta_0)}+(R_i-R_j), \\
R_i &=&  \frac{\varepsilon_i[\sigma(X_i,\theta_0)-\sigma(X_i,\hat{\theta}_n)]^2}{\sigma(X_i,\tilde{\theta}_0)
	\sigma(X_i,\hat{\theta}_n)}+\frac{m(X_i,{\beta}_0)-m(X_i,\hat{\beta}_n)}{\sigma^2(X_i,\theta_0)}
[\sigma(X_i,\theta_0)-\sigma(X_i,\hat{\theta}_n)] \\
&&  +\frac{m(X_i,{\beta}_0)-m(X_i,\hat{\beta}_n)}{\sigma^2(X_i,\theta_0)}\frac{[\sigma(X_i,\theta_0)
	-\sigma(X_i,\hat{\theta}_n)]^2}{\sigma(X_i,\hat{\theta}_n)}.
\end{eqnarray*}	
Similar to the arguments for (\ref{6.5}) in the proof of Theorem 3.1, $ \hat{U}_n$ can be decomposed as
\begin{eqnarray}\label{6.13}
\hat{U}_n  \nonumber
&=&   \frac{1}{C_n^4}\sum\limits_{i< j< k< l} \left( \frac{1}{6} \sum_{s< t,u< v}^{(i,j,k,l)} |\varepsilon_{st}|
(\|X_{st}\|+\|X_{uv}\|) -\frac{1}{12} \sum_{(s,t,u)}^{(i,j,k,l)}|\varepsilon_{st}|\|X_{su}\| \right) \\  \nonumber
&&    + \frac{1}{C_n^4}\sum\limits_{i< j< k< l} \left( \frac{1}{6} \sum_{s< t,u< v}^{(i,j,k,l)}\delta_{1st}
(\|X_{st}\|+\|X_{uv}\|) - \frac{1}{12} \sum_{(s,t,u)}^{(i,j,k,l)}\delta_{1st}\|X_{su}\| \right) \\  \nonumber
&&    + \frac{1}{C_n^4}\sum\limits_{i< j< k< l} \left( \frac{1}{6} \sum_{s< t,u< v}^{(i,j,k,l)}\delta_{2st}
(\|X_{st}\|+\|X_{uv}\|) - \frac{1}{12} \sum_{(s,t,u)}^{(i,j,k,l)}\delta_{2st}\|X_{su}\| \right) \\ \nonumber
&&    + \frac{1}{C_n^4}\sum\limits_{i< j< k< l} \left( \frac{1}{6} \sum_{s< t,u< v}^{(i,j,k,l)}\delta_{3st}
(\|X_{st}\|+\|X_{uv}\|) - \frac{1}{12} \sum_{(s,t,u)}^{(i,j,k,l)}\delta_{3st}\|X_{su}\| \right)\\
&=:& \hat{U}_{n0}+\hat{U}_{n1}+\hat{U}_{n2}+\hat{U}_{n3},
\end{eqnarray}
where
\begin{eqnarray*}
	\delta_{1st}
	&=& -[\frac{\varepsilon_s({\sigma(X_s,\hat{\theta}_n)}-{\sigma(X_s,\theta_0)})}{\sigma(X_s,\theta_0)}
	-\frac{\varepsilon_t({\sigma(X_t,\hat{\theta}_n)}-{\sigma(X_t,\theta_0)})}{\sigma(X_t,\theta_0)}
	+\frac{m(X_s,\hat{\beta}_n)-m(X_s,{\beta}_0)}{\sigma(X_s,\theta_0)} \\
	&&  -\frac{m(X_t,\hat{\beta}_n)-m(X_t,{\beta}_0)}{\sigma(X_t,\theta_0)}+(R_s-R_t)]\{\mathbb{I}(\varepsilon_s>\varepsilon_t)
	-\mathbb{I}(\varepsilon_s<\varepsilon_t)\}  \\
\delta_{2st}
&=&  -\{\frac{\varepsilon_s s(X_s)({\sigma(X_s,\hat{\theta}_n)}-{\sigma(X_s, \theta_0)})}{2\sqrt{n} \sigma^3(X_s, \theta_0)}
     -\frac{\varepsilon_t s(X_t)({\sigma(X_t,\hat{\theta}_n)}-{\sigma(X_t, \theta_0)})}{2\sqrt{n} \sigma^3(X_t, \theta_0)} \} [\mathbb{I}(\varepsilon_s>\varepsilon_t)-\mathbb{I}(\varepsilon_s < \varepsilon_t)].   \\
\delta_{3st}
&=&  2 \int_{0}^{P_{st}+\frac{\varepsilon_s s(X_s)({\sigma(X_s,\hat{\theta}_n)}-{\sigma(X_s, \theta_0)})}
     {2\sqrt{n}\sigma^3(X_s, \theta_0)}-\frac{\varepsilon_t s(X_t)({\sigma(X_t, \hat{\theta}_n)}
     -{\sigma(X_t, \theta_0)})}{2\sqrt{n}\sigma^3(X_t, \theta_0)}} [\mathbb{I}(\varepsilon_s-\varepsilon_t \leq z)
     -\mathbb{I}(\varepsilon_s \leq \varepsilon_t) ] dz .\\
\end{eqnarray*}

First we deal with the term $\hat{U}_{n2}$. By Taylor expansion, $\hat{U}_{n2}$ can be decomposed as
\begin{eqnarray*}
 \hat{U}_{n2}
&=&  \frac{1}{\sqrt{n}} (\hat{\theta}_n- \theta_0)^T \frac{1}{C_n^4}\sum\limits_{i< j< k< l}h_{21}(Z_i,Z_j,Z_k,Z_l) \\
&&   + \frac{1}{2\sqrt{n}}(\hat{\theta}_n- \theta_0)^T \frac{1}{C_n^4}\sum\limits_{i< j< k< l}h_{22}(Z_i,Z_j,Z_k,Z_l)
     (\hat{\theta}_n- \theta_0) \\
&&   + \frac{1}{2\sqrt{n}}(\hat{\theta}_n- \theta_0)^T \frac{1}{C_n^4}\sum\limits_{i< j< k< l}h_{23}(Z_i,Z_j,Z_k,Z_l)
     (\hat{\theta}_n- \theta_0) \\
&=:& I_{21} + I_{22} + I_{23},
\end{eqnarray*}
where $Z_i=(\varepsilon_i, X_i)$,
\begin{eqnarray*}
h_{2m}(Z_i,Z_j,Z_k,Z_l )
&=&  -6^{-1}\sum_{s< t,u< v}^{(i,j,k,l)}\delta_{2mst}\{\mathbb{I}(\varepsilon_s>\varepsilon_t)
   	 -\mathbb{I}(\varepsilon_s<\varepsilon_t)\}(\|X_{st}\|+\|X_{uv}\|)\\
&&   +12^{-1}\sum_{(s,t,u)}^{(i,j,k,l)}\delta_{2mst}\{\mathbb{I}(\varepsilon_s>\varepsilon_t)
	 -\mathbb{I}(\varepsilon_s<\varepsilon_t)\}\|X_{su}\|,  \quad {\rm for} \ m=1, 2, 3
\end{eqnarray*}
and
\begin{eqnarray*}
\delta_{21st}
&=&  \frac{\varepsilon_s s(X_s) \dot{\sigma}(X_s, \theta_0)}{2\sigma^3(X_s, \theta_0)}-\frac{\varepsilon_t
     s(X_t) \dot{\sigma} (X_t, \theta_0)}{2\sigma^3(X_t, \theta_0)} \\
\delta_{22st}
&=&  \frac{\varepsilon_s s(X_s)\ddot{\sigma}(X_s,(\hat{\theta}_n- \theta_0)\zeta+\theta_0)}{2
     \sigma^3(X_s, \theta_0)} -
 \frac{\varepsilon_s s(X_s) \ddot{\sigma}(X_s, \theta_0)}{2 \sigma^3(X_s, \theta_0)}  \\
&&- \frac{\varepsilon_t s(X_t) \ddot{\sigma}(X_t,(\hat{\theta}_n-\theta_0)\zeta+\theta_0)}
     {2 \sigma^3(X_t, \theta_0)}
	 -\frac{\varepsilon_t s(X_t) \ddot{\sigma}(X_t, \theta_0)}{2\sigma^3(X_t, \theta_0)} \\
\delta_{23st}
&=&  \frac{\varepsilon_s s(X_s) \ddot{\sigma}(X_s, \theta_0)}{2 \sigma^3(X_s, \theta_0)}
	 -\frac{\varepsilon_t s(X_t) \ddot{\sigma}(X_t, \theta_0)}{2\sigma^3(X_t, \theta_0)} ,
\end{eqnarray*}
for some $\zeta \in (0,1)$. For the term $I_{21}$, we have
\begin{eqnarray*}
n I_{21} &=& \sqrt{n}(\hat{\theta}_n- \theta_0)^T  \frac{1}{C_n^4}\sum\limits_{i< j< k< l}h_{21}(Z_i,Z_j,Z_k,Z_l),
\end{eqnarray*}
Note that
\begin{eqnarray*}
&&  E[h_{21}(Z_i,Z_j,Z_k,Z_l)] \\
&=& -E[\{\frac{\varepsilon_1s(X_1)\dot{\sigma}(X_1,\theta_0)}{2\sigma^3(X_1,\theta_0)}-\frac{\varepsilon_2s(X_2)
    \dot{\sigma}(X_2,\theta_0)}{2\sigma^3(X_2,\theta_0)} \} \{\mathbb{I}(\varepsilon_1>\varepsilon_2)-\mathbb{I}(\varepsilon_1<\varepsilon_2)\}C_x(X_1,X_2)]\\
&=& -2E[\{\frac{\varepsilon_1s(X_1)\dot{\sigma}(X_1,\theta_0)}{2\sigma^3(X_1,\theta_0)}\}
    \{ \mathbb{I}(\varepsilon_1>\varepsilon_2)-\mathbb{I}(\varepsilon_1<\varepsilon_2)\} C_x(X_1,X_2)]\\
&=& -2E[\varepsilon_1\{\mathbb{I}(\varepsilon_1>\varepsilon_2)-\mathbb{I}(\varepsilon_1<\varepsilon_2)\}]
    E[\frac{s(X_1) \dot{\sigma}(X_1,\theta_0)}{2\sigma^3(X_1,\theta_0)}E\{ C_x(X_1,X_2)|X_1\}]=0.
\end{eqnarray*}
where $C_x(X_1,X_2) = \|X_1-X_2\|-E(\|X_1-X_2\||X_1)-E(\|X_1-X_2\||X_2)+E(\|X_1-X_2\|)$.
Since $\sqrt{n}(\hat{\theta}_n- \theta_0) = O_p(1)$, it follows that $n I_{21} = o_p(1)$.
For the terms $I_{22}$ and $I_{23}$, similar to the arguments for $I_{12}$ and $I_{13}$ in the proof of Theorem 3.1, we can show
$n I_{22} = o_p(1)$  and $ n I_{23} = o_p(1) $. Consequently, we obtain that
$$ n \hat{U}_{n2} = n I_{21} +n I_{22} +n I_{23} = o_p(1). $$

Next we deal with the term $\hat{U}_{n3}$, decomposed it as
\begin{eqnarray*}
\hat{U}_{n3}
&=&  \frac{1}{C_n^4}\sum\limits_{i< j< k< l}h_{31}(Z_i,Z_j,Z_k,Z_l)
	 +\frac{1}{C_n^4}\sum\limits_{i< j< k< l}h_{32}(Z_i,Z_j,Z_k,Z_l) \\
&=:& \hat{U}_{n31} + \hat{U}_{n32},
\end{eqnarray*}
where
\begin{eqnarray*}
h_{31}(Z_i,Z_j,Z_k,Z_l)
&=& \frac{1}{6} \sum_{s< t,u< v}^{(i,j,k,l)}E(\delta_{3st}|X_s,X_t) (\|X_{st}\|+\|X_{uv}\|)
	-\frac{1}{12}\sum_{(s,t,u)}^{(i,j,k,l)}E(\delta_{3st}|X_s,X_t) \|X_{su}\|,  \\
	h_{32}(Z_i,Z_j,Z_k,Z_l)&=& \frac{1}{6} \sum_{s< t,u< v}^{(i,j,k,l)}
    [\delta_{3st}-E(\delta_{3st}|X_s,X_t)](\|X_{st}\|+\|X_{uv}\|)
	-\frac{1}{12} \sum_{(s,t,u)}^{(i,j,k,l)}[\delta_{3st}-E(\delta_{3st}|X_s,X_t)]\|X_{su}\|.
\end{eqnarray*}
For the term $\hat{U}_{n32}$, similar to arguments for the term $\hat{U}_{n22}$ in the proof of Theorem 3.1, we have $ n \hat{U}_{n32} =o_p(1).$
For the term $\hat{U}_{n31} $, following the same line as Theorem 1 of Xu and Cao (2021), we can show that uniformly over $ 1 \leq s, t \leq n$,
\begin{eqnarray*}
&&   E(\delta_{3st}|X_s,X_t) \\
&=&  2E[ \int_{0}^{(P_{st}+\frac{\varepsilon_s s(X_s)
		({\sigma(X_s,\hat{\theta}_n)}-{\sigma(X_s,\theta_0)})}
     {2\sqrt{n}\sigma^3(X_s,\theta_0)}-\frac{\varepsilon_t s(X_t)
     	({\sigma(X_t,\hat{\theta}_n)}
     -{\sigma(X_t,\theta_0)})}{2\sqrt{n}\sigma^3(X_t,\theta_0)})+R_s-R_t}
      [\mathbb{I}(\varepsilon_s
     -\varepsilon_t\leq z)-\mathbb{I}(\varepsilon_s\leq \varepsilon_t)]dz |X_s,X_t]\\	
&=&  2 E[ \int_{0}^{\frac{m(X_s,\hat{\beta}_n)-m(X_s,{\beta}_0)}{\sigma(X_s,\theta_0)}
	-\frac{m(X_t,\hat{\beta}_n)- m(X_t,{\beta}_0)}{\sigma(X_t,\theta_0)}} \mathbb{I}(\varepsilon_s - \varepsilon_t \leq  z)
	 - \mathbb{I}
	 (\varepsilon_s-\varepsilon_t\leq 0) dz |X_s,X_t]  \\
&&   + 2E[\int_{\frac{m(X_s,\hat{\beta}_n)-m(X_s,{\beta}_0)}{\sigma(X_s,\theta_0)}
	 -\frac{m(X_t,\hat{\beta}_n)-m(X_t,{\beta}_0)}{\sigma(X_t,\theta_0)}}^{\frac{m(X_s,\hat{\beta}_n)-m(X_s,{\beta}_0)}
	 {\sigma(X_s,\theta_0)}-\frac{m(X_t,\hat{\beta}_n)-m(X_t,{\beta}_0)}{\sigma(X_t,\theta_0)}
	 + \frac{\varepsilon_s({\sigma(X_s,\hat{\theta}_n)}-{\sigma(X_s,\theta_0)})}
	 {\sigma(X_s,\theta_0)}- \frac{\varepsilon_t({\sigma(X_t,\hat{\theta}_n)}-{\sigma(X_t,\theta_0)})}{\sigma(X_t,\theta_0)}}\\
&&   [\mathbb{I}(\varepsilon_s-\varepsilon_t \leq  z)-\mathbb{I}(\varepsilon_s-\varepsilon_t\leq 0)]dz|X_s,X_t] \\	
&&   + 2E[\int_{\frac{m(X_s,\hat{\beta}_n)-m(X_s,{\beta}_0)}{\sigma(X_s,\theta_0)}
	-\frac{m(X_t,\hat{\beta}_n)-m(X_t,{\beta}_0)}{\sigma(X_t,\theta_0)}
	+\frac{\varepsilon_s({\sigma(X_s,\hat{\theta}_n)}-{\sigma(X_s,\theta_0)})}{\sigma(X_s,\theta_0)} -\frac{\varepsilon_t({\sigma(X_t,\hat{\theta}_n)}-{\sigma(X_t,\theta_0)})}{\sigma(X_t,\theta_0)}}
     ^{\frac{m(X_s,\hat{\beta}_n)-m(X_s,\beta_0)}
	 {\sigma(X_s,\theta_0)}-\frac{m(X_t,\hat{\beta}_n)-m(X_t,{\beta}_0)}{\sigma(X_t,\theta_0)}
	 +\frac{\varepsilon_s({\sigma(X_s,\hat{\theta}_n)}-{\sigma(X_s,\theta_0)})}{\sigma(X_s,\theta_0)}
	 -\frac{\varepsilon_t({\sigma(X_t,\hat{\theta}_n)}-{\sigma(X_t,\theta_0)})}{\sigma(X_t,\theta_0)}+R_s-R_t}\\
&&   [\mathbb{I}(\varepsilon_s-\varepsilon_t \leq  z)-\mathbb{I}(\varepsilon_s-\varepsilon_t\leq 0)]dz|X_s,X_t] \\	
&&   + 2E[\int_{P_{st}}^{(P_{st}+\frac{\varepsilon_s
		s(X_s)({\sigma(X_s,\hat{\theta}_n)}-{\sigma(X_s,\theta_0)})}
     {2\sqrt{n}\sigma^3(X_s,\theta_0)}-\frac{\varepsilon_t s(X_t)({\sigma(X_t,\hat{\theta}_n)}
     - {\sigma(X_t,\theta_0)})}{2\sqrt{n}\sigma^3(X_t,\theta_0)})}\{\mathbb{I}(\varepsilon_s-\varepsilon_t
     \leq z) - \mathbb{I}(\varepsilon_s \leq \varepsilon_t)\}dz|X_s,X_t]  \\
&=&  Q_{\varepsilon} \left( [\frac{\dot{m}(X_s,{\beta}_0)}{\sigma(X_s,\theta_0)}-\frac{\dot{m}(X_t,{\beta}_0)}
	 {\sigma(X_t,\theta_0)}]^T (\hat{\beta}_n-\beta_0) \right)^2
	 + 2A_{\varepsilon}(\hat{\theta}_n-\theta_0)^T [\frac{\dot{\sigma}(X_s,\theta_0)}
	 {\sigma(X_s,\theta_0)} +\frac{\dot{\sigma}(X_t,\theta_0)}{\sigma(X_t,\theta_0)}] \\
&&   + 2A_{\varepsilon} \left([\frac{\dot{\sigma}(X_s,\theta_0)}{\sigma(X_s,\theta_0)}
     +\frac{\dot{\sigma}(X_t,\theta_0)}{\sigma(X_t,\theta_0)}]^T
	 (\hat{\theta}_n-\theta_0) \right) ^2 + 2A_{\varepsilon}(\hat{\theta}_n-\theta_0)^T
     [\frac{s(X_s) \dot{\sigma} (X_s,\theta_0)}{2\sqrt{n}\sigma^3(X_s,{\theta}_0)}
	 + \frac{s(X_t) \dot{\sigma} (X_t,\theta_0)} {2\sqrt{n}\sigma^3(X_t,\theta_0)}] \\
&&   2A_{\varepsilon} \left([\frac{s(X_s)\dot{\sigma}(X_s,\theta_0)}
     {2\sqrt{n}\sigma^3(X_s,\theta_0)}+\frac{s(X_t)\dot{\sigma}(X_t,\theta_0)}{2\sqrt{n}
     \sigma^3(X_t,\theta_0)}]^T (\hat{\theta}_n-\theta_0) \right) ^2 + o_p(\frac{1}{n}),
\end{eqnarray*}
where $Q_{\varepsilon} = E[f_{\varepsilon}(\varepsilon)]$ and $A_{\varepsilon}=E[\varepsilon F_{\varepsilon}(\varepsilon)]$.
Consequently, we obtain that
\begin{eqnarray}\label{6.14}
\nonumber
\hat{U}_{n31}
&=&   Q_{\varepsilon}(\hat{\beta}_n-\beta_0)^T
\frac{1}{C_n^4}\sum\limits_{i< j< k< l}h_{311}(Z_i,Z_j,Z_k,Z_l)
     (\hat{\beta}_n-\beta_0) \\ \nonumber
&&   +2A_{\varepsilon} (\hat{\theta}_n-\theta_0)^T
\frac{1}{C_n^4}\sum\limits_{i< j< k< l}h_{312}(Z_i,Z_j,Z_k,Z_l)
     \\ \nonumber
&&   +2A_{\varepsilon}(\hat{\theta}_n-\theta_0)^T
\frac{1}{C_n^4}\sum\limits_{i< j< k< l}h_{313}(Z_i,Z_j,Z_k,Z_l)
     (\hat{\theta}_n-\theta_0)  \\ \nonumber
&&    +2A_{\varepsilon} (\hat{\theta}_n-\theta_0)^T
\frac{1}{\sqrt{n}}  \frac{1}{C_n^4}\sum\limits_{i< j< k< l}h_{314}(Z_i,Z_j,Z_k,Z_l)
     \\ \nonumber
&&   +2A_{\varepsilon}(\hat{\theta}_n-\theta_0)^T
      \frac{1}{n}  \frac{1}{C_n^4}\sum\limits_{i< j< k< l}h_{315}(Z_i,Z_j,Z_k,Z_l)
     (\hat{\theta}_n-\theta_0)  + o_p(\frac{1}{n}) \\
&=:& Q_{\varepsilon} \hat{U}_{n311} + 2A_{\varepsilon} \hat{U}_{n312} + 2A_{\varepsilon} \hat{U}_{n313} + 2A_{\varepsilon}
     \hat{U}_{n314} + 2A_{\varepsilon} \hat{U}_{n315}  +o_p(\frac{1}{n}),
\end{eqnarray}
where
\begin{eqnarray*}
&&  h_{311}(Z_i,Z_j,Z_k,Z_l) \\
&=& 6^{-1}\sum_{s< t,u< v}^{(i,j,k,l)}\{\frac{\dot{m}(X_s,{\beta}_0)}{\sigma(X_s,\theta_0)}
-\frac{\dot{m}(X_t,{\beta}_0)}{\sigma(X_t,\theta_0)} \} \{\frac{\dot{m}(X_s,{\beta}_0)}{\sigma(X_s,\theta_0)}
-\frac{\dot{m}(X_t,{\beta}_0)}{\sigma(X_t,\theta_0)}\}^T (\|X_{st}\|+\|X_{uv}\|)   \\
&&  -12^{-1}\sum_{(s,t,u)}^{(i,j,k,l)}\{\frac{\dot{m}(X_s,{\beta}_0)}{\sigma(X_s,\theta_0)}-\frac{\dot{m}(X_t,{\beta}_0)}
   {\sigma(X_t,\theta_0)}\} \{\frac{\dot{m}(X_s,{\beta}_0)}{\sigma(X_s,\theta_0)}-\frac{\dot{m}(X_t,{\beta}_0)}
    {\sigma(X_t,\theta_0)}\}^T \|X_{su}\|, \\
&&  h_{312}(Z_i,Z_j,Z_k,Z_l) \\
&=& 6^{-1}\sum_{s< t,u<v}^{(i,j,k,l)}\{\frac{\dot{\sigma}(X_s,\theta_0)}{\sigma(X_s,\theta_0)}
    + \frac{\dot{\sigma}(X_t,\theta_0)}{\sigma(X_t,\theta_0)} \}(\|X_{st}\|+\|X_{uv}\|)
    - 12^{-1}\sum_{(s,t,u)}^{(i,j,k,l)}\{\frac{\dot{\sigma}(X_s,\theta_0)}{\sigma(X_s,\theta_0)}
    + \frac{\dot{\sigma}(X_t,\theta_0)}{\sigma(X_t,\theta_0)} \} \|X_{su}\|, \\
&&  h_{213}(Z_i,Z_j,Z_k,Z_l) \\
&=& 6^{-1}\sum_{s< t,u<v}^{(i,j,k,l)}\{\frac{\dot{\sigma}(X_s,\theta_0)}{\sigma(X_s,\theta_0)}
    + \frac{\dot{\sigma}(X_t,\theta_0)}{\sigma(X_t,\theta_0)} \}\{\frac{\dot{\sigma}(X_s,\theta_0)}{\sigma(X_s,\theta_0)}
    + \frac{\dot{\sigma}(X_t,\theta_0)}{\sigma(X_t,\theta_0)} \}^T(\|X_{st}\|+\|X_{uv}\|)   \\
&&  - 12^{-1}\sum_{(s,t,u)}^{(i,j,k,l)}\{\frac{\dot{\sigma}(X_s,\theta_0)}{\sigma(X_s,\theta_0)}
    + \frac{\dot{\sigma}(X_t,\theta_0)}{\sigma(X_t,\theta_0)} \}\{\frac{\dot{\sigma}(X_s,\theta_0)}{\sigma(X_s,\theta_0)}
    + \frac{\dot{\sigma}(X_t,\theta_0)}{\sigma(X_t,\theta_0)} \}^T \|X_{su}\|,\\
&&  h_{314}(Z_i,Z_j,Z_k,Z_l) \\
&=& 6^{-1}\sum_{s< t,u<v}^{(i,j,k,l)}\{\frac{s(X_s)\dot{\sigma}(X_s,\theta_0)}{2\sigma^3(X_s,\theta_0)}
	+ \frac{s(X_t)\dot{\sigma}(X_t,\theta_0)}{2 \sigma^3(X_t,\theta_0)} \}(\|X_{st}\|+\|X_{uv}\|)   \\
&&  - 12^{-1}\sum_{(s,t,u)}^{(i,j,k,l)}\{\frac{s(X_s)\dot{\sigma}(X_s,\theta_0)}{2 \sigma^3(X_s,\theta_0)}
	+ \frac{s(X_t)\dot{\sigma}(X_t,\theta_0)}{2 \sigma^3(X_t,\theta_0)} \} \|X_{su}\|, \\
&&  h_{315}(Z_i,Z_j,Z_k,Z_l) \\
&=& 6^{-1}\sum_{s< t,u<v}^{(i,j,k,l)}\{\frac{s(X_s)\dot{\sigma}(X_s,\theta_0)}{2\sigma^3(X_s,\theta_0)}
    + \frac{s(X_t)\dot{\sigma}(X_t,\theta_0)}{2\sigma^3(X_t,\theta_0)} \}
    \{\frac{s(X_s)\dot{\sigma}(X_s,\theta_0)}{2\sigma^3(X_s,\theta_0)}
    + \frac{s(X_t)\dot{\sigma}(X_t,\theta_0)}{2\sigma^3(X_t,\theta_0)} \}^T(\|X_{st}\|+\|X_{uv}\|)   \\
&&  - 12^{-1}\sum_{(s,t,u)}^{(i,j,k,l)}\{\frac{s(X_s)\dot{\sigma}(X_s,\theta_0)}{2\sigma^3(X_s,\theta_0)}
+ \frac{s(X_t)\dot{\sigma}(X_t,\theta_0)}{2\sigma^3(X_t,\theta_0)} \}
\{\frac{s(X_s)\dot{\sigma}(X_s,\theta_0)}{2\sigma^3(X_s,\theta_0)}
+ \frac{s(X_t)\dot{\sigma}(X_t,\theta_0)}{2\sigma^3(X_t,\theta_0)} \}^T \|X_{su}\|.
\end{eqnarray*}
For the terms $\hat{U}_{n311}$, $\hat{U}_{n312}$ and $\hat{U}_{n313}$, similar to the arguments for (\ref{6.8}), (\ref{6.9}) and (\ref{6.10}) in the proof of Theorem 3.1, we have
\begin{eqnarray*}
n \hat{U}_{n311} &=& \sqrt{n}(\hat{\beta}_n-\beta_0)^T M_1 \sqrt{n}(\hat{\beta}_n-\beta_0)+o_p(1).\\
n \hat{U}_{n312} &=& \sqrt{n}(\hat{\theta}_n-\tilde{\theta}_0) \frac{1}{\sqrt{n}}\sum_{i=1}^{n}4
                     E[h_{312}(Z_i,Z_j,Z_k,Z_l)|Z_i] +o_p(1).\\
n \hat{U}_{n313} &=& \sqrt{n}(\hat{\theta}_n-\tilde{\theta}_0)^T M_2 \sqrt{n}(\hat{\theta}_n-\theta_0)+o_p(1),
\end{eqnarray*}
where $E[h_{312}(Z_i,Z_j,Z_k,Z_l)|Z_i] = E[\{\frac{\dot{\sigma}(X_i,{\beta}_0)} {\sigma(X_i,\theta_0)} +\frac{\dot{\sigma}(X,{\beta}_0)}{\sigma(X,\theta_0)}\}C_x(X_i,X)|X_i]$, $M_1$ and $M_2$ are given in Theorem 3.1.

For the term $\hat{U}_{n314}$, recall that
\begin{eqnarray*}
n\hat{U}_{n314}
&=& \sqrt{n}(\hat{\theta}_n-\theta_0)^T \frac{1}{C_n^4}\sum\limits_{i< j< k< l} h_{314}(Z_i,Z_j,Z_k,Z_l) .
\end{eqnarray*}
By some elementary calculations, we have
$$ E[h_{314}(Z_i,Z_j,Z_k,Z_l)] = E[\{\frac{s(X_1){\dot{\sigma}(X_1,\theta_0)}}{\sigma^3(X_1,\theta_0)}+\frac{s(X_2){\dot{\sigma}
(X_2,\theta_0)}}{\sigma^3(X_2,\theta_0)}\}^TC_x(X_1,X_2)] =0 $$
with $C_x(X_1,X_2) = \|X_1-X_2\|-E(\|X_1-X_2\||X_1)-E(\|X_1-X_2\||X_2)+E(\|X_1-X_2\|)$.
Consequently, we obtain that $n \hat{U}_{n314} = o_p(1) $.
Similarly, we have $ n \hat{U}_{n315} = o_p(1) $. Hence we obtain
\begin{eqnarray*}
n \hat{U}_{n3}
&=& Q_{\varepsilon} \sqrt{n}(\hat{\beta}_n-\beta_0)^T M_1 \sqrt{n}(\hat{\beta}_n-\beta_0)  + 2A_{\varepsilon}
    \sqrt{n}(\hat{\theta}_n-\theta_0)^T M_2 \sqrt{n}(\hat{\theta}_n-\theta_0) \\
&&  + 2A_{\varepsilon} \sqrt{n}(\hat{\theta}_n-\theta_0) \frac{1}{\sqrt{n}}\sum_{i=1}^{n}4
    E[h_{312}(Z_i,Z_j,Z_k,Z_l)|Z_i] + o_p(1).
\end{eqnarray*}

According to the arguments in Theorem 3.1, we have
\begin{eqnarray*}
n \hat{U}_{n0}
&=& \frac{6}{n-1}\sum_{i=1}^{n}\sum_{j\neq i}^{n}E\{h_0(Z_i,Z_j,Z_k,Z_l)|Z_i,Z_j \}+o_p(1) \\
n \hat{U}_{n1}
&=& \sqrt{n}(\hat{\beta}_n-\beta_0)^T \frac{4}{\sqrt{n}}\sum_{i=1}^{n}E\{h_{11}(Z_i,Z_j,Z_k,Z_l)|Z_i\}+o_p(1)\\
&&  +\sqrt{n}(\hat{\theta}_n-\theta_0)^T \frac{4}{\sqrt{n}}\sum_{i=1}^{n}E\{h_{14}(Z_i,Z_j,Z_k,Z_l)|Z_i\} + o_p(1),
\end{eqnarray*}
with $E\{h_0(Z_1,Z_2,Z_3,Z_4)|Z_1,Z_2 \} =6^{-1} C_{\varepsilon}(\varepsilon_1,\varepsilon_2)C_x(X_1,X_2) $ and
\begin{eqnarray*}
E[h_{11}(Z_1,Z_2,Z_3,Z_4)|Z_1]
&=&  -(F_{\varepsilon}(\varepsilon_1)-1/2)E[\{\frac{\dot{m}(X_1,{\beta}_0)}{\sigma(X_1, \theta_0)}
     -\frac{\dot{m}(X_2,{\beta}_0)}{\sigma(X_2, \theta_0)}\}C_x(X_1,X_2)|X_1] \\
E[h_{14}(Z_1,Z_2,Z_3,Z_4)|Z_1 ]
&=&  -\frac{1}{2}E\{[\frac{\varepsilon_1\dot{\sigma}(x_1, \theta_0)}{\sigma(X_1, \theta_0)}
     -\frac{\varepsilon_2\dot{\sigma}(X_2, \theta_0)}{\sigma(X_2, \theta_0)}][\mathbb{I}(\varepsilon_1>\varepsilon_2)
     -\mathbb{I}(\varepsilon_1<\varepsilon_2)] C_x(X_1,X_2)|Z_1 \}.
\end{eqnarray*}
Altogether we obtain that
\begin{eqnarray*}
n\hat{U}_n
&=& \frac{6}{n-1}\sum_{i=1}^{n}\sum_{j\neq i}^{n}E\{h_0(Z_i,Z_j,Z_k,Z_l)|Z_i,Z_j \} \\
&&  \sqrt{n}(\hat{\beta}_n-\beta_0)^T \frac{4}{\sqrt{n}}\sum_{i=1}^{n}E\{h_{11}(Z_i,Z_j,Z_k,Z_l)|Z_i\}+o_p(1)\\
&&  +\sqrt{n}(\hat{\theta}_n-\theta_0)^T \frac{4}{\sqrt{n}}\sum_{i=1}^{n}E\{h_{14}(Z_i,Z_j,Z_k,Z_l)|Z_i\}   \\
&&  Q_{\varepsilon} \sqrt{n}(\hat{\beta}_n-\beta_0)^T M_1 \sqrt{n}(\hat{\beta}_n-\beta_0)  + 2A_{\varepsilon}
    \sqrt{n}(\hat{\theta}_n-\tilde{\theta}_0)^T M_2 \sqrt{n}(\hat{\theta}_n-\theta_0) \\
&&  + 2A_{\varepsilon} \sqrt{n}(\hat{\theta}_n-\theta_0) \frac{1}{\sqrt{n}}\sum_{i=1}^{n}4
    E[h_{312}(Z_i,Z_j,Z_k,Z_l)|Z_i] + o_p(1).
\end{eqnarray*}
By Assumption 1 and Proposition 4 of Tan et al. (2022), we have
\begin{eqnarray*}
\sqrt{n}(\hat{\beta}_n-\beta_0) &=&  \frac{1}{\sqrt{n}}\sum_{i=1}^{n}l(Y_i,X_i,\beta_0)+o_p(1) \\
\sqrt{n} (\hat{\theta}_n-\theta_0)
&=&  \frac{1}{\sqrt{n}} \sum_{i=1}^{n} [\sigma^2(X_i, \theta_0) (\varepsilon_i^2 - 1)] \Sigma_{\sigma}^{-1}
     \dot{\sigma}^2(X_i, \theta_0) + \Sigma_{\sigma}^{-1} E[s(X)\dot{\sigma}^2(X_i,{\theta}_0)] +o_p(1).
\end{eqnarray*}
Consequently,
\begin{eqnarray*}
&&   n\hat{U}_n \\
&=&  \frac{6}{n-1}\sum_{i=1}^{n}\sum_{j\neq i}^{n}E\{h_0(Z_i,Z_j,Z_k,Z_l)|Z_i,Z_j\}   \\
&&   +4 \frac{1}{\sqrt{n}}\sum_{i=1}^{n}l(Y_i,X_i,\beta_0)^T  \frac{1}{\sqrt{n}}\sum_{i=1}^{n}
     E[h_{11}(Z_i,Z_j,Z_k,Z_l)|Z_i]  \\
&&   +4 \frac{1}{\sqrt{n}}\sum_{i=1}^{n}[\sigma^2(X_i,\theta_0)(\varepsilon_i^2-1)]\dot{\sigma}^2
	(X_i,\theta_0) ^T \Sigma_{\sigma}^{-1} \frac{1}{\sqrt{n}}\sum_{i=1}^{n} E[h_{14}(Z_i,Z_j,Z_k,Z_l)|Z_i] \\
&&  +8 A_{\varepsilon} \frac{1}{\sqrt{n}}\sum_{i=1}^{n}[\sigma^2(X_i,\theta_0)(\varepsilon_i^2-1)]\dot{\sigma}^2
    (X_i,\theta_0)^T \Sigma_{\sigma}^{-1} E[h_{312}(Z_i,Z_j,Z_k,Z_l)|Z_i] \\
&&  +8A_\varepsilon E[s(X)\dot{\sigma}^2(X_i,\theta_0)]^T \Sigma_{\sigma}^{-1}
    \frac{1}{\sqrt{n}}\sum_{i=1}^{n}E[h_{312}(Z_i,Z_j,Z_k,Z_l|Z_i)] \\
&&  +4 E[s(X)\dot{\sigma}^2(X_i,\theta_0)]^T \Sigma_{\sigma}^{-1} \frac{1}{\sqrt{n}}\sum_{i=1}^{n}
    E[h_{14}(Z_i,Z_j,Z_k,Z_l|Z_i)] \\
&&  +2A_{\varepsilon} \frac{1}{\sqrt{n}}\sum_{i=1}^{n}[\sigma^2(X_i,\theta_0)(\varepsilon_i^2-1)]\dot{\sigma}^2
    (X_i,\theta_0)^T \Sigma_{\sigma}^{-1} M_2 \Sigma_{\sigma}^{-1} \frac{1}{\sqrt{n}}\sum_{i=1}^{n}[\sigma^2(X_i,\theta_0)
	(\varepsilon_i^2-1)]\dot{\sigma}^2(X_i,\theta_0) \\
	&&   +4A_{\varepsilon} \frac{1}{\sqrt{n}}\sum_{i=1}^{n}[\sigma^2(X_i,\theta_0)(\varepsilon_i^2-1)]
	\dot{\sigma}^2(X_i,\theta_0)^T\Sigma_{\sigma} ^{-1} M_2 \Sigma_{\sigma} ^{-1} E[s(X)\dot{\sigma}^2(X_i,{\theta}_0)]
	\\
	&&   +2A_{\varepsilon} E[s(X)\dot{\sigma}^2(X_i,{\theta}_0)]^T \Sigma_{\sigma} ^{-1}
	M_2 \Sigma_{\sigma} ^{-1} E[s(X)\dot{\sigma}^2(X_i,{\theta}_0)]\\
	&&  + Q_{\varepsilon} \frac{1}{\sqrt{n}}\sum_{i=1}^{n}l(Y_i,X_i,\beta_0)^T M_1 \frac{1}{\sqrt{n}}\sum_{i=1}^{n}l(Y_i,X_i,\beta_0)
    + o_p(1).
\end{eqnarray*}
Since $E[h_0(Z_i,Z_j,Z_k,Z_l)]^2\leq CE\|X\|^2E(\varepsilon^2)<\infty$, it follows that
\begin{eqnarray*}
n \hat{U}_n
&\longrightarrow&  \sum_{k=1}^{\infty} \lambda_k (\mathcal{Z}_k^2-1) + 4\mathcal{N}^{\top} \mathcal{P}_1 + 4\mathcal{W}^{\top}
    \Sigma_{\sigma}^{-1} \mathcal{P}_2 +8A_\varepsilon \mathcal{W}^{\top} \Sigma_{\sigma}^{-1}   \mathcal{P}_3+ 2A_\varepsilon \mathcal{W}^{\top} \Sigma_{\sigma}^{-1}  M_2 \Sigma_{\sigma}^{-1} \mathcal{W} \\
&&  +Q_\varepsilon\mathcal{N}^{\top} M_1 \mathcal{N} + 4E[s(X)\dot{\sigma}^2 (X_i,{\theta}_0)]^{\top} \Sigma_{\sigma}^{-1}
    \mathcal{P}_2 + 8A_\varepsilon E[s(X) \dot{\sigma}^2(X_i,{\theta}_0)]^{\top} \Sigma_{\sigma}^{-1} \mathcal{P}_3\\
&&   +2A_{\varepsilon} E[s(X)\dot{\sigma}^2(X_i,{\theta}_0)]^T \Sigma_{\sigma} ^{-1}
M_2 \Sigma_{\sigma} ^{-1} E[s(X)\dot{\sigma}^2(X_i,{\theta}_0)]  \\
&&   +4A_{\varepsilon} \mathcal{W}^T  M_2 \Sigma_{\sigma} ^{-1} E[s(X)\dot{\sigma}^2(X_i,{\theta}_0)]
\end{eqnarray*}
where $\Sigma_{\sigma} = E[\dot{\sigma}^2 (X_i,{\theta}_0)\dot{\sigma}^2 (X_i,{\theta}_0)^T]$, $A_\varepsilon$, $Q_\varepsilon$, $\lambda_i, \mathcal{Z}_i, \mathcal{N}, \mathcal{W}, \mathcal{P}_1, \mathcal{P}_2, \mathcal{P}_3 $, $M_1$ and $M_2$ are defined in Theorem 3.1. Hence we complete the proof of the first part of Theorem 3.2.

(2) Recall that $\eta_i=\frac{Y_i-m(X_i, \beta_0)}{\sigma(X_i, \tilde{\theta}_0)}$ and $\hat{\eta}_i = \frac{Y_i - m(X_i,\hat{\beta}_n)}{\sigma(X_i, \hat{\theta}_n)} $ in nonlinear cases. Under the global alternative $H_1$, we have $\eta_i=\frac{\varepsilon_i\sigma(X_i)}{\sigma(X_i,\tilde{\theta}_0)}$.
It follows from (\ref{6.1})  and (\ref{6.2}) in the proof of Theorem 3.1 that
\begin{eqnarray}\label{6.15}
&&  |\hat{\eta}_i-\hat{\eta}_j|  \nonumber \\
&&  =|\eta_i-\eta_j|-[\frac{m(X_i,\hat{\beta}_n)-m(X_i,\beta_0)}{\sigma(X_i,\tilde{\theta}_0)}-\frac{m(X_j,\hat{\beta}_n)
    -m(X_j,\beta_0)}{\sigma(X_j,\tilde{\theta}_0)}+\frac{\varepsilon_i\sigma(X_i)({\sigma(X_i,\hat{\theta}_n)}
    -{\sigma(X_i,\tilde{\theta}_0)})}{\sigma^2(X_i,\tilde{\theta}_0)} \nonumber \\
&&  -\frac{\varepsilon_j\sigma(X_j)({\sigma(X_j,\hat{\theta}_n)}-{\sigma(X_j,\tilde{\theta}_0)})}{\sigma^2(X_j,\tilde{\theta}_0)}
    +(R_i-R_j)] [\mathbb{I}(\eta_i>\eta_j)-\mathbb{I}(\eta_i<\eta_j)] \nonumber \\
&&  +2\int_{0}^{\frac{m(X_i,\hat{\beta}_n)-m(X_i,{\beta}_0)}{\sigma(X_i,\tilde{\theta}_0)}-\frac{m(X_i,\hat{\beta}_n)
    -m(X_j,{\beta}_0)}{\sigma(X_j,\tilde{\theta}_0)}+\frac{\varepsilon_i\sigma(X_i)({\sigma(X_i,\hat{\theta}_n)}	-{\sigma(X_i,\tilde{\theta}_0)})}{\sigma^2(X_i,\tilde{\theta}_0)}-\frac{\varepsilon_j\sigma(X_j)({\sigma(X_j,\hat{\theta}_n)}
    -{\sigma(X_j,{\theta}_0)})}{\sigma^2(X_j,\tilde{\theta}_0)}+R_i-R_j} \nonumber \\
&&  [\mathbb{I}(\eta_i-\eta_j \leq z)-\mathbb{I}(\eta_i \leq \eta_j)]dz.
\end{eqnarray}
where
\begin{equation*}
\begin{split}
R_i
=&  \frac{\varepsilon_i[\sigma(X_i,\tilde{\theta}_0)-\sigma(X_i,\hat{\theta}_n)]^2}{\sigma(X_i,\tilde{\theta}_0)
	\sigma(X_i,\hat{\theta}_n)}+\frac{m(X_i,{\beta}_0)-m(X_i,\hat{\beta}_n)}{\sigma^2(X_i,\tilde{\theta}_0)}
[\sigma(X_i,\tilde{\theta}_0)-\sigma(X_i,\hat{\theta}_n)] \\
&  +\frac{m(X_i,{\beta}_0)-m(X_i,\hat{\beta}_n)}{\sigma^2(X_i,\tilde{\theta}_0)}\frac{[\sigma(X_i,\tilde{\theta}_0)
	-\sigma(X_i,\hat{\theta}_n)]^2}{\sigma(X_i,\hat{\theta}_n)}.
\end{split}
\end{equation*}
By the analog to (6.5) in the proof of Theorem 3.1, $\hat{U}_n$ can be decomposed into three parts:
\begin{eqnarray}\label{6.16}
\hat{U}_n  \nonumber
&=&   \frac{1}{C_n^4}\sum\limits_{i< j< k< l} \left( \frac{1}{6} \sum_{s< t,u< v}^{(i,j,k,l)} |\eta_{st}|
(\|X_{st}\|+\|X_{uv}\|) -\frac{1}{12} \sum_{(s,t,u)}^{(i,j,k,l)}|\eta_{st}|\|X_{su}\| \right) \\  \nonumber
&&    + \frac{1}{C_n^4}\sum\limits_{i< j< k< l} \left( \frac{1}{6} \sum_{s< t,u< v}^{(i,j,k,l)}\delta_{5st}
(\|X_{st}\|+\|X_{uv}\|) - \frac{1}{12} \sum_{(s,t,u)}^{(i,j,k,l)}\delta_{5st}\|X_{su}\| \right) \\  \nonumber
&&    + \frac{1}{C_n^4}\sum\limits_{i< j< k< l} \left( \frac{1}{6} \sum_{s< t,u< v}^{(i,j,k,l)}\delta_{6st}
(\|X_{st}\|+\|X_{uv}\|) - \frac{1}{12} \sum_{(s,t,u)}^{(i,j,k,l)}\delta_{6st}\|X_{su}\| \right) \\
&=:& \hat{U}_{n4}+\hat{U}_{n5}+\hat{U}_{n6},
\end{eqnarray}
where $\eta_{st} = \eta_s - \eta_t$,
\begin{eqnarray*}
\delta_{5st}
&=&  -[\frac{\varepsilon_s\sigma(X_s)({\sigma(X_s,\hat{\theta}_n)}-{\sigma(X_s,\tilde{\theta}_0)})}
     {\sigma^2(X_s,\tilde{\theta}_0)}-\frac{\varepsilon_t\sigma(X_t)({\sigma(X_t,\hat{\theta}_n)}-{\sigma(x_t,\tilde{\theta}_0)})}
     {\sigma^2(X_t,\tilde{\theta}_0)} \\
&&   +\frac{m(X_s,\hat{\beta}_n)-m(X_s,{\beta}_0)}{\sigma(X_s,\tilde{\theta}_0)} -\frac{m(X_t,\hat{\beta}_n)
     -m(X_t,{\beta}_0)}{\sigma(X_t,\tilde{\theta}_0)}+(R_s-R_t)] [\mathbb{I}(\eta_s>\eta_t)-\mathbb{I}(\eta_s<\eta_t)], \\
\delta_{6st}
&=&2 \int_{0}^{\frac{m(X_s,\hat{\beta}_n)-m(X_s,{\beta}_0)}{\sigma(X_s,\tilde{\theta}_0)}-\frac{m(X_t,\hat{\beta}_n)
     -m(X_t,{\beta}_0)}{\sigma(X_t,\tilde\tilde{\theta}_0)}+\frac{\varepsilon_s\sigma(x_s)({\sigma(X_s,\hat{\theta}_n)}
     -{\sigma(X_s,\tilde{\theta}_0)})}{\sigma^2(X_s,\tilde{\theta}_0)}-\frac{\varepsilon_t\sigma(X_t)({\sigma(X_t,\hat{\theta}_n)}
     -{\sigma(X_t,\tilde{\theta}_0)})}{\sigma^2(X_t,\tilde{\theta}_0)}+R_s-R_t}  \\
&&   [\mathbb{I}(\eta_s-\eta_t \leq z)-\mathbb{I}(\eta_s  \leq \eta_t)]dz.
\end{eqnarray*}

For the term $\hat{U}_{n5}$, similar to the arguments for $\hat{U}_{n1}$ in the proof of Theorem 3.1, we have
\begin{eqnarray*}
\hat{U}_{n5}
&=& (\hat{\beta}_n-\beta_0)^T \frac{1}{C_n^4} \sum\limits_{i< j< k< l} h_{51}(Z_i,Z_j,Z_k,Z_l) \\
&&  + (\hat{\theta}_n-\tilde{\theta}_0)^T \frac{1}{C_n^4}\sum\limits_{i< j< k< l}h_{52}(Z_i,Z_j,Z_k,Z_l) +o_p(\frac{1}{n})
\end{eqnarray*}
where
\begin{eqnarray*}
&&   h_{51}(Z_i,Z_j,Z_k,Z_l)  \\
&=&  -6^{-1}\sum_{s< t,u< v}^{(i,j,k,l)}[\frac{\dot{m}(X_s,{\beta}_0)}{\sigma(X_s,\tilde{\theta}_0)}
     -\frac{\dot{m}(X_t,{\beta}_0)}{\sigma(X_t,\tilde{\theta}_0)} ] [\mathbb{I}(\eta_s>\eta_t)
     -\mathbb{I}(\eta_s<\eta_t)] (\|X_{st}\|+\|X_{uv}\|)\\
&&   +12^{-1}\sum_{(s,t,u)}^{(i,j,k,l)} [\frac{\dot{m}(X_s,{\beta}_0)}{\sigma(X_s,\tilde{\theta}_0)}
     -\frac{\dot{m}(X_t,{\beta}_0)}{\sigma(X_t,\tilde{\theta}_0)}] [\mathbb{I}(\eta_s>\eta_t)
     -\mathbb{I}(\eta_s<\eta_t)] \|X_{su}\| \\
&&   h_{52}(Z_i,Z_j,Z_k,Z_l) \\
&=&  -6^{-1}\sum_{s< t,u< v}^{(i,j,k,l)} [\frac{\varepsilon_s\sigma (X_s)\dot{\sigma}(X_s,\tilde{\theta}_0)}
     {\sigma^2(X_s,\tilde{\theta}_0)}-\frac{\varepsilon_t\sigma (X_t)\dot{\sigma}(X_t,\tilde{\theta}_0)}
     {\sigma^2(X_t,\tilde{\theta}_0)}] [\mathbb{I}(\eta_s>\eta_t)-\mathbb{I}(\eta_s<\eta_t)] (\|X_{st}\|+\|X_{uv}\|) \\
&&   +12^{-1}\sum_{(s,t,u)}^{(i,j,k,l)} [\frac{\varepsilon_s\sigma (X_s)
     \dot{\sigma}(X_s,\tilde{\theta}_0)}{\sigma^2(X_s,\tilde{\theta}_0)}-\frac{\varepsilon_t\sigma (X_t) \dot{\sigma}(X_t,\tilde{\theta}_0)}{\sigma^2(X_t,\tilde{\theta}_0)}] [\mathbb{I}(\eta_s>\eta_t) -\mathbb{I}(\eta_s<\eta_t)] \|X_{su}\|.
\end{eqnarray*}
It is easy to see that $\frac{1}{C_n^4}\sum\limits_{i< j< k< l}h_{51}(Z_i,Z_j,Z_k,Z_l)$ and $\frac{1}{C_n^4}\sum\limits_{i< j< k< l}h_{52}(Z_i,Z_j,Z_k,Z_l)$ are non-degenerate $U$-statistic of order $4$.
By some elementary calculations, we have
\begin{eqnarray*}
E[h_{51}(Z_i,Z_j,Z_k,Z_l)]
&=&   -2E[(\frac{\dot{m}(X_1,{\beta}_0)}{\sigma(X_1,\tilde{\theta}_0)}-\frac{\dot{m} (x_2,{\beta}_0)}
      {\sigma(X_2,\tilde{\theta}_0)})\mathbb{I}(\eta_1>\eta_2)C_x(X_1,X_2)] \overset{def}{=}2K_1 \\
E[h_{52}(Z_i,Z_j,Z_k,Z_l)]
&=&   -2E[(\frac{\varepsilon_1\sigma (X_1)\dot{\sigma}(X_1,\tilde{\theta}_0)}{\sigma^2(X_1,\tilde{\theta}_0)}
      -\frac{\varepsilon_2\sigma(X_2)\dot{\sigma}(X_2,\tilde{\theta}_0)}{\sigma^2(X_2,\tilde{\theta}_0)})
      \mathbb{I}(\eta_1>\eta_2)C_x(X_1,X_2)] \overset{def}{=}2K_2,
\end{eqnarray*}
where $C_x(X_1,X_2) = \|X_1-X_2\|-E(\|X_1-X_2\||X_1)-E(\|X_1-X_2\||X_2)+E(\|X_1-X_2\|)$.
Thus we obtain that
\begin{eqnarray*}
\hat{U}_{n5}
= (\hat{\beta}_n-\beta_0)^T K_1 + (\hat{\theta}_n-\tilde{\theta}_0)^T K_2 + O_p(\frac{1}{n}).
\end{eqnarray*}
Following the same line for the term $\hat{U}_{n2}$ in the proof of Theorem 3.1, we can show that $\sqrt{n}\hat{U}_{n6} = o_p(1)$. Altogether we obtain that
\begin{eqnarray*}
\hat{U}_{n}
&=&  \hat{U}_{n4} + 2(\hat{\beta}_n-\beta_0)^T K_1 + 2(\hat{\theta}_n-\tilde{\theta}_0)^T K_2 + o_p(\frac{1}{\sqrt n}) \\
&=:& \frac{1}{C_n^4}\sum\limits_{i< j< k< l} h_4(Z_i,Z_j,Z_k,Z_l) + 2(\hat{\beta}_n-\beta_0)^T K_1 +
     2(\hat{\theta}_n-\tilde{\theta}_0)^T K_2 + o_p(\frac{1}{\sqrt n}),
\end{eqnarray*}
where $Z_i = (\eta_i, X_i)$ and $h_4(Z_i,Z_j,Z_k,Z_l) = \frac{1}{6} \sum_{s< t,u< v}^{(i,j,k,l)} |\eta_{st}| (\|X_{st}\|+\|X_{uv}\|) -\frac{1}{12} \sum_{(s,t,u)}^{(i,j,k,l)}|\eta_{st}|\|X_{su}\| $.
Consequently,
\begin{eqnarray*}
\sqrt{n} [\hat{U}_n-dCov^2(\eta,X)] = \sqrt{n} [\hat{U}_{n4}- dCov^2(\eta,X)]  +2 \sqrt{n}(\hat{\beta}_n-\beta_0)^T K_1
+ 2\sqrt{n}(\hat{\theta}_n-\tilde{\theta}_0)^T K_2 + o_p(1).
\end{eqnarray*}
By Assumption 1 and Proposition 3 of Tan et al. (2022), we have
\begin{eqnarray*}
	\sqrt{n}(\hat{\beta}_n-\beta_0)
	&=&  \frac{1}{\sqrt{n}}\sum_{i=1}^{n}l(Y_i,X_i,\beta_0)+o_p(1) \\
	\sqrt{n} (\hat{\theta}_n-\tilde{\theta}_0)
	&=&  \frac{1}{\sqrt{n}} \sum_{i=1}^{n} [\sigma^2(X_i)\varepsilon_i^2 - \sigma^2(X_i, \tilde{\theta}_0)] \Sigma^{-1} \dot{\sigma}^2(X_i, \tilde{\theta}_0) + o_p(1).
\end{eqnarray*}
It follows that
\begin{eqnarray*}
&&  \sqrt{n} [\hat{U}_n-dCov^2(\eta,X)]  \\
&=& \sqrt{n} \frac{1}{C_n^4}\sum\limits_{i< j< k< l} [h_4(Z_i,Z_j,Z_k,Z_l)- dCov(\eta,X)] \\
&&  + \frac{2}{\sqrt{n}}\sum_{i=1}^{n} K_1^T l(Y_i,X_i,\beta_0) + \frac{2}{\sqrt{n}} \sum_{i=1}^{n} [\sigma^2(X_i)\varepsilon_i^2 - \sigma^2(X_i, \tilde{\theta}_0)] \Sigma^{-1} K_2^T \dot{\sigma}^2(X_i, \tilde{\theta}_0) + o_p(1).
\end{eqnarray*}
It is easy to verify that $E[h_4(Z_i,Z_j,Z_k,Z_l)] = dCov^2(\eta,X)$ and $\hat{U}_{n4}$ is non-degenerate. According to technical appendix 1.1 of Yao et al. (2018), we can obtain
\begin{equation*}
E[h_4(Z_1,Z_2,Z_3,Z_4|Z_1)]=\frac{1}{2}\{ E[C_\eta(\eta_1,\eta_2)C_x(X_1,X_2)|Z_1]+dCov^2(\eta,X)\}
\end{equation*}
where $ dCov^2(\eta,X)=E[C_\eta(\eta_i, \eta_j)C_x(X_i, X_j)]$,
\begin{eqnarray*}
C_{\eta}(\eta_i,\eta_j)&=& |\eta_i-\eta_j|-E(|\eta_i-\eta_j\|\eta_i)-E(|\eta_i-\eta_j\|\eta_j) +E(|\eta_i-\eta_j|) \\
C_x(X_i,X_j) &=& \|X_i-X_j\|-E(\|X_i-X_j\| |X_i)-E(\|X_i-X_j\| |X_j)+E(\|X_i-X_j\|).
\end{eqnarray*}
According to the formula (2) in Section 5.3.4 of Serfling (1984), we have
\begin{eqnarray*}
&& \sqrt{n}[\hat{U}_n-dCov^2(\eta,X)] \\
&=&\frac{1}{\sqrt{n}}\sum_{i=1}^{n} 2 \{\mathcal{G}(\eta_i,X_i) + K_1^Tl(Y_i,X_i,\beta_0) +[\sigma ^2(X_i)\varepsilon^2_i
   -\sigma ^2(X_i,\tilde{\theta}_0) ]K_2^T\Sigma^{-1}\dot{\sigma}^2(X_i, \tilde{\theta}_0) \}+o_p(1),
\end{eqnarray*}
%where $\mathcal{G}(\eta_1,X_1)= E[C_\eta(\eta_1,\eta_2)C_x(X_1,X_2)|Z_1]-dCov^2(\eta_1,X_1)$.
it follows that
$$ \sqrt{n} [\hat{U}_n-dCov^2(\eta, X)] \longrightarrow N(0, \sigma_1^2), $$
where $ \sigma_1^2 = 4 var\{\mathcal{G}(\eta, X)+ K_1^{T} l(Y, X, \beta_0) + [\sigma ^2(X) \varepsilon^2 -\sigma^2(X, \tilde{\theta}_0)] K_2^{T} \Sigma^{-1} \dot{\sigma}^2(X, \tilde{\theta}_0) \} $ with
\begin{eqnarray*}
	K_1 &=& -E[(\frac{\dot{m}(X_1,{\beta}_0)}{\sigma(X_1,\tilde{\theta}_0)}-\frac{\dot{m}(X_2,{\beta}_0)}
	{\sigma(X_2,\tilde{\theta}_0)}) I(\eta_1 > \eta_2) C_x(X_1,X_2)], \\
	K_2 &=& -E[(\frac{\eta_1 \dot{\sigma}(X_1,\tilde{\theta}_0)}{\sigma(X_1,\tilde{\theta}_0)}- \frac{\eta_2
		\dot{\sigma}(X_2,\tilde{\theta}_0)}{\sigma(X_2,\tilde{\theta}_0)})I(\eta_1 > \eta_2) C_x(X_1,X_2)], \\
%	\mathcal{G}(\eta_i, X_i) &=& E[C_\eta(\eta_i, \eta_j) C_x(X_i,X_j)|\eta_i, X_i]-dCov^2(\eta,X), \\
	\mathcal{G}(\eta_1,X_1) &=&  E[C_\eta(\eta_1,\eta_2)C_x(X_1,X_2)|Z_1]-dCov^2(\eta_1,X_1)\\
	C_{\eta}(\eta_i, \eta_j) &=& |\eta_i - \eta_j | -E(|\eta_i - \eta_j| | \eta_i) - E(|\eta_i - \eta_j| |\eta_j) + E(|\eta_i - \eta_j |),
\end{eqnarray*}
Hence we complete the proof of Theorem 3.2.   \hfill$\Box$

{\bf Proof of Theorem 3.3.} (1) First we discuss the asymptotic properties of $n\hat{U}_n$ under the null hypothesis in the nonparametric models. Recall that $\eta_i=\frac{Y_i-m(X_i)}{\sigma(X_i, \tilde{\theta}_0)}$ and $\hat{\eta}_i = \frac{Y_i - \hat{m}(X_i)}{\sigma(X_i, \hat{\theta}_n)} $ in nonparametric cases. It follows from (\ref{6.1}) and (\ref{6.2}) in the proof of Theorem 3.1 that
\begin{equation}\label{6.14}
\begin{split}
& |\hat{\eta}_i-\hat{\eta}_j|\\
=&|\varepsilon_i-\varepsilon_j|-[\frac{\hat{m}(X_i)-m(X_i)}{\sigma(X_i,\tilde{\theta}_0)}-\frac{\hat{m}(X_j)-m(X_j)}
  {\sigma(X_j,\tilde{\theta}_0)}+\frac{\varepsilon_i({\sigma(X_i,\hat{\theta}_n)}-{\sigma(X_i,\tilde{\theta}_0)})}
  {\sigma(X_i,\tilde{\theta}_0)}\\
& -\frac{\varepsilon_j({\sigma(X_j,\hat{\theta}_n)}-{\sigma(X_j,\tilde{\theta}_0)})}{\sigma(X_j,\tilde{\theta}_0)}+(R_i-R_j)]
  [\mathbb{I}(\varepsilon_i<\varepsilon_j)-\mathbb{I}(\varepsilon_i>\varepsilon_j)]\\
& +2\int_{0}^{\frac{\hat{m}(X_i)-m(X_i)}{\sigma(X_i,\tilde{\theta}_0)}-{\frac{\hat{m}(X_j)-m(X_j)}{\sigma(X_j,\tilde{\theta}_0)}
  +\frac{\varepsilon_i({\sigma(X_i,\hat{\theta}_n)}-{\sigma(X_i,\tilde{\theta}_0)})}{\sigma(X_i,\tilde{\theta}_0)}
  -\frac{\varepsilon_j({\sigma(X_j,\hat{\theta}_n)}-{\sigma(X_j,\tilde{\theta}_0)})}{\sigma(X_j,\tilde{\theta}_0)}+R_i-R_j}}\\
& [\mathbb{I}(\varepsilon_i-\varepsilon_j \leq z)-\mathbb{I}(\varepsilon_i \leq \varepsilon_j)]dz.
\end{split}
\end{equation}
where
\begin{equation*}
\begin{split}
R_i
=& \frac{\varepsilon_i[\sigma(X_i,\tilde{\theta}_0)-\sigma(X_i,\hat{\theta}_n)]^2}{\sigma(X_i,\tilde{\theta}_0)
   \sigma(X_i,\hat{\theta}_n)}+\frac{\hat{m}(X_i)-m(X_i)}{\sigma^2(X_i,\tilde{\theta}_0)}
   [\sigma(X_i,\tilde{\theta}_0)-\sigma(X_i,\hat{\theta}_n)] \\
&  +\frac{\hat{m}(X_i)-m(X_i)}{\sigma^2(X_i,\tilde{\theta}_0)}\frac{[\sigma(X_i,\tilde{\theta}_0)
   -\sigma(X_i,\hat{\theta}_n)]^2}{\sigma(X_i,\hat{\theta}_n)}.
\end{split}
\end{equation*}
Similar to the arguments in Theorem 3.1, we can rewrite $\hat{U}_n$ as
$$ \hat{U}_n=\frac{1}{C_n^4} \sum\limits_{i< j< k< l}\tilde{h}_0(\hat{Z}_i,\hat{Z}_j,\hat{Z}_k,\hat{Z}_l), $$
where
\begin{equation}\label{6.18}
\tilde{h}_0 (\hat{Z}_i,\hat{Z}_j,\hat{Z}_k,\hat{Z}_l)=\frac{1}{6} \sum_{s< t,u< v}^{(i,j,k,l)} |\hat{\eta}_{st}|
(\|X_{st}\| + \|X_{uv}\|)-\frac{1}{12} \sum_{(s,t,u)}^{(i,j,k,l)} |\hat{\eta}_{st}| \| X_{su} \|,
\end{equation}
$\hat{Z}_i=(\hat{\eta}_i, X_i)$, $X_{st}=X_s-X_t$, and $\hat{\eta}_{st}=\hat{\eta}_s-\hat{\eta}_t$. Here the summation in (\ref{6.18}) is over all permutations of the 4-tuples of indices $(i,j,k,l)$.
By the analog to (\ref{6.5}) in the proof of Theorem 3.1, we have	
\begin{eqnarray}\label{6.19}
\hat{U}_n  \nonumber
&=&   \frac{1}{C_n^4}\sum\limits_{i< j< k< l} \left( \frac{1}{6} \sum_{s< t,u< v}^{(i,j,k,l)} |\varepsilon_{st}|
(\|X_{st}\|+\|X_{uv}\|) -\frac{1}{12} \sum_{(s,t,u)}^{(i,j,k,l)}|\varepsilon_{st}|\|X_{su}\| \right) \\  \nonumber
&&    + \frac{1}{C_n^4}\sum\limits_{i< j< k< l} \left( \frac{1}{6} \sum_{s< t,u< v}^{(i,j,k,l)}\delta^*_{1st}
(\|X_{st}\|+\|X_{uv}\|) - \frac{1}{12} \sum_{(s,t,u)}^{(i,j,k,l)}\delta^*_{1st}\|X_{su}\| \right) \\  \nonumber
&&    + \frac{1}{C_n^4}\sum\limits_{i< j< k< l} \left( \frac{1}{6} \sum_{s< t,u< v}^{(i,j,k,l)}\delta^*_{2st}
(\|X_{st}\|+\|X_{uv}\|) - \frac{1}{12} \sum_{(s,t,u)}^{(i,j,k,l)}\delta^*_{2st}\|X_{su}\| \right) \\
&=:& \hat{U}^*_{n0}+\hat{U}^*_{n1}+\hat{U}^*_{n2},
\end{eqnarray}
where
\begin{eqnarray*}
	\delta^*_{1st}
	&=& -[\frac{\varepsilon_s({\sigma(X_s,\hat{\theta}_n)}-{\sigma(X_s,\tilde{\theta}_0)})}{\sigma(X_s,\tilde{\theta}_0)}
	-\frac{\varepsilon_t({\sigma(X_t,\hat{\theta}_n)}-{\sigma(X_t,\tilde{\theta}_0)})}{\sigma(X_t,\tilde{\theta}_0)}
	+\frac{\hat{m}(X_s)-m(X_s)}{\sigma(X_s,\tilde{\theta}_0)} \\
	&&  -\frac{\hat{m}(X_t)-m(X_t)}{\sigma(X_t,\tilde{\theta}_0)}+(R_s-R_t)]\{\mathbb{I}(\varepsilon_s>\varepsilon_t)
	-\mathbb{I}(\varepsilon_s<\varepsilon_t)\}  \\
	\delta^*_{2st}
	&=&  2\int_{0}^{\frac{\hat{m}(X_s)-m(X_s)}{\sigma(X_s,\tilde{\theta}_0)}-\frac{\hat{m}(X_t)
			-m(X_t)}{\sigma(X_t,\tilde{\theta}_0)}+\frac{\varepsilon_s{\sigma(X_s,\hat{\theta}_n)}-{\sigma(X_s,\tilde{\theta}_0)})}
		{\sigma(X_s,\tilde{\theta}_0)}-\frac{\varepsilon_t({\sigma(X_t,\hat{\theta}_n)}-{\sigma(X_t,\tilde{\theta}_0)})}
		{\sigma(X_t,\tilde{\theta}_0)}+R_s-R_t} \\
	&&   \{\mathbb{I}(\varepsilon_s-\varepsilon_t\leq z)-\mathbb{I}(\varepsilon_s\leq \varepsilon_t)\}dz.
\end{eqnarray*}

First we deal with the term $\hat{U}^*_{n1}$.  Recall that $\hat{g}(X_t)=\frac{1}{n-1}\sum_{j=1,j\neq t}^{n} K_h(X_t-X_j)Y_j $,  $\hat{f}_X(X_t)=\frac{1}{n-1}\sum_{j=1,j\neq t}^{n} K_h(X_t-X_j)$, and $\hat{m}(X_t)=\hat{g}(X_t)/\hat{f}_X(X_t)$, it follows that
\begin{eqnarray}\label{6.20}
\hat{m}(X_t)-m(X_t)
&=& \frac{\hat{g}(X_t)-g(X_t)}{f_X(X_t)}-m(X_t)\frac{\hat{f}_X(X_t)-f_X(X_t)}{f_X(X_t)} \nonumber \\
&&  -\frac{[\hat{g}(X_t)-g(X_t)][\hat{f}_X(X_t)-f_X(X_t)]}{f_X(X)\hat{f}_X(X_t)}+\frac{m(X_t)[\hat{f}_X(X_t)-f_X(X_t)]^2}
    {f_X(X_t)\hat{f}_X(X_t)}
\end{eqnarray}
where $h$ is the bandwidth, $K_h(\cdot)=K(\cdot/h)/h^p$, and $K(\cdot)$ is a kernel function.

For $\hat{g}(X_t)-g(X_t)$, we have
\begin{eqnarray}\label{6.21}
&&  \hat{g}(X_t)-g(X_t)  \nonumber \\
&=& \frac{1}{n-1}\sum_{j=1,j\neq t}^{n} K_h(X_t-X_j)Y_j-g(X_t) \nonumber \\
&=& \frac{1}{n-1}\sum_{j=1,j\neq t}^{n} K_h(X_t-X_j)m(X_j) -g(X_t)+ \frac{1}{n-1}\sum_{j=1,j\neq t}^{n}
    K_h(X_t-X_j)\sigma(X_j)\varepsilon_j \nonumber \\
&=& \frac{1}{n-1}\sum_{j=1,j\neq t}^{n} K_h(X_t-X_j)m(X_j)  -E[K_h(X_t-X_j)m(X_j)|X_t] \nonumber \\
&&  +E[K_h(X_t-X_j)m(X_j)|X_t]-g(X_t)+\frac{1}{n-1}\sum_{j=1,j\neq t}^{n} K_h(X_t-X_j)\sigma(X_j)\varepsilon_j.
\end{eqnarray}
Using Taylor expansion, we obtain that
\begin{eqnarray*}
	&&  E[K_h(X_t-X_j)m(X_j)|X_t]-g(X_t) \\
	&=& \int K_h(X_t-X)m(X)f_X(X)dx-g(X_t) \\
	&=& \int \frac{K(u)g(X_t+uh)h^p}{h^p}du-g(X_t)\\
	&=& C\frac{h^k}{k!}\int \frac{\partial g^{(k)}(X_t)}{\partial x_1^{l_1}\partial x_2^{l_2} \cdots \partial x_p^{l_p}}
	u_1^{l_1}u_2^{l_2} \cdots u_p^{l_p}K(u)du+ o_p(h^k) \\
	&=& h^kD_1(X_t)+o_p(h^k).
\end{eqnarray*}
Similarly, we have
\begin{eqnarray*}
	&&  \hat{f}_X(X_t)-f_X(X_t) \\
	&=&\frac{1}{n-1}\sum_{j=1,j \neq t}^{n}\{K_h(X_t-X_j)-E[K_h(X_t-X_j)|X_t] \} +E[K_h(X_t-X_j)|X_t]-f_X(X_t)  \\
	&=& C\frac{h^k}{k!}\int \frac{\partial f_X^{(k)}(X_t)}{\partial x_1^{l_1}\partial x_2^{l_2} \cdots \partial x_p^{l_p}}
	u_1^{l_1}u_2^{l_2} \cdots u_p^{l_p}K(u)du+\frac{1}{n-1}\sum_{j=1,j\neq t}^{n}\{K_h(X_t-X_j) \\
	&&  -E[K_h(X_t-X_j)|X_t]\}+ o_p(h^k)  \\
	&=& \frac{1}{n-1}\sum_{j=1,j\neq t}^{n}\{K_h(X_t-X_j)-E[K_h(X_t-X_j)|X_t]\}+h^kD_2(X_t)+o_p(h^k).
\end{eqnarray*}
Consequently,
\begin{eqnarray*}
        	&&  \hat{m}(X_t)-m(X_t)  \\
	        &=& \frac{1}{n-1}\sum_{j=1,j\neq t}^{n}\{ \omega_{t,j}m(X_j)-E[\omega_{t,j}m(X_j)|X_t] \}
                +\frac{1}{n-1} \sum_{j=1,j \neq t}^{n}(\omega_{t,j}-E[\omega_{t,j}|X_t])m(X_t)\\
	        &&  +\frac{1}{n-1}\sum_{j=1,j\neq t}^{n}\omega_{t,j}\sigma(X_j)\varepsilon_j + h^kD(X_t) +o_p(h^k),
\end{eqnarray*}
where $D(X_t)=\frac{D_1(X_t) + D_2(X_t)m(X_t)}{f_X(X_t)}$ and $ \omega_{i,j}= \frac{K_h(X_i-X_j)}{f_X(X_i)}$.
By Taylor expansion and the decomposition of $ \hat{m}(X_t)-m(X_t)$, we can decompose $\hat{U}^*_{n1}$ as
\begin{eqnarray*}
\nonumber
	\hat{U}^*_{n1}
&=&  \frac{1}{C_n^4} \sum\limits_{i< j< k< l} h^k h^*_{11}(Z_i,Z_j,Z_k,Z_l)
     + \frac{1}{C_n^4} \sum\limits_{i< j< k< l} h^*_{12}(Z_i,Z_j,Z_k,Z_l) \\
&&   + \frac{1}{C_n^4} \sum\limits_{i< j< k< l} h^*_{13}(Z_i,Z_j,Z_k,Z_l)
     + \frac{1}{C_n^4} \sum\limits_{i< j< k< l} h^*_{14}(Z_i,Z_j,Z_k,Z_l) \\
&&   + (\hat{\theta}_n-\tilde{\theta}_0)^T \frac{1}{C_n^4}\sum\limits_{i< j< k< l}h_{15}^*(Z_i,Z_j,Z_k,Z_l) \\
&&   + 2^{-1}(\hat{\theta}_n-\tilde{\theta}_0)^T \frac{1}{C_n^4}\sum\limits_{i< j< k< l}h_{16}^*(Z_i,Z_j,Z_k,Z_l)
	 (\hat{\theta}_n-\tilde{\theta}_0) \\
&&   + 2^{-1}(\hat{\theta}_n-\tilde{\theta}_0)^T \frac{1}{C_n^4}\sum\limits_{i< j< k< l}h_{17}^*(Z_i,Z_j,Z_k,Z_l)
	 (\hat{\theta}_n-\tilde{\theta}_0) \\
&&   + \frac{1}{C_n^4}\sum\limits_{i< j< k< l}h_{18}^*(Z_i,Z_j,Z_k,Z_l) \\
&=:& I^*_{11} + I^*_{12}+ I^*_{13}+ I^*_{14}+ I^*_{15}+ I^*_{16}+ I^*_{17} + I^*_{18},
\end{eqnarray*}
where $Z_i=(\varepsilon_i, X_i)$ and
\begin{eqnarray*}
&&  h^*_{1m}(Z_i,Z_j,Z_k,Z_l) \\
&=& -6^{-1}\sum_{s < t,u < v}^{(i,j,k,l)}{\delta}^*_{1mst} \{\mathbb{I}(\varepsilon_s>\varepsilon_t)-
	\mathbb{I}(\varepsilon_s<\varepsilon_t)\}(\|X_{st}\|+\|X_{uv}\|) \\
&&  +12^{-1}\sum_{(s,t,u)}^{(i,j,k,l)}{\delta}^*_{1mst}\{\mathbb{I}(\varepsilon_s>\varepsilon_t)
    -\mathbb{I}(\varepsilon_s<\varepsilon_t)\}\|X_{su}\|, \quad {\rm for} \ m=1, 2, \cdots, 8
\end{eqnarray*}
with
\begin{eqnarray*}
{\delta}^*_{11st}
&=&  \frac{D(X_s)}{\sigma(X_s,\tilde{\theta}_0)}-\frac{D(X_t)}{\sigma(X_t,\tilde{\theta}_0)} \\
{\delta}^*_{12st}
&=&  \frac{\frac{1}{n-1}\sum_{p=1,p\neq s}^{n}\omega_{s,p}\sigma(X_p,\tilde{\theta}_0)\varepsilon_p}
     {\sigma(X_s,\tilde{\theta}_0)}-\frac{\frac{1}{n-1}\sum_{q=1,q\neq t}^{n}\omega_{t,q}
     \sigma(X_q,\tilde{\theta}_0)\varepsilon_q}{\sigma(X_t,\tilde{\theta}_0)},\\
{\delta}^*_{13st}
&=&  \frac{\frac{1}{n-1}\sum_{p=1,p\neq s}^{n}(\omega_{s,p}m(X_p)-E[\omega_{s,p} m(X_p)|X_s])}{\sigma(X_s,\tilde{\theta}_0)}
     -\frac{(\frac{1}{n-1}\sum_{q=1,q\neq t}^{n}\omega_{t,q}m(X_q)-E[\omega_{t,q} m(X_q)|X_t])}{\sigma(X_t,\tilde{\theta}_0)},\\
{\delta}^*_{14st}
&=&  \frac{\frac{m(X_s)}{n-1}\sum_{p=1,p\neq s}^{n}(\omega_{s,p}-E[\omega_{s,p} |X_s])}{\sigma(X_s,\tilde{\theta}_0)}
     -\frac{\frac{m(X_t)}{n-1}\sum_{q=1,q\neq t}^{n}(\omega_{t,q}-E[\omega_{t,q} |X_t])}{\sigma(X_t,\tilde{\theta}_0)}, \\
\delta^*_{15st}
&=&  \frac{\varepsilon_s \dot{\sigma}(X_s,\tilde{\theta}_0)}{\sigma(X_s,\tilde{\theta}_0)}-\frac{\varepsilon_t\dot{\sigma}
	 (X_t,\tilde{\theta}_0)}{\sigma(X_t,\tilde{\theta}_0)} \\
\delta^*_{16st}
&=&  \frac{\varepsilon_s \ddot{\sigma}(X_s,(\hat{\theta}_n-\tilde{\theta}_0)\zeta+\theta_0)}{\sigma(X_s,\tilde{\theta}_0)}
	 -\frac{\varepsilon_s\ddot{\sigma}(X_s,\tilde{\theta}_0)}{\sigma(X_s,\tilde{\theta}_0)}
     -[\frac{\varepsilon_t\ddot{\sigma}(X_t,(\hat{\theta}_n-\tilde{\theta}_0)\zeta+\tilde{\theta}_0)}{\sigma(X_t,\tilde{\theta}_0)}
 	 -\frac{\varepsilon_t \ddot{\sigma}(X_t,\tilde{\theta}_0)}{\sigma(X_t,\tilde{\theta}_0)}] \\
\delta^*_{17st}
&=&  \frac{\varepsilon_s \ddot{\sigma}(X_s,\tilde{\theta}_0)}{\sigma(X_s,\tilde{\theta}_0)}-\frac{\varepsilon_t\ddot{\sigma}
	 (X_t,\tilde{\theta}_0)}{\sigma(X_t,\tilde{\theta}_0)} \\
\delta^*_{18st} &=& R_s-R_t.
\end{eqnarray*}

For the term $I^*_{11}$, it is easy to see that $E[ h^*_{11}(Z_i,Z_j,Z_k,Z_l)] = 0$ and $\frac{1}{C_n^4} \sum\limits_{i< j< k< l} h^*_{11}(Z_i,Z_j,Z_k,Z_l)$ is non-degenerate. Combining this with the assumption 6(d), we have
$$n I^*_{11} = \sqrt{n}h^k \frac{\sqrt{n}}{C_n^4} \sum\limits_{i< j< k< l} h^*_{11}(Z_i,Z_j,Z_k,Z_l)=o_p(1).$$
For the term $I^*_{12}$, decomposed it as
\begin{eqnarray*}
&&  I^*_{12} \\
&=& \frac{1}{n(n-1)^3(n-2)(n-3)}\sum\limits_{i=1}^{n}\sum\limits_{j=1.j\neq i}^{n}\sum\limits_{k=1,k\neq i,j }^{n}
    \sum\limits_{l=1,l\neq i,j,k }^{n}\sum\limits_{p=1,p\neq i}^{n}\sum\limits_{q=1,q\neq j}^{n}
    \{ \frac{\omega_{p,i}\sigma(X_p,\tilde{\theta}_0)\varepsilon_p}{\sigma(X_i,\tilde{\theta}_0)}\\
&&  -\frac{\omega_{q,j}\sigma(X_q,\tilde{\theta}_0)\varepsilon_q}{\sigma(X_j,\tilde{\theta}_0)} \}
    \{\mathbb{I}(\varepsilon_i>\varepsilon_j)-\mathbb{I}(\varepsilon_i<\varepsilon_j)\}(\|X_{ij}\|+\|X_{kl}\|-2\|X_{ik}\|)\\
&=& \frac{1}{n(n-1)^3(n-2)(n-3)}\sum_{i=1}^{n}\sum_{j\neq i}^{n}\sum_{k\neq i,j }^{n} \sum_{l\neq i,j,k }^{n}
    \sum_{p=q\neq i,j}^{n}\{ \frac{\omega_{p,i}\sigma(X_p,\tilde{\theta}_0)\varepsilon_p}{\sigma(X_i,\tilde{\theta}_0)}
    -\frac{\omega_{q,j} \sigma(X_q,\tilde{\theta}_0)\varepsilon_q}{\sigma(X_j,\tilde{\theta}_0)} \}\\
&&  \times \{\mathbb{I}(\varepsilon_i>\varepsilon_j)-\mathbb{I}(\varepsilon_i<\varepsilon_j)\}(\|X_{ij}\|+\|X_{kl}\|
    -2\|X_{ik}\|)\\
&&  +\frac{1}{n(n-1)^3(n-2)(n-3)}\sum_{i=1}^{n}\sum_{j\neq i}^{n}\sum_{k\neq i,j }^{n} \sum_{l\neq i,j,k }^{n}
    \sum_{p=j}\sum_{q=i}\{ \frac{\omega_{j,i}\sigma(X_j,\tilde{\theta}_0)\varepsilon_j}{\sigma(X_i,\tilde{\theta}_0)}
    -\frac{\omega_{i,j} \sigma(X_i,\tilde{\theta}_0)\varepsilon_i}{\sigma(X_j,\tilde{\theta}_0)} \}\\
&&  \times \{\mathbb{I}(\varepsilon_i>\varepsilon_j)-\mathbb{I}(\varepsilon_i<\varepsilon_j)\}
    (\|X_{ij}\|+\|X_{kl}\|-2\|X_{ik}\|)\\
&&  +\frac{1}{n(n-1)^3(n-2)(n-3)}\sum_{i=1}^{n}\sum_{j\neq i}^{n}\sum_{k\neq i,j }^{n} \sum_{l\neq i,j,k }^{n}
    \sum_{p=j}^{n}\sum_{q\neq i,j}^{n}\{ \frac{\omega_{p,i}\sigma(X_p,\tilde{\theta}_0)\varepsilon_p}
    {\sigma(X_i,\tilde{\theta}_0)}-\frac{\omega_{p,j}\sigma(X_p,\tilde{\theta}_0)\varepsilon_p}
    {\sigma(X_j,\tilde{\theta}_0)} \}\\
&&  \times \{\mathbb{I}(\varepsilon_i>\varepsilon_j)-\mathbb{I}(\varepsilon_i<\varepsilon_j)\}
    (\|X_{ij}\|+\|X_{kl}\|-2\|X_{ik}\|)\\
&&  +\frac{1}{n(n-1)^3(n-2)(n-3)}\sum_{i=1}^{n}\sum_{j\neq i}^{n}\sum_{k\neq i,j }^{n} \sum_{l\neq i,j,k }^{n}
    \sum_{p\neq i,j}^{n}\sum_{q=i}^{n}\{ \frac{\omega_{p,i}\sigma(X_p,\tilde{\theta}_0)\varepsilon_p}
    {\sigma(X_i,\tilde{\theta}_0)}-\frac{\omega_{p,j}\sigma(X_p,\tilde{\theta}_0)\varepsilon_p}
    {\sigma(X_j,\tilde{\theta}_0)} \}\\
&&  \times \{\mathbb{I}(\varepsilon_i>\varepsilon_j)-\mathbb{I}(\varepsilon_i<\varepsilon_j)\}
    (\|X_{ij}\|+\|X_{kl}\|-2\|X_{ik}\|)  \\
&&  +\frac{1}{n(n-1)^3(n-2)(n-3)}\sum_{i=1}^{n}\sum_{j\neq i}^{n}\sum_{k\neq i,j }^{n}
    \sum_{l\neq i,j,k }^{n}\sum_{p\neq i\neq j\neq k\neq l}^{n}\sum_{q\neq i\neq j\neq k\neq l \neq p}^{n}\{ \frac{\omega_{p,i}\sigma(X_p,\tilde{\theta}_0)\varepsilon_p}{\sigma(X_i,\tilde{\theta}_0)} \\
&&  -\frac{\omega_{q,j}\sigma(X_q,\tilde{\theta}_0)\varepsilon_q}{\sigma(X_j,\tilde{\theta}_0)} \}
    \{\mathbb{I}(\varepsilon_i>\varepsilon_j)-\mathbb{I}(\varepsilon_i<\varepsilon_j)\}(\|X_{ij}\|+\|X_{kl}\|-2\|X_{ik}\|)\\
&=:&I^*_{121} + I^*_{122}+ I^*_{123}+ I^*_{124}+ I^*_{125}
\end{eqnarray*}
For the term $I^*_{121}$, we decompose it as
\begin{eqnarray*}
I^*_{121}
&=&\frac{1}{n(n-1)^3(n-2)(n-3)}\sum_{i=1}^{n}\sum_{j\neq i}^{n}\sum_{k\neq i,j }^{n} \sum_{l\neq i,j,k }^{n}\sum_{p=q=k}^{}\{ \frac{\omega_{p,i}\sigma(X_p,\tilde{\theta}_0)\varepsilon_p}{\sigma(X_i,\tilde{\theta}_0)}\\
&&-\frac{\omega_{q,j}\sigma(X_q,\tilde{\theta}_0)\varepsilon_q}{\sigma(X_j,\tilde{\theta}_0)} \} \{\mathbb{I}(\varepsilon_i>\varepsilon_j)-\mathbb{I}(\varepsilon_i<\varepsilon_j)\}(\|X_{ij}\|+\|X_{kl}\|-2\|X_{ik}\|)\\
&&+\frac{1}{n(n-1)^3(n-2)(n-3)}\sum_{i=1}^{n}\sum_{j\neq i}^{n}\sum_{k\neq i,j }^{n} \sum_{l\neq i,j,k }^{n}\sum_{p=q=l}^{}\{ \frac{\omega_{p,i}\sigma(X_p,\tilde{\theta}_0)\varepsilon_p}{\sigma(X_i,\tilde{\theta}_0)}\\
&&-\frac{\omega_{q,j}\sigma(X_q,\tilde{\theta}_0)\varepsilon_q}{\sigma(X_j,\tilde{\theta}_0)} \} \{\mathbb{I}(\varepsilon_i>\varepsilon_j)-\mathbb{I}(\varepsilon_i<\varepsilon_j)\}(\|X_{ij}\|+\|X_{kl}\|-2\|X_{ik}\|)\\
&&+\frac{1}{n(n-1)^3(n-2)(n-3)}\sum_{i=1}^{n}\sum_{j\neq i}^{n}\sum_{k\neq i,j }^{n} \sum_{l\neq i,j,k }^{n}\sum_{p=q\neq i,j,l,k}^{n}\{ \frac{\omega_{p,i}\sigma(X_p,\tilde{\theta}_0)\varepsilon_p}{\sigma(X_i,\tilde{\theta}_0)}\\
&&-\frac{\omega_{q,j}\sigma(X_q,\tilde{\theta}_0)\varepsilon_q}{\sigma(X_j,\tilde{\theta}_0)} \} \{\mathbb{I}(\varepsilon_i>\varepsilon_j)-\mathbb{I}(\varepsilon_i<\varepsilon_j)\}(\|X_{ij}\|+\|X_{kl}\|-2\|X_{ik}\|)\\
&=:&I^*_{1211}+I^*_{1212}+I^*_{1213}.
\end{eqnarray*}
For the term $I^*_{1211}$,
\begin{eqnarray*}
nI^*_{1211}
&=&  \frac{1}{(n-1)^3(n-2)(n-3)}\sum\limits_{i=1}^{n}\sum\limits_{j\neq i}^{n}\sum\limits_{k\neq i,j }^{n}
    \sum \limits_{l\neq i,j,k }^{n}\{ \frac{\omega_{k,i}\sigma(X_k,\tilde{\theta}_0)\varepsilon_k}
    {\sigma(X_i,\tilde{\theta}_0)}-\frac{\omega_{k,j}\sigma(X_k,\tilde{\theta}_0)\varepsilon_k}{\sigma(X_j,\tilde{\theta}_0)}\}\\
&&  \times \{\mathbb{I}(\varepsilon_i>\varepsilon_j)-\mathbb{I}(\varepsilon_i< \varepsilon_j)\}
    (\|X_{ij}\|+\|X_{kl}\|-2\|X_{ik}\|)\\
&=& \frac{n}{(n-1)^2}\frac{1}{C_n^4}\mathop{\sum\sum\sum\sum}_{1 \leq i< j< k< l \leq n} {\delta}^*_{1211}(Z_i,Z_j,Z_k,Z_l),
\end{eqnarray*}
where
\begin{eqnarray*}
\delta^*_{1211}(Z_i,Z_j,Z_k,Z_l)&=&\frac{1}{24}\sum_{(s,t,u,v)}^{(i,j,k,l)}\{ \frac{\omega_{u,s}\sigma(X_u,\tilde{\theta}_0)
\varepsilon_u}{\sigma(X_s,\tilde{\theta}_0)}-\frac{\omega_{u,t}\sigma(X_u,\tilde{\theta}_0)\varepsilon_u}
{\sigma(X_t,\tilde{\theta}_0)} \} \{\mathbb{I}(\varepsilon_s>\varepsilon_t)-\mathbb{I}(\varepsilon_s<\varepsilon_t)\} \\
&& (\|X_{st}\| + \|X_{uv}\| - 2\|X_{su}\|).
\end{eqnarray*}
Since $E |\delta^*_{1211}(Z_i,Z_j,Z_k,Z_l)| < \infty$, it follows from the law of large numbers for $U$-statistics that $nI^*_{1211}=o_p(1)$.
Similarly, we can show that $ nI^*_{1212}=o_p(1)$ and $ nI^*_{1213}=o_p(1)$.
%For the term $I^*_{1213}$,
%\begin{eqnarray*}
%I^*_{1213}&=&\frac{1}{(n-1)^3(n-2)(n-3)}\sum_{i=1}^{n}\sum_{j\neq i}^{n}\sum_{k\neq i,j }^{n} \sum_{l\neq i,j,k }^{n}\sum_{p=q\neq i,j,l,k}^{n}\{ \frac{\omega_{p,i}\sigma(X_p,\tilde{\theta}_0)\varepsilon_p}{\sigma(X_i,\tilde{\theta}_0)}-\frac{\omega_{p,j}\sigma(X_p,\tilde{\theta}_0)\varepsilon_p}{\sigma(X_j,\tilde{\theta}_0)} \}\\
%&& \times \{\mathbb{I}(\varepsilon_i>\varepsilon_j)-\mathbb{I}(\varepsilon_i<\varepsilon_j)\}(\|X_{ij}\|+\|X_{kl}\|-2\|X_{ik}\|)\\
%&=&\frac{n(n-4)}{(n-1)^2}\frac{1}{C_n^5}\mathop{\sum\sum\sum\sum\sum}_{1 \leq i< j< k< l< p \leq n}{\delta}^*_{1213}(Z_i,Z_j,Z_k,Z_l,Z_p),
%\end{eqnarray*}
%where
%\begin{eqnarray*}
%{\delta}^*_{1213}(Z_i,Z_j,Z_k,Z_l,Z_p)&=&\frac{1}{120}\sum_{(s,t,u,v,r)}^{(i,j,k,l,p)}
%\{ \frac{\omega_{r,s}\sigma(X_r,\tilde{\theta}_0)\varepsilon_r}{\sigma(X_s,\tilde{\theta}_0)}-\frac{\omega_{r,t}\sigma(X_r,\tilde{\theta}_0)\varepsilon_r}{\sigma(X_t,\tilde{\theta}_0)} \} \{\mathbb{I}(\varepsilon_s>\varepsilon_t)-\mathbb{I}(\varepsilon_s<\varepsilon_t) \}(\|X_{st}\|\\
%&&+\|X_{uv}\|-2\|X_{su}\|).
%\end{eqnarray*}
%By the law of large numbers for U-statistics,  we have $ nI^*_{1213}=o_p(1)$.
Consequently,
$$n I^*_{121} = n I^*_{1211}+ n I^*_{1212} + n I^*_{1213}= o_p(1).$$

For the  term $I^*_{122}$, recall that
\begin{eqnarray*}
nI^*_{122}
&=&  \frac{1}{(n-1)^3(n-2)(n-3)}\sum_{i=1}^{n}\sum_{j\neq i}^{n}\sum_{k\neq i,j }^{n} \sum_{l\neq i,j,k }^{n}\{
     \frac{\omega_{j,i}\sigma(X_j,\tilde{\theta}_0)\varepsilon_j}{\sigma(X_i,\tilde{\theta}_0)}-
     \frac{\omega_{i,j}\sigma(X_i,\tilde{\theta}_0)\varepsilon_i}{\sigma(X_j,\tilde{\theta}_0)} \}  \\
&&   \times \{\mathbb{I}(\varepsilon_i>\varepsilon_j)-\mathbb{I}(\varepsilon_i<\varepsilon_j)\}
     (\|X_{ij}\|+\|X_{kl}\|-2\|X_{ik}\|)
\end{eqnarray*}
By the law of large numbers for $U$-statistics, it is readily seen that $nI^*_{122}=o_p(1) $.

For the  term $I^*_{123}$,  recall that
\begin{eqnarray*}
	I^*_{123}&=&\frac{1}{n(n-1)^3(n-2)(n-3)}\sum_{i=1}^{n}\sum_{j\neq i}^{n}\sum_{k\neq i,j }^{n} \sum_{l\neq i,j,k }^{n}\sum_{p=j}^{}\sum_{q\neq i,j}^{n}\{ \frac{\omega_{p,i}\sigma(X_p,\tilde{\theta}_0)\varepsilon_p}{\sigma(X_i,\tilde{\theta}_0)}
	-\frac{\omega_{q,j}\sigma(X_q,\tilde{\theta}_0)\varepsilon_q}{\sigma(X_j,\tilde{\theta}_0)} \} \\
	&& \times \{\mathbb{I}(\varepsilon_i>\varepsilon_j)-\mathbb{I}(\varepsilon_i<\varepsilon_j)\}(\|X_{ij}\|+\|X_{kl}\|-2\|X_{ik}\|).
\end{eqnarray*}
Decompose $I^*_{123}$ as follows,
\begin{eqnarray*}
I^*_{123}
&=&\frac{1}{n(n-1)^3(n-2)(n-3)}\sum_{i=1}^{n}\sum_{j\neq i}^{n}\sum_{k\neq i,j }^{n} \sum_{l\neq i,j,k }^{n}\sum_{p=j}^{}\sum_{q=k}^{}\{ \frac{\omega_{p,i}\sigma(X_p,\tilde{\theta}_0)\varepsilon_p}{\sigma(X_i,\tilde{\theta}_0)}
-\frac{\omega_{q,j}\sigma(X_q,\tilde{\theta}_0)\varepsilon_q}{\sigma(X_j,\tilde{\theta}_0)} \}   \\
&& \times \{\mathbb{I}(\varepsilon_i>\varepsilon_j)-\mathbb{I}(\varepsilon_i<\varepsilon_j)\}(\|X_{ij}\|+\|X_{kl}\|-2\|X_{ik}\|)\\
&&+\frac{1}{n(n-1)^3(n-2)(n-3)}\sum_{i=1}^{n}\sum_{j\neq i}^{n}\sum_{k\neq i,j }^{n} \sum_{l\neq i,j,k }^{n}\sum_{p=j}^{}\sum_{q=l}^{}\{ \frac{\omega_{p,i}\sigma(X_p,\tilde{\theta}_0)\varepsilon_p}{\sigma(X_i,\tilde{\theta}_0)}
-\frac{\omega_{q,j}\sigma(X_q,\tilde{\theta}_0)\varepsilon_q}{\sigma(X_j,\tilde{\theta}_0)} \}  \\
&& \times \{\mathbb{I}(\varepsilon_i>\varepsilon_j)-\mathbb{I}(\varepsilon_i<\varepsilon_j)\}(\|X_{ij}\|+\|X_{kl}\|-2\|X_{ik}\|)\\
&&+\frac{1}{n(n-1)^3(n-2)(n-3)}\sum_{i=1}^{n}\sum_{j\neq i}^{n}\sum_{k\neq i,j }^{n} \sum_{l\neq i,j,k }^{n}\sum_{p=j}\sum_{q\neq i,j,k,l}\{ \frac{\omega_{p,i}\sigma(X_p,\tilde{\theta}_0)\varepsilon_p}{\sigma(X_i,\tilde{\theta}_0)}
-\frac{\omega_{q,j}\sigma(X_q,\tilde{\theta}_0)\varepsilon_q}{\sigma(X_j,\tilde{\theta}_0)} \}  \\
&& \times \{\mathbb{I}(\varepsilon_i>\varepsilon_j)-\mathbb{I}(\varepsilon_i<\varepsilon_j)\}(\|X_{ij}\|+\|X_{kl}\|-2\|X_{ik}\|)\\
&=:&I^*_{1231}+I^*_{1232}+I^*_{1233}
\end{eqnarray*}
Similar to the arguments for $I^*_{122}$, we have $nI^*_{1231}=o_p(1) $ and $nI^*_{1232}=o_p(1)$.
For the term $I^*_{1233}$,
\begin{eqnarray*}
nI^*_{1233}
&=&  \frac{1}{n(n-1)^3(n-2)(n-3)}\sum_{i=1}^{n}\sum_{j\neq i}^{n}\sum_{k\neq i,j }^{n} \sum_{l\neq i,j,k }^{n}
     \sum_{p=j}\sum_{q\neq i,j,k,l}\{ \frac{\omega_{p,i}\sigma(X_p,\tilde{\theta}_0)\varepsilon_p}
     {\sigma(X_i,\tilde{\theta}_0)}  \\
&&   -\frac{\omega_{q,j}\sigma(X_q,\tilde{\theta}_0)\varepsilon_q}{\sigma(X_j,\tilde{\theta}_0)} \}
     \{\mathbb{I}(\varepsilon_i>\varepsilon_j)-\mathbb{I}(\varepsilon_i<\varepsilon_j)\}(\|X_{ij}\|+\|X_{kl}\|-2\|X_{ik}\|)\\
&=&  \frac{n(n-4)}{(n-1)^2}\frac{1}{C_n^5}\mathop{\sum\sum\sum\sum\sum}_{1 \leq i< j< k< l< r \leq n }
     {\delta}^*_{1233}(Z_i,Z_j,Z_k,Z_l,Z_r),
\end{eqnarray*}
where
\begin{eqnarray*}
\delta^*_{1233}(Z_i,Z_j,Z_k,Z_l,Z_r)
&=& \frac{1}{120}\sum\limits_{(s,t,u,v,r)}^{(i,j,k,l,q)} \{ \frac{\omega_{t,s}\sigma(X_t,\tilde{\theta}_0)\varepsilon_t}
    {\sigma(X_s,\tilde{\theta}_0)}-\frac{\omega_{r,t}\sigma(X_r,\tilde{\theta}_0)\varepsilon_r}{\sigma(X_t,\tilde{\theta}_0)}\}\\
&&  \times \{\mathbb{I}(\varepsilon_s>\varepsilon_t)-\mathbb{I}(\varepsilon_s<\varepsilon_t) \}
    (\|X_{st}\|+\|X_{uv}\|-2\|X_{su}\|).
\end{eqnarray*}
By the law of large numbers for U-statistics, it follows that
$ n I^*_{1233} \rightarrow 2A_\varepsilon E \|X_1 - X_2\|, $
where $A_{\varepsilon}=E[\varepsilon F_{\varepsilon}(\varepsilon)]$.
Consequently, we obtain that $ n I^*_{123} \rightarrow 2A_\varepsilon E \|X_1 - X_2\|. $
Similarly, we can show that
$$ n I^*_{124} \longrightarrow 2A_\varepsilon E \|X_1 - X_2\| .$$

For the term  $I^*_{125}$, recall that
\begin{eqnarray*}
I^*_{125}&=&\frac{1}{n(n-1)^3(n-2)(n-3)}\sum_{i=1}^{n}\sum_{j\neq i}^{n}\sum_{k\neq i,j }^{n} \sum_{l\neq i,j,k }^{n}\sum_{p\neq i\neq j\neq k\neq l}^{n}\sum_{q\neq i\neq j\neq k\neq l \neq p}^{n}\{ \frac{\omega_{p,i}\sigma(X_p,\tilde{\theta}_0)\varepsilon_p}{\sigma(X_i,\tilde{\theta}_0)}\\
&&-\frac{\omega_{q,j}\sigma(X_q,\tilde{\theta}_0)\varepsilon_q}{\sigma(X_j,\tilde{\theta}_0)} \} \{\mathbb{I}(\varepsilon_i>\varepsilon_j)-\mathbb{I}(\varepsilon_i<\varepsilon_j)\}(\|X_{ij}\|+\|X_{kl}\|-2\|X_{ik}\|)\\
&=&\frac{(n-4)(n-5)}{(n-1)^2}\frac{1}{C_n^6}\sum_{i=1}^{n}\sum_{j\neq i}^{n}\sum_{k\neq i,j }^{n} \sum_{l\neq i,j,k }^{n}\sum_{p\neq i\neq j\neq k\neq l}^{n}\sum_{q\neq i\neq j\neq k\neq l \neq p}^{n}\{ \frac{\omega_{p,i}\sigma(X_p,\tilde{\theta}_0)\varepsilon_p}{\sigma(X_i,\tilde{\theta}_0)}\\
&&-\frac{\omega_{q,j}\sigma(X_q,\tilde{\theta}_0)\varepsilon_q}{\sigma(X_j,\tilde{\theta}_0)} \} \{\mathbb{I}(\varepsilon_i>\varepsilon_j)-\mathbb{I}(\varepsilon_i<\varepsilon_j)\}(\|X_{ij}\|+\|X_{kl}\|-2\|X_{ik}\|)\\
&=&\frac{(n-4)(n-5)}{(n-1)^2}\frac{1}{C_n^6}\mathop{\sum\sum\sum\sum\sum\sum}_{1 \leq i< j< k< l< r< m \leq n } {\delta}^*_{125}(Z_i,Z_j,Z_k,Z_l,Z_r,Z_m),
\end{eqnarray*}
where
\begin{eqnarray*}
{\delta}^*_{125}(Z_i,Z_j,Z_k,Z_l,Z_r,Z_m)
&=& \frac{1}{6!}\sum_{(s,t,u,v,r,m)}^{(i,j,k,l,p,q)} \{\frac{\omega_{r,s}\sigma(X_r,\tilde{\theta}_0)\varepsilon_r}
    {\sigma(X_s,\tilde{\theta}_0)}-\frac{\omega_{m,t}\sigma(X_m,\tilde{\theta}_0)\varepsilon_m}{\sigma(X_t,\tilde{\theta}_0)}\} \\
&& \times \{\mathbb{I}(\varepsilon_s>\varepsilon_t)-\mathbb{I}(\varepsilon_s< \varepsilon_t)\}
   (\|X_{st}\|+\|X_{uv}\|-2\|X_{su}\|).
\end{eqnarray*}
Some elementary calculations show that $I^*_{125}$ is degenerate of order $1$. By the arguments in Section 5.3.4 of  Serfling (2009), we can obtain
\begin{equation*}
nI^*_{125}=\frac{2}{(n-1)}\mathop{\sum\sum}_{1 \leq i< j \leq n }h^*_{1}(Z_i,Z_j)+o_p(1),
\end{equation*}
where $h^*_{1}(Z_i,Z_j)=\frac{1}{2}(H_{1ij}+H_{2ij})$,
\begin{eqnarray*}	
H_{1ij}
&=&  \frac{1}{4!}E[\sum_{(k_1,l_1,p_1,q_1)}^{(k,l,p,q)}(\frac{\omega_{p_1,q_1}\sigma(X_{p_1},\tilde{\theta}_0)
     \varepsilon_{p_1}}{\sigma(X_{q_1},\tilde{\theta}_0)}-\frac{\omega_{i,j}\sigma(X_i,\tilde{\theta}_0)\varepsilon_i}
     {\sigma(X_j,\tilde{\theta}_0)})\{\mathbb{I}(\varepsilon_{q_1}>\varepsilon_j)-\mathbb{I}(\varepsilon_{q_1}<\varepsilon_j)\}\\
&&   \times (\|X_{q_1j}\|+\|X_{kl}\|-2\|X_{q_1k}\|)|Z_i,Z_j],\\
H_{2ij}
&=&  \frac{1}{4!}E[\sum_{(k_1,l_1,p_1,q_1)}^{(k,l,p,q)}(\frac{\omega_{j,i}\sigma(X_j,\tilde{\theta}_0)\varepsilon_j}
     {\sigma(X_i,\tilde{\theta}_0)}-\frac{\omega_{q_1,p_1}\sigma(X_{q_1},\tilde{\theta}_0)\varepsilon_{q_1}}
     {\sigma(X_{p_1},\tilde{\theta}_0)})\{\mathbb{I}(\varepsilon_i>\varepsilon_{p_1})
     -\mathbb{I}(\varepsilon_i<\varepsilon_{p_1})\} \\
&&   \times (\|X_{ip_1}\|+\|X_{k_1l_1}\| -2\|X_{ik_1}\|)|Z_i,Z_j].
\end{eqnarray*}
Hence we obtain that
$$nI^*_{12} = \frac{2}{(n-1)}\mathop{\sum\sum}_{1 \leq i< j \leq n }h^*_{1}(Z_i,Z_j) + 4A_\varepsilon E \|X_1 - X_2\| + o_p(1).$$

For the term $I^*_{13}$ and $I^*_{14}$, similar to the arguments for $I^*_{12}$, we can show that
\begin{eqnarray*}
n I^*_{13}=\frac{2}{n-1}\mathop{\sum\sum}_{1 \leq i< j \leq n }h^*_{2}(Z_i,Z_j)+o_p(1), \\
n I^*_{14}=\frac{2}{n-1}\mathop{\sum\sum}_{1 \leq i< j \leq n }h^*_{3}(Z_i,Z_j)+o_p(1),
\end{eqnarray*}
where $h^*_{2}(Z_i,Z_j)=\frac{1}{2}(H_{3ij}+H_{4ij})$, $ h^*_{3}(Z_i,Z_j)=\frac{1}{2}(H_{5ij}+H_{6ij})$,
\begin{eqnarray*}
H_{3ij}
&=&   \frac{1}{4!}\sum_{(k_1,l_1,p_1,q_1)}^{(k,l,p,q)}E[ (\frac{(\omega_{j,i}m(X_j)-E[\omega_{j,i} m(X_j)|X_i])}
      {\sigma(X_i,\tilde{\theta}_0)} - \frac{(\omega_{q_1,p_1} m(X_{q_1})-E[\omega_{q_1,p_1} m(X_{q_1})|X_{p_1}])}{\sigma(X_j,\tilde{\theta}_0)}) \\
&&    \times \{\mathbb{I}(\varepsilon_i>\varepsilon_{p_1})-\mathbb{I}(\varepsilon_i<\varepsilon_{p_1})\}(\|X_{(i)}^{(p_1)}\|
      +\|X_{(k_1)}^{(l_1)}\| - 2\|X_{(i)}^{(k_1)}\|)|Z_i,Z_j],\\
H_{4ij}
&=&  \frac{1}{4!}\sum_{(k_1,l_1,p_1,q_1)}^{(k,l,p,q)}E[ (\frac{(\omega_{p_1,q_1}m(X_{p_1})-E[\omega_{p_1,q_1}
     m(X_{p_1})|X_{q_1}])}{\sigma(X_{q_1},\tilde{\theta}_0)} - \frac{(\omega_{j,i} m(X_j)-E[\omega_{j,i} m(X_j)|X_i])}
     {\sigma(X_i,\tilde{\theta}_0)} )\\
&&   \times \{\mathbb{I}(\varepsilon_{q_1}>\varepsilon_j) - \mathbb{I}(\varepsilon_{q_1}<\varepsilon_j)\}
     (\|X_{(q_1)}^{(j)}\|+\|X_{(k_1)}^{(l_1)}\|-2\|X_{(q_1)}^{(k_1)}\|)|Z_i,Z_j], \\
H_{5ij}
&=& \frac{1}{4!} E[\sum_{(k_1,l_1,p_1,q_1)}^{(k,l,p,q)}\{\frac{(\omega_{j,i}-E[\omega_{j,i}|X_i])m(X_i)}
    {\sigma(X_i,\tilde{\theta}_0)}-\frac{(\omega_{q_1,p_1}
    -E[\omega_{q_1,p_1} |X_{p_1}])m(X_j)}{\sigma(X_j,\tilde{\theta}_0)} \} \\
&&	\{\mathbb{I}(\varepsilon_i>\varepsilon_{p_1}) - \mathbb{I}(\varepsilon_i<\varepsilon_{p_1})\}
	(\|X_{ip_1}\|+\|X_{k_1l_1}\|-2\|X_{ik_1}\|)|Z_i,Z_j],\\
H_{6ij}
&=& \frac{1}{4!} E[\sum_{(k_1,l_1,p_1,q_1)}^{(k,l,p,q)} \{\frac{(\omega_{p_1,q_1}-E[\omega_{p_1,q_1}
    |X_{q_1}])m(X_{q_1})}{\sigma(X_{q_1},\tilde{\theta}_0)} - \frac{(\omega_{j,i} -E[\omega_{j,i} |X_i])m(X_{i})}
    {\sigma(X_i,\tilde{\theta}_0)} \} \\
&&  \{\mathbb{I}(\varepsilon_{q_1}>\varepsilon_j) - \mathbb{I}(\varepsilon_{q_1}<\varepsilon_j)\}
    (\|X_{q_1j}\|+\|X_{k_1l_1}\|-2\|X_{q_1k_1}\|)|Z_i,Z_j].
\end{eqnarray*}
For the term $I^*_{15}$, $I^*_{16}$, $I^*_{17}$, $I^*_{18}$, similar to the arguments for $I_{14}$, $I_{15}$, $I_{16}$, $I_{17}$ in the proof of Theorem 3.1, we can show that $ n I^*_{16} = o_p(1), n I^*_{17} = o_p(1), n I^*_{18} = o_p(1)$, and
\begin{eqnarray*}
I^*_{15} =
(\hat{\theta}_n-\tilde{\theta}_0)^T \frac{1}{C_n^4} \sum\limits_{i< j< k< l} h^*_{15} (Z_i,Z_j,Z_k,Z_l) + o_p(1).
\end{eqnarray*}
Hence we obtain that
\begin{eqnarray*}
n \hat{U}^*_{n1}
&=& \frac{2}{n-1}\mathop{\sum\sum}_{1 \leq i< j \leq n }[h^*_1(Z_i,Z_j)+h^*_2(Z_i,Z_j)+h^*_3(Z_i,Z_j)] \\
&&  + n(\hat{\theta}_n-\tilde{\theta}_0)^T \frac{1}{C_n^4} \sum\limits_{i< j< k< l} h^*_{15} (Z_i,Z_j,Z_k,Z_l)+ 4A_\varepsilon E \|X_1 - X_2\| + o_p(1)
\end{eqnarray*}

Now we consider the term $\hat{U}^*_{n2}$. Recall that
\begin{eqnarray*}
\hat{U}^*_{n2}
&=&
\frac{1}{C_n^4}\sum\limits_{i< j< k< l} \left( \frac{1}{6} \sum_{s< t,u< v}^{(i,j,k,l)}\delta^*_{2st}
(\|X_{st}\|+\|X_{uv}\|) - \frac{1}{12} \sum_{(s,t,u)}^{(i,j,k,l)}\delta^*_{2st}\|X_{su}\| \right),
\end{eqnarray*}
where
\begin{eqnarray*}
	\delta^*_{2st}
	&=&  2\int_{0}^{\frac{\hat{m}(X_s)-m(X_s)}{\sigma(X_s,\tilde{\theta}_0)}-\frac{\hat{m}(X_t)
	-m(X_t)}{\sigma(X_t,\tilde{\theta}_0)}+\frac{\varepsilon_s{\sigma(X_s,\hat{\theta}_n)}-{\sigma(X_s,\tilde{\theta}_0)})}
	{\sigma(X_s,\tilde{\theta}_0)}-\frac{\varepsilon_t({\sigma(X_t,\hat{\theta}_n)}-{\sigma(X_t,\tilde{\theta}_0)})}
	{\sigma(X_t,\tilde{\theta}_0)}+R_s-R_t} \\
	&&   \{\mathbb{I}(\varepsilon_s-\varepsilon_t\leq z)-\mathbb{I}(\varepsilon_s\leq \varepsilon_t)\}dz.
\end{eqnarray*}
Following the same line as that for the term $\hat{U}_{n2}$ in Theorem 3.1, it can be decomposed as
\begin{eqnarray*}
     \hat{U}^*_{n2}
     &=& \frac{1}{C_n^4}\sum\limits_{i< j< k< l}{h}^*_{21}(Z_i,Z_j,Z_k,Z_l)+\frac{1}{C_n^4}\sum\limits_{i< j< k< l}{h}^*_{22}(Z_i,Z_j,Z_k,Z_l)\\
     &=:& \hat{U}^*_{n21}+\hat{U}^*_{n22},
\end{eqnarray*}
where
\begin{eqnarray*}
	{h}^*_{21}(Z_i,Z_j,Z_k,Z_l)
	&=&\frac{1}{6}\sum_{s< t,u< v}^{(i,j,k,l)}E(\delta^*_{2st}|X_s,X_t )(\|X_{st}\|+\|X_{uv}\|)-\frac{1}{12}\sum_{(s,t,u)}^{(i,j,k,l)}E(\delta^*_{2st}|X_s,X_t )\|X_{su}\|,\\
	{h}^*_{22}(Z_i,Z_j,Z_k,Z_l)&=&\frac{1}{6}\sum_{s< t,u< v}^{(i,j,k,l)}[{\delta}^*_{2st}-E(\delta^*_{2st}|X_s,X_t )](\|X_{st}\|+\|X_{uv}\|)-\frac{1}{12}\sum_{(s,t,u)}^{(i,j,k,l)}[{\delta}^*_{2st}-E(\delta^*_{2st}|X_s,X_t )]\|X_{su}\|.
\end{eqnarray*}
For $ \hat{U}^*_{n21}$, similar to the arguments in $U_{21}$ in Theorem 3.1, we have uniformly over $1\leq s,t \leq n$,
\begin{eqnarray*}
	&&E[{\delta}^*_{2st}|X_s,X_t ]\\
	&=&2E[\int_{0}^{\frac{\hat{m}(X_s)-m(X_s)}{\sigma(X_s,\tilde{\theta}_0)}-\frac{\hat{m}(X_t)-m(X_t)}{\sigma(X_t,\tilde{\theta}_0)}+\frac{\varepsilon_s({\sigma(X_s,\hat{\theta}_n)}-{\sigma(X_s,\tilde{\theta}_0)})}{\sigma(X_s,\tilde{\theta}_0)}-\frac{\varepsilon_t({\sigma(X_t,\hat{\theta}_n)}-{\sigma(X_t,\tilde{\theta}_0)})}{\sigma(X_t,\tilde{\theta}_0)}+R_s-R_t}\{\mathbb{I}(\varepsilon_s-\varepsilon_t\leq z)\\
	&&-\mathbb{I}(\varepsilon_s\leq \varepsilon_t)\}dz|X_s,X_t ]\\
	&=&Q_{\varepsilon}h^{2k}(\frac{D(X_s)}{\sigma(X_s,\tilde{\theta}_0)}-\frac{D(X_t)}{\sigma(X_t,\tilde{\theta}_0)})^T(\frac{D(X_s)}{\sigma(X_s,\tilde{\theta}_0)}-\frac{D(X_t)}{\sigma(X_t,\tilde{\theta}_0)})\\
	&&+Q_{\varepsilon}h^{k}(\frac{D(X_s)}{\sigma(X_s,\tilde{\theta}_0)}-\frac{D(X_t)}{\sigma(X_t,\tilde{\theta}_0)})^T(\frac{\frac{1}{n-1}\sum_{p=1,p\neq s,t}^{n}\omega_{s,p}\sigma(X_p,\tilde{\theta}_0)\varepsilon_p}{\sigma(X_s,\tilde{\theta}_0)}-\frac{\frac{1}{n-1}\sum_{q=1,q\neq s,t}^{n}\omega_{t,q}\sigma(X_q,\tilde{\theta}_0)\varepsilon_q}{\sigma(X_t,\tilde{\theta}_0)} )\\
	&&+Q_{\varepsilon}(\frac{\frac{1}{n-1}\sum_{p=1,p\neq s,t}^{n}\omega_{s,p}\sigma(X_p,\tilde{\theta}_0)\varepsilon_p}{\sigma(X_s,\tilde{\theta}_0)}-\frac{\frac{1}{n-1}\sum_{q=1,q\neq s,t}^{n}\omega_{t,q}\sigma(X_q,\tilde{\theta}_0)\varepsilon_q}{\sigma(X_t,\tilde{\theta}_0)} )^T\\
	&&(\frac{\frac{1}{n-1}\sum_{p=1,p\neq s,t}^{n}\omega_{s,p}\sigma(X_p,\tilde{\theta}_0)\varepsilon_p}{\sigma(X_s,\tilde{\theta}_0)}
	-\frac{\frac{1}{n-1}\sum_{q=1,q\neq s,t}^{n}\omega_{t,q}\sigma(X_q,\tilde{\theta}_0)\varepsilon_q}{\sigma(X_t,\tilde{\theta}_0)} )\\
	&&+Q_{\varepsilon}h^{k}(\frac{D(X_s)}{\sigma(X_s,\tilde{\theta}_0)}-\frac{D(X_t)}{\sigma(X_t,\tilde{\theta}_0)})^T(\frac{\frac{1}{n-1}\sum_{p=1,p\neq s}^{n}\omega_{s,p}m(X_p)-E[\omega_{s,p}m(X_p)|X_s]}{\sigma(X_s,\tilde{\theta}_0)}\\
	&&-\frac{\frac{1}{n-1}\sum_{q=1,q\neq t}^{n}\omega_{t,q}m(X_q)-E[\omega_{t,q}m(X_q)|X_t]}{\sigma(X_t,\tilde{\theta}_0)} )\\
	&&+Q_{\varepsilon}h^{k}(\frac{D(X_s)}{\sigma(X_s,\tilde{\theta}_0)}-\frac{D(X_t)}{\sigma(X_t,\tilde{\theta}_0)})^T(\frac{\frac{m(X_s)}{n-1}\sum_{p=1,p\neq s}^{n}(\omega_{s,p}-E[\omega_{s,p}|X_s])}{\sigma(X_s,\tilde{\theta}_0)}-\frac{\frac{m(X_t)}{n-1}\sum_{q=1,q\neq t}^{n}(\omega_{t,q}-E[\omega_{t,q}|X_t])}{\sigma(X_t,\tilde{\theta}_0)} )\\
	&&+Q_{\varepsilon}(\frac{\frac{1}{n-1}\sum_{p=1,p\neq s}^{n}\omega_{s,p}m(X_p)-E[\omega_{s,p}m(X_p)|X_s]}{\sigma(X_s,\tilde{\theta}_0)}-\frac{\frac{1}{n-1}\sum_{q=1,q\neq t}^{n}\omega_{t,q}m(X_q)-E[\omega_{t,q}m(X_q)|X_t]}{\sigma(X_t,\tilde{\theta}_0)} )^T\\
	&& \times (\frac{\frac{1}{n-1}\sum_{p=1,p\neq s}^{n}\omega_{s,p}m(X_p)-E[\omega_{s,p}m(X_p)|X_s]}{\sigma(X_s,\tilde{\theta}_0)}-\frac{\frac{1}{n-1}\sum_{q=1,q\neq t}^{n}\omega_{t,q}m(X_q)-E[\omega_{t,q}m(X_q)|X_t]}{\sigma(X_t,\tilde{\theta}_0)})\\     \nonumber
	&&+Q_{\varepsilon}(\frac{\frac{m(X_s)}{n-1}\sum_{p=1,p\neq s}^{n}(\omega_{s,p}-E[\omega_{s,p}|X_s])}{\sigma(X_s,\tilde{\theta}_0)}-\frac{\frac{m(X_t)}{n-1}\sum_{q=1,q\neq t}^{n}(\omega_{t,q}-E[\omega_{t,q}|X_t])}{\sigma(X_t,\tilde{\theta}_0)} )^T\\    \nonumber
	&& \times (\frac{\frac{m(X_s)}{n-1}\sum_{p=1,p\neq s}^{n}(\omega_{s,p}-E[\omega_{s,p}|X_s])}{\sigma(X_s,\tilde{\theta}_0)}-\frac{\frac{m(X_t)}{n-1}\sum_{q=1,q\neq t}^{n}(\omega_{t,q}-E[\omega_{t,q}|X_t])}{\sigma(X_t,\tilde{\theta}_0)} )
	\\
	&&+Q_{\varepsilon}(\frac{\frac{1}{n-1}\sum_{p=1,p\neq s}^{n}\omega_{s,p}m(X_p)-E[\omega_{s,p}m(X_p)|X_s]}{\sigma(X_s,\tilde{\theta}_0)}-\frac{\frac{1}{n-1}\sum_{q=1,q\neq t}^{n}\omega_{t,q}m(X_q)-E[\omega_{t,q}m(X_q)|X_t]}{\sigma(X_t,\tilde{\theta}_0)} )^T\\
	&& \times (\frac{\frac{m(X_s)}{n-1}\sum_{p=1,p\neq s}^{n}(\omega_{s,p}-E[\omega_{s,p}|X_s])}{\sigma(X_s,\tilde{\theta}_0)}-\frac{\frac{m(X_t)}{n-1}\sum_{q=1,q\neq t}^{n}(\omega_{t,q}-E[\omega_{t,q}|X_t])}{\sigma(X_t,\tilde{\theta}_0)} )
	\\
	&&+Q_{\varepsilon}(\frac{\frac{1}{n-1}\sum_{p=1,p\neq s}^{n}\omega_{s,p}m(X_p)-E[\omega_{s,p}m(X_p)|X_s]}{\sigma(X_s,\tilde{\theta}_0)}-\frac{\frac{1}{n-1}\sum_{q=1,q\neq t}^{n}\omega_{t,q}m(X_q)-E[\omega_{t,q}m(X_q)|X_t]}{\sigma(X_t,\tilde{\theta}_0)} )^T\\
	&& \times (\frac{\frac{1}{n-1}\sum_{p=1,p\neq s,t}^{n}\omega_{s,p}\sigma(X_p,\tilde{\theta}_0)\varepsilon_p}{\sigma(X_s,\tilde{\theta}_0)}-\frac{\frac{1}{n-1}\sum_{q=1,q\neq s,t}^{n}\omega_{t,q}\sigma(X_q,\tilde{\theta}_0)\varepsilon_q}{\sigma(X_t,\tilde{\theta}_0)} )
	\\
	&&+Q_{\varepsilon}(\frac{\frac{m(X_s)}{n-1}\sum_{p=1,p\neq s}^{n}(\omega_{s,p}-E[\omega_{s,p}|X_s])}{\sigma(X_s,\tilde{\theta}_0)}-\frac{\frac{m(X_t)}{n-1}\sum_{q=1,q\neq t}^{n}(\omega_{t,q}-E[\omega_{t,q}|X_t])}{\sigma(X_t,\tilde{\theta}_0)} )^T\\
	&& \times (\frac{\frac{1}{n-1}\sum_{p=1,p\neq s,t}^{n}\omega_{s,p}\sigma(X_p,\tilde{\theta}_0)\varepsilon_p}{\sigma(X_s,\tilde{\theta}_0)}-\frac{\frac{1}{n-1}\sum_{q=1,q\neq s,t}^{n}\omega_{t,q}\sigma(X_q,\tilde{\theta}_0)\varepsilon_q}{\sigma(X_t,\tilde{\theta}_0)})\\
	&&   + 2A_{\varepsilon} \left([\frac{\dot{\sigma}(X_s,\tilde{\theta}_0)}
	{\sigma(X_s,\tilde{\theta}_0)}+\frac{\dot{\sigma}(X_t,\tilde{\theta}_0)}{\sigma(X_t,\tilde{\theta}_0)}]^T
	(\hat{\theta}_n-\tilde{\theta}_0) \right) ^2+2A_{\varepsilon}(\hat{\theta}_n-\tilde{\theta}_0)^T [\frac{\dot{\sigma}(X_s,\tilde{\theta}_0)}
	{\sigma(X_s,\tilde{\theta}_0)} +\frac{\dot{\sigma}(X_t,\tilde{\theta}_0)}{\sigma(X_t,\tilde{\theta}_0)}]+o_p(\frac{1}{n}),
\end{eqnarray*}
where $Q_{\varepsilon} = E[f_{\varepsilon}(\varepsilon)]$ and $A_{\varepsilon}=E[\varepsilon F_{\varepsilon}(\varepsilon)]$.
Consequently, we obtain that
\begin{eqnarray}\label{6.22}
     \nonumber
	\hat{U}^*_{n21}   \nonumber
	&=&  h^{2k}  Q_{\varepsilon} \frac{1}{C_n^4}\sum\limits_{i< j< k< l}h^*_{211}(Z_i,Z_j,Z_k,Z_l)  + h^{k}  Q_{\varepsilon} \frac{1}{C_n^4}\sum\limits_{i< j< k< l}h^*_{212}(Z_i,Z_j,Z_k,Z_l) \\  \nonumber
	&& + h^{k} Q_{\varepsilon} \frac{1}{C_n^4}\sum\limits_{i< j< k< l}h^*_{213}(Z_i,Z_j,Z_k,Z_l) + h^{k}  Q_{\varepsilon} \frac{1}{C_n^4}\sum\limits_{i< j< k< l}h^*_{214}(Z_i,Z_j,Z_k,Z_l) \\
	\nonumber
	&& +   Q_{\varepsilon} \frac{1}{C_n^4}\sum\limits_{i< j< k< l}h^*_{215}(Z_i,Z_j,Z_k,Z_l)  +  Q_{\varepsilon} \frac{1}{C_n^4}\sum\limits_{i< j< k< l}h^*_{216}(Z_i,Z_j,Z_k,Z_l) \\
	\nonumber
	&&+  Q_{\varepsilon} \frac{1}{C_n^4}\sum\limits_{i< j< k< l}h^*_{217}(Z_i,Z_j,Z_k,Z_l) +  Q_{\varepsilon} \frac{1}{C_n^4}\sum\limits_{i< j< k< l}h^*_{218}(Z_i,Z_j,Z_k,Z_l)\\
	\nonumber
	&& +  Q_{\varepsilon} \frac{1}{C_n^4}\sum\limits_{i< j< k< l}h^*_{219}(Z_i,Z_j,Z_k,Z_l) +  Q_{\varepsilon} \frac{1}{C_n^4}\sum\limits_{i< j< k< l}h^*_{2110}(Z_i,Z_j,Z_k,Z_l) \\
	\nonumber
&& +2A_{\varepsilon} (\hat{\theta}_n-\tilde{\theta}_0)^T \frac{1}{C_n^4}\sum\limits_{i< j< k< l}h_{2111}^*(Z_i,Z_j,Z_k,Z_l)  \\
	\nonumber
	&&    +2A_{\varepsilon}(\hat{\theta}_n-\tilde{\theta}_0)^T \frac{1}{C_n^4}\sum\limits_{i< j< k< l} h_{2112}^*(Z_i,Z_j,Z_k,Z_l)(\hat{\theta}_n-\tilde{\theta}_0)  + o_p(\frac{1}{n}), \\ \nonumber
	&=:& h^{2k} Q_{\varepsilon} \hat{U}^*_{n211} + h^{k}Q_{\varepsilon} \hat{U}^*_{n212}
	+ h^{k} Q_{\varepsilon} \hat{U}^*_{n213} + h^{k} Q_{\varepsilon} \hat{U}^*_{n214}
	+ Q_{\varepsilon} \hat{U}^*_{n215} +Q_{\varepsilon} \hat{U}^*_{n216}
	+Q_{\varepsilon} \hat{U}^*_{n217}  \\
&&  +Q_{\varepsilon} \hat{U}^*_{n218}+Q_{\varepsilon} \hat{U}^*_{n219} +Q_{\varepsilon} \hat{U}^*_{n2110}
	+2A_{\varepsilon} \hat{U}^*_{n2111} + 2A_{\varepsilon} \hat{U}^*_{n2112}  +o_p(\frac{1}{n}),
\end{eqnarray}
where
\begin{eqnarray*}
	&&   h^*_{21m}(Z_i,Z_j,Z_k,Z_l) \\
	&=&  6^{-1}\sum_{s< t,u< v}^{(i,j,k,l)}\delta^*_{21mst}(\|X_{st}\|+\|X_{uv}\|)-12^{-1}\sum_{(s,t,u)}^{(i,j,k,l)}\delta^*_{21mst}\|X_{su}\|,  \quad {\rm for} \ m=1, \cdots, 12  \\
\end{eqnarray*}
and
\begin{eqnarray*}
\delta^*_{211st}
&=&  (\frac{D(X_s)}{\sigma(X_s,\tilde{\theta}_0)}-\frac{D(X_t)}{\sigma(X_t,\tilde{\theta}_0)})(\frac{D(X_s)}
     {\sigma(X_s,\tilde{\theta}_0)}-\frac{D(X_t)}{\sigma(X_t,\tilde{\theta}_0)})^T \\
\delta^*_{212st}
&=&  \{\frac{\frac{1}{n-1}\sum_{p=1,p\neq s,t}^{n}\omega_{s,p}\sigma(X_p,\tilde{\theta}_0)\varepsilon_p}
     {\sigma(X_s,\tilde{\theta}_0)}-\frac{\frac{1}{n-1}\sum_{q=1,q\neq s,t}^{n}
     \omega_{t,q}\sigma(X_q,\tilde{\theta}_0)\varepsilon_q}{\sigma(X_t,\tilde{\theta}_0)} \}
     \{\frac{D(X_s)}{\sigma(X_s,\tilde{\theta}_0)}-\frac{D(X_t)}{\sigma(X_t,\tilde{\theta}_0)}\}^T \\
\delta^*_{213st}
&=&  (\frac{D(X_s)}{\sigma(X_s,\tilde{\theta}_0)}-\frac{D(X_t)}{\sigma(X_t,\tilde{\theta}_0)})^T(\frac{\frac{m(X_p)}{n-1}
     \sum_{p=1,p\neq s}^{n}(\omega_{s,p}m(X_p)-E[\omega_{s,p}m(X_p)|X_s])}{\sigma(X_s,\tilde{\theta}_0)}\\
&&   -\frac{\frac{m(X_t)}{n-1}\sum_{q=1,q\neq t}^{n}(\omega_{t,q}m(X_q)-E[\omega_{t,q}m(X_q)|X_t])}
     {\sigma(X_t,\tilde{\theta}_0)} )\\
\delta^*_{214st}
&=&  (\frac{D(X_s)}{\sigma(X_s,\tilde{\theta}_0)}-\frac{D(X_t)}{\sigma(X_t,\tilde{\theta}_0)})^T(\frac{\frac{m(X_s)}{n-1}
     \sum_{p=1,p\neq s}^{n}(\omega_{s,p}-E[\omega_{s,p}|X_s])}{\sigma(X_s,\tilde{\theta}_0)}
	 -\frac{\frac{m(X_t)}{n-1}\sum_{q=1,q\neq t}^{n}(\omega_{t,q}-E[\omega_{t,q}|X_t])}{\sigma(X_t,\tilde{\theta}_0)} )\\
\delta^*_{215st}
&=&  \{\frac{\frac{1}{n-1}\sum_{p=1,p\neq s,t}^{n}\omega_{s,p}\sigma(X_p,\tilde{\theta}_0)
	\varepsilon_{p}}{\sigma(X_{s},\tilde{\theta}_0)}-\frac{\frac{1}{n-1}\sum_{q=1,q\neq s,t}^{n}
	\omega_{t,q}\sigma(X_q,\tilde{\theta}_0)\varepsilon_{q}}{\sigma(X_{t},\tilde{\theta}_0)}  \} \\
&&   \times \{\frac{\frac{1}{n-1}\sum_{p'=1,p'\neq s,t}^{n}\omega_{s,p'}\sigma(X_{p'},\tilde{\theta}_0)
	\varepsilon_{p'}}{\sigma(X_{s},\tilde{\theta}_0)}-\frac{\frac{1}{n-1}\sum_{q'=1,q'\neq s,t}^{n}
	\omega_{t,q'}\sigma(X_{q'},\tilde{\theta}_0) \varepsilon_{q'}}{\sigma(X_{t},\tilde{\theta}_0)}  \}^T \\	
\delta^*_{216st}
&=&  (\frac{\frac{1}{n-1}\sum_{p=1,p\neq s}^{n}\omega_{s,p}m(X_p)-E[\omega_{s,p}m(X_p)|X_s]}{\sigma(X_s,\tilde{\theta}_0)}
	 -\frac{\frac{1}{n-1}\sum_{q=1,q\neq t}^{n}\omega_{t,q}m(X_q)-E[\omega_{t,q}m(X_q)|X_t]}{\sigma(X_t,\tilde{\theta}_0)} )^T\\
&&   \times (\frac{\frac{1}{n-1}\sum_{p=1,p\neq s}^{n}\omega_{s,p}m(X_p)-E[\omega_{s,p}m(X_p)|X_s]}{\sigma(X_s,\tilde{\theta}_0)}
	 -\frac{\frac{1}{n-1}\sum_{q=1,q\neq t}^{n}\omega_{t,q}m(X_q)-E[\omega_{t,q}m(X_q)|X_t]}{\sigma(X_t,\tilde{\theta}_0)})\\
\delta^*_{217st}
&=&  (\frac{\frac{m(X_s)}{n-1}\sum_{p=1,p\neq s}^{n}(\omega_{s,p}-E[\omega_{s,p}|X_s])}
     {\sigma(X_s,\tilde{\theta}_0)}-\frac{\frac{m(X_t)}{n-1}\sum_{q=1,q\neq t}^{n}(\omega_{t,q}
     -E[\omega_{t,q}|X_t])}{\sigma(X_t,\tilde{\theta}_0)} )^T\\
&&   \times (\frac{\frac{m(X_s)}{n-1}\sum_{p=1,p\neq s}^{n}(\omega_{s,p}-E[\omega_{s,p}|X_s])}
     {\sigma(X_s,\tilde{\theta}_0)}-\frac{\frac{m(X_t)}{n-1}\sum_{q=1,q\neq t}^{n}
     (\omega_{t,q}-E[\omega_{t,q}|X_t])}{\sigma(X_t,\tilde{\theta}_0)} ) \\
\delta^*_{218st}
&=&  (\frac{\frac{1}{n-1}\sum_{p=1,p\neq s}^{n}\omega_{s,p}m(X_p)-E[\omega_{s,p}m(X_p)|X_s]}
     {\sigma(X_s,\tilde{\theta}_0)}-\frac{\frac{1}{n-1}\sum_{q=1,q\neq t}^{n}\omega_{t,q}m(X_q)
     -E[\omega_{t,q}m(X_q)|X_t]}{\sigma(X_t,\tilde{\theta}_0)} )^T\\
&&   \times (\frac{\frac{m(X_s)}{n-1}\sum_{p=1,p\neq s}^{n}(\omega_{s,p}-E[\omega_{s,p}|X_s])}
     {\sigma(X_s,\tilde{\theta}_0)}-\frac{\frac{m(X_t)}{n-1}\sum_{q=1,q\neq t}^{n}(\omega_{t,q}-E[\omega_{t,q}|X_t])}
     {\sigma(X_t,\tilde{\theta}_0)} ) \\
\delta^*_{219st}
&=&  (\frac{\frac{1}{n-1}\sum_{p=1,p\neq s}^{n}\omega_{s,p}m(X_p)-E[\omega_{s,p}m(X_p)|X_s]}
     {\sigma(X_s,\tilde{\theta}_0)}-\frac{\frac{1}{n-1}\sum_{q=1,q\neq t}^{n}\omega_{t,q}m(X_q)
     -E[\omega_{t,q}m(X_q)|X_t]}{\sigma(X_t,\tilde{\theta}_0)} )^T\\
&&  \times (\frac{\frac{1}{n-1}\sum_{p=1,p\neq s,t}^{n}\omega_{s,p}\sigma(X_p,\tilde{\theta}_0)\varepsilon_p}
    {\sigma(X_s,\tilde{\theta}_0)}-\frac{\frac{1}{n-1}\sum_{q=1,q\neq s,t}^{n}\omega_{t,q}
    \sigma(X_q,\tilde{\theta}_0)\varepsilon_q}{\sigma(X_t,\tilde{\theta}_0)} ) \\
\delta^*_{2110st}
&=&  (\frac{\frac{m(X_s)}{n-1}\sum_{p=1,p\neq s}^{n}(\omega_{s,p}-E[\omega_{s,p}|X_s])}
     {\sigma(X_s,\tilde{\theta}_0)}-\frac{\frac{m(X_t)}{n-1}\sum_{q=1,q\neq t}^{n}(\omega_{t,q}-E[\omega_{t,q}|X_t])}
     {\sigma(X_t,\tilde{\theta}_0)} )^T\\
&&   \times (\frac{\frac{1}{n-1}\sum_{p=1,p\neq s,t}^{n}\omega_{s,p}\sigma(X_p,\tilde{\theta}_0)\varepsilon_p}
     {\sigma(X_s,\tilde{\theta}_0)}-\frac{\frac{1}{n-1}\sum_{q=1,q\neq s,t}^{n}\omega_{t,q}
     \sigma(X_q,\tilde{\theta}_0)\varepsilon_q}{\sigma(X_t,\tilde{\theta}_0)})	\\
\delta^*_{2111st}
&=&  \frac{\dot{\sigma}(X_s,\tilde{\theta}_0)}{\sigma(X_s,\tilde{\theta}_0)}
	 + \frac{\dot{\sigma}(X_t,\tilde{\theta}_0)}{\sigma(X_t,\tilde{\theta}_0)}  \\
\delta^*_{2112st}
&=&  \{\frac{\dot{\sigma}(X_s,\tilde{\theta}_0)}{\sigma(X_s,\tilde{\theta}_0)}
	 + \frac{\dot{\sigma}(X_t,\tilde{\theta}_0)}{\sigma(X_t,\tilde{\theta}_0)} \}\{\frac{\dot{\sigma}(X_s,\tilde{\theta}_0)}
     {\sigma(X_s,\tilde{\theta}_0)} + \frac{\dot{\sigma}(X_t,\tilde{\theta}_0)}{\sigma(X_t,\tilde{\theta}_0)} \}^T .
\end{eqnarray*}

For the term $\hat{U}^*_{n211}$,  by the law of large numbers, we have
\begin{eqnarray*}
\frac{1}{C_n^4}\sum\limits_{i< j< k< l}h^*_{211}(Z_i,Z_j,Z_k,Z_l) \longrightarrow E[h^*_{211}(Z_i,Z_j,Z_k,Z_l)], \quad {\rm in \ probability,}
\end{eqnarray*}
where
\begin{equation*}
E[h^*_{211}(Z_i,Z_j,Z_k,Z_l)] = E[(\frac{D(X_1)}{\sigma(X_1,\tilde{\theta}_0)}-\frac{D(X_2)}{\sigma(X_2,\tilde{\theta}_0)})(\frac{D(X_1)}{\sigma(X_1,\tilde{\theta}_0)}-\frac{D(X_2)}{\sigma(X_2,\tilde{\theta}_0)})^T C_x(X_1,X_2)],
\end{equation*}
and $C_x(X_1,X_2) = \|X_1-X_2\|-E(\|X_1-X_2\||X_1)-E(\|X_1-X_2\||X_2)+E(\|X_1-X_2\|)$.
Together with the assumption 6(d),  we have
$$n h^{2k} Q_{\varepsilon} \hat{U}^*_{n211}= n h^{2k} Q_{\varepsilon} \frac{1}{C_n^4}\sum\limits_{i< j< k< l}h^*_{211}(Z_i,Z_j,Z_k,Z_l)= o_p(1).$$

For the term $\hat{U}^*_{n212}$, it can be decomposed as
\begin{eqnarray*}
\hat{U}^*_{n212}
&=& \frac{1}{n(n-1)^2(n-2)^2(n-3)}\sum_{i=1}^{n}\sum_{j=1.j\neq i}^{n}\sum_{k=1,k\neq i,j }^{n} \sum_{l=1,l\neq i,j,k }^{n}
    \sum_{p=1,p\neq i,j}^{n}\sum_{q=1,q\neq i,j}^{n} \{ \frac{\omega_{p,i}\sigma(X_p,\tilde{\theta}_0)
    \varepsilon_p}{\sigma(X_i,\tilde{\theta}_0)}\\
&&  -\frac{\omega_{q,j}\sigma(X_q,\tilde{\theta}_0)\varepsilon_q}{\sigma(X_j,\tilde{\theta}_0)} \}\{\frac{D(X_s)}
    {\sigma(X_s,\tilde{\theta}_0)}-\frac{D(X_t)}{\sigma(X_t,\tilde{\theta}_0)}\}^T(\|X_{ij}\|+\|X_{kl}\|-\|X_{ik}\|)\\
&=& \frac{1}{n(n-1)^2(n-2)^2(n-3)}\sum_{i=1}^{n}\sum_{j\neq i}^{n}\sum_{k\neq i,j }^{n} \sum_{l\neq i,j,k }^{n}
    \sum_{p=q\neq i,j}^{n} \{ \frac{\omega_{p,i}\sigma(X_p,\tilde{\theta}_0)\varepsilon_p}{\sigma(X_i,\tilde{\theta}_0)}
    -\frac{\omega_{q,j}\sigma(X_q,\tilde{\theta}_0)\varepsilon_q}{\sigma(X_j,\tilde{\theta}_0)}\}\\
&&  \times \{\frac{D(X_s)}{\sigma(X_s,\tilde{\theta}_0)}
	-\frac{D(X_t)}{\sigma(X_t,\tilde{\theta}_0)}\}^T(\|X_{ij}\|+\|X_{kl}\|-2\|X_{ik}\|)\\
&&  +\frac{1}{n(n-1)^2(n-2)^2(n-3)}\sum_{i=1}^{n}\sum_{j\neq i}^{n}\sum_{k\neq i,j }^{n} \sum_{l\neq i,j,k }^{n}\sum_{p=k }
    \sum_{q\neq i,j,k}^{n}\{ \frac{\omega_{k,i}\sigma(X_k,\tilde{\theta}_0)\varepsilon_k}{\sigma(X_i,\tilde{\theta}_0)}-\frac{\omega_{q,j}\sigma(X_q,\tilde{\theta}_0)\varepsilon_q}{\sigma(X_j,\tilde{\theta}_0)} \} \\
&& \times \{\frac{D(X_s)}{\sigma(X_s,\tilde{\theta}_0)}-\frac{D(X_t)}{\sigma(X_t,\tilde{\theta}_0)}\}^T(\|X_{ij}\|
    +\|X_{kl}\|-2\|X_{ik}\|)\\
&&  +\frac{1}{n(n-1)^2(n-2)^2(n-3)}\sum_{i=1}^{n}\sum_{j\neq i}^{n}\sum_{k\neq i,j }^{n} \sum_{l\neq i,j,k }^{n}
   \sum_{p=l}\sum_{q\neq i,j,l}^{n}\{ \frac{\omega_{p,i}\sigma(X_p,\tilde{\theta}_0)\varepsilon_p}{\sigma(X_i,\tilde{\theta}_0)}
   -\frac{\omega_{p,j}\sigma(X_p,\tilde{\theta}_0)\varepsilon_p}{\sigma(X_j,\tilde{\theta}_0)} \} 	\\
&& \times \{\frac{D(X_s)}{\sigma(X_s,\tilde{\theta}_0)}-\frac{D(X_t)}{\sigma(X_t,\tilde{\theta}_0)}\}^T
    (\|X_{ij}\|+\|X_{kl}\|-2\|X_{ik}\|)\\
&& +\frac{1}{n(n-1)^2(n-2)^2(n-3)}\sum_{i=1}^{n}\sum_{j\neq i}^{n}\sum_{k\neq i,j }^{n} \sum_{l\neq i,j,k }^{n}\sum_{q=k }
   \sum_{p\neq i,j,k }^{n}\{ \frac{\omega_{p,i}\sigma(X_p,\tilde{\theta}_0)\varepsilon_p}{\sigma(X_i,\tilde{\theta}_0)}
	-\frac{\omega_{q,j}\sigma(X_q,\tilde{\theta}_0)\varepsilon_q}{\sigma(X_j,\tilde{\theta}_0)} \}	\\
&& \times \{\frac{D(X_s)}{\sigma(X_s,\tilde{\theta}_0)}-\frac{D(X_t)}{\sigma(X_t,\tilde{\theta}_0)}\}^T
    (\|X_{ij}\|+\|X_{kl}\|-2\|X_{ik}\|)\\
&& +\frac{1}{n(n-1)^2(n-2)^2(n-3)}\sum_{i=1}^{n}\sum_{j\neq i}^{n}\sum_{k\neq i,j }^{n} \sum_{l\neq i,j,k }^{n}\sum_{q=l }
   \sum_{p\neq i,j,l }^{n}\{ \frac{\omega_{p,i}\sigma(X_p,\tilde{\theta}_0)\varepsilon_p}{\sigma(X_i,\tilde{\theta}_0)}
   -\frac{\omega_{q,j}\sigma(X_q,\tilde{\theta}_0)\varepsilon_q}{\sigma(X_j,\tilde{\theta}_0)} \} \\
&& \times \{\frac{D(X_s)}{\sigma(X_s,\tilde{\theta}_0)}-\frac{D(X_t)}{\sigma(X_t,\tilde{\theta}_0)}\}^T
    (\|X_{ij}\|+\|X_{kl}\|-2\|X_{ik}\|) \\
&&  +\frac{1}{n(n-1)^2(n-2)^2(n-3)}\sum_{i=1}^{n}\sum_{j\neq i}^{n}\sum_{k\neq i,j }^{n} \sum_{l\neq i,j,k }^{n}
    \sum_{p\neq i\neq j\neq k\neq l}^{n}\sum_{q\neq i\neq j\neq k\neq l \neq p}^{n}\{ \frac{\omega_{p,i}\sigma(X_p,\tilde{\theta}_0)\varepsilon_p}{\sigma(X_i,\tilde{\theta}_0)}\\
&&  -\frac{\omega_{q,j}\sigma(X_q,\tilde{\theta}_0)\varepsilon_q}{\sigma(X_j,\tilde{\theta}_0)} \}
    \{\frac{D(X_s)}{\sigma(X_s,\tilde{\theta}_0)}-\frac{D(X_t)}{\sigma(X_t,\tilde{\theta}_0)}\}^T
    (\|X_{ij}\|+\|X_{kl}\|-2\|X_{ik}\|) \\
&=:& O_p(\frac{1}{n}) +\hat{U}^*_{n2121},
\end{eqnarray*}
where
\begin{eqnarray*}
\hat{U}^*_{n2121}
&=& \frac{1}{n(n-1)^2(n-2)^2(n-3)}\sum_{i=1}^{n}\sum_{j\neq i}^{n}\sum_{k\neq i,j }^{n} \sum_{l\neq i,j,k }^{n}
    \sum_{p\neq i\neq j\neq k\neq l}^{n}\sum_{q\neq i\neq j\neq k\neq l \neq p}^{n}\{ \frac{\omega_{p,i}\sigma(X_p,\tilde{\theta}_0)\varepsilon_p}{\sigma(X_i,\tilde{\theta}_0)}\\
&&  -\frac{\omega_{q,j}\sigma(X_q,\tilde{\theta}_0)\varepsilon_q}{\sigma(X_j,\tilde{\theta}_0)} \}
    \{\frac{D(X_s)}{\sigma(X_s,\tilde{\theta}_0)}-\frac{D(X_t)}{\sigma(X_t,\tilde{\theta}_0)}\}^T
    (\|X_{ij}\|+\|X_{kl}\|-2\|X_{ik}\|)\\
&=& \frac{(n-4)(n-5)}{(n-1)(n-2)}\frac{1}{C_n^6}\mathop{\sum\sum\sum\sum\sum\sum}_{1 \leq i< j< k< l< p< q \leq n }
    {\delta}^*_{2121}(Z_i,Z_j,Z_k,Z_l,Z_p,Z_q),
\end{eqnarray*}
with
\begin{eqnarray*}
&&   {\delta}^*_{2121}(Z_i,Z_j,Z_k,Z_l,Z_p,Z_q) \\
&=&  \frac{1}{6!}\sum_{(s,t,u,v,r,m)}^{(i,j,k,l,p,q)} \{ \frac{\omega_{r,s}\sigma(X_r,\tilde{\theta}_0)\varepsilon_r}
     {\sigma(X_s,\tilde{\theta}_0)}-\frac{\omega_{m,t}\sigma(X_m,\tilde{\theta}_0)\varepsilon_m}
     {\sigma(X_t,\tilde{\theta}_0)} \} \{\frac{D(X_s)}{\sigma(X_s,\tilde{\theta}_0)}-\frac{D(X_t)}
     {\sigma(X_t,\tilde{\theta}_0)}\}^T \\
&&   (\|X_{st}\|+\|X_{uv}\|-2\|X_{su}\|).
\end{eqnarray*}
Note that $E[\delta^*_{2121}(Z_i,Z_j,Z_k,Z_l,Z_p,Z_q) ] = 0$, it follows that $n h^k \hat{U}^*_{n2121}=o_p(1)$.
Consequently, we obtain that
$$ nh^k\hat{U}^*_{n212}=o_p(1). $$
Similarly, we can show that
$$ nh^k\hat{U}^*_{n213}=o_p(1) \quad  {\rm and} \quad nh^k\hat{U}^*_{n214}=o_p(1). $$

For the term $\hat{U}^*_{n215}$, recall that
\begin{eqnarray*}
\hat{U}^*_{n215}  \nonumber
&=&    \frac{1}{C_n^4}\sum\limits_{i< j< k< l} \left( \frac{1}{6} \sum_{s< t,u< v}^{(i,j,k,l)}{\delta}^*_{215}
(\|X_{st}\|+\|X_{uv}\|) - \frac{1}{12} \sum_{(s,t,u)}^{(i,j,k,l)}{\delta}^*_{215}\|X_{su}\| \right)
\end{eqnarray*}
with
\begin{eqnarray*}
{\delta}^*_{215}
&=& \{\frac{\frac{1}{n-1}\sum_{p=1,p\neq s,t}^{n}
    \omega_{s,p}\sigma(X_p,\tilde{\theta}_0)\varepsilon_{p}}{\sigma(X_{s},\tilde{\theta}_0)}-\frac{\frac{1}{n-1}
    \sum_{q=1,q\neq s,t}^{n}\omega_{t,q}\sigma(X_q,\tilde{\theta}_0)\varepsilon_{q}}{\sigma(X_{t},\tilde{\theta}_0)}  \} \\
&& \times \{\frac{\frac{1}{n-1}\sum_{p'=1,p'\neq s,t}^{n}\omega_{s,p'}\sigma(X_{p'},\tilde{\theta}_0)
   \varepsilon_{p'}}{\sigma(X_{s},\tilde{\theta}_0)}-\frac{\frac{1}{n-1}\sum_{q'=1,q'\neq s,t}^{n} \omega_{t,q'}\sigma(X_{q'},\tilde{\theta}_0)\varepsilon_{q'}}{\sigma(X_{t},\tilde{\theta}_0)}  \}^T
\end{eqnarray*}
By some tedious calculations, we have
\begin{eqnarray*}
&&\hat{U}^*_{n215} \\
&=&  \frac{1}{n(n-1)^3(n-2)^3(n-3)}\sum_{i=1}^{n}\sum_{j=1,j\neq i}^{n}\sum_{k=1,k\neq i,j }^{n} \sum_{l=1,l\neq i,j,k }^{n}
     \sum_{p=1,p\neq i,j}^{n}\sum_{q=1,q\neq i,j}^{n}\sum_{p'=1,p'\neq i,j}^{n}\sum_{q'=1,q'\neq i,j}^{n}\\
&&   \{ \frac{\omega_{p,i}\sigma(X_p,\tilde{\theta}_0)\varepsilon_p}{\sigma(X_i,\tilde{\theta}_0)}-\frac{\omega_{q,j}
    \sigma(X_q,\tilde{\theta}_0)\varepsilon_q}{\sigma(X_j,\tilde{\theta}_0)} \}
    \{ \frac{\omega_{p',i}\sigma(X_{p'},\tilde{\theta}_0)\varepsilon_{p'}}{\sigma(X_i,\tilde{\theta}_0)}
    -\frac{\omega_{q',j}\sigma(X_{q'},\tilde{\theta}_0)\varepsilon_{q'}}{\sigma(X_j,\tilde{\theta}_0)}\}^T
    (\|X_{ij}\|+\|X_{kl}\|-2\|X_{ik}\|)\\
&=&\frac{(n-4)(n-5)(n-6)(n-7)}{(n-1)^2(n-2)^2}\frac{1}{C_n^8}\mathop{\sum\sum\sum\sum\sum\sum\sum\sum}_{1\leq i< j< k< l< p< p'<
   q < q'\leq n} {\delta}^*_{215}(Z_i,Z_j,Z_k,Z_l,Z_p,Z_{p'},Z_q,Z_{q'})+o_p(\frac{1}{n})\\
&=:&\hat{U}^*_{n2151}+o_p(\frac{1}{n}),
\end{eqnarray*}
where
\begin{eqnarray*}
&&{\delta}^*_{215}(Z_i,Z_j,Z_k,Z_l,Z_p,Z_{p'},Z_q,Z_{q'})\\
&=&\frac{1}{8!}\sum_{(s,t,u,v,r,m,n,e)}^{(i,j,k,l,p,p',q,q')}\{ \frac{\omega_{r,s}\sigma(X_r,\tilde{\theta}_0)\varepsilon_r}{\sigma(X_s,\tilde{\theta}_0)}
-\frac{\omega_{m,t}\sigma(X_m,\tilde{\theta}_0)\varepsilon_m}{\sigma(X_t,\tilde{\theta}_0)}\} \{ \frac{\omega_{n,s}\sigma(X_{n},\tilde{\theta}_0)\varepsilon_{n}}{\sigma(X_s,\tilde{\theta}_0)}
-\frac{\omega_{e,t}\sigma(X_{e},\tilde{\theta}_0)\varepsilon_{e}}{\sigma(X_t,\tilde{\theta}_0)}\}\\
&& \times (\|X_{st}\|+\|X_{uv}\|-2\|X_{su}\|).
\end{eqnarray*}
Some elementary calculations show that $E[\delta^*_{215}(Z_i,Z_j,Z_k,Z_l,Z_p,Z_{p'},Z_q,Z_{q'})]  =0 $ and $\hat{U}^*_{n2151}$ is degenerate of order $1$. By the arguments in Section 5.3.4 of Serfling (2009), we can obtain
$$
n\hat{U}^*_{n2151}=\frac{2}{(n-1)}\mathop{\sum\sum}_{1 \leq i< j \leq n } \tilde{h}^{*}_5(Z_i,Z_j)+o_p(1),
$$
where  $\tilde{h}^*_5(Z_i,Z_j)=\frac{1}{2}(H_{7ij}+H_{8ij})$ and
\begin{eqnarray*}
H_{7ij}&=&\frac{1}{6!}E[\sum_{(k_1,l_1,p_1,q_1,{{p}_1'},{{q}_1'})}^{(k,l,p,q,{p'},{q'})}\{ \frac{\omega_{i,p_1}\sigma(X_i,\tilde{\theta}_0)\varepsilon_i}{\sigma(X_{p_1},\tilde{\theta}_0)}
	-\frac{\omega_{q_1,p_1'}\sigma(X_{q_1},\tilde{\theta}_0)\varepsilon_{q_1}}{\sigma(X_{p_1'},\tilde{\theta}_0)}\}
   \{ \frac{\omega_{j,p_1}\sigma(X_j,\tilde{\theta}_0)\varepsilon_{j}}{\sigma(X_{p_1},\tilde{\theta}_0)}
   -\frac{\omega_{q_1',p_1'}\sigma(X_{q_1'},\tilde{\theta}_0)\varepsilon_{q_1'}}{\sigma(X_{p_1}',\tilde{\theta}_0)}\}\\
&&  (\|X_{(p_1)}^{(p_1')}\|+\|X_{(k_1)}^{(l_1)}\|-2\|X_{(p_1)}^{(k_1)}\|)|Z_i,Z_j],\\
H_{8ij} &=&\frac{1}{6!}E[\sum_{(k_1,l_1,p_1,q_1,{{p}_1'},{{q}_1'})}^{(k,l,p,q,{p'},{q'})}
      \{ \frac{\omega_{i,p_1}\sigma(X_i,\tilde{\theta}_0)\varepsilon_i}{\sigma(X_{p_1},\tilde{\theta}_0)}
	  -\frac{\omega_{q_1,q_1'}\sigma(X_{q_1},\tilde{\theta}_0)\varepsilon_{q_1}}{\sigma(X_{q_1'},\tilde{\theta}_0)}\}
      \{ \frac{\omega_{p_1',p_1}\sigma(X_{p_1'},\tilde{\theta}_0)\varepsilon_{p_1'}}{\sigma(X_{p_1},\tilde{\theta}_0)}
      -\frac{\omega_{j,q_1'}\sigma(X_j,\tilde{\theta}_0)\varepsilon_{j}}{\sigma(X_{q_1'},\tilde{\theta}_0)}\} \\
&&    (\|X_{(p_1)}^{(q_1')}\|+\|X_{(k_1)}^{(l_1)}\|-2\|X_{(p_1)}^{(k_1)}\|)|Z_i,Z_j].
\end{eqnarray*}
For the term $\hat{U}^*_{n21m},m=6,\cdots,10$,  similar to the arguments for $ \hat{U}^*_{n215} $, we have
\begin{eqnarray*}
	n \hat{U}^*_{n216}=\frac{2}{n-1}\mathop{\sum\sum}_{1 \leq i< j \leq n }\tilde{h}^*_{6}(Z_i,Z_j)+o_p(1), \\
	n \hat{U}^*_{n217}=\frac{2}{n-1}\mathop{\sum\sum}_{1 \leq i< j \leq n }\tilde{h}^*_{7}(Z_i,Z_j)+o_p(1),\\
	n \hat{U}^*_{n218}=\frac{2}{n-1}\mathop{\sum\sum}_{1 \leq i< j \leq n }\tilde{h}^*_{8}(Z_i,Z_j)+o_p(1),\\
	n \hat{U}^*_{n219}=\frac{2}{n-1}\mathop{\sum\sum}_{1 \leq i< j \leq n }\tilde{h}^*_{9}(Z_i,Z_j)+o_p(1),\\
	n \hat{U}^*_{n2110}=\frac{2}{n-1}\mathop{\sum\sum}_{1 \leq i< j \leq n }\tilde{h}^*_{10}(Z_i,Z_j)+o_p(1),
\end{eqnarray*}
where $\tilde{h}^*_6(Z_i,Z_j)=\frac{1}{2}(H_{9ij}+H_{10ij})$, $\tilde{h}^*_7(Z_i,Z_j)=\frac{1}{2}(H_{11ij}+H_{12ij})$, $\tilde{h}^*_8(Z_i,Z_j)=\frac{1}{2}(H_{13ij}+H_{14ij})$,
$\tilde{h}^*_9(Z_i,Z_j)=\frac{1}{2}(H_{15ij}+H_{16ij})$,
$\tilde{h}^*_{10}(Z_i,Z_j)=\frac{1}{2}(H_{17ij}+H_{18ij})$,
\begin{eqnarray*}
	H_{9ij}&=&\frac{1}{6!}E[\sum_{(k_1,l_1,p_1,q_1,{{p}_1'},{{q}_1'})}^{(k,l,p,q,{p'},{q'})}\{ \frac{\omega_{i,p_1}m(X_i)-E[\omega_{i,p_1}m(X_i)|X_{p_1}]}{\sigma(X_{p_1},\tilde{\theta}_0)}
	-\frac{\omega_{p_1',q_1}m(X_{p_1'})-E[\omega_{p_1',q_1}m(X_{p_1'})|X_{q_1}]}{\sigma(X_{q_1},\tilde{\theta}_0)}\}\\
	&&  \times  \{ \frac{\omega_{j,p_1}m(X_j)-E[\omega_{j,p_1}m(X_j)|X_{p_1}]}{\sigma(X_{p_1},\tilde{\theta}_0)}-\frac{\omega_{q'_1,q_1}m(X_{q'_1})
-E[\omega_{q'_1,q_1}m(X_{q'_1})|X_{q_1}]}{\sigma(X_{q_1},\tilde{\theta}_0)}\}^T(\|X_{p_1q_1}\|+\|X_{k_1l_1}\|\\
	&&-2\|X_{p_1k_1}\|)|Z_i,Z_j],\\
	H_{10ij}&=&\frac{1}{6!}E[\sum_{(k_1,l_1,p_1,q_1,{{p}_1'},{{q}_1'})}^{(k,l,p,q,{p'},{q'})}\{ \frac{\omega_{i,p_1}m(X_i)-E[\omega_{i,p_1}m(X_i)|X_{p_1}]}{\sigma(X_{p_1},\tilde{\theta}_0)}
	-\frac{\omega_{p_1',q_1}m(X_{p_1'})-E[\omega_{p_1',q_1}m(X_{p_1'})|X_{q_1}]}{\sigma(X_{q_1},\tilde{\theta}_0)}\}\\
	&& \times  \{ \frac{\omega_{q'_1,p_1}m(X_{q'_1})-E[\omega_{q'_1,p_1}m(X_{q'_1})|X_{p_1}]}{\sigma(X_{p_1},\tilde{\theta}_0)}
  -\frac{\omega_{j,q_1}m(X_j)-E[\omega_{j,q_1}m(X_j)|X_{q_1}]}{\sigma(X_{q_1},\tilde{\theta}_0)}\}^T(\|X_{p_1q_1}\|+\|X_{k_1l_1}\|\\
	&&-2\|X_{p_1k_1}\|)|Z_i,Z_j],\\
	H_{11ij}&=&\frac{1}{6!}E[\sum_{(k_1,l_1,p_1,q_1,{{p}_1'},{{q}_1'})}^{(k,l,p,q,{p'},{q'})}\{ \frac{(\omega_{i,p_1}-E[\omega_{i,p_1}|X_{p_1}])m(X_{p_1})}{\sigma(X_{p_1},\tilde{\theta}_0)}
	-\frac{(\omega_{p_1',q_1}-E[\omega_{p_1',q_1}|X_{q_1}])m(X_{q_1})}{\sigma(X_{q_1},\tilde{\theta}_0)}\}
	\\ && \times
	\{ \frac{(\omega_{j,p_1}-E[\omega_{j,p_1}|X_{p_1}])m(X_{p_1})}{\sigma(X_{p_1},\tilde{\theta}_0)}
	-\frac{(\omega_{q'_1,q_1} -E[\omega_{q'_1,q_1}|X_{q_1}])m(X_{q_1})}{\sigma(X_{q_1},\tilde{\theta}_0)}\}^T(\|X_{p_1q_1)}\|
	\\ &&
	+\|X_{k_1l_1}\|-2\|X_{p_1k_1}\|)|Z_i,Z_j],\\
	H_{12ij}&=&\frac{1}{6!}E[\sum_{(k_1,l_1,p_1,q_1,{{p}_1'},{{q}_1'})}^{(k,l,p,q,{p'},{q'})}\{ \frac{(\omega_{i,p_1}-E[\omega_{i,p_1}|X_{p_1}])m(X_{p_1})}{\sigma(X_{p_1},\tilde{\theta}_0)}
	-\frac{(\omega_{p_1',q_1}-E[\omega_{p_1',q_1}|X_{q_1}])m(X_{q_1})}{\sigma(X_{q_1},\tilde{\theta}_0)}\}\\
	&& \times \{ \frac{(\omega_{q'_1,p_1}-E[\omega_{q'_1,p_1}|X_{p_1}])m(X_{p_1})}{\sigma(X_{p_1},\tilde{\theta}_0)}
	-\frac{(\omega_{j,q_1}-E[\omega_{j,q_1}|X_{q_1}])m(X_{q_1})}{\sigma(X_{q_1},\tilde{\theta}_0)}\}^T(\|X_{p_1q_1}\|\\
	&&+\|X_{k_1l_1}\|-2\|X_{p_1k_1}\|)|Z_i,Z_j],\\
	H_{13ij}&=&\frac{1}{6!}E[\sum_{(k_1,l_1,p_1,q_1,{{p}_1'},{{q}_1'})}^{(k,l,p,q,{p'},{q'})}\{ \frac{\omega_{i,p_1}m(X_i)-E[\omega_{i,p_1}m(X_i)|X_{p_1}]}{\sigma(X_{p_1},\tilde{\theta}_0)}
	-\frac{\omega_{p_1',q_1}m(X_{p_1'})-E[\omega_{p_1',q_1}m(X_{p_1'})|X_{q_1}]}{\sigma(X_{q_1},\tilde{\theta}_0)}\}\\
	&&  \times  \{ \frac{(\omega_{j,p_1}-E[\omega_{j,p_1}|X_{p_1}])m(X_{p_1})}{\sigma(X_{p_1},\tilde{\theta}_0)}-\frac{(\omega_{q'_1,q_1}
-E[\omega_{q'_1,q_1}|X_{q_1}])m(X_{q_1})}{\sigma(X_{q_1},\tilde{\theta}_0)}\}^T(\|X_{p_1q_1}\|+\|X_{k_1l_1}\|\\
	&&-2\|X_{p_1k_1}\|)|Z_i,Z_j],\\
	H_{14ij}&=&\frac{1}{6!}E[\sum_{(k_1,l_1,p_1,q_1,{{p}_1'},{{q}_1'})}^{(k,l,p,q,{p'},{q'})}\{ \frac{\omega_{i,p_1}m(X_i)-E[\omega_{i,p_1}m(X_i)|X_{p_1}]}{\sigma(X_{p_1},\tilde{\theta}_0)}
	-\frac{\omega_{p_1',q_1}m(X_{p_1'})-E[\omega_{p_1',q_1}m(X_{p_1'})|X_{q_1}]}{\sigma(X_{q_1},\tilde{\theta}_0)}\}\\
	&&  \times  \{ \frac{(\omega_{q'_1,p_1}-E[\omega_{q'_1,p_1}|X_{p_1}])m(X_{p_1})}{\sigma(X_{p_1},\tilde{\theta}_0)}-\frac{(\omega_{j,q_1}
-E[\omega_{j,q_1}|X_{q_1}])m(X_{q_1})}{\sigma(X_{q_1},\tilde{\theta}_0)}\}^T(\|X_{p_1q_1}\|+\|X_{k_1l_1}\| \\
	&&-2\|X_{p_1k_1}\|)|Z_i,Z_j],\\
	H_{15ij}&=&\frac{1}{6!}E[\sum_{(k_1,l_1,p_1,q_1,{{p}_1'},{{q}_1'})}^{(k,l,p,q,{p'},{q'})}\{ \frac{\omega_{i,p_1}m(X_i)-E[\omega_{i,p_1}m(X_i)|X_{p_1}]}{\sigma(X_{p_1},\tilde{\theta}_0)}
	-\frac{\omega_{p_1',q_1}m(X_{p_1'})-E[\omega_{p_1',q_1}m(X_{p_1'})|X_{q_1}]}{\sigma(X_{q_1},\tilde{\theta}_0)}\}\\
	&& \times  \{ \frac{\omega_{j,p_1}\sigma(X_j,\tilde{\theta}_0)\varepsilon_j}{\sigma(X_{p_1},\tilde{\theta}_0)}
    -\frac{\omega_{q'_1,q_1}\sigma(X_{q'_1},\tilde{\theta}_0)\varepsilon_{q'_1}}{\sigma(X_{q_1},\tilde{\theta}_0)}\}^T
    (\|X_{p_1q_1}\| +\|X_{k_1l_1}\|-2\|X_{p_1k_1}\|)|Z_i,Z_j],\\
	H_{16ij}&=&\frac{1}{6!}E[\sum_{(k_1,l_1,p_1,q_1,{{p}_1'},{{q}_1'})}^{(k,l,p,q,{p'},{q'})}\{ \frac{\omega_{i,p_1}m(X_i)-E[\omega_{i,p_1}m(X_i)|X_{p_1}]}{\sigma(X_{p_1},\tilde{\theta}_0)}
	-\frac{\omega_{p_1',q_1}m(X_{p_1'})-E[\omega_{p_1',q_1}m(X_{p_1'})|X_{q_1}]}{\sigma(X_{q_1},\tilde{\theta}_0)}\}\\
	&& \times  \{ \frac{\omega_{q'_1,p_1}\sigma(X_{q'_1},\tilde{\theta}_0)\varepsilon_{q'_1}}{\sigma(X_{p_1},\tilde{\theta}_0)}
   -\frac{\omega_{j,q_1}\sigma(X_{j},\tilde{\theta}_0)\varepsilon_{j}}{\sigma(X_{q_1},\tilde{\theta}_0)}\}^T
   (\|X_{p_1q_1}\|+\|X_{k_1l_1}\|-2\|X_{p_1k_1}\|)|Z_i,Z_j],\\
	H_{17ij}&=&\frac{1}{6!}E[\sum_{(k_1,l_1,p_1,q_1,{{p}_1'},{{q}_1'})}^{(k,l,p,q,{p'},{q'})}\{
    \frac{(\omega_{i,p_1}-E[\omega_{i,p_1}|X_{p_1}])m(X_{p_1})}{\sigma(X_{p_1},\tilde{\theta}_0)}
	-\frac{(\omega_{p_1',q_1}-E[\omega_{p_1',q_1}|X_{q_1}])m(X_{q_1})}{\sigma(X_{q_1},\tilde{\theta}_0)}\}\\
&& \times \{ \frac{\omega_{j,p_1}\sigma(X_j,\tilde{\theta}_0)\varepsilon_j}{\sigma(X_{p_1},\tilde{\theta}_0)}
    -\frac{\omega_{q'_1,q_1}\sigma(X_{q'_1},\tilde{\theta}_0)\varepsilon_{q'_1}}{\sigma(X_{q_1},\tilde{\theta}_0)}\}^T
    (\|X_{p_1q_1}\|+\|X_{k_1l_1}\|-2\|X_{p_1k_1}\|)|Z_i,Z_j],\\
H_{18ij}&=&\frac{1}{6!}E[\sum_{(k_1,l_1,p_1,q_1,{{p}_1'},{{q}_1'})}^{(k,l,p,q,{p'},{q'})}
    \{ \frac{(\omega_{i,p_1}-E[\omega_{i,p_1}|X_{p_1}])m(X_{p_1})}{\sigma(X_{p_1},\tilde{\theta}_0)}
	-\frac{(\omega_{p_1',q_1}-E[\omega_{p_1',q_1}|X_{q_1}])m(X_{q_1})}{\sigma(X_{q_1},\tilde{\theta}_0)}\} \\
&& \times \{ \frac{\omega_{q'_1,p_1}\sigma(X_{q'_1},\tilde{\theta}_0)\varepsilon_{q'_1}}{\sigma(X_{p_1},\tilde{\theta}_0)}
    -\frac{\omega_{j,q_1}\sigma(X_{j},\tilde{\theta}_0)\varepsilon_{j}}{\sigma(X_{q_1},\tilde{\theta}_0)}\}^T(\|X_{p_1q_1}\|
    +\|X_{k_1l_1}\|-2\|X_{p_1k_1}\|)|Z_i,Z_j].
\end{eqnarray*}

For the last two terms $\hat{U}^*_{n2111}$ and $\hat{U}^*_{n2112}$, similar to the arguments for $ \hat{U}_{n212}$ and $ \hat{U}_{n213}$  in
the proof of Theorem 3.1, we can obtain that
\begin{eqnarray*}
	n \hat{U}^*_{n2111} &=&  \sqrt{n}(\hat{\theta}_n-\tilde{\theta}_0) \frac{1}{\sqrt{n}}\sum_{i=1}^{n}4E[h^*_{2111}(Z_i,Z_j,Z_k,Z_l|Z_i)]+o_p(1).\\
	n \hat{U}^*_{n2112} &=& \sqrt{n}(\hat{\theta}_n-\tilde{\theta}_0)^T M_2 \sqrt{n}(\hat{\theta}_n-\theta_0)+o_p(1).
\end{eqnarray*}
where
\begin{eqnarray*}
&&E[h^*_{2111}(Z_i,Z_j,Z_k,Z_l)|Z_i]=
E[\{\frac{\dot{\sigma}(X_i,{\beta}_0)} {\sigma(X_i,\theta_0)} +\frac{\dot{\sigma}(X,{\beta}_0)}{\sigma(X,\theta_0)}\}C_x(X_i,X)|X_i]
\\
&&M_2 = E[\{\frac{\dot{\sigma}(X_1,\tilde{\theta}_0)}{\sigma(X_1,\tilde{\theta}_0)}+
\frac{\dot{\sigma}(X_2,\tilde{\theta}_0)}{\sigma(X_2,\tilde{\theta}_0)} \} \{\frac{\dot{\sigma}(X_1,\tilde{\theta}_0)}{\sigma(X_1,\tilde{\theta}_0)}
+ \frac{\dot{\sigma}(X_2,\tilde{\theta}_0)}{\sigma(X_2,\tilde{\theta}_0)} \}^T C_x(X_1,X_2)].
\end{eqnarray*}
with $C_x(X_1,X_2) = \|X_1-X_2\|-E(\|X_1-X_2\||X_1)-E(\|X_1-X_2\||X_2)+E(\|X_1-X_2\|)$.
Hence we obtain that
\begin{eqnarray*}
&&   n U^*_{n21} \\
&=&  \frac{2Q_{\varepsilon}}{n-1}\mathop{\sum\sum}_{1 \leq i< j \leq n }
     [\tilde{h}^*_5(Z_i,Z_j)+\tilde{h}^*_6(Z_i,Z_j)+\tilde{h}^*_7(Z_i,Z_j)+\tilde{h}^*_8(Z_i,Z_j)+\tilde{h}^*_9(Z_i,Z_j)+\tilde{h}^*_{10}(Z_i,Z_j)]  \\
&&   +\sqrt{n}(\hat{\theta}_n-\tilde{\theta}_0) \frac{1}{\sqrt{n}}\sum_{i=1}^{n}4E[h^*_{2111}(Z_i,Z_j,Z_k,Z_l)|Z_i]+	
	 \sqrt{n}(\hat{\theta}_n-\tilde{\theta}_0)^T M_2 \sqrt{n}(\hat{\theta}_n-\tilde{\theta}_0) + o_p(1).
\end{eqnarray*}
For the term ${U}^*_{n22}$, similar to the arguments in Theorem 1 of Xu and Cao (2021), we can also show that $n {U}^*_{n22} =o_p(1).$ Hence we obtain that
\begin{eqnarray*}
	n \hat{U}^*_{n2}
&=& \frac{2Q_{\varepsilon}}{n-1}\mathop{\sum\sum}_{1 \leq i< j \leq n } \sum_{k=5}^{10} \tilde{h}^*_k(Z_i,Z_j)
    +\sqrt{n}(\hat{\theta}_n-\tilde{\theta}_0) \frac{1}{\sqrt{n}}\sum_{i=1}^{n}4E[h^*_{2111}(Z_i,Z_j,Z_k,Z_l)|Z_i] \\
&&  +\sqrt{n}(\hat{\theta}_n-\tilde{\theta}_0)^T M_2 \sqrt{n}(\hat{\theta}_n-\tilde{\theta}_0) + o_p(1).
\end{eqnarray*}

For the term $\hat{U}_{n0}^*$, note that it is the same as the term $\hat{U}_{n0}$ in the proof of Theorem 3.1. Consequently,
\begin{eqnarray*}
n \hat{U}_{n0}^*
&=& \frac{6}{n-1}\sum_{i=1}^{n}\sum_{j\neq i}^{n}E\{h_0(Z_i,Z_j,Z_k,Z_l)|Z_i,Z_j \}+o_p(1) \\
&=& \frac{1}{n-1}\sum_{i=1}^{n}\sum_{j\neq i}^{n}C_{\varepsilon}(\varepsilon_i,\varepsilon_j)C_x(X_i,X_j) + o_p(1).
\end{eqnarray*}
Here $h_0(Z_i,Z_j,Z_k,Z_l) = \frac{1}{6} \sum_{s< t,u< v}^{(i,j,k,l)} |\varepsilon_{st}|
(\|X_{st}\|+\|X_{uv}\|) -\frac{1}{12} \sum_{(s,t,u)}^{(i,j,k,l)}|\varepsilon_{st}|\|X_{su}\|.$
Hence we obtain that
\begin{eqnarray*}
n \hat{U}^*_{n}
&=& \frac{1}{n-1}\sum_{i=1}^{n}\sum_{j\neq i}^{n}C_{\varepsilon}(\varepsilon_i,\varepsilon_j)C_x(X_i,X_j)
    + \frac{2}{n-1}\mathop{\sum\sum}_{1 \leq i< j \leq n }\sum_{k=1}^3 h^*_k(Z_i,Z_j) \\
&&  + n(\hat{\theta}_n-\tilde{\theta}_0)^T \frac{1}{C_n^4} \sum\limits_{i< j< k< l} h^*_{15} (Z_i,Z_j,Z_k,Z_l)
    + \frac{2Q_{\varepsilon}}{n-1}\mathop{\sum\sum}_{1 \leq i< j \leq n } \sum_{k=5}^{10} \tilde{h}^*_k(Z_i,Z_j)  \\
&&  +\sqrt{n}(\hat{\theta}_n-\tilde{\theta}_0) \frac{1}{\sqrt{n}}\sum_{i=1}^{n}4E[h^*_{2111}(Z_i,Z_j,Z_k,Z_l)|Z_i] \\
&&  +\sqrt{n}(\hat{\theta}_n-\tilde{\theta}_0)^T M_2 \sqrt{n}(\hat{\theta}_n-\tilde{\theta}_0)
    + 4A_\varepsilon E \|X_1 - X_2\| + o_p(1).
\end{eqnarray*}
To obtain the limiting distribution of $n\hat{U}_n$, it remains to derive the asymptotic expansion of  $\hat{\theta}_n - \tilde{\theta}_0$.  By proof of theorem 1 in Appendix A of Dette et al. (2007), we have
\begin{eqnarray*}
	\sqrt{n} (\hat{\theta}_n-\tilde{\theta}_0)
	&=&  \frac{1}{\sqrt{n}} \sum_{i=1}^{n} [\sigma^2(X_i)\varepsilon_i^2 - \sigma^2(X_i, \tilde{\theta}_0)] \Sigma^{-1} \dot{\sigma}^2(X_i, \tilde{\theta}_0) + o_p(1).
\end{eqnarray*}
Altogether we obtain that
\begin{eqnarray*}
	&&   n\hat{U}_n \\
	&=&  \frac{1}{n-1}\sum_{i=1}^{n}\sum_{j\neq i}^{n}H(Z_i,Z_j)
	+4 \frac{1}{\sqrt{n}}\sum_{i=1}^{n}[\sigma^2(X_i,\tilde{\theta}_0)(\varepsilon_i^2-1)]\dot{\sigma}^2
	(X_i,\tilde{\theta}_0) ^T \Sigma^{-1} \frac{1}{\sqrt{n}}\sum_{i=1}^{n} E[h^*_{15}(Z_i,Z_j,Z_k,Z_l)|Z_i] \\
	&&   +8 A_{\varepsilon} \frac{1}{\sqrt{n}}\sum_{i=1}^{n}[\sigma^2(X_i,\tilde{\theta}_0)(\varepsilon_i^2-1)]\dot{\sigma}^2
	(X_i,\tilde{\theta}_0)^T \Sigma^{-1} E[h_{2111}^*(Z_i,Z_j,Z_k,Z_l)|Z_i] \\
	&&   +2A_{\varepsilon} \frac{1}{\sqrt{n}}\sum_{i=1}^{n}[\sigma^2(X_i,\tilde{\theta}_0)(\varepsilon_i^2-1)]\dot{\sigma}^2
	(X_i,\tilde{\theta}_0)^T\Sigma^{-1} M_2 \Sigma^{-1} \frac{1}{\sqrt{n}}\sum_{i=1}^{n}[\sigma^2(X_i,\tilde{\theta}_0)
	(\varepsilon_i^2-1)]\dot{\sigma}^2(X_i,\tilde{\theta}_0)\\
	&&+4 A_\varepsilon E\|X_1-X_2\| + o_p(1),
\end{eqnarray*}
where $H(Z_i,Z_j)=C_{\varepsilon}(\varepsilon_i,\varepsilon_j)C_x(X_i,X_j)+\sum_{k=1}^{3}h^*_k(Z_i,Z_j)
+Q_{\varepsilon}\sum_{k=5}^{10} \tilde{h}^*_k(Z_i,Z_j)$.
Finally, we obtain that
\begin{eqnarray*}
n\hat{U}_n
& \longrightarrow& \sum_{k=1}^{\infty}\lambda_k (\mathcal{Z}_k^2-1)+4\mathcal{W}^T\Sigma^{-1}\mathcal{P}_2
                   + 8A_{\varepsilon}\mathcal{W}^T\Sigma^{-1}\mathcal{P}_3 \\
&&                 + 2A_\varepsilon \mathcal{W}^{\top} \Sigma^{-1}M_2\Sigma^{-1}\mathcal{W}+4 A_\varepsilon E\|X_1-X_2\|,
\end{eqnarray*}
where $\mathcal{Z}_1, \mathcal{Z}_2,..$  are independent standard normal random variables, the eigenvalues
$\{\lambda_q \}_{q=1}^\infty$ are the solutions of the integral equation
$$ \int H(Z_i,Z_j)\psi_q(Z_j)dF(Z_j) = \lambda_q\psi_q(Z_i),$$
with $\{ \psi_i(\cdot) \}_{i=1}^\infty$ being the orthonormal eigenfunctions and $F_Z(\cdot)$ being the cumulative distribution function of $Z$,
and $(\mathcal{Z}_i,\mathcal{W},\mathcal{P}_2,\mathcal{P}_3) \in \mathbb{R}^{3p+1}$ is a Gaussian random vector with zero-mean and the covariance matrix satisfying
\begin{eqnarray*}
var(\mathcal{Z}_i)&=&1\\
var(\mathcal{P}_2)&=&var(E\{h^*_{15}(Z_i,Z_j,Z_k,Z_l)|Z_i\})\\
var(\mathcal{P}_3)&=&var(E\{h_{2111}^*(Z_i,Z_j,Z_k,Z_l)|Z_i\})\\
var(\mathcal{W})&=&var([\sigma^2(X_i,{\theta}_0)(\varepsilon_i^2-1)]\dot{\sigma}^2(X_i,{\theta}_0))\\
cov(\mathcal{Z}_i,\mathcal{P}_2)&=&cov(\psi_i(Z_i), E\{h_{15}^*(Z_i,Z_j,Z_k,Z_l)|Z_i\})\\
cov(\mathcal{Z}_i,\mathcal{P}_3)&=&cov(\psi_i(Z_i), E\{h_{2111}^*(Z_i,Z_j,Z_k,Z_l)|Z_i\})\\
cov(\mathcal{Z}_i,\mathcal{W})&=&cov(\psi_i(Z_i),[\sigma^2(X_i,{\theta}_0)(\varepsilon_i^2-1)]\dot{\sigma}^2(X_i,{\theta}_0))\\		cov(\mathcal{P}_2,\mathcal{W})&=&cov(E\{h_{15}^*(Z_i,Z_j,Z_k,Z_l)|Z_i\},[\sigma^2(X_i,{\theta}_0)(\varepsilon_i^2-1)]
                                 \dot{\sigma}^2(X_i,{\theta}_0))\\	
cov(\mathcal{P}_3,\mathcal{W})&=&cov(E\{h_{2111}^*(Z_i,Z_j,Z_k,Z_l)|Z_i\},[\sigma^2(X_i,{\theta}_0)(\varepsilon_i^2-1)]
                                 \dot{\sigma}^2(X_i,{\theta}_0))\\
cov(\mathcal{P}_2,\mathcal{P}_3)&=&cov(E\{h_{15}^*(Z_i,Z_j,Z_k,Z_l)|Z_i\},E\{h_{2111}^*(Z_i,Z_j,Z_k,Z_l)|Z_i\}).
\end{eqnarray*}
Hence we complete the proof of the first part of Theorem 3.3.

(2) In this part we discuss the asymptotic properties of $n\hat{U}_n$ under the local alternatives $H_{1n}$. Recall that $\eta_i=\frac{Y_i-m(X_i)}{\sigma(X_i, \theta_0)}$ and $\hat{\eta}_i = \frac{Y_i - \hat{m}(X_i)}{\sigma(X_i, \hat{\theta}_n)} $ in nonparametric cases. It follows from (\ref{6.1}) and (\ref{6.2}) that
\begin{equation}\label{6.23}
\begin{split}
&  | \hat{\eta}_i-\hat{\eta}_j|  \\
&= |\varepsilon_i-\varepsilon_j| \\
&  -\{P^*_{ij}+ \frac{\varepsilon_i s(X_i) [\sigma(X_i,\hat{\theta}_n)
   - \sigma(X_i, \theta_0)]}{2 \sqrt{n}	\sigma^3(X_i, \theta_0)}-\frac{\varepsilon_j s(X_j)
   [ \sigma(X_j,\hat{\theta}_n) - \sigma(X_j, \theta_0)]}{2\sqrt{n}\sigma^3(X_j, \theta_0)}\} [\mathbb{I}(\varepsilon_i>\varepsilon_j)-\mathbb{I}(\varepsilon_i < \varepsilon_j)] \\
&  +2\int_{0}^{P^*_{ij} + \frac{\varepsilon_i s(X_i)[{\sigma(X_i, \hat{\theta}_n)}-{\sigma(X_i, \theta_0)}]}
	{2\sqrt{n}\sigma^3(X_i, \theta_0)}-\frac{\varepsilon_j s(X_j)
    [{\sigma(X_j,\hat{\theta}_n)}-{\sigma(X_j, \theta_0)}]}{2\sqrt{n}\sigma^3(X_j, \theta_0)}} \{\mathbb{I}(\varepsilon_i-\varepsilon_j \leq z)-\mathbb{I}(\varepsilon_i \leq \varepsilon_j)\}dz,
\end{split}
\end{equation}
where
\begin{eqnarray*}
P^*_{ij}
&=& \frac{\hat{m}(X_i)-m(X_i)} {\sigma(X_i,\theta_0)} -\frac{\hat{m}(X_j)-m(X_j)} {\sigma(X_j,\theta_0)}
    +\frac{\varepsilon_i({\sigma(X_i,\hat{\theta}_n)}-{\sigma(X_i,\theta_0)})}{\sigma(X_i,\theta_0)}
    -\frac{\varepsilon_j({\sigma(X_j,\hat{\theta}_n)}-{\sigma(X_j,\theta_0)})}{\sigma(X_j,\theta_0)}\\
&&  +(R_i-R_j).  \\
R_i
&=& \frac{\varepsilon_i[\sigma(X_i,\theta_0)-\sigma(X_i,\hat{\theta}_n)]^2}{\sigma(X_i,\theta_0)
     	\sigma(X_i,\hat{\theta}_n)}+\frac{\hat{m}(X_i)-m(X_i)}{\sigma^2(X_i,\theta_0)}
        [\sigma(X_i,\theta_0)-\sigma(X_i,\hat{\theta}_n)] \\
&&  +\frac{\hat{m}(X_i)-m(X_i)}{\sigma^2(X_i,\theta_0)} \frac{[\sigma(X_i,\theta_0)
	-\sigma(X_i,\hat{\theta}_n)]^2}{\sigma(X_i,\hat{\theta}_n)}.
\end{eqnarray*}
Similar to the arguments for (\ref{6.5}) in the proof of Theorem 3.1, $n\hat{U}_n$ can be decomposed as
\begin{eqnarray}\label{6.24}
\hat{U}_n  \nonumber
&=&   \frac{1}{C_n^4}\sum\limits_{i< j< k< l} \left( \frac{1}{6} \sum_{s< t,u< v}^{(i,j,k,l)}
       |\varepsilon_{st}|  (\|X_{st}\|+\|X_{uv}\|) -\frac{1}{12} \sum_{(s,t,u)}^{(i,j,k,l)}
       |\varepsilon_{st}|\|X_{su}\| \right) \\  \nonumber
&&    + \frac{1}{C_n^4}\sum\limits_{i< j< k< l} \left( \frac{1}{6} \sum_{s< t,u< v}^{(i,j,k,l)}
       \delta^*_{1st} (\|X_{st}\|+\|X_{uv}\|) - \frac{1}{12} \sum_{(s,t,u)}^{(i,j,k,l)}
       \delta^*_{1st}\|X_{su}\| \right) \\  \nonumber
&&    + \frac{1}{C_n^4}\sum\limits_{i< j< k< l} \left( \frac{1}{6} \sum_{s< t,u< v}^{(i,j,k,l)}
      \delta^*_{2st} (\|X_{st}\|+\|X_{uv}\|) - \frac{1}{12} \sum_{(s,t,u)}^{(i,j,k,l)}
      \delta^*_{2st}\|X_{su}\| \right) \\ \nonumber
&&    + \frac{1}{C_n^4}\sum\limits_{i< j< k< l} \left( \frac{1}{6} \sum_{s< t,u< v}^{(i,j,k,l)}
      \delta^*_{3st} (\|X_{st}\|+\|X_{uv}\|) - \frac{1}{12} \sum_{(s,t,u)}^{(i,j,k,l)}
      \delta^*_{3st}\|X_{su}\| \right)\\
&=:& \hat{U}^*_{n0}+\hat{U}^*_{n1}+\hat{U}^*_{n2}+\hat{U}^*_{n3},
\end{eqnarray}
where
\begin{eqnarray*}
\delta^*_{1st}
&=&   -[\frac{\varepsilon_s({\sigma(X_s,\hat{\theta}_n)}-{\sigma(X_s,\theta_0)})}{\sigma(X_s,\theta_0)}
	  -\frac{\varepsilon_t({\sigma(X_t,\hat{\theta}_n)} -{\sigma(X_t,\theta_0)})}{\sigma(X_t,\theta_0)}
	  +\frac{\hat{m}(X_s)-m(X_s)}{\sigma(X_s,\theta_0)} -\frac{\hat{m}(X_t)-m(X_t)}{\sigma(X_t,\theta_0)}\\
&&    +(R_s-R_t)]\{\mathbb{I}(\varepsilon_s>\varepsilon_t)-\mathbb{I}(\varepsilon_s<\varepsilon_t)\}  \\
\delta^*_{2st}
&=&    -\{\frac{\varepsilon_s s(X_s)({\sigma(X_s,\hat{\theta}_n)}-{\sigma(X_s, \theta_0)})}
	   {2\sqrt{n} \sigma^3(X_s, \theta_0)}-\frac{\varepsilon_t s(X_t)
	   ({\sigma(X_t,\hat{\theta}_n)}-{\sigma(X_t, \theta_0)})}{2\sqrt{n} \sigma^3(X_t, \theta_0)} \} [\mathbb{I}(\varepsilon_s>\varepsilon_t)-\mathbb{I}(\varepsilon_s < \varepsilon_t)].   \\
\delta^*_{3st}
&=&  2 \int_{0}^{P^*_{st}+\frac{\varepsilon_s s(X_s)({\sigma(X_s,\hat{\theta}_n)}
	 -{\sigma(X_s, \theta_0)})}{2\sqrt{n}\sigma^3(X_s, \theta_0)}
	 -\frac{\varepsilon_t s(X_t)({\sigma(X_t, \hat{\theta}_n)}-{\sigma(X_t, \theta_0)})}
	 {2\sqrt{n}\sigma^3(X_t, \theta_0)}} [\mathbb{I}(\varepsilon_s-\varepsilon_t \leq z)
	 -\mathbb{I}(\varepsilon_s \leq \varepsilon_t) ] dz .\\
\end{eqnarray*}

For the term $\hat{U}^*_{n3}$, decomposed it as
\begin{eqnarray*}
	\hat{U}^*_{n3}
	&=&  \frac{1}{C_n^4}\sum\limits_{i< j< k< l}h^*_{31}(Z_i,Z_j,Z_k,Z_l)
	+\frac{1}{C_n^4}\sum\limits_{i< j< k< l}h^*_{32}(Z_i,Z_j,Z_k,Z_l) \\
	&=:& \hat{U}^*_{n31} + \hat{U}^*_{n32},
\end{eqnarray*}
where
\begin{eqnarray*}
	h^*_{31}(Z_i,Z_j,Z_k,Z_l)&=& \frac{1}{6} \sum_{s< t,u< v}^{(i,j,k,l)}E(\delta^*_{3st}|X_s,X_t) (\|X_{st}\|+\|X_{uv}\|)
	-\frac{1}{12}\sum_{(s,t,u)}^{(i,j,k,l)}E(\delta^*_{3st}|X_s,X_t) \|X_{su}\|,  \\
	h^*_{32}(Z_i,Z_j,Z_k,Z_l)&=& \frac{1}{6} \sum_{s< t,u< v}^{(i,j,k,l)}[\delta^*_{3st}-E(\delta^*_{3st}|X_s,X_t)](\|X_{st}\|+\|X_{uv}\|)
	-\frac{1}{12} \sum_{(s,t,u)}^{(i,j,k,l)}[\delta^*_{3st}-E(\delta^*_{3st}|X_s,X_t)]\|X_{su}\|.
\end{eqnarray*}
For the term $\hat{U}^*_{n31}$, similar to the arguments for $\hat{U}^*_{n21}$ in the proof of Theorem 3.2(1), we have uniformly over $1\leq s,t \leq n$,
\begin{eqnarray*}
&&   E[{\delta}^*_{3st}|X_s,X_t ]\\	
&=&  2E[\int_{0}^{\frac{\hat{m}(X_s)-m(X_s)}{\sigma(X_s,\theta_0)}-\frac{\hat{m}(X_t)-m(X_t)}{\sigma(X_t,\theta_0)}
     +\frac{\varepsilon_s({\sigma(X_s,\hat{\theta}_n)}-{\sigma(X_s,\theta_0)})}
	 {\sigma(X_s,\theta_0)}-\frac{\varepsilon_t({\sigma(X_t,\hat{\theta}_n)}-{\sigma(X_t,\theta_0)})}
	 {\sigma(X_t,\theta_0)}+R_s-R_t}\{\mathbb{I}(\varepsilon_s-\varepsilon_t\leq z)\\
&&   -\mathbb{I}(\varepsilon_s\leq \varepsilon_t)\}dz|X_s,X_t ]\\
&&  +2E[\int_{0}^{(P_{st}+\frac{\varepsilon_is(X_i)({\sigma(X_i,\hat{\theta}_n)}-{\sigma(X_i,\theta_0)})}
    {2\sqrt{n}\sigma^3(X_i,\theta_0)} -\frac{\varepsilon_js(X_j)({\sigma(X_j,\hat{\theta}_n)}-{\sigma(X_j,\tilde{\theta}_0)})}
    {2\sqrt{n}\sigma^3(X_j,\theta_0)})} \{\mathbb{I}(\varepsilon_s-\varepsilon_t \leq z )
    -\mathbb{I}(\varepsilon_s \leq \varepsilon_t)\}dz |X_s,X_t]\\
&=& Q_{\varepsilon}h^{2k}(\frac{D(X_s)}{\sigma(X_s,\theta_0)}-\frac{D(X_t)}{\sigma(X_t,\tilde{\theta}_0)})^T
	(\frac{D(X_s)}{\sigma(X_s,\theta_0)}-\frac{D(X_t)}{\sigma(X_t,\theta_0)}) \\
&&  +Q_{\varepsilon}h^{k}(\frac{D(X_s)}{\sigma(X_s,\tilde{\theta}_0)}-\frac{D(X_t)}{\sigma(X_t,\theta_0)})^T
    (\frac{\frac{1}{n-1}\sum_{p=1,p\neq s,t}^{n}\omega_{s,p}\sigma(X_p,\theta_0)\varepsilon_p}
    {\sigma(X_s,\theta_0)}-\frac{\frac{1}{n-1}\sum_{q=1,q\neq s,t}^{n}\omega_{t,q}
    \sigma(X_q,\theta_0)\varepsilon_q}{\sigma(X_t,\theta_0)} )\\
&&  +Q_{\varepsilon}(\frac{\frac{1}{n-1}\sum_{p=1,p\neq s,t}^{n}\omega_{s,p}\sigma(X_p,\theta_0)\varepsilon_p}
    {\sigma(X_s,\theta_0)}-\frac{\frac{1}{n-1} \sum_{q=1,q\neq s,t}^{n}\omega_{t,q}
    \sigma(X_q,\theta_0)\varepsilon_q}{\sigma(X_t,\theta_0)} )^T\\
&&  (\frac{\frac{1}{n-1}\sum_{p=1,p\neq s,t}^{n}\omega_{s,p}\sigma(X_p,\theta_0)\varepsilon_p}{\sigma(X_s,\theta_0)}
	-\frac{\frac{1}{n-1}\sum_{q=1,q\neq s,t}^{n}\omega_{t,q}\sigma(X_q,\theta_0)\varepsilon_q}{\sigma(X_t,\theta_0)} )\\
&&  +Q_{\varepsilon}h^{k}(\frac{D(X_s)}{\sigma(X_s,\theta_0)}-\frac{D(X_t)}{\sigma(X_t,\theta_0)})^T
	(\frac{\frac{1}{n-1}\sum_{p=1,p\neq s}^{n}\omega_{s,p}m(X_p)-E[\omega_{s,p}m(X_p)|X_s]}{\sigma(X_s,\theta_0)}\\
&&  -\frac{\frac{1}{n-1}\sum_{q=1,q\neq t}^{n}\omega_{t,q}m(X_q)-E[\omega_{t,q}m(X_q)|X_t]}{\sigma(X_t,\theta_0)} )\\
&&  +Q_{\varepsilon}h^{k}(\frac{D(X_s)}{\sigma(X_s,\theta_0)}-\frac{D(X_t)}{\sigma(X_t,\theta_0)})^T
	(\frac{\frac{m(X_s)}{n-1}\sum_{p=1,p\neq s}^{n}(\omega_{s,p}-E[\omega_{s,p}|X_s])}{\sigma(X_s,\theta_0)}
	-\frac{\frac{m(X_t)}{n-1}\sum_{q=1,q\neq t}^{n}(\omega_{t,q}-E[\omega_{t,q}|X_t])}{\sigma(X_t,\theta_0)} )\\
&&  +Q_{\varepsilon}(\frac{\frac{1}{n-1}\sum_{p=1,p\neq s}^{n}\omega_{s,p}m(X_p)
    -E[\omega_{s,p}m(X_p)|X_s]}{\sigma(X_s,\theta_0)}-\frac{\frac{1}{n-1}\sum_{q=1,q\neq t}^{n}
    \omega_{t,q}m(X_q)-E[\omega_{t,q}m(X_q)|X_t]}{\sigma(X_t,\theta_0)} )^T\\
&&  \times (\frac{\frac{1}{n-1}\sum_{p=1,p\neq s}^{n}\omega_{s,p}m(X_p)-E[\omega_{s,p}m(X_p)|X_s]}
    {\sigma(X_s,\theta_0)}-\frac{\frac{1}{n-1}\sum_{q=1,q\neq t}^{n}\omega_{t,q}m(X_q)
    -E[\omega_{t,q}m(X_q)|X_t]}{\sigma(X_t,\theta_0)})\\     \nonumber
&&  +Q_{\varepsilon}(\frac{\frac{m(X_s)}{n-1}\sum_{p=1,p\neq s}^{n}(\omega_{s,p}-E[\omega_{s,p}|X_s])}
    {\sigma(X_s,\theta_0)}-\frac{\frac{m(X_t)}{n-1}\sum_{q=1,q\neq t}^{n}(\omega_{t,q}
    -E[\omega_{t,q}|X_t])}{\sigma(X_t,\theta_0)} )^T\\    \nonumber
&&  \times (\frac{\frac{m(X_s)}{n-1}\sum_{p=1,p\neq s}^{n}(\omega_{s,p}-E[\omega_{s,p}|X_s])}
    {\sigma(X_s,\theta_0)}-\frac{\frac{m(X_t)}{n-1}\sum_{q=1,q\neq t}^{n}(\omega_{t,q}-E[\omega_{t,q}|X_t])}
    {\sigma(X_t,\theta_0)} ) \\
&&  +Q_{\varepsilon}(\frac{\frac{1}{n-1}\sum_{p=1,p\neq s}^{n}\omega_{s,p}m(X_p)
    -E[\omega_{s,p}m(X_p)|X_s]}{\sigma(X_s,\theta_0)}-\frac{\frac{1}{n-1}\sum_{q=1,q\neq t}^{n}
    \omega_{t,q}m(X_q)-E[\omega_{t,q}m(X_q)|X_t]}{\sigma(X_t,\theta_0)} )^T\\
&&  \times (\frac{\frac{m(X_s)}{n-1}\sum_{p=1,p\neq s}^{n}(\omega_{s,p}-E[\omega_{s,p}|X_s])}
    {\sigma(X_s,\theta_0)}-\frac{\frac{m(X_t)}{n-1}\sum_{q=1,q\neq t}^{n}(\omega_{t,q}
    -E[\omega_{t,q}|X_t])}{\sigma(X_t,\theta_0)}) \\
&&  +Q_{\varepsilon}(\frac{\frac{1}{n-1}\sum_{p=1,p\neq s}^{n}\omega_{s,p}m(X_p)
    -E[\omega_{s,p}m(X_p)|X_s]}{\sigma(X_s,\theta_0)}-\frac{\frac{1}{n-1}\sum_{q=1,q\neq t}^{n}
    \omega_{t,q}m(X_q)-E[\omega_{t,q}m(X_q)|X_t]}{\sigma(X_t,\theta_0)} )^T\\
&&  \times (\frac{\frac{1}{n-1}\sum_{p=1,p\neq s,t}^{n}\omega_{s,p}\sigma(X_p,\theta_0)\varepsilon_p}
	{\sigma(X_s,\theta_0)}-\frac{\frac{1}{n-1}\sum_{q=1,q\neq s,t}^{n}\omega_{t,q}\sigma(X_q,\theta_0)\varepsilon_q}
	{\sigma(X_t,\theta_0)} ) \\
&&  +Q_{\varepsilon}(\frac{\frac{m(X_s)}{n-1}\sum_{p=1,p\neq s}^{n}(\omega_{s,p}
    -E[\omega_{s,p}|X_s])}{\sigma(X_s,\theta_0)}-\frac{\frac{m(X_t)}{n-1}\sum_{q=1,q\neq t}^{n}
    (\omega_{t,q}-E[\omega_{t,q}|X_t])}{\sigma(X_t,\theta_0)} )^T\\
&&  \times (\frac{\frac{1}{n-1}\sum_{p=1,p\neq s,t}^{n}\omega_{s,p}\sigma(X_p,\theta_0)\varepsilon_p}
    {\sigma(X_s,\theta_0)}-\frac{\frac{1}{n-1}\sum_{q=1,q\neq s,t}^{n}\omega_{t,q}\sigma(X_q,\theta_0)\varepsilon_q}
    {\sigma(X_t,\theta_0)})\\
&&  + 2A_{\varepsilon} \left([\frac{\dot{\sigma}(X_s,\theta_0)}{\sigma(X_s,\theta_0)}
	+\frac{\dot{\sigma}(X_t,\theta_0)}{\sigma(X_t,\theta_0)}]^T
	(\hat{\theta}_n-\theta_0) \right) ^2+2A_{\varepsilon}(\hat{\theta}_n-\theta_0)^T [\frac{\dot{\sigma}(X_s,\theta_0)}
	{\sigma(X_s,\theta_0)}+\frac{\dot{\sigma}(X_t,\theta_0)}{\sigma(X_t,\theta_0)}]\\
&&  + 2A_{\varepsilon}(\hat{\theta}_n-\theta_0)^T
	[\frac{s(X_s) \dot{\sigma} (X_s,\theta_0)}{2\sqrt{n}\sigma^3(X_s,{\theta}_0)}
	+\frac{s(X_t) \dot{\sigma} (X_t,\theta_0)} {2\sqrt{n}\sigma^3(X_t,\theta_0)}]\\
&&  +2A_{\varepsilon} \left([\frac{s(X_s)\dot{\sigma}(X_s,\theta_0)}
	{2\sqrt{n}\sigma^3(X_s,\theta_0)}+\frac{s(X_t)\dot{\sigma}(X_t,\theta_0)}{2\sqrt{n}
	\sigma^3(X_t,\theta_0)}]^T (\hat{\theta}_n-\theta_0) \right) ^2 + o_p(\frac{1}{n})
\end{eqnarray*}
where $Q_{\varepsilon} = E[f_{\varepsilon}(\varepsilon)]$ and $A_{\varepsilon}=E[\varepsilon F_{\varepsilon}(\varepsilon)]$.
Consequently,
\begin{eqnarray*}
\nonumber
\hat{U}^*_{n31}   \nonumber
&=&  h^{2k}  Q_{\varepsilon} \frac{1}{C_n^4}\sum\limits_{i< j< k< l}h^*_{311}(Z_i,Z_j,Z_k,Z_l)  + h^{k}  Q_{\varepsilon} \frac{1}{C_n^4}\sum\limits_{i< j< k< l}h^*_{312}(Z_i,Z_j,Z_k,Z_l) \\  \nonumber
&& + h^{k} Q_{\varepsilon} \frac{1}{C_n^4}\sum\limits_{i< j< k< l}h^*_{313}(Z_i,Z_j,Z_k,Z_l) + h^{k}  Q_{\varepsilon} \frac{1}{C_n^4}\sum\limits_{i< j< k< l}h^*_{314}(Z_i,Z_j,Z_k,Z_l) \\
\nonumber
&& +   Q_{\varepsilon} \frac{1}{C_n^4}\sum\limits_{i< j< k< l}h^*_{315}(Z_i,Z_j,Z_k,Z_l)  +  Q_{\varepsilon} \frac{1}{C_n^4}\sum\limits_{i< j< k< l}h^*_{316}(Z_i,Z_j,Z_k,Z_l) \\
\nonumber
&&+  Q_{\varepsilon} \frac{1}{C_n^4}\sum\limits_{i< j< k< l}h^*_{317}(Z_i,Z_j,Z_k,Z_l) +  Q_{\varepsilon} \frac{1}{C_n^4}\sum\limits_{i< j< k< l}h^*_{318}(Z_i,Z_j,Z_k,Z_l)\\
\nonumber
&& +  Q_{\varepsilon} \frac{1}{C_n^4}\sum\limits_{i< j< k< l}h^*_{319}(Z_i,Z_j,Z_k,Z_l) +  Q_{\varepsilon} \frac{1}{C_n^4}\sum\limits_{i< j< k< l}h^*_{3110}(Z_i,Z_j,Z_k,Z_l) \\
\nonumber
&& +2A_{\varepsilon} (\hat{\theta}_n-\theta_0)^T \frac{1}{C_n^4}\sum\limits_{i< j< k< l}h^*_{3111}(Z_i,Z_j,Z_k,Z_l)  \\
\nonumber
&&    +2A_{\varepsilon}(\hat{\theta}_n-\theta_0)^T \frac{1}{C_n^4}\sum\limits_{i< j< k< l}h^*_{3112}(Z_i,Z_j,Z_k,Z_l)(\hat{\theta}_n-\theta_0)  \\ \nonumber
&& +2A_{\varepsilon} (\hat{\theta}_n-\theta_0)^T \frac{1}{\sqrt{n}} \frac{1}{C_n^4}\sum\limits_{i< j< k< l}h^*_{3113}(Z_i,Z_j,Z_k,Z_l)  \\
\nonumber
&&    +2A_{\varepsilon}(\hat{\theta}_n-\theta_0)^T \frac{1}{n} \frac{1}{C_n^4}\sum\limits_{i< j< k< l}h^*_{3114}(Z_i,Z_j,Z_k,Z_l)(\hat{\theta}_n-\theta_0)  + o_p(\frac{1}{n}), \\ \nonumber
&=:&  h^{2k}Q_{\varepsilon} \hat{U}^*_{n311} + h^{k}Q_{\varepsilon} \hat{U}^*_{n312}
+h^{k} Q_{\varepsilon} \hat{U}^*_{n313} +h^{k} Q_{\varepsilon} \hat{U}^*_{n314}
+ Q_{\varepsilon} \sum_{k=5}^{10} \hat{U}^*_{n31k} \\
&& +2A_{\varepsilon} \sum_{k=11}^{14} \hat{U}^*_{n31k} +o_p(\frac{1}{n}),
\end{eqnarray*}
where
\begin{eqnarray*}
&&   h^*_{31m}(Z_i,Z_j,Z_k,Z_l) \\
&=&  6^{-1}\sum_{s< t,u< v}^{(i,j,k,l)}\delta_{31mst}(\|X_{st}\|+\|X_{uv}\|)
     -12^{-1}\sum_{(s,t,u)}^{(i,j,k,l)}\delta_{31mst}\|X_{su}\|,  \quad {\rm for} \ m=1, \cdots, 14  \\
\end{eqnarray*}
and
\begin{eqnarray*}
	\delta^*_{311st}
	&=&  (\frac{D(X_s)}{\sigma(X_s,\theta_0)}-\frac{D(X_t)}{\sigma(X_t,\theta_0)})
	(\frac{D(X_s)}{\sigma(X_s,\theta_0)}-\frac{D(X_t)}{\sigma(X_t,\theta_0)})^T
	\\
	\delta^*_{312st}
	&=& \{\frac{\frac{1}{n-1}\sum_{p=1,p\neq s,t}^{n} \omega_{s,p}\sigma(X_p,\theta_0)\varepsilon_p}{\sigma(X_s,\theta_0)}-\frac{\frac{1}{n-1}\sum_{q=1,q\neq s,t}^{n} \omega_{t,q}\sigma(X_q,\theta_0)\varepsilon_q}{\sigma(X_t,\theta_0)} \} \{\frac{D(X_s)}{\sigma(X_s,\theta_0)}-\frac{D(X_t)}{\sigma(X_t,\theta_0)}\}^T \\
	\delta^*_{313st}
	&=& (\frac{D(X_s)}{\sigma(X_s,\theta_0)}-\frac{D(X_t)}{\sigma(X_t,\theta_0)})^T(\frac{\frac{1}{n-1}\sum_{p=1,p\neq s}^{n} \omega_{s,p}m(X_p)-E[\omega_{s,p}m(X_p)|X_s]}{\sigma(X_s,\theta_0)}\\
&&  -\frac{\frac{1}{n-1}\sum_{q=1,q\neq t}^{n}\omega_{t,q}m(X_q)-E[\omega_{t,q}m(X_q)|X_t]}{\sigma(X_t,\theta_0)} )\\
	\delta^*_{314st}
	&=& (\frac{D(X_s)}{\sigma(X_s,\tilde{\theta}_0)}-\frac{D(X_t)}{\sigma(X_t,\tilde{\theta}_0)})^T(\frac{\frac{m(X_s)}{n-1}\sum_{p=1,p\neq s}^{n}(\omega_{s,p}-E[\omega_{s,p}|X_s])}{\sigma(X_s,\theta_0)}-\frac{\frac{m(X_t)}{n-1}\sum_{q=1,q\neq t}^{n}(\omega_{t,q}-E[\omega_{t,q}|X_t])}{\sigma(X_t,\theta_0)} )\\
	\delta^*_{315st}
	&=& \{\frac{\frac{1}{n-1}\sum_{p=1,p\neq s,t}^{n} \omega_{s,p}\sigma(X_p,\theta_0)\varepsilon_{p}}{\sigma(X_{s},\theta_0)}-\frac{\frac{1}{n-1}\sum_{q=1,q\neq s,t}^{n} \omega_{t,q}\sigma(X_q,\theta_0)\varepsilon_{q}}{\sigma(X_{t},\theta_0)}  \} \\
	&&\times
	\{\frac{\frac{1}{n-1}\sum_{p'=1,p'\neq s,t}^{n}\omega_{s,p'}\sigma(X_{p'},\theta_0)\varepsilon_{p'}}{\sigma(X_{s},\theta_0)}
	-\frac{\frac{1}{n-1}\sum_{q'=1,q'\neq s,t}^{n}\omega_{t,q'} \sigma(X_{q'},\theta_0)\varepsilon_{q'}}{\sigma(X_{t},\theta_0)}\}^T  \\
	\delta^*_{316st}
	&=&  (\frac{\frac{1}{n-1}\sum_{p=1,p\neq s}^{n} \omega_{s,p}m(X_p)-E[\omega_{s,p}m(X_p)|X_s]}{\sigma(X_s,\theta_0)}-\frac{\frac{1}{n-1}\sum_{q=1,q\neq t}^{n} \omega_{t,q}m(X_q)-E[\omega_{t,q}m(X_q)|X_t]}{\sigma(X_t,\theta_0)} )^T\\
	&& \times (\frac{\frac{1}{n-1}\sum_{p=1,p\neq s}^{n} \omega_{s,p}m(X_p)-E[\omega_{s,p}m(X_p)|X_s]}{\sigma(X_s,\theta_0)}-\frac{\frac{1}{n-1}\sum_{q=1,q\neq t}^{n} \omega_{t,q}m(X_q)-E[\omega_{t,q}m(X_q)|X_t]}{\sigma(X_t,\theta_0)})\\
	\delta^*_{317st}
	&=& (\frac{\frac{m(X_s)}{n-1}\sum_{p=1,p\neq s}^{n} (\omega_{s,p}-E[\omega_{s,p}|X_s])}{\sigma(X_s,\theta_0)}-\frac{\frac{m(X_t)}{n-1}\sum_{q=1,q\neq t}^{n} (\omega_{t,q}-E[\omega_{t,q}|X_t])}{\sigma(X_t,\theta_0)} )^T\\
	&& \times (\frac{\frac{m(X_s)}{n-1}\sum_{p=1,p\neq s}^{n} (\omega_{s,p}-E[\omega_{s,p}|X_s])}{\sigma(X_s,\theta_0)}-\frac{\frac{m(X_t)}{n-1}\sum_{q=1,q\neq t}^{n} (\omega_{t,q}-E[\omega_{t,q}|X_t])}{\sigma(X_t,\theta_0)} )
	\\
	\delta^*_{318st}
	&=&
	(\frac{\frac{1}{n-1}\sum_{p=1,p\neq s}^{n} \omega_{s,p}m(X_p)-E[\omega_{s,p}m(X_p)|X_s]}{\sigma(X_s,\theta_0)}-\frac{\frac{1}{n-1}\sum_{q=1,q\neq t}^{n} \omega_{t,q}m(X_q)-E[\omega_{t,q}m(X_q)|X_t]}{\sigma(X_t,\theta_0)} )^T\\
	&& \times (\frac{\frac{m(X_s)}{n-1}\sum_{p=1,p\neq s}^{n} (\omega_{s,p}-E[\omega_{s,p}|X_s])}{\sigma(X_s,\theta_0)}-\frac{\frac{m(X_t)}{n-1}\sum_{q=1,q\neq t}^{n} (\omega_{t,q}-E[\omega_{t,q}|X_t])}{\sigma(X_t,\theta_0)} )
	\\
	\delta^*_{319st}
	&=& (\frac{\frac{1}{n-1}\sum_{p=1,p\neq s}^{n} \omega_{s,p}m(X_p)-E[\omega_{s,p}m(X_p)|X_s]}{\sigma(X_s,\theta_0)}-\frac{\frac{1}{n-1}\sum_{q=1,q\neq t}^{n}
\omega_{t,q}m(X_q)-E[\omega_{t,q}m(X_q)|X_t]}{\sigma(X_t,\theta_0)} )^T\\
	&& \times (\frac{\frac{1}{n-1}\sum_{p=1,p\neq s,t}^{n}\omega_{s,p}\sigma(X_p,\theta_0)\varepsilon_p}
	{\sigma(X_s,\theta_0)}
	-\frac{\frac{1}{n-1}\sum_{q=1,q\neq s,t}^{n}\omega_{t,q}\sigma(X_q,\theta_0)\varepsilon_q}
	{\sigma(X_t,\theta_0)} )\\
	\delta^*_{3110st}
	&=&
	(\frac{\frac{m(X_s)}{n-1}\sum_{p=1,p\neq s}^{n} (\omega_{s,p}-E[\omega_{s,p}|X_s])}{\sigma(X_s,\theta_0)}-\frac{\frac{m(X_t)}{n-1}\sum_{q=1,q\neq t}^{n} (\omega_{t,q}-E[\omega_{t,q}|X_t])}{\sigma(X_t,\theta_0)} )^T\\
	&& \times (\frac{\frac{1}{n-1}\sum_{p=1,p\neq s,t}^{n} \omega_{s,p}\sigma(X_p,\theta_0)\varepsilon_p}{\sigma(X_s,\theta_0)}-\frac{\frac{1}{n-1}\sum_{q=1,q\neq s,t}^{n} \omega_{t,q}\sigma(X_q,\theta_0)\varepsilon_q}{\sigma(X_t,\theta_0)})
	\\
	\delta^*_{3111st}
	&=&\{\frac{\dot{\sigma}(X_s,\theta_0)}{\sigma(X_s,\theta_0)}
	+ \frac{\dot{\sigma}(X_t,\theta_0)}{\sigma(X_t,\theta_0)}\}\\
	\delta^*_{3112st}
	&=&\{\frac{\dot{\sigma}(X_s,\theta_0)}{\sigma(X_s,\theta_0)}
	+ \frac{\dot{\sigma}(X_t,\theta_0)}{\sigma(X_t,\theta_0)}\}\{\frac{\dot{\sigma}(X_s,\theta_0)}{\sigma(X_s,\theta_0)}
	+ \frac{\dot{\sigma}(X_t,\theta_0)}{\sigma(X_t,\theta_0)}\}^T\\	
	\delta^*_{3113st}
	&=& \frac{s(X_s)\dot{\sigma}(X_s,\theta_0)}{2 \sigma^3(X_s,\theta_0)}
	+ \frac{s(X_t)\dot{\sigma}(X_t,\theta_0)}{2 \sigma^3(X_t,\theta_0)} \\	
	\delta^*_{3114st}
		&=& \frac{s(X_s)\dot{\sigma}(X_s,\theta_0)}{2 \sigma^3(X_s,\theta_0)}
	+ \frac{s(X_t)\dot{\sigma}(X_t,\theta_0)}{2 \sigma^3(X_t,\theta_0)}
	\frac{s(X_s)\dot{\sigma}(X_s,\theta_0)}{2 \sigma^3(X_s,\theta_0)}
	+ \frac{s(X_t)\dot{\sigma}(X_t,\theta_0)}{2 \sigma^3(X_t,\theta_0)}^T.
\end{eqnarray*}
	
For the term $\hat{U}^*_{n311}$,  by the law of large numbers and some elementary calculations, we have
\begin{eqnarray*}
\frac{1}{C_n^4}\sum\limits_{i< j< k< l}h^*_{311}(Z_i,Z_j,Z_k,Z_l) \longrightarrow E[(\frac{D(X_1)}{\sigma(X_1, \theta_0)}-\frac{D(X_2)}{\sigma(X_2, \theta_0)})(\frac{D(X_1)}
{\sigma(X_1, \theta_0)}-\frac{D(X_2)}{\sigma(X_2, \theta_0)})^T C_x(X_1,X_2)],
\end{eqnarray*}
in probability, where
\begin{equation*}
E[h^*_{311}(Z_i,Z_j,Z_k,Z_l)] = E[(\frac{D(X_1)}{\sigma(X_1, \theta_0)}-\frac{D(X_2)}{\sigma(X_2, \theta_0)})(\frac{D(X_1)}{\sigma(X_1, \theta_0)}-\frac{D(X_2)}{\sigma(X_2, \theta_0)})^T C_x(X_1,X_2)]
\end{equation*}
and $C_x(X_1,X_2) = \|X_1-X_2\|-E(\|X_1-X_2\||X_1)-E(\|X_1-X_2\||X_2)+E(\|X_1-X_2\|)$.
Together with the assumption 6(d), we have
$$n h^{2k} \hat{U}^*_{n311}= n h^{2k} \frac{1}{C_n^4}\sum\limits_{i< j< k< l}h^*_{311}(Z_i,Z_j,Z_k,Z_l)= o_p(1).$$
For the term $\{ \hat{U}^*_{n31m} \}_{m=2}^{10}$, similar to the arguments for $\{ \hat{U}^*_{n21m}\}_{m=2}^{10}$ in the proof of Theorem 3.3(1), we have $n h^k \hat{U}^*_{n31m} = o_p(1)$ for $m=2, 3, 4$ and
\begin{eqnarray*}
n \hat{U}^*_{n31m}=\frac{2}{n-1}\mathop{\sum\sum}_{1 \leq i< j \leq n }\tilde{h}^*_{m}(Z_i,Z_j)+o_p(1), \quad m = 5, \cdots, 10,
\end{eqnarray*}
where $\tilde{h}^*_{m}(Z_i,Z_j)$ is given in the first part of this proof.

For the terms $\hat{U}^*_{n3111}$ and $\hat{U}^*_{n3112}$, similar to the arguments for $\hat{U}^*_{n2111}$ and $\hat{U}^*_{n2112}$ in the proof of Theorem 3.3(1), we have
\begin{eqnarray*}
	n \hat{U}^*_{n3111} &=&  \sqrt{n}(\hat{\theta}_n-\tilde{\theta}_0) \frac{1}{\sqrt{n}}\sum_{i=1}^{n}4E[h^*_{3111}(Z_i,Z_j,Z_k,Z_l|Z_i)]+o_p(1).\\
	n \hat{U}^*_{n3112} &=& \sqrt{n}(\hat{\theta}_n-\tilde{\theta}_0)^T M_2 \sqrt{n}(\hat{\theta}_n-\theta_0)+o_p(1).
\end{eqnarray*}
where
\begin{eqnarray*}
	&&E[h^*_{3111}(Z_i,Z_j,Z_k,Z_l)|Z_i]=
	E[\{\frac{\dot{\sigma}(X_i,{\beta}_0)} {\sigma(X_i,\theta_0)} +\frac{\dot{\sigma}(X,{\beta}_0)}{\sigma(X,\theta_0)}\}C_x(X_i,X)|X_i]
	\\
	&&M_2 = E[\{\frac{\dot{\sigma}(X_1,\theta_0)}{\sigma(X_1,\theta_0)}+
	\frac{\dot{\sigma}(X_2,\theta_0)}{\sigma(X_2,\theta_0)} \} \{\frac{\dot{\sigma}(X_1,\theta_0)}{\sigma(X_1,\theta_0)}
	+ \frac{\dot{\sigma}(X_2,\theta_0)}{\sigma(X_2,\theta_0)} \}^T C_x(X_1,X_2)].
\end{eqnarray*}
with $C_x(X_1,X_2) = \|X_1-X_2\|-E(\|X_1-X_2\||X_1)-E(\|X_1-X_2\||X_2)+E(\|X_1-X_2\|)$.

For the term $\hat{U}^*_{n3113}$ and $\hat{U}^*_{n3114}$ similar to the arguments for $\hat{U}^*_{n314}$ and $\hat{U}^*_{n315}$ in the proof of Theorem 3.2(1), we have
$$n \hat{U}^*_{n3113}=o_p(1), \quad {\rm and } \quad n \hat{U}^*_{n3114}=o_p(1). $$
Thus we obtain that
\begin{eqnarray*}
n \hat{U}^*_{n31}
&=& Q_{\varepsilon} \frac{2}{n-1}\mathop{\sum\sum}_{1 \leq i< j \leq n } \sum_{m=5}^{10}\tilde{h}^*_{m}(Z_i,Z_j) \\
&&  + \sqrt{n}(\hat{\theta}_n-\theta_0) \frac{1}{\sqrt{n}}\sum_{i=1}^{n}4E[h^*_{3111}(Z_i,Z_j,Z_k,Z_l)|Z_i] \\
&&  + \sqrt{n}(\hat{\theta}_n-\theta_0)^T M_2 \sqrt{n}(\hat{\theta}_n-\theta_0) + o_p(1).
\end{eqnarray*}
Following the same line as $\hat{U}^*_{n22}$ in the first part of this proof, we can show that $\hat{U}^*_{n32}=o_p(1)$.
It follows that
\begin{eqnarray*}
n \hat{U}^*_{n3}
&=& Q_{\varepsilon} \frac{2}{n-1}\mathop{\sum\sum}_{1 \leq i< j \leq n } \sum_{m=5}^{10}\tilde{h}^*_{m}(Z_i,Z_j) \\
&&  + 8 A_{\varepsilon} \sqrt{n}(\hat{\theta}_n-\theta_0) \frac{1}{\sqrt{n}}\sum_{i=1}^{n} E[h^*_{3111}(Z_i,Z_j,Z_k,Z_l)|Z_i] \\
&&  + \sqrt{n}(\hat{\theta}_n-\theta_0)^T M_2 \sqrt{n}(\hat{\theta}_n-\theta_0) + o_p(1).
\end{eqnarray*}

Note that $\hat{U}^*_{n0}$ and $\hat{U}^*_{n1}$ are the same as the terms $\hat{U}^*_{n0}$ and $\hat{U}^*_{n1}$ in the first part of this proof. Consequently,
\begin{eqnarray*}
n\hat{U}^*_{n0}&=& \frac{1}{n-1}\sum_{i=1}^{n}\sum_{j\neq i}^{n}C_{\varepsilon}(\varepsilon_i,\varepsilon_j)C_x(X_i,X_j)
                   +o_p(1), \\
n\hat{U}^*_{n1}&=& \frac{2}{n-1}\mathop{\sum\sum}_{1 \leq i< j \leq n }[h^*_1(Z_i,Z_j)+h^*_2(Z_i,Z_j)+h^*_3(Z_i,Z_j)] \\
&&                 + n(\hat{\theta}_n-\theta_0)^T \frac{1}{C_n^4} \sum\limits_{i< j< k< l} h^*_{15} (Z_i,Z_j,Z_k,Z_l)
                   + 4A_\varepsilon E \|X_1 - X_2\| + o_p(1).
\end{eqnarray*}
For the term $\hat{U}^*_{n2}$,  similar to the arguments in Theorem 1 of Xu and Cao (2021), we have
$$n \hat{U}^*_{n2} = o_p(1). $$
Altogether we obtain that
\begin{eqnarray*}
n\hat{U}^*_{n}
&=& \frac{1}{n-1}\sum_{i=1}^{n}\sum_{j\neq i}^{n}C_{\varepsilon}(\varepsilon_i,\varepsilon_j)C_x(X_i,X_j)  \\
&&  \frac{2}{n-1}\mathop{\sum\sum}_{1 \leq i< j \leq n }[h^*_1(Z_i,Z_j)+h^*_2(Z_i,Z_j)+h^*_3(Z_i,Z_j)] \\
&&  + n(\hat{\theta}_n-\theta_0)^T \frac{1}{C_n^4} \sum\limits_{i< j< k< l} h^*_{15} (Z_i,Z_j,Z_k,Z_l)
    + 4A_\varepsilon E \|X_1 - X_2\|  \\
&&  + Q_{\varepsilon} \frac{2}{n-1}\mathop{\sum\sum}_{1 \leq i< j \leq n } \sum_{m=5}^{10}\tilde{h}^*_{m}(Z_i,Z_j) \\
&&  + 8A_\varepsilon  \sqrt{n}(\hat{\theta}_n-\theta_0) \frac{1}{\sqrt{n}}\sum_{i=1}^{n}E[h^*_{3111}(Z_i,Z_j,Z_k,Z_l|Z_i)] \\
&&  + 2A_\varepsilon  \sqrt{n}(\hat{\theta}_n-\theta_0)^T M_2 \sqrt{n}(\hat{\theta}_n-\theta_0) + o_p(1).
\end{eqnarray*}
According to the proof of theorem 1 in Appendix A of Dette et al. (2007), we have
\begin{eqnarray*}
	\sqrt{n} (\hat{\theta}_n-\theta_0)
	&=&  \frac{1}{\sqrt{n}} \sum_{i=1}^{n} [\sigma^2(X_i, \theta_0) (\varepsilon_i^2 - 1)] \Sigma_{\sigma}^{-1}
	\dot{\sigma}^2(X_i, \theta_0) + \Sigma_{\sigma}^{-1} E[s(X)\dot{\sigma}^2(X_i,{\theta}_0)] +o_p(1),
\end{eqnarray*}
where $\Sigma_{\sigma} = E[\dot{\sigma}^2 (X_i,{\theta}_0)\dot{\sigma}^2 (X_i,{\theta}_0)^T]$.
Hence we obtain that
\begin{eqnarray*}
	&&   n\hat{U}_n \\
	&=&   \frac{1}{n-1}\sum_{i=1}^{n}\sum_{j\neq i}^{n}H(Z_i,Z_j)
	+4 \frac{1}{\sqrt{n}}\sum_{i=1}^{n}[\sigma^2(X_i,\theta_0)(\varepsilon_i^2-1)]\dot{\sigma}^2
	(X_i,\theta_0) ^T \Sigma_{\sigma}^{-1} \frac{1}{\sqrt{n}}\sum_{i=1}^{n} E[h^*_{15}(Z_i,Z_j,Z_k,Z_l)|Z_i] \\
	&&   +8 A_{\varepsilon} \frac{1}{\sqrt{n}}\sum_{i=1}^{n}[\sigma^2(X_i,\theta_0)(\varepsilon_i^2-1)]\dot{\sigma}^2
	(X_i,\theta_0)^T \Sigma_{\sigma} ^{-1} E[h^*_{3111}(Z_i,Z_j,Z_k,Z_l)|Z_i] \\
	&&   +2A_{\varepsilon} \frac{1}{\sqrt{n}}\sum_{i=1}^{n}[\sigma^2(X_i,\theta_0)(\varepsilon_i^2-1)]\dot{\sigma}^2
	(X_i,\theta_0)^T\Sigma_{\sigma} ^{-1} M_2 \Sigma_{\sigma} ^{-1} \frac{1}{\sqrt{n}}\sum_{i=1}^{n}[\sigma^2(X_i,\theta_0)
	(\varepsilon_i^2-1)]\dot{\sigma}^2(X_i,\theta_0)\\
	&&   +4A_{\varepsilon} \frac{1}{\sqrt{n}}\sum_{i=1}^{n}[\sigma^2(X_i,\theta_0)(\varepsilon_i^2-1)]\dot{\sigma}^2
	(X_i,\theta_0)^T\Sigma_{\sigma} ^{-1} M_2 \Sigma_{\sigma} ^{-1} E[s(X)\dot{\sigma}^2(X_i,{\theta}_0)]
	\\
	&&   +2A_{\varepsilon} E[s(X)\dot{\sigma}^2(X_i,{\theta}_0)]^T \Sigma_{\sigma} ^{-1}
	M_2 \Sigma_{\sigma} ^{-1} E[s(X)\dot{\sigma}^2(X_i,{\theta}_0)]
	\\
	&&+ 8A_{\varepsilon} E[s(X)\dot{\sigma}^2(X_i,{\theta}_0)]^T \Sigma_{\sigma} ^{-1}
	\{
	\frac{1}{\sqrt{n}}\sum_{i=1}^{n}E\{h^*_{3111}(Z_i,Z_j,Z_k,Z_l|Z_i)\}+4 A_\varepsilon E\|X_1-X_2\| \\
	&&+ 4E[s(X)\dot{\sigma}^2(X_i,{\theta}_0)]^T \Sigma_{\sigma} ^{-1}
	\frac{1}{\sqrt{n}}\sum_{i=1}^{n}E\{h^*_{15}(Z_i,Z_j,Z_k,Z_l|Z_i)\}+ o_p(1),
\end{eqnarray*}
whence
\begin{eqnarray*}
n\hat{U}_n
&\longrightarrow&  \sum_{k=1}^{\infty}\lambda_k (\mathcal{Z}_k^2-1)
    +4\mathcal{W}^T\Sigma^{-1}\mathcal{P}_2
    +8A_{\varepsilon}\mathcal{W}^T\Sigma^{-1}\mathcal{P}_3
    + 2A_\varepsilon \mathcal{W}^{\top} \Sigma^{-1}M_2\Sigma^{-1}\mathcal{W}
    +4 A_\varepsilon E\|X_1-X_2\|\\
&&  + 4E[s(X)\dot{\sigma}^2 (X_i,{\theta}_0)]^{\top} \Sigma^{-1} \mathcal{P}_2
    +8A_\varepsilon E[s(X) \dot{\sigma}^2(X_i,{\theta}_0)]^{\top} \Sigma^{-1}\mathcal{P}_3 +4A_{\varepsilon} \mathcal{W}^T  M_2 \Sigma_{\sigma} ^{-1} E[s(X)\dot{\sigma}^2(X_i,{\theta}_0)]\\
    \\
&&   +2A_{\varepsilon} E[s(X)\dot{\sigma}^2(X_i,{\theta}_0)]^T \Sigma_{\sigma} ^{-1}
    M_2 \Sigma_{\sigma} ^{-1} E[s(X)\dot{\sigma}^2(X_i,{\theta}_0)],
\end{eqnarray*}
where $H(Z_i,Z_j)$, $A_\varepsilon$, $Q_\varepsilon$, $\lambda_i, \mathcal{Z}_i, \mathcal{N}, \mathcal{W}, \mathcal{P}_2, \mathcal{P}_3 $ and $M_2$ are defined in Theorem 3.3(1).
Hence we complete the proof of the second part of Theorem 3.2.

(3) Now we discuss the asymptotic properties of $n \hat{U}_n$ under the global alternative $H_1$ in nonparametric models.
Recall that $\eta_i=\frac{Y_i-m(X_i)}{\sigma(X_i, \tilde{\theta}_0)}=\frac{\varepsilon_i\sigma(X_i)}{\sigma(X_i,\theta_0)}$ and $\hat{\eta}_i = \frac{Y_i - \hat{m}(X_i)}{\sigma(X_i, \hat{\theta}_n)} $ in nonparametric cases.
Under the global alternative $H_1$, applying (\ref{6.1}) and (\ref{6.2}) in the proof of Theorem 3.1 again, we have
\begin{equation}\label{6.25}
\begin{split}
&| \hat{\eta}_i-\hat{\eta}_j| \\
=&| \eta_i-\eta_j|
    -[\frac{\hat{m}(X_i)-m(X_i)}{\sigma(X_i,\tilde{\theta}_0)}
    -\frac{\hat{m}(X_j)-m(X_j)}{\sigma(X_j,\tilde{\theta}_0)}
    +\frac{\varepsilon_i\sigma(X_i)({\sigma(X_i,\hat{\theta}_n)}
	-{\sigma(X_i,\tilde{\theta}_0)})}{\sigma^2(X_i,\tilde{\theta}_0)}\\
&-\frac{\varepsilon_j\sigma(X_j)({\sigma(X_j,\hat{\theta}_n)}
	-{\sigma(X_j,\tilde{\theta}_0)})}{\sigma^2(X_j,\tilde{\theta}_0)}
    +(R_i-R_j)]\{\mathbb{I}(\eta_i>\eta_j)-\mathbb{I}(\eta_i<\eta_j)\}\\
&+2\int_{0}^{\frac{\hat{m}(X_i)-m(X_i)}{\sigma(X_i,\tilde{\theta}_0)}
	-\frac{\hat{m}(X_j)-m(X_j)}{\sigma(X_j,\tilde{\theta}_0)}
    +\frac{\varepsilon_i\sigma(X_i)({\sigma(X_i,\hat{\theta}_n)}
  	-{\sigma(X_i,\tilde{\theta}_0)})}{\sigma^2(X_i,\tilde{\theta}_0)}
    -\frac{\varepsilon_j\sigma(X_j)({\sigma(X_j,\hat{\theta}_n)}
  	-{\sigma(X_j,\tilde{\theta}_0)})}{\sigma^2(X_j,\tilde{\theta}_0)}
    +R_i-R_j}\\
& \{\mathbb{I}(\eta_i-\eta_j \leq z)-\mathbb{I}(\eta_i \leq \eta_j)\}dz,
\end{split}
\end{equation}
By the analog to (\ref{6.13}), $\hat{U}_n$ can be decomposed into three parts:
\begin{eqnarray}\label{6.26}
\hat{U}_n  \nonumber
&=&   \frac{1}{C_n^4}\sum\limits_{i< j< k< l} \left( \frac{1}{6} \sum_{s< t,u< v}^{(i,j,k,l)} |\eta_{st}|
(\|X_{st}\|+\|X_{uv}\|) -\frac{1}{12} \sum_{(s,t,u)}^{(i,j,k,l)}|\eta_{st}|\|X_{su}\| \right) \\  \nonumber
&&    + \frac{1}{C_n^4}\sum\limits_{i< j< k< l} \left( \frac{1}{6} \sum_{s< t,u< v}^{(i,j,k,l)}\delta_{5st}^*
(\|X_{st}\|+\|X_{uv}\|) - \frac{1}{12} \sum_{(s,t,u)}^{(i,j,k,l)}\delta_{5st}^* \|X_{su}\| \right) \\  \nonumber
&&    + \frac{1}{C_n^4}\sum\limits_{i< j< k< l} \left( \frac{1}{6} \sum_{s< t,u< v}^{(i,j,k,l)}\delta_{6st}^*
(\|X_{st}\|+\|X_{uv}\|) - \frac{1}{12} \sum_{(s,t,u)}^{(i,j,k,l)}\delta_{6st}^* \|X_{su}\| \right) \\
&=:& \hat{U}^*_{n4}+\hat{U}^*_{n5}+\hat{U}^*_{n6},
\end{eqnarray}
where
\begin{eqnarray*}
	\delta^*_{5st}&=&-[\frac{\varepsilon_s\sigma(X_s)({\sigma(X_s,\hat{\theta}_n)}
		-{\sigma(X_s,\tilde{\theta}_0)})}{\sigma^2(X_s,{\theta}_0)}-
	\frac{\varepsilon_t\sigma(X_t)({\sigma(X_t,\hat{\theta}_n)}-{\sigma(X_t,\tilde{\theta}_0)})}
	{\sigma^2(X_t,{\theta}_0)}+\frac{\hat{m}(X_i)-m(X_i)}{\sigma(X_i,\tilde{\theta}_0)}\\
	&&-\frac{\hat{m}(X_j)-m(X_j)}{\sigma(X_j,\tilde{\theta}_0)}
	+(R_s-R_t)]\{\mathbb{I}(\eta_s>\eta_t)-\mathbb{I}(\eta_s<\eta_t)\}.\\
	\delta^*_{6st}
	&=&
    2\int_{0}^{\frac{\hat{m}(X_i)-m(X_i)}{\sigma(X_i,{\theta}_0)}
    	-\frac{\hat{m}(X_j)-m(X_j)}{\sigma(X_j,\tilde{\theta}_0)}
    	+\frac{\varepsilon_s\sigma(X_s)({\sigma(X_s,\hat{\theta}_n)}-{\sigma(X_s,\tilde{\theta}_0)})}
    	{\sigma^2(X_s,\tilde{\theta}_0)}-\frac{\varepsilon_t\sigma(X_t)
    	({\sigma(X_t,\hat{\theta}_n)}-{\sigma(X_t,\tilde{\theta}_0)})}{\sigma^2(X_t,\tilde{\theta}_0)}
        +R_s-R_t}\\
    && \{\mathbb{I}(\eta_s-\eta_t \leq z)-\mathbb{I}(\eta_s  \leq \eta_t)\}dz.
\end{eqnarray*}
For the term $\hat{U}^*_{n5}$, similar to the arguments for $\hat{U}^*_{n1}$ in the proof of Theorem 3.3(1), we have
\begin{eqnarray*}
	\hat{U}^*_{n5}
	&=&  \frac{1}{C_n^4} \sum\limits_{i< j< k< l} h^k h^*_{51}(Z_i,Z_j,Z_k,Z_l)  + \frac{1}{C_n^4} \sum\limits_{i< j< k< l} h^*_{52}(Z_i,Z_j,Z_k,Z_l) \\
	&&  + \frac{1}{C_n^4} \sum\limits_{i< j< k< l} h^*_{53}(Z_i,Z_j,Z_k,Z_l)   + \frac{1}{C_n^4} \sum\limits_{i< j< k< l} h^*_{54}(Z_i,Z_j,Z_k,Z_l) \\
	&&  + (\hat{\theta}_n-\tilde{\theta}_0)^T \frac{1}{C_n^4}\sum\limits_{i< j< k< l}h^*_{55}(Z_i,Z_j,Z_k,Z_l) +o_p(\frac{1}{\sqrt{n}})\\
	&=&:h^k \hat{U}^*_{n51} + \hat{U}^*_{n52} +\hat{U}^*_{n53} +\hat{U}^*_{n54} +\hat{U}^*_{n55}  +o_p(\frac{1}{\sqrt{n}})
\end{eqnarray*}
where
\begin{eqnarray*}
	&&  h^*_{5m}(Z_i,Z_j,Z_k,Z_l) \\
	&=& -6^{-1}\sum_{s < t,u < v}^{(i,j,k,l)}{\delta}^*_{5mst} \{\mathbb{I}(\eta_s>\eta_t)-
	\mathbb{I}(\eta_s<\eta_t)\}(\|X_{st}\|+\|X_{uv}\|) \\
	&&  +12^{-1}\sum_{(s,t,u)}^{(i,j,k,l)}{\delta}^*_{5mst}\{\mathbb{I}(\eta_s>\eta_t)
	-\mathbb{I}(\eta_s<\eta_t)\}\|X_{su}\|, \quad {\rm for} \ m=1, 2, 3, 4
\end{eqnarray*}
with
\begin{eqnarray*}
	{\delta}^*_{51st}
	&=&  \frac{D(X_s)}{\sigma(X_s,\tilde{\theta}_0)}-\frac{D(X_t)}
	{\sigma(X_t,\tilde{\theta}_0)} \\
	{\delta}^*_{52st}
	&=&  \frac{\frac{1}{n-1}\sum_{p=1,p\neq s}^{n}\omega_{s,p}\sigma(X_p,\tilde{\theta}_0)\varepsilon_p}
	{\sigma(X_s,\tilde{\theta}_0)}-\frac{\frac{1}{n-1}\sum_{q=1,q\neq t}^{n}\omega_{t,q}
		\sigma(X_q,\tilde{\theta}_0)\varepsilon_q}{\sigma(X_t,\tilde{\theta}_0)},\\
	{\delta}^*_{53st}
	&=&  \frac{\frac{1}{n-1}\sum_{p=1,p\neq s}^{n}(\omega_{s,p}m(X_p)-E[\omega_{s,p} m(X_p)|X_s])}{\sigma(X_s,\tilde{\theta}_0)}-\frac{(\frac{1}{n-1}\sum_{q=1,q\neq t}^{n}\omega_{t,q}m(X_q)-E[\omega_{t,q} m(X_q)|X_t])}
	{\sigma(X_t,\tilde{\theta}_0)},\\
	{\delta}^*_{54st}
	&=&  \frac{\frac{m(X_s)}{n-1}\sum_{p=1,p\neq s}^{n}(\omega_{s,p}-E[\omega_{s,p} |X_s])}{\sigma(X_s,\tilde{\theta}_0)}
	-\frac{\frac{m(X_t)}{n-1}\sum_{q=1,q\neq t}^{n}(\omega_{t,q}-E[\omega_{t,q} |X_t])}{\sigma(X_t,\tilde{\theta}_0)},\\
		{\delta}^*_{55st}
	&=&  \{\frac{\varepsilon_s\sigma   (X_s)\dot{\sigma}(X_s,\tilde{\theta}_0)}{\sigma^2(X_s,\tilde{\theta}_0)}
	-\frac{\varepsilon_t\sigma(X_t)\dot{\sigma}(X_t,\tilde{\theta}_0)}
	    {\sigma^2(X_t,\tilde{\theta}_0)} \}.
\end{eqnarray*}
For the term $\hat{U}^*_{n51}$, by the law of large numbers for U-statistics and the assumption 6(d),we have
$$\sqrt{n} h^k \hat{U}^*_{n51}=\sqrt{n} h^k E[ h^*_{51}(Z_i,Z_j,Z_k,Z_l)]= o_p(1).$$
For the term $\hat{U}^*_{n52}$, similar to the arguments for $I^*_{12}$ in the proof of Theorem 3.3(1),  $\hat{U}^*_{n52}$  can be decomposed as
\begin{eqnarray*}
	&&  \hat{U}^*_{n52} \\
	&=& \frac{1}{n(n-1)^3(n-2)(n-3)}\sum\limits_{i=1}^{n}\sum\limits_{j=1.j\neq i}^{n}\sum\limits_{k=1,k\neq i,j }^{n}
	\sum\limits_{l=1,l\neq i,j,k }^{n}\sum\limits_{p=1,p\neq i}^{n}\sum\limits_{q=1,q\neq j}^{n}
	\{ \frac{\omega_{p,i}\sigma(X_p,\tilde{\theta}_0)\varepsilon_p}{\sigma(X_i,\tilde{\theta}_0)}\\
	&&  -\frac{\omega_{q,j}\sigma(X_q,\tilde{\theta}_0)\varepsilon_q}{\sigma(X_j,\tilde{\theta}_0)} \}
	\{\mathbb{I}(\varepsilon_i>\varepsilon_j)-\mathbb{I}(\varepsilon_i<\varepsilon_j)\}(\|X_{ij}\|+\|X_{kl}\|-2\|X_{ik}\|)\\
	&=& \frac{1}{n(n-1)^3(n-2)(n-3)}\sum_{i=1}^{n}\sum_{j\neq i}^{n}\sum_{k\neq i,j }^{n} \sum_{l\neq i,j,k }^{n}
	\sum_{p=q\neq i,j}^{n}\{ \frac{\omega_{p,i}\sigma(X_p,\tilde{\theta}_0)\varepsilon_p}{\sigma(X_i,\tilde{\theta}_0)}
	-\frac{\omega_{q,j} \sigma(X_q,\tilde{\theta}_0)\varepsilon_q}{\sigma(X_j,\tilde{\theta}_0)} \}\\
	&&  \times \{\mathbb{I}(\varepsilon_i>\varepsilon_j)-\mathbb{I}(\varepsilon_i<\varepsilon_j)\}(\|X_{ij}\|+\|X_{kl}\|
	-2\|X_{ik}\|)\\
	&&  +\frac{1}{n(n-1)^3(n-2)(n-3)}\sum_{i=1}^{n}\sum_{j\neq i}^{n}\sum_{k\neq i,j }^{n} \sum_{l\neq i,j,k }^{n}
	\sum_{p=j}\sum_{q=i}\{ \frac{\omega_{j,i}\sigma(X_j,\tilde{\theta}_0)\varepsilon_j}{\sigma(X_i,\tilde{\theta}_0)}
	-\frac{\omega_{i,j} \sigma(X_i,\tilde{\theta}_0)\varepsilon_i}{\sigma(X_j,\tilde{\theta}_0)} \}\\
	&&  \times \{\mathbb{I}(\varepsilon_i>\varepsilon_j)-\mathbb{I}(\varepsilon_i<\varepsilon_j)\}
	(\|X_{ij}\|+\|X_{kl}\|-2\|X_{ik}\|)\\
	&&  +\frac{1}{n(n-1)^3(n-2)(n-3)}\sum_{i=1}^{n}\sum_{j\neq i}^{n}\sum_{k\neq i,j }^{n} \sum_{l\neq i,j,k }^{n}
	\sum_{p=j}^{n}\sum_{q\neq i,j}^{n}\{ \frac{\omega_{p,i}\sigma(X_p,\tilde{\theta}_0)\varepsilon_p}
	{\sigma(X_i,\tilde{\theta}_0)}-\frac{\omega_{p,j}\sigma(X_p,\tilde{\theta}_0)\varepsilon_p}
	{\sigma(X_j,\tilde{\theta}_0)} \}\\
	&&  \times \{\mathbb{I}(\varepsilon_i>\varepsilon_j)-\mathbb{I}(\varepsilon_i<\varepsilon_j)\}
	(\|X_{ij}\|+\|X_{kl}\|-2\|X_{ik}\|)\\
	&&  +\frac{1}{n(n-1)^3(n-2)(n-3)}\sum_{i=1}^{n}\sum_{j\neq i}^{n}\sum_{k\neq i,j }^{n} \sum_{l\neq i,j,k }^{n}
	\sum_{p\neq i,j}^{n}\sum_{q=i}^{n}\{ \frac{\omega_{p,i}\sigma(X_p,\tilde{\theta}_0)\varepsilon_p}
	{\sigma(X_i,\tilde{\theta}_0)}-\frac{\omega_{p,j}\sigma(X_p,\tilde{\theta}_0)\varepsilon_p}
	{\sigma(X_j,\tilde{\theta}_0)} \}\\
	&&  \times \{\mathbb{I}(\varepsilon_i>\varepsilon_j)-\mathbb{I}(\varepsilon_i<\varepsilon_j)\}
	(\|X_{ij}\|+\|X_{kl}\|-2\|X_{ik}\|)  \\
	&&  +\frac{1}{n(n-1)^3(n-2)(n-3)}\sum_{i=1}^{n}\sum_{j\neq i}^{n}\sum_{k\neq i,j }^{n}
	\sum_{l\neq i,j,k }^{n}\sum_{p\neq i\neq j\neq k\neq l}^{n}\sum_{q\neq i\neq j\neq k\neq l \neq p}^{n}\{ \frac{\omega_{p,i}\sigma(X_p,\tilde{\theta}_0)\varepsilon_p}{\sigma(X_i,\tilde{\theta}_0)} \\
	&&  -\frac{\omega_{q,j}\sigma(X_q,\tilde{\theta}_0)\varepsilon_q}{\sigma(X_j,\tilde{\theta}_0)} \}
	\{\mathbb{I}(\varepsilon_i>\varepsilon_j)-\mathbb{I}(\varepsilon_i<\varepsilon_j)\}(\|X_{ij}\|+\|X_{kl}\|-2\|X_{ik}\|)\\
	&=:&\hat{U}^*_{n521} + \hat{U}^*_{n522}+ \hat{U}^*_{n523}+ \hat{U}^*_{n524}+ \hat{U}^*_{n525}
\end{eqnarray*}
For the term $\{ \hat{U}^*_{n52m} \}_{m=2}^{4}$, similar to the arguments for $\{ I^*_{12m}\}_{m=2}^{4}$ in the proof of Theorem 3.3(1),  we obtain that $\sqrt{n}\hat{U}^*_{n52m}=o_p(1),m=2,3,4$.

For the term $\hat{U}^*_{n525}$, we have
\begin{equation*}
\hat{U}^*_{n525}=\frac{(n-4)(n-5)}{(n-1)^2}\frac{1}{C_n^6}
\mathop{\sum\sum\sum\sum\sum\sum}_{1 \leq i< j< k< l< r< m \leq n }
{\delta}^*_{525}(Z_i,Z_j,Z_k,Z_l,Z_r,Z_m)
\end{equation*}
where
\begin{eqnarray*}
&&{\delta}^*_{525}(Z_i,Z_j,Z_k,Z_l,Z_r,Z_m) \\
&=&\frac{1}{6!}\sum_{(s,t,u,v,r,m)}^{(i,j,k,l,p,q)}
    \{\frac{\omega_{r,s}\sigma(X_r,\tilde{\theta}_0)\varepsilon_r}{\sigma(X_s,\tilde{\theta}_0)}
     -\frac{\omega_{m,t}\sigma(X_m,\tilde{\theta}_0)\varepsilon_m}{\sigma(X_t,\tilde{\theta}_0)} \} \{\mathbb{I}(\eta_s>\eta_t)-\mathbb{I}(\eta_s<\eta_t)\}(\|X_{st}\|+\|X_{uv}\|-2\|X_{su}\|).
\end{eqnarray*}
Some elementary calculations show that $\hat{U}^*_{n525}$ is non-degenerate.
By the standard theory of $U$-statistics (see Section 5.3 in Serfling (2009),
for instance), we have
\begin{eqnarray*}
	\sqrt{n}\hat{U}^*_{n525}
	&=&  \frac{1}{\sqrt{n}}\sum_{i=1}^{n}E[{\delta}^*_{525}(Z_i,Z_j,Z_k,Z_l,Z_r,Z_m)|Z_i]+o_p(1)\\
	&=& \frac{1}{\sqrt{n}}\sum_{i=1}^{n}(H_{11i}+H_{12i})+o_p(1)
\end{eqnarray*}
where
\begin{eqnarray*}
	H_{11i}&=&\frac{1}{5!}\sum_{(j_1,k_1,l_1,p_1,q_1)}^{(j,k,l,p,q)}
	E[\{\frac{\omega_{i,p_1}\sigma(X_i,{\theta}_0)\varepsilon_i}{\sigma(X_{p_1},{\theta}_0)}
	-\frac{\omega_{q_1,j_1}\sigma(X_{q_1},{\theta}_0)\varepsilon_{q_1}}{\sigma(X_{j_1},{\theta}_0)}\}  \{\mathbb{I}(\eta_{p_1}>\eta_{j_1})-\mathbb{I}(\eta_{p_1}<\eta_{j_1})\}(\|X_{p_1j_1}\|\\
	&&+\|X_{k_1l_1}\|-2\|X_{p_1k_1}\|)|Z_i] \\
	H_{12i}&=&\frac{1}{5!}\sum_{(j_1,k_1,l_1,p_1,q_1)}^{(j,k,l,p,q)}
	E[\{\frac{\omega_{p_1,q_1}\sigma(X_{p_1},{\theta}_0)\varepsilon_{p_1}}{\sigma(X_{q_1},{\theta}_0)}
	-\frac{\omega_{i,j_1}\sigma(X_i,{\theta}_0)\varepsilon_i}{\sigma(X_{j_1},{\theta}_0)} \} \{\mathbb{I}(\eta_{q_1}>\eta_{j_1})-\mathbb{I}(\eta_{q_1}<\eta_{j_1})\}(\|X_{q_1j_1}\|\\
	&&+\|X_{k_1l_1}\|-2\|X_{q_1k_1}\|)|Z_i].
\end{eqnarray*}
Similarly, we have
\begin{eqnarray*}
	\sqrt{n} \hat{U}^*_{n53}=\frac{1}{\sqrt{n}}\sum_{i=1}^{n}
	E[{\delta}^*_{535}(Z_i,Z_j,Z_k,Z_l,Z_r,Z_m)|Z_i]+o_p(1)
	=\frac{1}{\sqrt{n}}\sum_{i=1}^{n}(H_{13i}+H_{14i})+o_p(1)\\
    \sqrt{n}  \hat{U}^*_{n54}=\frac{1}{\sqrt{n}}\sum_{i=1}^{n}
	E[{\delta}^*_{545}(Z_i,Z_j,Z_k,Z_l,Z_r,Z_m)|Z_i]+o_p(1)
	=\frac{1}{\sqrt{n}}\sum_{i=1}^{n}(H_{15i}+H_{16i})+o_p(1)
\end{eqnarray*}
where
\begin{eqnarray*}
	H_{13i}&=&\frac{1}{5!}\sum_{(j_1,k_1,l_1,p_1,q_1)}^{(j,k,l,p,q)}
	E[\{\frac{\omega_{i,p_1}m(X_i)-E[\omega_{i,p_1}m(X_i)|X_{p_1}]}{\sigma(X_{p_1},{\theta}_0)}
	-\frac{\omega_{q_1,j_1}m(X_{q_1})-E[\omega_{q_1,j_1}m(X_{q_1})|X_{j_1}]}{\sigma(X_{j_1},{\theta}_0)} \},\\
	&& \times \{\mathbb{I}(\eta_{p_1}>\eta_{j_1})-\mathbb{I}(\eta_{p_1}<\eta_{j_1})\}
	(\|X_{p_1j_1}\|+\|X_{k_1l_1}\|-2\|X_{p_1k_1}\|)|Z_i],\\
	H_{14i}&=&\frac{1}{5!}\sum_{(j_1,k_1,l_1,p_1,q_1)}^{(j,k,l,p,q)}
	E[\{\frac{\omega_{p_1,q_1}m(X_{p_1})-E[\omega_{p_1,q_1}m(X_{p_1})|X_{q_1}]}{\sigma(X_{q_1},{\theta}_0)}
	-\frac{\omega_{i,j_1}m(X_i)-E[\omega_{i,j_1}m(X_i)|X_{j_1}]}{\sigma(X_{j_1},{\theta}_0)} \} \\
	&& \times
	\{\mathbb{I}(\eta_{q_1}>\eta_{j_1})-\mathbb{I}(\eta_{q_1}<\eta_{j_1})\}
	   (\|X_{q_1j_1}\|+\|X_{k_1l_1}\|-2\|X_{q_1k_1}\|)|Z_i],\\
	H_{15i}&=&\frac{1}{5!}\sum_{(j_1,k_1,l_1,p_1,q_1)}^{(j,k,l,p,q)}
	E[\{\frac{\omega_{i,p_1}-E[\omega_{i,p_1}|X_{p_1}]}{\sigma(X_{p_1},{\theta}_0)}
	-\frac{\omega_{q_1,j_1}m(X_{q_1})-E[\omega_{q_1,j_1}|X_{j_1}]}{\sigma(X_{j_1},{\theta}_0)} \} \{\mathbb{I}(\eta_{p_1}>\eta_{j_1})\\
	&&-\mathbb{I}(\eta_{p_1}<\eta_{j_1})\}(\|X_{p_1j_1}\|+\|X_{k_1l_1}\|-2\|X_{p_1k_1}\|)|Z_i],\\
	H_{16i}&=&\frac{1}{5!}\sum_{(j_1,k_1,l_1,p_1,q_1)}^{(j,k,l,p,q)}
	E[\{\frac{\omega_{p_1,q_1}m(X_{p_1})-E[\omega_{p_1,q_1}|X_{q_1}]}{\sigma(X_{q_1},{\theta}_0)}
	-\frac{\omega_{i,j_1}-E[\omega_{i,j_1}|X_{j_1}]}{\sigma(X_{j_1},{\theta}_0)} \} \\
	&& \times
	\{\mathbb{I}(\eta_{q_1}>\eta_{j_1})-\mathbb{I}(\eta_{q_1}<\eta_{j_1})\}
	(\|X_{q_1j_1}\|+\|X_{k_1l_1}\|-2\|X_{q_1k_1}\|)|Z_i].
\end{eqnarray*}

%For $\hat{U}^*_{n6}$, following the same line as that for the term  $\hat{U}^*_{n2}$ in Theorem 3.3(1), it can be decomposed as
%\begin{eqnarray*}
%	\hat{U}^*_{n6}
%	&=& \frac{1}{C_n^4}\sum\limits_{i< j< k< l}{h}^*_{61}(Z_i,Z_j,Z_k,Z_l)+\frac{1}{C_n^4}\sum\limits_{i< j< k< l}{h}^*_{62}(Z_i,Z_j,Z_k,Z_l)\\
%	&=:& \hat{U}^*_{n61}+\hat{U}^*_{n62},
%\end{eqnarray*}
%where
%\begin{eqnarray*}
%	{h}^*_{61}(Z_i,Z_j,Z_k,Z_l)
%	&=&\frac{1}{6}\sum_{s< t,u< v}^{(i,j,k,l)}E(\delta^*_{6st}|X_s,X_t )(\|X_{st}\|+\|X_{uv}\|)-\frac{1}{12}\sum_{(s,t,u)}^{(i,j,k,l)}E(\delta^*_{6st}|X_s,X_t )\|X_{su}\|,\\
%	{h}^*_{62}(Z_i,Z_j,Z_k,Z_l)&=&\frac{1}{6}\sum_{s< t,u< v}^{(i,j,k,l)}[{\delta}^*_{6st}-E(\delta^*_{6st}|X_s,X_t )](\|X_{st}\|+\|X_{uv}\|)-\frac{1}{12}\sum_{(s,t,u)}^{(i,j,k,l)}[{\delta}^*_{6st}-E(\delta^*_{6st}|X_s,X_t )]\|X_{su}\|.
%\end{eqnarray*}
For the term $\hat{U}^*_{n55}$, it is easy to see that $\frac{1}{C_n^4}\sum\limits_{i< j< k< l}h_{55}(Z_i,Z_j,Z_k,Z_l)$ is non-degenerate $U$-statistic of order $4$.
By some elementary calculations, we have
\begin{eqnarray*}
	E[h_{55}(Z_i,Z_j,Z_k,Z_l)]
	&=&   -2E[(\frac{\varepsilon_1\sigma (X_1)\dot{\sigma}(X_1,\tilde{\theta}_0)}{\sigma^2(X_1,\tilde{\theta}_0)}
	-\frac{\varepsilon_2\sigma(X_2)\dot{\sigma}(X_2,\tilde{\theta}_0)}{\sigma^2(X_2,\tilde{\theta}_0)})
	\mathbb{I}(\eta_1>\eta_2)C_x(X_1,X_2)] \overset{def}{=}2K_2,
\end{eqnarray*}
where $C_x(X_1,X_2) = \|X_1-X_2\|-E(\|X_1-X_2\||X_1)-E(\|X_1-X_2\||X_2)+E(\|X_1-X_2\|)$.
Thus we obtain that
\begin{eqnarray*}
	\sqrt{n} \hat{U}^*_{n5}
	= \frac{1}{\sqrt{n}}\sum_{i=1}^{n}\sum_{k=1}^{6}H_{1ki} + 2 \sqrt{n} (\hat{\theta}_n-\tilde{\theta}_0)^T K_2 + o_p(1).
\end{eqnarray*}
Following the same line for the term $\hat{U}^*_{n2}$ in the proof of Theorem 3.3(1), we can show that $\sqrt{n}\hat{U}^*_{n6} = o_p(1)$. Altogether we obtain that
\begin{eqnarray*}
	\sqrt{n}  \hat{U}_{n}
	&=&  \sqrt{n} \hat{U}^*_{n4}
	     + \frac{1}{\sqrt{n}}\sum_{i=1}^{n}\sum_{k=1}^{6}H_{1ki}
	     + 2 \sqrt{n} (\hat{\theta}_n-\tilde{\theta}_0)^T K_2 + o_p(1)\\
	&=:& \frac{\sqrt{n}}{C_n^4}\sum\limits_{i< j< k< l} h^*_4(Z_i,Z_j,Z_k,Z_l)
	     + \frac{1}{\sqrt{n}}\sum_{i=1}^{n}\sum_{k=1}^{6}H_{1ki}
	     + 2 \sqrt{n} (\hat{\theta}_n-\tilde{\theta}_0)^T K_2 + o_p(1),
\end{eqnarray*}
where $Z_i = (\eta_i, X_i)$ and $h^*_4(Z_i,Z_j,Z_k,Z_l) = \frac{1}{6} \sum_{s< t,u< v}^{(i,j,k,l)} |\eta_{st}| (\|X_{st}\|+\|X_{uv}\|) -\frac{1}{12} \sum_{(s,t,u)}^{(i,j,k,l)}|\eta_{st}|\|X_{su}\| $.
Consequently,
\begin{eqnarray*}
	\sqrt{n} [\hat{U}_n-dCov^2(\eta,X)] = \sqrt{n} [\hat{U}_{n4}- dCov^2(\eta,X)]  + \frac{1}{\sqrt{n}}\sum_{i=1}^{n}\sum_{k=1}^{6}H_{1ki}
	+ 2\sqrt{n}(\hat{\theta}_n-\tilde{\theta}_0)^T K_2 + o_p(1).
\end{eqnarray*}
To obtain the limiting distribution of $n\hat{U}_n$, it remains to derive the asymptotic expansion of  $\hat{\theta}_n - \theta_0$.  According to the arguments of Theorem 1 in Appendix A of  Dette et al. (2007), we have
\begin{eqnarray*}
	\sqrt{n} (\hat{\theta}_n-\tilde{\theta}_0)
	&=&  \frac{1}{\sqrt{n}} \sum_{i=1}^{n} [\sigma^2(X_i)\varepsilon_i^2 - \sigma^2(X_i, \tilde{\theta}_0)] \Sigma^{-1} \dot{\sigma}^2(X_i, \tilde{\theta}_0) + o_p(1).
\end{eqnarray*}
Altogether we obtain that
\begin{eqnarray*}
&&\sqrt{n}[\hat{U}_n-dCov^2(\eta_i,X_i)]\\
&=&\frac{1}{\sqrt{n}}\sum_{i=1}^{n}2\{\mathcal{G}(\eta_i,X_i)
    +\mathcal{I}_i(\eta_i,X_i)+[\sigma ^2(X_i)\varepsilon^2_i -\sigma ^2(X_i, \tilde{\theta}_0) ]K_2^T \Sigma^{-1}\dot{\sigma}^2(X_i,\tilde{\theta}_0)\}+o_p(1),
\end{eqnarray*}
it follows that
$$ \sqrt{n}[\hat{U}_n-dCov^2(\eta, X)] \longrightarrow  N(0, \sigma_2^2), $$
where $\sigma_2^2 = 4var\{\mathcal{G}(\eta, X)+\mathcal{I}_1(\eta, X)+[\sigma ^2(X)\varepsilon^2 -\sigma ^2(X, \tilde{\theta}_0) ]K_2^T \Sigma^{-1}\dot{\sigma}^2(X,\tilde{\theta}_0)\}$ with
\begin{eqnarray*}
	\mathcal{G}(\eta_1,X_1)&=&
	E[C_\eta(\eta_1,\eta_2)C_x(X_1,X_2)|Z_1]-dCov^2(\eta_1,X_1)\\
	C_{\eta}(\eta_i,\eta_j)&=& |\eta_i-\eta_j|-E(|\eta_i-\eta_j\|\eta_i)-E(|\eta_i-\eta_j\|\eta_j)
	+E(|\eta_i-\eta_j|) \\
	C_x(X_i,X_j) &=& \|X_i-X_j\|-E(\|X_i-X_j\| |X_i)-E(\|X_i-X_j\| |X_j)+E(\|X_i-X_j\|)\\
	dCov^2(\eta,X)&=&E[C_\eta(\eta_i,\eta_j)C_x(X_i,X_j)],
\end{eqnarray*}
$\mathcal{I}_1(\eta_i,X_i)= \frac{1}{2} \sum_{k=1}^{6}H_{1ki}$ and $K_2$ are defined in Theorem 3.2(2). Hence we complete the proof of Theorem 3.3. \hfill$\Box$

{\bf Proof of Theorem 4.1.}
This proof follows the same line as Theorem 4 in Xu and Cao (2021). Thus we omit the details here. \hfill$\Box$

\end{document}